\documentclass[journal]{IEEEtran}
\ifCLASSINFOpdf
  \usepackage[pdftex]{graphicx}
\else
  \usepackage[dvips]{graphicx}
\fi
\usepackage{array}
\usepackage{amsmath}
\usepackage{amssymb}
\usepackage{amsthm}
\usepackage{amsfonts}
\usepackage[ruled,vlined]{algorithm2e}
\usepackage[caption=false,font=normalsize]{subfig}
\usepackage{multirow}
\usepackage{booktabs}
\usepackage{url}
\usepackage{cite}
\usepackage{balance}
\usepackage{diagbox}
\usepackage{comment}

\newcommand{\argmin}{\mathop{\rm argmin}\limits}

\graphicspath{{./images/}}
\def\x{{\mathbf x}}

\def\y{{\mathbf y}}

\def\v{{\mathbf v}}

\def\g{{\mathbf g}}

\def\q{{\mathbf q}}
\def\n{{\mathbf n}}

\def\u{{\mathbf u}}

\def\A{{\mathbf A}}
\def\B{{\mathbf B}}

\def\D{{\mathbf D}}

\def\L{{\mathbf L}}
\def\R{{\mathbb R}}
\def\bR{{\mathbf R}}

\def\U{{\mathbf U}}

\def\M{{\mathbf M}}

\def\S{{\mathbf S}}

\def\Hmath{{\mathcal H}}

\def\Bmath{{\mathcal B}}

\DeclareMathOperator{\prox}{prox}

\DeclareMathOperator{\HSSTV}{HSSTV}
\DeclareMathOperator{\sgn}{sgn}

\begin{document}
%
\title{Robust Hyperspectral Image Fusion with Simultaneous Guide Image Denoising via Constrained Convex Optimization}
%
%
%

\author{Saori~Takeyama~\IEEEmembership{Member,~IEEE}~and~Shunsuke~Ono,~\IEEEmembership{Member,~IEEE}
\thanks{S. Takeyama is with the Department of Information and Communications Engineering, School of Engineering, Tokyo Institute of Technology, Kanagawa, 226-8503, Japan, e-mail: takeyama.s.aa@m.titech.ac.jp (see https://sites.google.com/view/saori-takeyama/home).}
\thanks{S. Ono is with the Department of Computer Science, School of Computing, Tokyo Institute of technology, Kanagawa, 226-8503, Japan.}
\thanks{Manuscript received xxx xx, 2021; revised xxxx xx, 2021.}}

%
%

\markboth{Journal of \LaTeX\ Class Files,~Vol.~XX, No.~X, April~2022}%
{Takeyama \MakeLowercase{\textit{et al.}}: Bare Demo of IEEEtran.cls for IEEE Journals}
%



\maketitle

\begin{abstract}
The paper proposes a new high spatial resolution hyperspectral (HR-HS) image estimation method based on convex optimization.
The method assumes a low spatial resolution HS (LR-HS) image and a guide image as observations, where both observations are contaminated by noise.
Our method simultaneously estimates an HR-HS image and a noiseless guide image, so the method can utilize spatial information in a guide image even if it is contaminated by heavy noise. 
The proposed estimation problem adopts hybrid spatio-spectral total variation as regularization and evaluates the edge similarity between HR-HS and guide images to effectively use apriori knowledge on an HR-HS image and spatial detail information in a guide image.
To efficiently solve the problem, we apply a primal-dual splitting method.
Experiments demonstrate the performance of our method and the advantage over several existing methods.

\end{abstract}


%
\IEEEpeerreviewmaketitle

\section{Introduction}\label{sec:intro}
\IEEEPARstart{H}{igh-resolution} spectral information helps reveal the intrinsic characteristics of objects and environment lighting.
Hyperspectral (HS) imaging can capture spectral information from 400nm to 2500nm in the 5-10nm interval and is expected to resolve many problems in a wide range of fields, e.g., remote sensing, biomedical, industrial, and food safety \cite{HSI1, HSI2, HSI3}.
Owing to high-resolution spectral information, HS imaging is limited to the amount of incident energy in each band, so it has critical tradeoffs between the spatial resolution, the spectral resolution, and the signal-to-noise ratio.
Many applications require high spatial and spectral resolution HS (HR-HS) images without noise, so it is an essential task to resolve the problem.

In general, observed HS images have a high spectral resolution but a low spatial resolution (called LR-HS images) because the spatial detail information is lost through an imaging process.
Therefore, fusion techniques have been widely studied to estimate HR-HS images~\cite{Pansharpening1,Pansharpening2}.
This technique fuses two observations of an LR-HS image and a guide image that has high spatial resolution but low spectral one.
Here, the guide image is assumed a panchromatic (PAN) image or a multispectral (MS) image, and the fusion technique with a PAN image is especially called HS pansharpening.
This fusion technique has been known to effectively estimate HR-HS images because it can utilize both the high spatial resolution information of the guide image and the high spectral resolution information of the LR-HS image.

Many fusion methods have been proposed, which can be roughly classified into three groups: component analysis-based methods, optimization-based methods, and deep learning-based methods.
Component analysis-based methods~\cite{PCA_pan, GS, GSA, MTF_GLP, MTF_GLP_HPM, SFIM, GFPCA} separate spatial details from a guide image using a linear transformation, replace the spatial information of an LR-HS image with the extracted spatial details, and then apply the inverse transformation to the fused image.
They are easy implementation and have low computational cost but the performance is limited in especially noisy cases because they cannot directly handle apriori knowledge on HR-HS images.

In contrast, optimization-based approaches can evaluate apriori knowledge on HR-HS images using a regularization function.
Many methods~\cite{HySure, LTMR, yang2021regularizing, NED, CNMF, FUMI, Lanaras_fusion, NCTRF} assume the low rankness of HR-HS images and model an HR-HS image as the product of endmember spectra and abundance maps.
Here, endmember and abundance have spectral and spatial information of an HR-HS image, respectively.
Then, the methods estimate the endmember and the abundance by independently evaluating spatial and spectral apriori knowledge.
The methods achieve an estimation preserving spectral information but produce spatial artifacts especially when noisy cases.
This is because they do not directly evaluate spatial edge information in a guide image even though a guide image has high-resolution spatial detailed information.
Besides, the performance relies on the rank number of the endmember that cannot be known, and the setting error sufficiently affects their estimation performance.
In other words, it is an essential and a difficult tasks for the methods to decide the suitable rank number.

Some optimization-based methods~\cite{PSSGDNSP, VPECT} evaluate spatial detail information to estimate an HR-MS image, so they can estimate a spatial detail-preserving HR-MS image.
In \cite{PSSGDNSP, VPECT}, they can estimate a spatial detail-preserving HR-MS image from an MS image with less than 10 bands.
The two methods assume a noiseless guide image, so the performance would be degraded for a noisy one.
Moreover, they use a nonconvex function or do not directly evaluate spectral information, so they require improvement to achieve the estimation unaffected by the initial value and preserving spectral information if they are used for HR-HS image estimation.

\begin{figure}[t]
    \centering
    \begin{minipage}[t]{0.45\hsize}
        \includegraphics[width=\hsize]{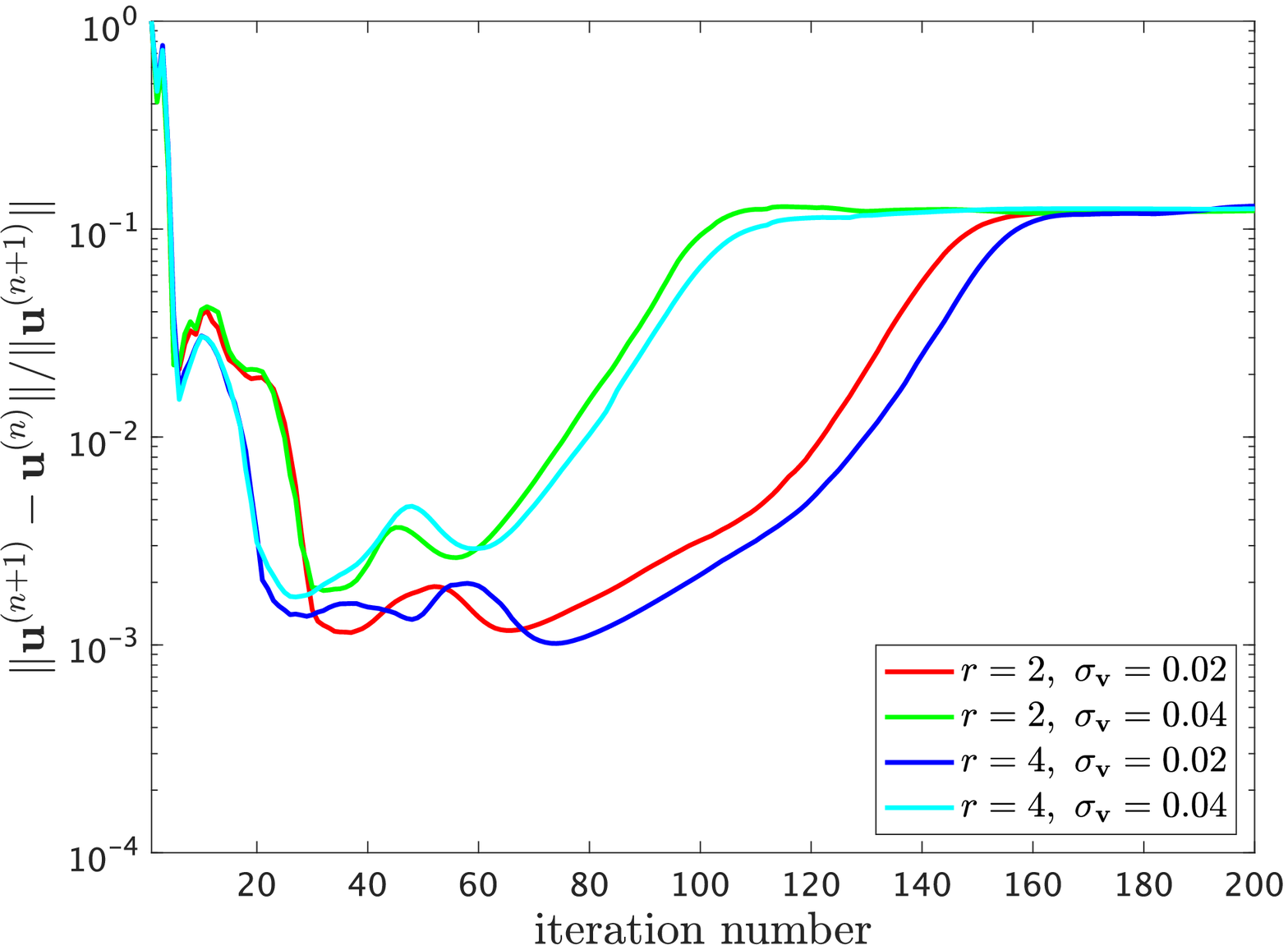}
    \end{minipage}
    \begin{minipage}[t]{0.45\hsize}
        \includegraphics[width=\hsize]{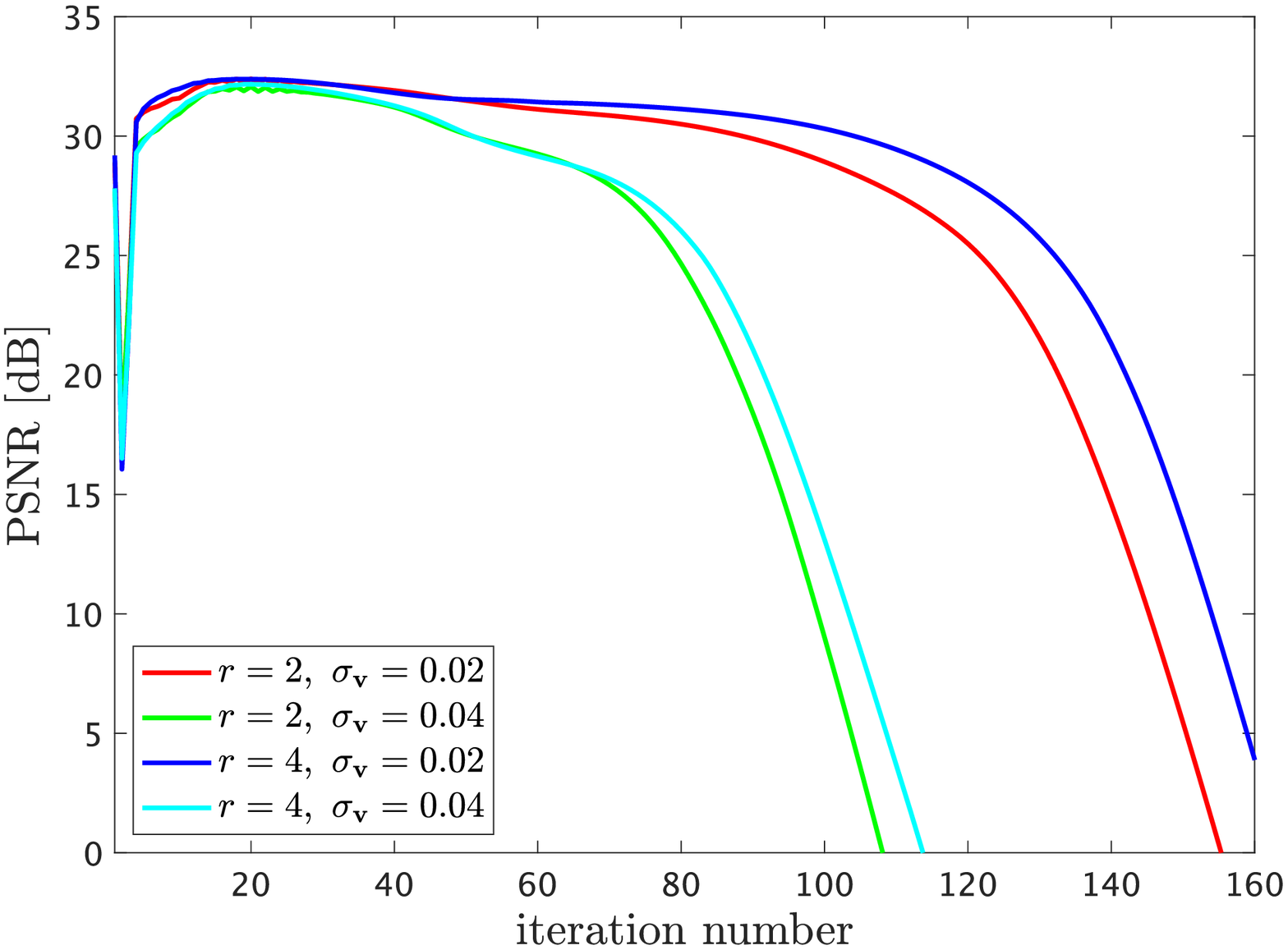}
    \end{minipage}
    
    \begin{minipage}[t]{0.45\hsize}
        \centerline{\footnotesize{Convergence}}
    \end{minipage}
    \begin{minipage}[t]{0.45\hsize}
        \centerline{\footnotesize{PSNR}}
    \end{minipage}
    \caption{PSNR and the error of the results by CNN-Fus at each iteration on HS pansharpening experiments (Salinas)}
    \label{fig:conv_CNN_Fus}
    \vspace{-5pt}
\end{figure}

Deep learning has been recently known as a powerful approach to image processing, so some deep learning-based fusion methods have been proposed~\cite{u2MDN, DHSIS, VaFuNet, wu2020new, MSSL}.
The methods train the intrinsic characteristics of HR-HS images and use the characteristics for HR-HS image estimation.
For the training process, they extract latent properties from a large number of HR-HS images, thus they achieve high estimation performance.
However, capturing a large number of HR-HS images is very difficult, and in fact, the number of HS datasets is limited.
Moreover, the training HR-HS images are required to have the same number of bands as an observed LR-HS image, so they need re-training when the methods are used on another LR-HS image with a different number of bands.
To solve the dilemma, CNN-Fus~\cite{CNN_Fus} applies a convolution neural network (CNN) pre-trained with grayscale images.
CNN-Fus can evaluate a spatial underlying characteristics of an HR-HS image and achieves high performance.
However, the performance depends on the iteration number as in Fig.~\ref{fig:conv_CNN_Fus}, so the setting is an essential task.
In \cite{CNN_Fus}, the authors manually determine the iteration number at which the quality measure is stable, but one cannot know the suitable iteration number in a real estimation.
Hence, CNN-Fus faces the problem of how determining the iteration number in real observations.

In this paper, we propose a new HR-HS image estimation method from LR-HS and guide images based on optimization, which preserves spatial edge and is robust to noise included in observations.
Our proposed method estimates both an HR-HS image and a noiseless guide image, so it is rarely affected by noise in a guide image.
Then, our method evaluates the edge similarity between HR-HS and guide images to take over the spatial edge in a guide image to an HR-HS image.
To effectively evaluate apriori knowledge on an HR-HS image, the method adopts hybrid spatio-spectral total variation (HSSTV)~\cite{HSSTV} as a regularization function. 
The optimization problem is convex, and it can be efficiently solved by a primal-dual splitting method. 
In experiments, we verified the performance of our proposed method and confirm the advantage over several existing fusion methods.

Our paper makes the following contributions:
\begin{itemize}
    \item The method simultaneously estimates an HR-HS image and a noiseless guide image to achieve robust HR-HS image estimation to noise included in an LR-HS image and an observed guide image. 
    The experiments illustrate that our method can estimate an HR-HS image without spatial artifacts when observations are contaminated by noise.
    \item To effectively utilize the spatial detailed information in a guide image, our method evaluates the edge similarity between an HR-HS image and a guide image.
    Thanks to the setting, the method achieves high performance for HS pansharpening and HS and MS image fusion.
\end{itemize}

The rest of this article is organized as follows.
Section ref{sec:RW} reviews several existing optimization-based fusion methods in detail.
In Section~\ref{sec:pre}, we introduce mathematical ingredients.
Our proposed method is described in Section~\ref{sec:PM}.
Experimental results and comparisons on HS pansharpening and HS and MS image fusion are presented in Section~\ref{sec:E}.
Section~\ref{sec:C} concludes this article.


\section{Related Works}\label{sec:RW}
\begin{table*}[t]
\centering
    \caption{The features of existing HR-HS image estimation methods based on optimization}
    \label{tab:feature}
    \begin{tabular}{|c||c|c|c|c|c|} \hline
         \multicolumn{5}{|c|}{HR-HS image estimation methods} \\ \hline
         method & guide image & noise & spatial evaluation & spectral evaluation \\ \hline
         CNMF~\cite{CNMF} & MS image & both images & - & low-rankness \\ \hline
         HySure~\cite{HySure} & MS or PAN image & both images & piecewise smoothness & low-rankness \\ \hline
         NED~\cite{NED} & MS image & both images & $\ell_2$ sparsity & low-rankness \\ \hline
         Yang's~\cite{yang2021regularizing} & MS image & both images & the similarity of non-local patches & low-rankness \\ \hline
         FUMI~\cite{FUMI} & MS image & both images & $\ell_1$ sparsity & low-rankness \\ \hline
         NCTRF~\cite{NCTRF} & MS image & both images & low-rankness & low-rankness \\ \hline
         Lanaras's~\cite{Lanaras_fusion} & MS or PAN image & both images & $\ell_1$ sparsity & low-rankness \\ \hline
         LTMR~\cite{LTMR} & MS image & both images & non-local similarity & low-rankness \\ \hline
         {\bf{proposed}} & MS or PAN image & both images & \begin{tabular}{c} piecewise smoothness \\ edge similarity \end{tabular} & piecewise smoothness \\ \hline
         \multicolumn{5}{|c|}{HR-MS image estimation methods} \\ \hline
         PSSGDNSP~\cite{PSSGDNSP} & PAN image & only MS image & edge similarity & edge similarity\\ \hline
         VPECT~\cite{VPECT} & PAN image & only MS image & \begin{tabular}{c} piecewise smoothness of cartoon \\ low-rankness of texture patches \end{tabular} & - \\ \hline
    \end{tabular}
\end{table*}
We explain several optimization-based existing HR-HS image estimation methods. 
Tab.~\ref{tab:feature} shows the features of several existing methods.


Many optimization-based fusion methods assume that an HS image has low-rank properties and can be decomposed into some bases and coefficients.
In HS unmixing, one of the low-rank modeling (LRM)-based approaches, the bases and the coefficients are handled as endmember spectra and abundance maps, respectively.
Here, the endmember and the abundance represent the intrinsic spectral response of materials and the existing ratio of each material.
From the characteristics, the abundance and the endmember are considered to have spatial and spectral information of an HS image, respectively. 
Based on the assumption, the methods estimate them from a pair of LR-HS and guide images.

All methods in Tab.~\ref{tab:feature} consider that a guide image is an MS image and HySure and Lanaras's also consider an HS pansharpening case. 
The methods~\cite{NED, yang2021regularizing, FUMI, NCTRF,LTMR} only consider an MS guide image, but they only use a guide image to evaluate the data fidelity in abundance estimation.
Therefore, by setting the suitable spectral response operator, they can estimate an HR-HS image in the HS pansharpening case.
Coupled NMF (CNMF)~\cite{CNMF} decomposes an MS image into an LR-endmember and an HR-abundance using nonnegative matrix factorization, and use an HR-abundance to generate an HR-HS image.
PAN images do not have spectral information, so estimating an LR-endmember is difficult.
However, according to \cite{CNMF_pan}, the method can be adopted for HS pansharpening by only estimating HR abundance from a PAN image using an estimated HR-endmember and a spectral response operator of a PAN image.

All methods consider that both observations are contaminated by additive Gaussian noises.
Here, in some papers~\cite{Lanaras_fusion, NED, LTMR, NCTRF}, the authors do not clearly model the noise, but all methods evaluate the data fidelity between observations and an HR-HS image.
Therefore, we regard them as methods considering noisy observations.

Some methods~\cite{HySure, LTMR, NED, yang2021regularizing} separately estimate the endmember and the abundance.
The methods first decompose an LR-HS image into HR-endmember and LR-abundance and use HR-endmember.
Then, they estimate abundance from the estimated endmember and observations.
Since an LR-HS image is generally contaminated by noise, the endmember estimation is required to be robust.
To reduce the influence of noise, the methods~\cite{HySure, LTMR, yang2021regularizing} use singular value decomposition for pre-denoising and/or endmember estimation, and the method~\cite{NED} estimates the endmember by solving a nonnegative matrix factorization (NMF)-based problem considering noise.
For the abundance estimation step, they formulate an optimization problem including regularization and data fidelity and effectively solve it by an iterative algorithm.
They adopt different regularization functions, i.e., they utilize the different spatial characteristics on an HR-HS image (See Tab.~\ref{tab:feature}).



The methods~\cite{CNMF, FUMI, Lanaras_fusion} simultaneously estimate the endmember and the abundance.
CNMF decomposes LR-HS and guide images into the endmembers and the abundances by NMF and generates an HR-HS image from the endmember of the LR-HS image and the abundance of the guide one.
This is because LR-HS and guide images have full spectral and spatial information, respectively.
CNMF uses the relationship between the observations for the decomposition and alternatively and iteratively estimates the endmember and the abundance.
The two methods~\cite{FUMI, Lanaras_fusion} formulate and solve optimization problems to simultaneously estimate the endmember and the abundance.
They evaluate the sparsity of the abundance by sum-to-one and nonnegative constraints.
The method~\cite{NCTRF} also assumes the low-rankness of HS images and considers three low-rank factors. 
It simultaneously estimates the factors from observations by solving an optimization problem that evaluates the data fidelity and spectral correlation.

The LRM-based methods achieve an HR-HS image estimation preserving spectral information, but they have problems.
Although a guide image has spatial edge information, they do not directly evaluate it.
Specifically, the methods use spatial information in a guide image when they estimate abundance or low-rank factors, but they only evaluate the data fidelity.
Therefore, the methods produce spatial artifacts or oversmoothing.
Besides, they require to set the number of the endmember, and the setting is affected by the estimation performance.
However, since the suitable number is unknown and different by objects, it is an essential and difficult task to search for it.

In \cite{PSSGDNSP, VPECT}, the authors propose an HR-MS image estimation based on other approaches.
Pansharpening with spatial and spectral gradient difference-induced nonconvex sparsity priors (PSSGDNSP)~\cite{PSSGDNSP} uses $\ell_{1/2}$ norm to evaluate the spatial gradient similarity between a guide image and an HR-MS image and the spectral gradient similarity between an LR-HS image and an HR-MS image.
This is because the authors experimentally confirm that the similarities are close to hyper-Laplacian.
Since the spatial gradient is almost the same as the  spatial edge information, PSSGDNSP can take over the spatial information in a guide image to an HR-MS image.
However, the method assumes that a guide image is not contaminated by noise, so the performance would significantly decrease when real observations.
Moreover, the $\ell_{1/2}$ norm can measure more sparsity than the $\ell_1$ norm but is non-convex. 
Therefore, the convergence is not guaranteed, and the performance of PSSGDNSP would depend on the initial value.
Tian et al. propose a cartoon-texture-based method for MS pansharpening in \cite{VPECT}, named a variational pansharpening method by exploiting
cartoon-texture similarities (VPECT).
VPECT decomposes a PAN image into the cartoon and the texture components, and then VPECT estimates an HR-MS image using the components.
From the design, the method can estimate an HR-MS image with spatial detail, but it cannot directly evaluate spectral information.
Therefore, the performance would be limited for HR-HS image estimation.
Moreover, VPECT assumes a noiseless PAN image to effectively extract the spatial texture component from a PAN image, but the situation is not realistic.

\section{Preliminaries}\label{sec:pre}
\subsection{Notations and Definitions}
Let $\R$ be the set of real numbers.
We shall use boldface lowercase and capital to represent vectors and matrices, respectively, and $:=$ to define something.
We denote the transpose of a vector/matrix by $(\cdot)^{\top}$ and the Euclidean norm (the $\ell_2$ norm) of a vector by $\|\cdot\|$.

For notational convenience,
we treat an HS image $\U \in \R^{N_v \times N_h \times B}$ as a vector $\u \in \R^{NB}$ ($N:=N_v N_h$ is the number of the pixels of each band, and $B$ is the number of the bands)
by stacking its columns on top of one another, i.e., the index of the component of the $i$th pixel in $k$th band is $i+(k-1)N$ (for $i=1,\ldots,N$ and $k = 1,\ldots, B$).

\subsection{Proximal Tools}\label{subsec:prox}
A function $f:\R^N\rightarrow(-\infty, \infty]$ is called \textit{proper lower semicontinuous convex} if
$\mbox{dom}(f) := \{\x\in\R^N|\;f(\x)<\infty\}\neq\emptyset$,
$\mbox{lev}_{\leq\alpha}(f):=\{\x \in\R^N|\; f(\x)\leq\alpha\}$ is closed for every $\alpha\in\R$,
and $f(\lambda\x+(1-\lambda)\y)\leq\lambda f(\x)+(1-\lambda)f(\y)$
for every $\x, \y\in\R^N$ and $\lambda\in(0,1)$, respectively.
Let $\Gamma_0(\R^N)$ be the set of all proper lower semicontinuous convex functions on $\R^N$.

The \textit{proximity operator}\cite{Moreau} plays a central role in convex optimization based on proximal splitting.
The proximity operator of $f \in \Gamma_0 (\R^N)$ with a index $\gamma > 0$ is then defined by
\begin{equation}\label{prox}
\prox_{\gamma f} (\x) := \argmin_{\y} f(\y) + \frac{1}{2 \gamma} \| \y - \x\|^2,
\end{equation}
where the existence and uniqueness of the minimizer are guaranteed respectively
by the coercivity\footnote{A function $f\in\Gamma_0(\R^N)$ is called \textit{coercive} if $\|\x\|\rightarrow\infty\Rightarrow f(\x)\rightarrow\infty$. In this case, the existence of a minimizer of $f$ is guaranteed, that is, there exists $\x^\star\in\mbox{dom}(f)$ such that $f(\x^\star)=\inf_{\x\in\Hmath}f(\x)$ (see, e.g., \cite{Combettesbook}). }
and the strict convexity of $f(\cdot)+\frac{1}{2\gamma}\|\cdot-\x\|^2$.

We introduce the indicator function of a nonempty closed convex set $C \subset \R^N$, which is defined as follows:
\begin{align}\label{indicator_def}
\iota_{C}(\x) := \left\{
\begin{array}{l l}
0, &\mbox{if} ~ \x \in C, \\
\infty , &\mbox{otherwise}. \\
\end{array}
\right.
\end{align}
Then, for any $\gamma>0$, the proximity operator is defined by
\begin{equation}\label{indicator_prox}
\prox_{\gamma \iota_{C}}(\x) = P_C(\x) :=\argmin_{\y \in C} \|\x-\y\|,
\end{equation}
where $P_C(\x)$ is a metric projection onto $C$, and this proximity operator is reduced to the metric projection onto $C$.

\subsection{Primal-Dual Splitting Method}\label{subsec:PDS}
The \textit{primal-dual splitting method}~\cite{PDChambolle} is based on proximal splitting, and it is an algorithm that can solve convex optimization problems of the form:
\begin{equation}\label{PDSequation}
\min_{\x} g(\x) + h(\L\x),
\end{equation}
where $g\in\Gamma_0(\R^{a})$, $h\in\Gamma_0(\R^{b})$, and $\L$ is a linear operator.
Here, we assume that $g$ and $h$ are \textit{proximable}, i.e., the proximity operators of $g$ and $h$ are computable.
For arbitrarily chosen $\y^{(0)}$ and $\gamma_1, \gamma_2 >0$ satisfying $\gamma_1 \gamma_2 \|\L\|_{op} \leq 1$ ($\|\cdot\|_{op}$ is the operator norm), the primal-dual splitting method iterates the following steps:
\begin{align}
\left \lfloor
\begin{array}{l}\label{PDS_renew}
\x^{(n+1)} = \prox_{\gamma_1 g}(\x^{(n)} - \gamma_1\L^{\top} \y{(n)}), \\
\y^{(n+1)} = \prox_{\gamma_2 h^*}(\y^{(n)} + \gamma_2 \L(2\x^{(n+1)} - \x^{(n)})).
\end{array}
\right.
\end{align}
The function $h^{*}$ is the convex conjugate of $h$, and the proximity operator $h^{*}$ is available via that of $h$~\cite[Theorem 14.3 (ii)]{Combettesbook} as follows:
\begin{equation}\label{prox_conj}
\prox_{\gamma h^*}(\x) = \x - \gamma\prox_{\frac{1}{\gamma} h}\left( \frac{1}{\gamma} \x \right).
\end{equation}

\section{Proposed Method}\label{sec:PM}

We propose a new HR-HS image estimation method from a pair of an LR-HS image and a guide image.
Let $\bar{\u} \in \R^{NB}$ be an HR-HS image, and a LR-HS image $\v$ and a guide image $\g$ are given by
\begin{align}
&\v = \S \B \bar{\u} + \n_{\v} \in \R^{\frac{NB}{r^2}}, \label{model_v} \\
&\g = \bR \bar{\u} + \n_{\g}  \in \R^{Nb}, \label{model_g}
\end{align}
where $\S$ is a down-sampling matrix with the ratio of $r^2$, $\B$ is a blur matrix, $\n_{\v}$ and $\n_{\g}$ are additive white Gaussian noises, and $b$ is the band number of a guide image.
We assume that $NB$ is divisible by $r^2$ because the number of pixels in $\v$ must be a positive integer.
In the paper, the guide image is supposed to be an MS image or a PAN image, where the PAN image is a single-channel image and is equal to a grayscale image.
The matrix $\bR$ represents the spectral response of the guide image and depends on the imaging system. 
This model considers both observations contaminated by noise, which is a natural situation.
In general, since HS images are more affected by noise than MS and PAN images, the noise intensity $\sigma_{\v}$ and $\sigma_{\g}$ of $\n_{\v}$ and $\n_{\g}$ satisfy $\sigma_{\v} > \sigma_{\g}$.
This model is based on Wald's protocol~\cite{WaldsProtocol}, and existing methods are also based on the same or very similar model.

\subsection{HR-HS Image Estimation}
Our proposed method estimates an HR-HS image by solving a convex optimization problem.
For robust estimation, the problem simultaneously estimates an HR-HS image $\u \in \R^{NB}$ and a noiseless guide image $\q \in \R^{Nb}$.
Our method evaluates apriori knowledge on $\u$ and $\q$ and the edge similarity between the two images, so it can effectively use spatial and spectral information in observations.
We formulate the HR-HS image estimation problem as follows:
\begin{align} \label{prob:HRHSestimation}
&\min_{\u, \q} \HSSTV(\u) + \lambda\|\D\M_u\u - \D\M\q\|_{1,2} + \rho\|\D\q\|_{1,2} \nonumber \\
&\mbox{s.t. } \left[ \begin{array}{l}
\S \B \u \in \Bmath_{2,\varepsilon}^{\v} := \{ \x \in \R^{\frac{NB}{r}} | \|\x - \v \| \leq \varepsilon\}, \\
\q \in \Bmath_{2, \eta}^{\g} := \{ \x \in \R^{M} | \|\x - \g \| \leq \eta \}, \\
\u \in [\underline{\mu_{\u}}, \overline{\mu_{\u}}]^{NB}, \\
\q \in [\underline{\mu_{\q}}, \underline{\mu_{\q}}]^{Nb}, \\
\end{array} \right.
\end{align}
where 
$\D = (\D_v^{\top} \D_h^{\top})^{\top}$ is a spatial difference operator with $\D_v$ and $\D_h$ being vertical and horizontal difference operators, respectively.
A linear operator $\M_u \in \R^{NB \times NB'}$ extracts the same spectral range information as a guide image.
Specifically, when a guide image captures information with spectral range $[\lambda_{\g,\mathrm{st}}, \lambda_{\g,\mathrm{en}}]$ ($\lambda_{\g,\mathrm{st}} < \lambda_{\g,\mathrm{en}}$), and from $B_{\lambda_{\g,\mathrm{st}}}$th to $B_{\lambda_{\g,\mathrm{en}}}$th bands of $\v$ have the spectral range information, $\M_u$ only maintains the information in $[B_{\lambda_{\g,\mathrm{st}}}, B_{\lambda_{\g,\mathrm{en}}}]$ and removes the other information.
Moreover, $\M \in \R^{NB' \times Nb}$ is a normalized transpose matrix of $\mathbf{R}$ in the spectral range included in a guide image.
The matrix is calculated by transposing $\mathbf{R}$, normalizing it in the row direction, and extracting a part of the spectral range.
This is because the evaluation of the edge similarity in other than the spectral range is not essential and would produce spatial over-smoothing.
Since the method assumes a noisy guide image, it estimates both the HR-HS image and the desirable guide image for a noise-robust estimation.

\begin{algorithm}[t]
\footnotesize{
\LinesNumbered
\SetKwInOut{Input}{input}
\SetKwInOut{Output}{output}
\caption{Primal-dual splitting algorithm \\ for solving Prob.~\eqref{prob:HRHSestimation}}
\label{alg:PDSHRHSestimation}
\Input{$\u^{(0)}$, $\q^{(0)}$, $\y_1^{(0)}$, $\y_2^{(0)}$, $\y_3^{(0)}$, $\y_4^{(0)}$, $\y_5^{(0)}$}
\While{A stopping criterion is not satisfied}{
$\u^{(n+1)} = \prox_{\gamma_1 \iota_{[\underline{\mu_{\u}}, \overline{\mu_{\u}}]^{NB}}}(\u^{(n)} - \gamma_1(\A_{\omega}^{\top} \y_1^{(n)} + \M_u^{\top}\D^{\top}\y_2^{(n)} + \B^{\top}\S^{\top}\y_4^{(n)}))$\;
$\q^{(n+1)} = \prox_{\gamma_1 \iota_{[\underline{\mu_{\q}}, \overline{\mu_{\q}}]^{Nb}}} (\q^{(n)} - \gamma_1(\M^{\top}\D^{\top}\y_2^{(n)} + \D^{\top}\y_3^{(n)} + \y_5^{(n)}))$\; 
$\y_1^{(n+1)} \leftarrow \y_1 + \gamma_2 \A_{\omega}(2\u^{(n+1)} - \u^{(n)})$\;
$\y_2^{(n+1)} \leftarrow \y_2 + \gamma_2 (\D\M_u(2\u^{(n+1)} - \u^{(n)}) - \D\M(2\q^{(n+1)} - \q^{(n)}))$\;
$\y_3^{(n+1)} \leftarrow \y_3 + \gamma_2 \D(2\q^{(n+1)} - \q^{(n)})$\;
$\y_4^{(n+1)} \leftarrow \y_4 + \gamma_2 \S\B(2\u^{(n+1)} - \u^{(n)})$\;
$\y_5^{(n+1)} \leftarrow \y_5 + \gamma_2 (2\q^{(n+1)} - \q^{(n)})$\;
$\y_1^{(n+1)} = \y_1^{(n)} - \gamma_2 \prox_{\frac{1}{\gamma_2} \|\cdot\|_{1,p}}\left(\frac{\y_1^{(n)}}{\gamma_2}\right)$\;
$\y_2^{(n+1)} = \y_2^{(n)} - \gamma_2 \prox_{\frac{\lambda}{\gamma_2} \|\cdot\|_{1,2}}\left(\frac{\y_2^{(n)}}{\gamma_2}\right)$\;
$\y_3^{(n+1)} = \y_3^{(n)} - \gamma_2 \prox_{\frac{\rho}{\gamma_2} \|\cdot\|_{1,2}}\left(\frac{\y_3^{(n)}}{\gamma_2}\right)$\;
$\y_4^{(n+1)} = \y_4^{(n)} - \gamma_2 \prox_{\frac{1}{\gamma_2} \iota_{\Bmath_{2, \varepsilon}^{\v}}}\left(\frac{\y_4^{(n)}}{\gamma_2}\right)$\;
$\y_5^{(n+1)} = \y_5^{(n)} - \gamma_2 \prox_{\frac{1}{\gamma_2} \iota_{\Bmath_{2, \eta}^{\g}}}\left(\frac{\y_5^{(n)}}{\gamma_2}\right)$\;
$n\leftarrow n+1$\;
}
}
\end{algorithm}

The first term in Prob.~\eqref{prob:HRHSestimation} is a hybrid regularization technique for HS image restoration, named HSSTV~\cite{HSSTV}.
HSSTV can effectively utilize a-prior knowledge of HS images, and according to \cite{HSSTV, ICIP2019, IGARSS2020, ICIP2020}, it has achieved high performance for several HS image restorations.
HSSTV is defined by
\begin{equation}\label{def:HSSTV}
\HSSTV(\u) := \|\A_{\omega} \u\|_{1,p} \mbox{ s.t. } \A_{\omega} := \left [
\begin{array}{c}
\D\D_b \\
\omega\D
\end{array} \right ],
\end{equation}
where $\|\cdot\|_{1,p}$ is a mixed $\ell_{1,p}$ norm, and $\D_b$ is a spectral difference operator.
In \cite{HSSTV_ICASSP, HSSTV_arxiv}, $p$ is assumed to be $1$ or $2$.
Since $\omega > 0$ adjusts the relative importance of $\D$ to $\D\D_b$, HSSTV effectively utilizes a-prior knowledge on HS images by a suitable $\omega$.

The second term in Prob.~\eqref{prob:HRHSestimation} evaluates the edge similarity between the HR-HS image and the guide image, which compares the edges in the range of the common spectral range of the two images. 
The technique is based on \cite{chen2014image} and assumes that the non-zero differences of the HR-HS and guide images are sparse and correspond to edges and that their edge positions are similar.
Therefore, minimizing the mixed $\ell_{1,2}$ norm of the difference between the edge images of $\M_u \u$ and $\M \q$ helps to exploit spatial information on the guide image.
The third term in Prob.~\eqref{prob:HRHSestimation} is a TV for the guide image. 
The parameter $\lambda$ and $\rho$ balance the three terms.

The first and second constraints serve as data-fidelity of $\u$ and $\q$, respectively.
The sets $\Bmath_{2,\varepsilon}^{\v}$ and $\Bmath_{2, \eta}^{\g}$ are defined as the $\v$-centered $\ell_2$-norm ball with radius $\varepsilon>0$ and the $\g$-centered $\ell_2$-norm ball with radius $\eta>0$, respectively.
As mentioned in \cite{afonso2011augmented, EPIpre, TSP2015Involved, HSSTV_ICASSP, HSSTV_arxiv, ono2019efficient, ICIP2019, IGARSS2020, ICIP2020}, such hard constraints facilitate the parameter settings because $\varepsilon$ and $\eta$ can decide from the observation information.
The third and fourth constraints in Prob.~\eqref{prob:HRHSestimation} are the dynamic range of $\u$ and $\q$ with $\underline{\mu} < \overline{\mu}$, respectively.

\subsection{Optimization}
\begin{table*}[t]
\centering
\caption{Parameter settings for experiments (top: HS pansharpening experiments, bottom: HS and MS image fusion experiments)}
\label{parameter_settings}
\scalebox{0.8}{\begin{tabular}{|c|c||c|c|c|c|c|c|c|c|} \hline
    \multicolumn{10}{|c|}{HS pansharpening} \\ \hline
    method & \diagbox{parameters}{$(r,~\sigma_{\g})$} & $(2, 0)$ & $(2, 0.02)$ & $(2, 0.04)$ & $(4, 0)$ & $(4, 0.02)$ & $(4, 0.04)$ & $(8, 0.02)$ & $(16 ,0.02)$ \\ \hline
    HySure & $(\lambda_m, \lambda_{\varphi})$ & $(10, 0.11)$ & $(10, 0.11)$ & $(6, 0.15)$ & $(7, 0.11)$ & $(5, 0.11)$ & $(4, 0.11)$ & $(10, 0.12)$ & $(4, 0.12)$ \\ \hline
     Lanaras's & $p$ & 4 & 4 & 4 & 5 & 5 & 5 & 4 & 3 \\ \hline
    LTMR & $(L, \lambda)$ & $(2, 5\times 10^{-5})$ & $(3, 5 \times 10^{-5})$ & $(3, 5 \times 10^{-5})$ & $(2, 1 \times 10^{-4})$ & $(2, 5 \times 10^{-4})$ & $(2, 1 \times 10^{-3})$ & $(2, 5 \times 10^{-4})$ & $(2, 1\times 10^{-3})$ \\ \hline
    CNN-Fus & $(L, \lambda, T)$ & $(5, 0.1, 85)$ & $(3, 0.01, 25)$ & $(3, 0.01, 25)$ & $(3, 0.01, 25)$ & $(3, 0.01, 25)$ & $(3, 0.01, 25)$ & $(2, 0.01, 60)$ & $(2, 0.001, 10)$ \\ \hline
    \multirow{2}{*}{\begin{tabular}{c}proposed \\ ($p = 1$)\end{tabular}} & $\omega$ & 0.01 & 0.01 & 0.01 & 0.01 & 0.01 & 0.01 & 0.02 & 0.02 \\ 
    & $(\lambda, \rho)$ & $(0.07, 1)$ & $(0.04, 1)$ & $(0.04, 1)$ & $(0.05, 1)$ & $(0.07, 1)$ & $(0.08, 1)$ & $(0.05, 1)$ & $(0.06, 1)$ \\ \hline
    \multirow{2}{*}{\begin{tabular}{c}proposed \\ ($p = 2$)\end{tabular}} & $\omega$ & 0.01 & 0.01 & 0.01 & 0.01 & 0.02 & 0.02 & 0.02 & 0.03 \\ 
    & $(\lambda, \rho)$ & $(0.07, 1)$ & $(0.04, 1)$ & $(0.04, 1)$ & $(0.07, 1)$ & $(0.04, 1)$ & $(0.04, 1)$ & $(0.05, 1)$ & $(0.1, 1)$ \\ \hline
    %
    \multicolumn{10}{|c|}{HS and MS image fusion} \\ \hline
    method & \diagbox{parameters}{$(r,~\sigma_{\g})$} & $(2, 0)$ & $(2, 0.05)$ & $(2, 0.1)$ & $(4, 0)$ & $(4, 0.05)$ & $(4, 0.1)$ & $(8, 0.05)$ & $(16 ,0.05)$ \\ \hline
    HySure & $(\lambda_m, \lambda_{\varphi})$ & $(10, 0.24)$ & $(7, 0.3)$ & $(3, 0.34)$ & $(10, 0.17)$ & $(6, 0.32)$ & $(3, 0.3)$ & $(10, 0.4)$ & $(3, 0.19)$ \\ \hline
     Lanaras's & $p$ & 4 & 3 & 3 & 4 & 3 & 2 & 3 & 3 \\ \hline
    LTMR & $(L,\lambda)$ & $(2,5\times 10^{-5})$ & $(2,5\times 10^{-5})$ & $(2,5\times 10^{-5})$ & $(2,5\times 10^{-5})$ & $(2,5\times 10^{-5})$ & $(2,5\times 10^{-5})$ & $(2,5\times 10^{-5})$ & $(2,5\times 10^{-5})$ \\ \hline
    CNN-Fus & $(L, \lambda, T)$ & $(3, 0.1, 50)$ & $(3, 0.1, 25)$ & $(3, 0.1, 25)$ & $(2, 0.01, 10)$ & $(2, 0.01, 10)$ & $(2, 0.01, 10)$ & $(3, 0.01, 15)$ & $(4, 0.01, 15)$ \\ \hline
    \multirow{2}{*}{\begin{tabular}{c}proposed \\ ($p = 1$)\end{tabular}} & $\omega$ & 0 & 0 & 0 & 0.02 & 0 & 0 & 0 & 0.01 \\ 
    & $(\lambda, \rho)$ & $(0.07, 1)$ & $(0.03, 1)$ & $(0.02, 1)$ & $(0.1, 1)$ & $(0.08, 1)$ & $(0.05, 1)$ & $(0.08, 1)$ & $(0.1, 1)$ \\ \hline
    \multirow{2}{*}{\begin{tabular}{c}proposed \\ ($p = 2$)\end{tabular}} & $\omega$ & 0.02 & 0 & 0 & 0.02 & 0 & 0 & 0 & 0.03 \\ 
    & $(\lambda, \rho)$ & $(0.07, 1)$ & $(0.03, 1)$ & $(0.02, 1)$ & $(0.08, 1)$ & $(0.07, 1)$ & $(0.05, 1)$ & $(0.07, 1)$ & $(0.09, 1)$ \\ \hline
\end{tabular}}
\end{table*}

Prob.~\eqref{prob:HRHSestimation} is a non-smooth constrained convex optimization, and so a suitable iterative algorithm is required to solve it.
In this paper, we adopt the primal-dual splitting method, which is introduced in Sec.~\ref{subsec:PDS}.
By the definition of the primal-dual splitting method, we reformulate Prob.~\eqref{prob:HRHSestimation} into Prob.~\eqref{PDSequation}.

Since Prob.~\eqref{prob:HRHSestimation} has four constraints, we put them into the objective function using the indicator function.
The problem is rewritten as follows:
\begin{align}\label{prob:after_indicator}
&\min_{\u, \q} \HSSTV(\u) + \lambda\|\D\M_u\u - \D\M\q\|_{1,2} + \rho\|\D\q\|_{1,2} \nonumber \\
&+\iota_{\Bmath_{2,\varepsilon}^{\v}}(\S \B \u) + \iota_{\Bmath_{2, \eta}^{\g}}(\q) + \iota_{[\underline{\mu_{\u}}, \overline{\mu_{\u}}]^{NB}}(\u) + \iota_{[\underline{\mu_{\q}}, \overline{\mu_{\q}}]^{Nb}}(\q).
\end{align}
Note that Prob.~\eqref{prob:after_indicator} exactly equals Prob.~\eqref{prob:HRHSestimation} because of the definition of the indicator function (see \eqref{indicator_def}).

By letting
\begin{align}
&g:\R^{N(B+b)}\rightarrow\R^2:(\u, \q) \mapsto (\iota_{[\underline{\mu_{\u}}, \overline{\mu_{\u}}]^{NB}}(\u), \iota_{[\underline{\mu_{\q}}, \overline{\mu_{\q}}]^{Nb}}(\q)) \label{g} \\
&h:\R^{\left(6+\frac{1}{r^2}\right)NB + 2NB' + Nb}\rightarrow\R\cup\{\infty\}: (\y_1, \y_2, \y_3, \y_4, \y_5) \nonumber \\
&\mapsto \|\y_1\|_{1,p} + \|\y_2\|_{1,2} + \|\y_3\|_{1,2} + \iota_{\Bmath^{\v}_{2,\varepsilon}}(\y_4) + \iota_{\Bmath^{\g}_{2, \eta}}, \label{h} \\
&\L:\R^{N(B+b)}\rightarrow\R^{\left(6+\frac{1}{r^2}\right)NB + 2NB' + Nb}: \nonumber \\
&(\u, \q)\mapsto(\A_{\omega}\u, \D\M_u\u-\D\M\q, \D\q, \S\B\u, \q), \label{L}
\end{align}
Prob.~\eqref{prob:HRHSestimation} is reduced to Prob.~\eqref{PDSequation}.
The resulting algorithm based on the primal-dual splitting method is summarized in Alg.~\ref{alg:PDSHRHSestimation} with \eqref{prox_conj}.

We explain how to calculate the proximity operator of some functions.
First, in step 2 and 3, the proximity operator of the indicator function of box constraint is required and defined as follows: for $i=1,\ldots,NB$
\begin{equation*}
[P_{[\underline{\mu}, \overline{\mu}]^{NB}}(\x)]_{i} = \min\{\max\{x_i, \underline{\mu}\}, \overline{\mu}\}.
\end{equation*}
Second, step 9, 10, and 11 require the proximity operators of $\ell_1$ and the mixed $\ell_{1,2}$ norm, which are reduced to a simple softthresholding type operation:
for $\gamma > 0$ and for $i=1,\ldots,4NB$,
(i) in the case of the $\ell_1$ norm, 
\begin{equation*}
[\prox_{\gamma \|\cdot\|_1}(\x)]_i
= \sgn(x_i)\max \left \{ |x_i|-\gamma, 0 \right \},
\end{equation*}
where $\sgn$ is the sign function, and (ii) in the case of the $\ell_{1,2}$ norm, 
\begin{equation*}
[\prox_{\gamma \|\cdot\|_{1,2}}(\x)]_i
= \max \left \{ 1 - \gamma \left( \textstyle\sum_{j=0}^{3} x_{\tilde{i}+jNB}^2 \right )^{-\frac{1}{2}}, 0 \right \}
x_i,
\end{equation*}
where $\tilde{i} := ((i-1) \mod NB) + 1$.
Finally, we introduce the proximity operators of $\iota_{\Bmath_{2,\varepsilon}^{\v}}$ and $\iota_{\Bmath_{2,\eta}^{\g}}$.
Since the two sets $\Bmath_{2,\varepsilon}^{\v}$ and $\Bmath_{2,\eta}^{\g}$ are $\ell_2$-norm ball, the above proximity operators can be calculated by following forms replacing the center and the radius:
\begin{equation*}
P_{\mathcal{B}_{2,\varepsilon}^{\v}}(\x) = \left \{
\begin{array}{l l}
\x, & \mbox{if } \x \in \mathcal{B}_{2,\varepsilon}^{\v},\\
\v + \frac{\varepsilon(\x-\v)}{\|\x-\v\|}, & \mbox{otherwise}.
\end{array} \right.
\end{equation*}

The algorithm requires fast Fourier transform and sort function to calculate the blur operator and the $\ell_1$-ball projection.
Therefore, the computational complexity is $O(NB\log NB)$.

\section{Experiments}\label{sec:E}
\begin{figure}[t]
\centering
\begin{minipage}[t]{0.24\hsize}
\includegraphics[width=\hsize]{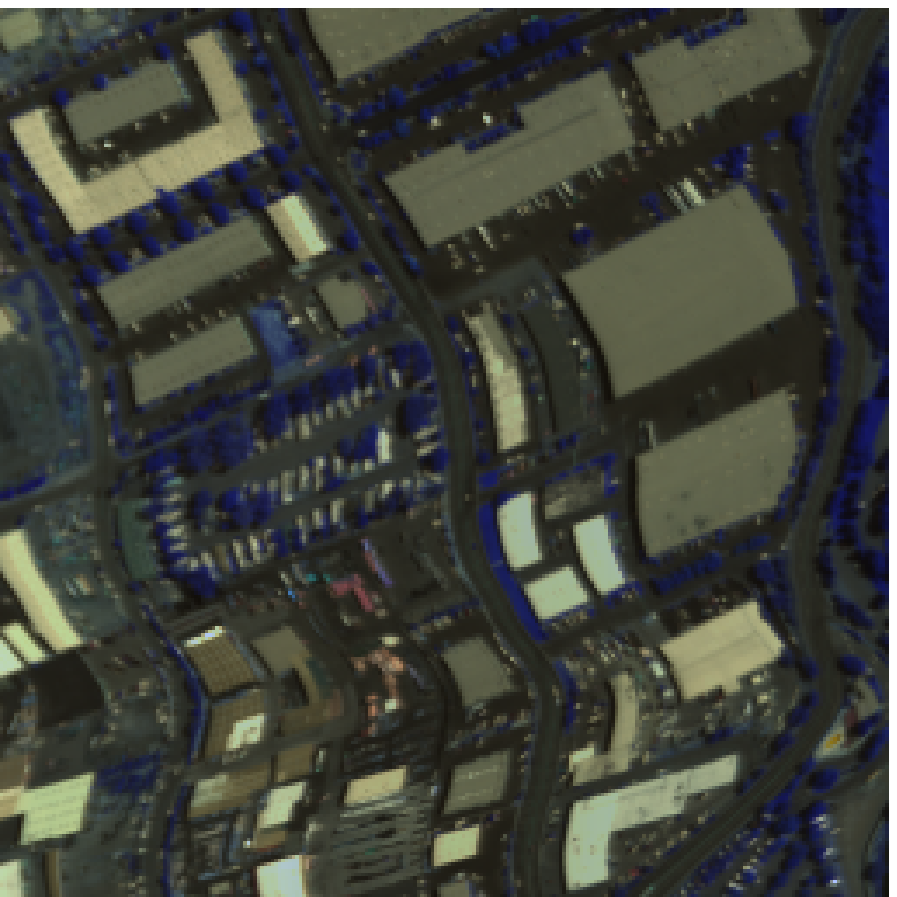}
\end{minipage}
\begin{minipage}[t]{0.24\hsize}
\includegraphics[width=\hsize]{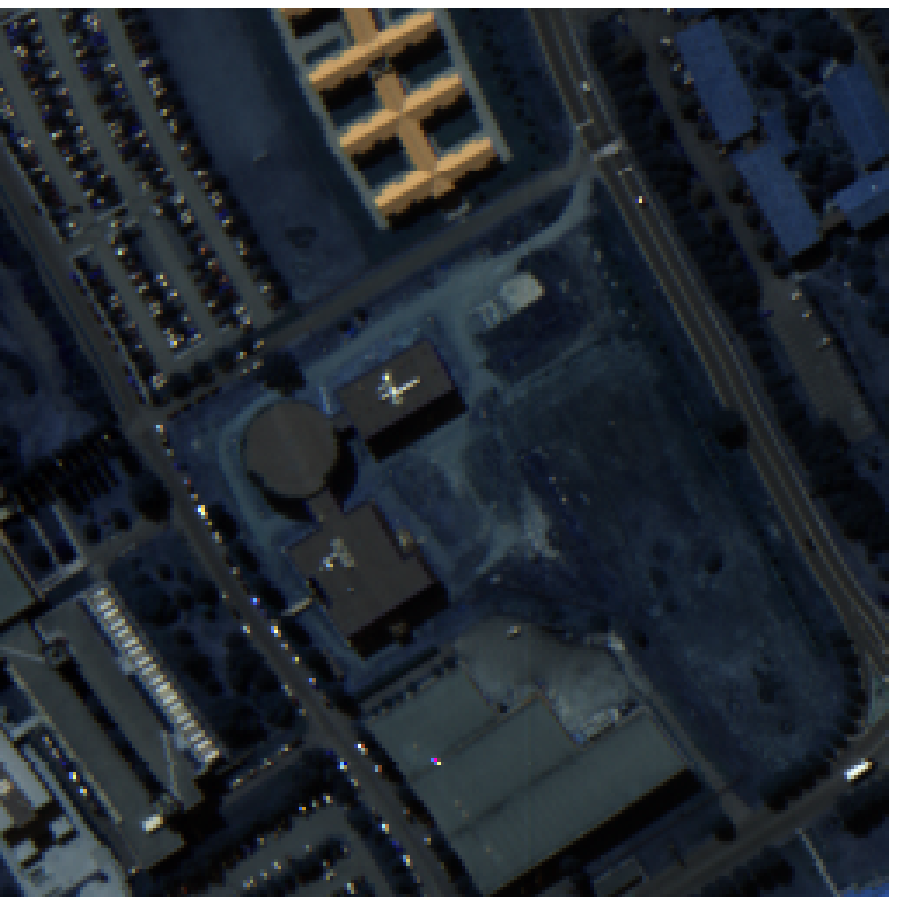}
\end{minipage}
\begin{minipage}[t]{0.24\hsize}
\includegraphics[width=\hsize]{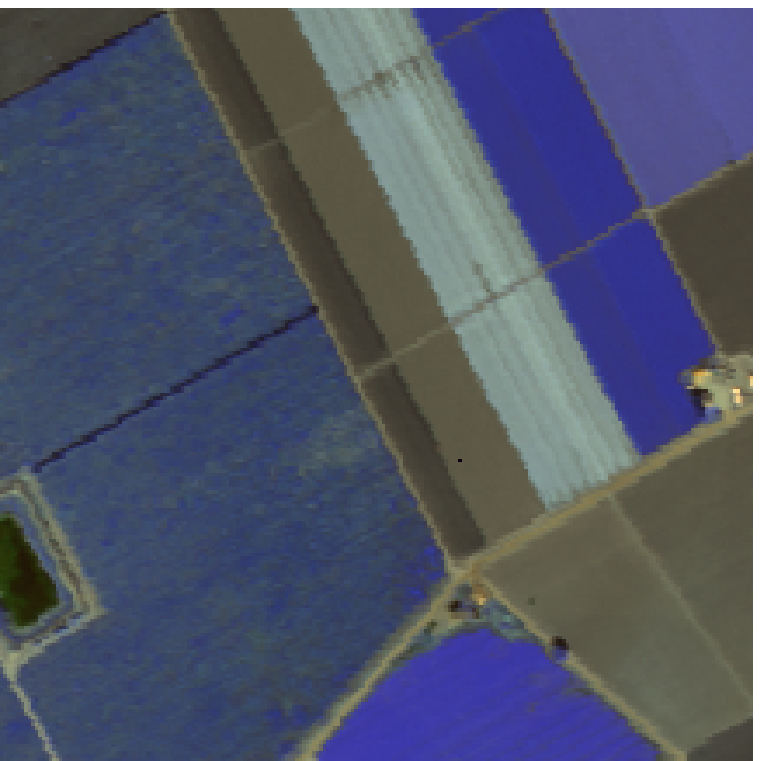}
\end{minipage}
\begin{minipage}[t]{0.24\hsize}
\includegraphics[width=\hsize]{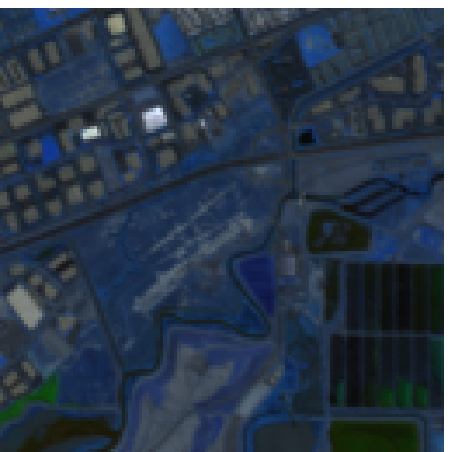}
\end{minipage}

\begin{minipage}[t]{0.24\hsize}
\centerline{\footnotesize{Reno}}
\end{minipage}
\begin{minipage}[t]{0.24\hsize}
\centerline{\footnotesize{Pavia U}}
\end{minipage}
\begin{minipage}[t]{0.24\hsize}
\centerline{\footnotesize{Salinas}}
\end{minipage}
\begin{minipage}[t]{0.24\hsize}
\centerline{\footnotesize{Moffett field}}
\end{minipage}

\begin{minipage}[t]{0.24\hsize}
\centerline{\footnotesize{$ 256 \times 256 \times 128 $}}
\end{minipage}
\begin{minipage}[t]{0.24\hsize}
\centerline{\footnotesize{$ 256 \times 256 \times 98 $}}
\end{minipage}
\begin{minipage}[t]{0.24\hsize}
\centerline{\footnotesize{$ 216 \times 216 \times 100 $}}
\end{minipage}
\begin{minipage}[t]{0.24\hsize}
\centerline{\footnotesize{$ 128 \times 128 \times 176 $}}
\end{minipage}

\caption{Test HS images. They are depicted as RGB images (R = 16th, G = 32nd, B = 64th bands).}
\label{fig:groundtruth}
\end{figure}

\begin{table*}[tp]
\begin{center}
	\caption{The quality measures  of the results on HS pansharpening experiments with $r = 2$ (boldface: the highest performance, underline: the second performance).}
	\label{tab:HSPan_2}
	\scalebox{0.88}{
	\begin{tabular}{|c|c||c|c|c|c||c|c|c|c||c|c|c|c|} \hline
	\multirow{2}{*}{image} & \multirow{2}{*}{method} & \multicolumn{4}{c||}{$\sigma_{\g} = 0$}& \multicolumn{4}{c||}{$\sigma_{\g} = 0.02$} & \multicolumn{4}{c|}{$\sigma_{\g} = 0.04$} \\ \cline{3-14
	} 
	& & PSNR[dB] & SAM & ERGAS & $Q2^n$ & PSNR[dB] & SAM & ERGAS & $Q2^n$ & PSNR[dB] & SAM & ERGAS & $Q2^n$ \\ \hline
	 \multirow{10}{*}{Reno} & GSA~\cite{GSA} & {\bf{35.22}} & \underline{4.588} & {\bf{5.573}} & 0.8363 & {\bf{34.09}} & {\bf{4.595}} & {\bf{6.109}} & 0.8095 & \underline{32.95} & {\bf{4.624}} & \underline{6.752} & 0.7674 \\ 
	 & MTF-GLP~\cite{MTF_GLP} & 34.26 & 4.666 & 6.007 & 0.7767 & {\bf{33.61}} & \underline{4.664} & \underline{6.345} & 0.7659 & {\bf{33.05}} & \underline{4.667} & {\bf{6.661}} & 0.7513 \\ 
	 & GFPCA~\cite{GFPCA} & 31.25 & 6.348 & 8.656 & 0.6420 & 30.87 & 6.352 & 8.933 & 0.6434 & 30.42 & 6.386 & 9.255 & 0.6395\\ 
	 & CNMF~\cite{CNMF} & 33.87 & 5.409 & 6.318 & 0.7781 & 31.70 & 5.841 & 7.643 & 0.7256 & 28.31 & 7.141 & 10.83 & 0.6849 \\ 
	 & HySure~\cite{HySure} & 33.24 & 4.707 & 6.687 & 0.7389 & 32.54 & 4.870 & 7.181 & 0.7380 & 30.94 & {\bf{4.894}} & 8.321 & 0.6862 \\
	 & Lanaras's~\cite{Lanaras_fusion} & \underline{34.73} & 5.772 & \underline{5.756} & 0.8528 & 32.21 & 6.127 & 7.186 & 0.7866 & 28.55 & 7.180 & 10.49 & 0.7245 \\ 
	 & LTMR~\cite{LTMR} & 32.36 & 6.382 & 7.629 & 0.7643 & 30.12 & 8.614 & 9.272 & 0.7096 & 28.12 & 9.841 & 11.00 & 0.5090 \\
	 & CNN-Fus~\cite{CNN_Fus} & 33.58 & {\bf{4.008}} & 6.115 & {\bf{0.9531}} & 31.48 & 5.332 & 7.770 & 0.9143 & 31.46 & 5.344 & 7.786 & {\bf{0.9141}} \\
	 & {\bf{proposed ($p = 1$)}} & 34.52 & 5.664 & 6.863 & 0.9403 & 32.99 & 5.939 & 7.885 & \underline{0.9209} & 32.19 & 6.033 & 8.345 & 0.9086 \\ 
	 & {\bf{proposed ($p = 2$)}} & 34.67 & 5.601 & 6.714 & \underline{0.9420} & 33.09 & 5.883 & 7.777 & {\bf{0.9220}} & 32.26 & 5.983 & 8.257 & \underline{0.9095} \\ \hline
	 \multirow{10}{*}{PaviaU} & GSA~\cite{GSA} & 31.44 & {\bf{5.409}} & 7.347 & 0.4752 & 30.84 & {\bf{5.521}} & 8.045 & 0.4809 & 30.15 & {\bf{5.674}} & 8.912 & 0.4537 \\ 
	 & MTF-GLP~\cite{MTF_GLP} & 30.95 & 5.700 & 8.037 & 0.4636 & 30.57 & \underline{5.719} & 8.461 & 0.4574 & 30.24 & \underline{5.716} & 8.847 & 0.4434 \\ 
	 & GFPCA~\cite{GFPCA} & 26.89 & 7.950 & 11.31 & 0.3203 & 26.75 & 7.958 & 11.61 & 0.3200 & 26.55 & 7.884 & 12.00 & 0.3185\\ 
	 & CNMF~\cite{CNMF} & 23.94 & 14.78 & 15.65 & 0.3905 & 23.25 & 16.02 & 17.23 & 0.3202 & 22.31 & 17.08 & 20.17 & 0.2752 \\ 
	 & HySure~\cite{HySure} & 29.21 & 6.052 & 8.912 & 0.4627 & 28.87 & 6.238 & 9.434 & 0.4718 & 27.97 & 6.734 & 10.92 & 0.4277 \\
	 & Lanaras's~\cite{Lanaras_fusion} & 30.60 & 7.204 & 7.982 & 0.5664 & 28.76 & 7.898 & 10.28 & 0.5275 & 25.56 & 10.27 & 15.36 & 0.4695 \\
	 & LTMR~\cite{LTMR} & 29.06 & 7.965 & 10.79 & 0.5339 & 28.47 & 9.405 & 11.69 & 0.5650 & 26.54 & 11.25 & 15.18 & 0.5461 \\
	 & CNN-Fus~\cite{CNN_Fus} & 30.63 & \underline{5.526} & 8.913 & 0.9151 & 29.02 & 6.531 & 10.48 & 0.8726 & 29.05 & 6.388 & 10.53 & 0.8710 \\
	 & {\bf{proposed ($p = 1$)}} & \underline{31.66} & 5.861 & \underline{7.222} & \underline{0.9413} & \underline{31.06} & 5.826 & \underline{7.966} & \underline{0.9239} & \underline{30.52} & 5.870 & \underline{8.650} & \underline{0.9121} \\ 
	 & {\bf{proposed ($p = 2$)}} & {\bf{31.80}} & 5.818 & {\bf{7.154}} & {\bf{0.9427}} & {\bf{31.13}} & 5.782 & {\bf{7.921}} & {\bf{0.9247}} & {\bf{30.58}} & 5.830 & {\bf{8.608}} & {\bf{0.9129}} \\ \hline
	 \multirow{10}{*}{Salinas} & GSA~\cite{GSA} & 34.57 & 2.751 & \underline{3.504} & 0.4749 & {\bf{34.25}} & 2.843 & {\bf{3.699}} & 0.4735 & {\bf{34.03}} & 2.898 & {\bf{3.840}} & 0.4707 \\ 
	 & MTF-GLP~\cite{MTF_GLP} & 33.27 & 3.074 & 3.990 & 0.4527 & 33.28 & 3.069 & 4.068 & 0.4665 & 33.23 & 3.064 & \underline{4.122} & 0.4583 \\ 
	 & GFPCA~\cite{GFPCA} & 32.43 & 2.730 & 4.083 & 0.4865 & 32.22 & \underline{2.810} & 4.234 & 0.4702 & 32.02 & \underline{2.882} & 4.384 & 0.4616 \\ 
	 & CNMF~\cite{CNMF} & 33.65 & 3.009 & 4.202 & 0.4302 & 32.08 & 3.469 & 5.146 & 0.4156 & 28.77 & 4.517 & 7.569 & 0.3470 \\ 
	 & HySure~\cite{HySure} & 32.93 & \underline{2.383} & 3.853 & 0.4820 & 32.73 & {\bf{2.372}}  & \underline{3.957} & 0.4798 & 32.31 & {\bf{2.460}} & 4.267 & 0.4692 \\
	 & Lanaras's~\cite{Lanaras_fusion} & 30.83 & 3.788 & 5.071 & 0.3685 & 28.81 & 4.731 & 6.475 & 0.3271 & 24.94 & 7.139 & 9.784 & 0.1887 \\
	 & LTMR~\cite{LTMR} & 31.05 & 4.446 & 5.491 & 0.3878 & 29.88 & 5.186 & 6.571 & 0.3325 & 28.55 & 6.066 & 7.919 & 0.3305 \\
	 & CNN-Fus~\cite{CNN_Fus} & {\bf{36.30}} & {\bf{1.919}} & {\bf{2.966}} & {\bf{0.8035}} & 33.38 & 3.267 & 4.355 & {\bf{0.6520}} & 33.38 & 3.293 & \underline{4.342} & {\bf{0.6591}} \\
	 & {\bf{proposed ($p = 1$)}} & 34.57 & 3.013 & 4.600 & 0.6878 & 33.69 & 3.315 & 5.351 & 0.6071 & 33.46 & 3.351 & 5.468 & 0.5987 \\ 
	 & {\bf{proposed ($p = 2$)}} & \underline{34.89} & 2.911 & 4.436 & \underline{0.6949} & \underline{33.97} & 3.200 & 5.171 & \underline{0.6159} & \underline{33.73} & 3.234 & 5.288 & \underline{0.6077} \\ \hline
	 \multirow{10}{*}{\begin{tabular}{c} Moffett\\field \end{tabular}} & GSA~\cite{GSA} & 24.94 & 7.820 & 14.06 & 0.3890 & 24.72 & 7.857 & 14.51 & 0.4532 & 24.25 & 7.949 & 15.40 & 0.3788 \\ 
	 & MTF-GLP~\cite{MTF_GLP} & 29.84 & 7.792 & 9.770 & 0.6018 & 29.42 & 7.818 & 10.16 & 0.6056 & 29.08 & 7.849 & 10.50 & 0.6004 \\ 
	 & GFPCA~\cite{GFPCA} & 28.55 & 6.957 & 9.583 & 0.6446 & 28.23 & 6.959 & 9.977 & 0.6446 & 27.74 & 7.113 & 10.63 & 0.5840 \\ 
	 & CNMF~\cite{CNMF} & 26.72 & 9.943 & 12.72 & 0.5837 & 24.24 & 10.58 & 15.94 & 0.4670 & 22.93 & 11.61 & 18.74 & 0.3325 \\
	 & HySure~\cite{HySure} & 29.83 & 5.949 & 8.388 & 0.6285 & 29.45 & 5.989 & 8.792  & 0.6284 & 28.52 & 6.294 & 9.913 & 0.5508 \\
	 & Lanaras's~\cite{Lanaras_fusion} & 26.32 & 8.749 & 12.48 & 0.4098 & 29.01 & 6.455 & 9.269 & 0.5821 & 25.99 & 7.996 & 13.39 & 0.5000 \\
	 & LTMR~\cite{LTMR} & 29.97 & 8.203 & 9.573 & 0.6226 & 30.14 & 7.715 & 8.737 & 0.5546 & 29.15 & 8.990 & 10.47 & 0.5542 \\
	 & CNN-Fus~\cite{CNN_Fus} & 32.07 & {\bf{5.175}} & 6.851 & 0.9185 & 30.48 & 5.745 & 8.170 & 0.8748 & 30.69 & 5.715 & 8.003 & 0.8816 \\
	 & {\bf{proposed ($p = 1$)}} & \underline{32.35} & 5.675 & \underline{6.311} & \underline{0.9308} & \underline{31.23} & \underline{5.668} & \underline{7.174} & \underline{0.9053} & \underline{30.90} & \underline{5.706} & \underline{7.542} & \underline{0.8958} \\ 
	 & {\bf{proposed ($p = 2$)}} & {\bf{32.60}} & \underline{5.599} & {\bf{6.154}} & {\bf{0.9339}} & {\bf{31.30}} & {\bf{5.625}} & {\bf{7.126}} & {\bf{0.9064}} & {\bf{30.96}} & {\bf{5.665}} & {\bf{7.500}} & {\bf{0.8968}} \\ \hline \hline
	 \multirow{10}{*}{Average} & GSA~\cite{GSA} & 31.54 & 5.142 & 7.622 & 0.5439 & 30.97 & 5.204 & 8.091 & 0.5543 & 30.34 & 5.286 & 8.725 & 0.5176 \\ 
	 & MTF-GLP~\cite{MTF_GLP} & 32.08 & 5.308 & 6.951 & 0.5737 & 31.72 & 5.317 & 7.258 & 0.5738 & 31.40 & 5.324 & 7.531 & 0.5633 \\
	 & GFPCA~\cite{GFPCA} & 29.78 & 5.996 & 8.409 & 0.5234 & 29.52 & 6.020 & 8.689 & 0.5196 & 29.18 & 6.066 & 9.066 & 0.5009 \\
	 & CNMF~\cite{CNMF} & 29.54 & 8.286 & 9.723 & 0.5456 & 28.27 & 8.791 & 10.96 & 0.5211 & 25.56 & 10.59 & 14.25 & 0.3668 \\
	 & HySure~\cite{HySure} & 31.30 & \underline{4.773} & 6.960 & 0.5780 & 30.90 & {\bf{4.925}} & 7.357 & 0.5801 & 29.93 & {\bf{5.111}} & 8.355 & 0.5477 \\
	 & Lanaras's~\cite{Lanaras_fusion} & 30.62 & 6.378 & 7.823 & 0.5494 & 29.70 & 6.302 & 8.302 & 0.5558 & 26.26 & 8.147 & 12.26 & 0.4706 \\
	 & LTMR~\cite{LTMR} & 30.61 & 6.749 & 8.370 & 0.5771 & 29.65 & 7.730 & 9.068 & 0.5404 & 28.09 & 9.036 & 11.14 & 0.4849 \\
	 & CNN-Fus~\cite{CNN_Fus} & 33.14 & {\bf{4.157}} & \underline{6.211} & {\bf{0.8975}} & 31.09 & 5.219 & 7.694 & 0.8284 & 31.15 & 5.185 & 7.665 & \underline{0.8315} \\
	 & {\bf{proposed ($p = 1$)}} & \underline{33.28} & 5.053 & 6.249 & 0.8751 & \underline{32.25} & 5.187 & \underline{7.094} & \underline{0.8393} & \underline{31.77} & 5.240 & \underline{7.501} & 0.8288 \\
	 & {\bf{proposed ($p = 2$)}} & {\bf{33.49}} & 4.982 & {\bf{6.114}} & \underline{0.8784} & {\bf{32.37}} & \underline{5.123} & {\bf{6.999}} & {\bf{0.8422}} & {\bf{31.88}} & \underline{5.178} & {\bf{7.413}} & {\bf{0.8318}} \\ \hline
	\end{tabular}}
\end{center}
\end{table*}

\begin{table*}[tp]
\begin{center}
	\caption{The quality measures  of the results on HS pansharpening experiments with $r = 4$ (boldface: the highest performance, underline: the second performance).}
	\label{tab:HSPan_4}
	\scalebox{0.88}{
	\begin{tabular}{|c|c||c|c|c|c||c|c|c|c||c|c|c|c|} \hline
	\multirow{2}{*}{image} & \multirow{2}{*}{method} & \multicolumn{4}{c||}{$\sigma_{\g} = 0$} & \multicolumn{4}{c||}{$\sigma_{\g} = 0.02$} & \multicolumn{4}{c|}{$\sigma_{\g} = 0.04$} \\ \cline{3-14}
	& & PSNR[dB] & SAM & ERGAS & $Q2^n$ & PSNR[dB] & SAM & ERGAS & $Q2^n$ & PSNR[dB] & SAM & ERGAS & $Q2^n$ \\ \hline
	 \multirow{10}{*}{Reno} & GSA~\cite{GSA} & 31.23 & \underline{6.436} & 4.568 & 0.7545 & 30.73 & 6.440 & 4.749 & 0.7440 & 30.16 & 6.460 & 4.964 & 0.7242 \\ 
	 & MTF-GLP~\cite{MTF_GLP} & 30.22 & 6.557 & 4.838 & 0.6295 & 29.88 & 6.560 & 4.980 & 0.6208 & 29.51 & 6.570 & 5.143 & 0.6341 \\ 
	 & GFPCA~\cite{GFPCA} & 27.79 & 6.782 & 5.954 & 0.4410 & 27.65 & 6.812 & 6.048 & 0.4378 & 27.57 & 6.830 & 6.082 & 0.4457 \\ 
	 & CNMF~\cite{CNMF} & 31.11 & 6.794 & 4.361 & 0.7384 & 30.18 & 6.668 & 4.604 & 0.6582 & 27.98 & 7.281 & 5.704 & 0.6173 \\ 
	 & HySure~\cite{HySure} & \underline{32.01} & {\bf{6.112}} & {\bf{4.047}} & 0.7835 & {\bf{31.66}} & {\bf{5.951}} & \underline{4.092} & 0.7511 & \underline{30.62} & {\bf{6.108}} & \underline{4.476} & 0.7102 \\ 
	 & Lanaras's~\cite{Lanaras_fusion} & 31.96 & 7.717 & 4.102 & 0.8446 & 30.44 & 8.130 & 4.630 & 0.7719 & 27.81 & 8.713 & 5.854 & 0.6505 \\
	 & LTMR~\cite{LTMR} & 31.64 & 7.175 & 4.245 & 0.8160 & 30.32 & 7.474 & 4.729 & 0.7423 & 27.07 & 10.91 & 6.419 & 0.5521 \\
	 & CNN-Fus~\cite{CNN_Fus} & 31.38 & 6.472 & \underline{4.100} & {\bf{0.9083}} & \underline{31.51} & \underline{6.242} & {\bf{4.043}} & {\bf{0.9094}} & {\bf{31.09}} & \underline{6.411} & {\bf{4.187}} & {\bf{0.9043}} \\
	 & {\bf{proposed ($p = 1$)}} & 31.72 & 7.482 & 4.958 & 0.8835 & 30.69 & 7.817 & 5.293 & 0.8659 & 29.95 & 7.829 & 5.431 & 0.8547 \\ 
	 & {\bf{proposed ($p = 2$)}} & {\bf{32.05}} & 7.247 & 4.603 & \underline{0.8963} & 31.17 & 7.197 & 4.638 & \underline{0.8860} & 30.23 & 7.220 & 4.940 & \underline{0.8700} \\ \hline
	 \multirow{10}{*}{Pavia U} & GSA~\cite{GSA} & 27.55 & 7.679 & 5.818 & 0.3582 & 27.29 & 7.736 & 6.040 & 0.3612 & 26.96 & 7.751 & 6.331 & 0.3421 \\ 
	 & MTF-GLP~\cite{MTF_GLP} & 26.78 & 8.326 & 6.757 & 0.3015 & 26.59 & 8.358 & 6.913 & 0.2945 & 26.37 & 8.371 & 7.097 & 0.2930 \\ 
	 & GFPCA~\cite{GFPCA} & 24.99 & 8.757 & 7.740 & 0.1962 & 24.95 & 8.718 & 7.782 & 0.1973 & 24.91 & 8.669 & 7.848 & 0.1962 \\ 
	 & CNMF~\cite{CNMF} & 23.59 & 15.82 & 8.167 & 0.3683 & 23.13 & 16.37 & 8.863 & 0.3256 & 22.16 & 17.63 & 10.45 & 0.2479 \\ 
	 & HySure~\cite{HySure} & 28.38 & 7.505 & 4.861 & 0.4083 & 28.03 & 7.691 & 5.181 & 0.4061 & 27.64 & 7.853 & 5.562 & 0.4029 \\
	 & Lanaras's~\cite{Lanaras_fusion} & 28.11 & 8.663 & 5.067 & 0.4927 & 26.90 & 9.236 & 6.008 & 0.4717 & 24.76 & 10.56 & 8.108 & 0.4332 \\
	 & LTMR~\cite{LTMR} & 28.04 & 8.620 & 5.555 & 0.5091 & 27.10 & 9.631 & 6.428 & 0.4825 & 25.26 & 12.33 & 8.390 & 0.4381 \\
	 & CNN-Fus~\cite{CNN_Fus} & 28.60 & 7.343 & 5.237 & 0.8634 & 28.60 & {\bf{7.282}} & 5.239 & 0.8628 & {\bf{28.32}} & \underline{7.571} & \underline{5.479} & {\bf{0.8516}} \\
	 & {\bf{proposed ($p = 1$)}} & \underline{29.29} & \underline{7.314} & \underline{4.420} & \underline{0.9063} & \underline{28.70} & 7.502 & {\bf{4.925}} & {\bf{0.8777}} & 28.05 & 7.695 & {\bf{5.464}} & {\underline{0.8509}} \\ 
	 & {\bf{proposed ($p = 2$)}} & {\bf{29.45}} & {\bf{7.310}} & {\bf{4.374}} & {\bf{0.9100}} & {\bf{28.76}} & {\underline{7.356}} & {\underline{4.944}} & \underline{0.8757} & {\underline{28.10}} & {\bf{7.427}} & 5.498 & 0.8490 \\ \hline
	 \multirow{10}{*}{Salinas} & GSA~\cite{GSA} & 31.08 & 4.216 & 2.633 & 0.4202 & 30.93 & 4.274 & 2.700 & 0.4178 & 30.83 & 4.310 & 2.748 & 0.4170 \\
	 & MTF-GLP~\cite{MTF_GLP} & 29.75 & 4.698 & 2.955 & 0.3264 & 29.74 & 4.695 & 2.989 & 0.3595 & 29.74 & 4.685 & 3.010 & 0.3547 \\ 
	 & GFPCA~\cite{GFPCA} & 31.00 & \underline{3.317} & 2.516 & 0.4383 & 30.87 & {\bf{3.405}} & 2.578 & 0.4395 & 30.72 & \underline{3.468} & 2.653 & 0.4438 \\ 
	 & CNMF~\cite{CNMF} & 31.07 & 3.638 & 2.588 & 0.3483 & 30.07 & 4.520 & 3.192 & 0.3380 & 28.23 & 5.635 & 4.033 & 0.2315 \\ 
	 & HySure~\cite{HySure} & 31.52 & 3.346 & \underline{2.368} & 0.4238 & 31.47 & \underline{3.410} & {\underline{2.438}} & 0.4213 & 31.37 & {\bf{3.274}} & {\bf{2.448}} & 0.4161 \\ 
	 & Lanaras's~\cite{Lanaras_fusion} & 29.37 & 4.522 & 2.979 & 0.3866 & 29.21 & 4.663 & 3.252 & 0.3610 & 27.44 & 5.607 & 4.185 & 0.2377 \\
	 & LTMR~\cite{LTMR} & 30.86 & 4.454 & 2.656 & 0.3789 & 29.79 & 4.929 & 3.116 & 0.3419 & 27.74 & 6.509 & 4.230 & 0.2752 \\
	 & CNN-Fus~\cite{CNN_Fus} & {\bf{33.17}} & {\bf{3.313}} & {\bf{2.144}} & {\bf{0.6926}} & {\bf{32.66}} & 3.541 & {\bf{2.276}} & {\bf{0.6683}} & \underline{31.67} & 4.022 & \underline{2.637} & {\bf{0.6282}} \\
	 & {\bf{proposed ($p = 1$)}} & 31.94 & 3.961 & 3.165 & \underline{0.5620} & 31.86 & 3.963 & 3.136 & 0.5485 & 31.62 & 4.007 & 3.201 & 0.5361  \\ 
	 & {\bf{proposed ($p = 2$)}} & \underline{32.13} & 3.935 & 3.144 & 0.5585 & {\underline{32.08}} & 3.824 & 2.950 & {\underline{0.5681}} & {\bf{31.81}} & 3.870 & 3.025 & {\underline{0.5562}} \\ \hline
	 \multirow{10}{*}{\begin{tabular}{c} Moffett\\field \end{tabular}} & GSA~\cite{GSA} & 24.31 & 9.924 & 7.575 & 0.3858 & 24.15 & 9.924 & 7.750 & 0.3861 & 23.81 & 9.924 & 8.107 & 0.3822 \\ 
	 & MTF-GLP~\cite{MTF_GLP} & 27.09 & 9.924 & 6.336 & 0.6588 & 26.80 & 9.924 & 6.524 & 0.5945 & 26.50 & 9.924 & 6.732 & 0.6563 \\ 
	 & GFPCA~\cite{GFPCA} & 26.35 & 7.952 & 6.435 & 0.5383 & 26.27 & 7.909 & 6.451 & 0.5298 & 26.13 & 7.928 & 6.581 & 0.5374 \\ 
	 & CNMF~\cite{CNMF} & 24.24 & 12.74 & 7.954 & 0.3923 & 23.54 & 13.14 & 8.608 & 0.3103 & 21.92 & 14.82 & 10.51 & 0.1723 \\
	 & HySure~\cite{HySure} & 28.56 & 7.329 & 4.833 & 0.6001 & 28.20 & \underline{7.367} & 5.048 & 0.5983 & 27.56 & 7.386 & 5.458 & 0.5974 \\
	 & Lanaras's~\cite{Lanaras_fusion} & 25.83 & 8.831 & 6.392 & 0.5760 & 25.40 & 8.969 & 6.851 & 0.5719 & 23.90 & 9.571 & 8.302 & 0.5669 \\
	 & LTMR~\cite{LTMR} & 28.17 & 9.316 & 5.410 & 0.5183 & 26.70 & 9.748 & 6.264 & 0.4382 & 26.70 & 10.21 & 6.543 & 0.2911 \\
	 & CNN-Fus~\cite{CNN_Fus} & 28.62 & {\bf{6.735}} & 4.866 & 0.8298 & \underline{28.72} & 6.938 & \underline{4.828} & \underline{0.8349} & {\bf{28.47}} & \underline{7.365} & \underline{5.083} & \underline{0.8185} \\
	 & {\bf{proposed ($p = 1$)}} & \underline{29.62} & 7.369 & \underline{4.276} & \underline{0.8789} & {\bf{28.91}} & 7.467 & {\bf{4.731}} & {\bf{0.8461}} & {\underline{28.42}} & 7.534 & {\bf{5.073}} & {\bf{0.8198}} \\ 
	 & {\bf{proposed ($p = 2$)}} & {\bf{29.76}} & \underline{7.265} & {\bf{4.229}} & {\bf{0.8824}} & 28.43 & {\bf{7.326}} & 4.982 & 0.8206 & 28.19 & {\bf{7.277}} & 5.181 & 0.8019 \\ \hline \hline
	 \multirow{10}{*}{Average} & GSA~\cite{GSA} & 28.54 & 7.064 & 5.148 & 0.4797 & 28.28 & 7.093 & 5.310 & 0.4773 & 27.94 & 7.111 & 5.537 & 0.4664 \\ 
	 & MTF-GLP~\cite{MTF_GLP} & 28.46 & 7.376 & 5.222 & 0.4791 & 28.25 & 7.384 & 5.351 & 0.4673 & 28.03 & 7.388 & 5.496 & 0.4845 \\
	 & GFPCA~\cite{GFPCA} & 27.53 & 6.702 & 5.661 & 0.4034 & 27.43 & 6.711 & 5.715 & 0.4011 & 27.33 & 6.724 & 5.791 & 0.4058 \\
	 & CNMF~\cite{CNMF} & 27.50 & 9.747 & 5.768 & 0.4618 & 26.71 & 9.903 & 6.261 & 0.4195 & 24.73 & 11.19 & 7.694 & 0.3022 \\
	 & HySure~\cite{HySure} & 30.12 & \underline{6.073} & {\bf{4.027}} & 0.5539 & 29.67 & \underline{6.260} & \underline{4.252} & 0.5367 & 29.16 & {\bf{6.241}} & \underline{4.525} & 0.5322 \\
	 & Lanaras's~\cite{Lanaras_fusion} & 28.82 & 7.433 & 4.635 & 0.575 & 27.99 & 7.750 & 5.185 & 0.5441 & 25.98 & 8.613 & 6.612 & 0.4721 \\
	 & LTMR~\cite{LTMR} & 29.67 & 7.391 & 4.466 & 0.5556 & 28.48 & 7.946 & 5.134 & 0.5012 & 26.30 & 9.512 & 6.627 & 0.3986 \\
	 & CNN-Fus~\cite{CNN_Fus} & 30.44 & {\bf{5.966}} & \underline{4.087} & {\bf{0.8235}} & {\bf{30.37}} & {\bf{6.001}} & {\bf{4.097}} & {\bf{0.8189}} & {\bf{29.89}} & \underline{6.342} & {\bf{4.347}} & {\bf{0.8007}} \\
	 & {\bf{proposed ($p = 1$)}} & \underline{30.64} & 6.531 & 4.205 & 0.8077 & 30.04 & 6.687 & 4.521 & 0.7846 & 29.51 & 6.766 & 4.792 & 0.7654 \\
	 & {\bf{proposed ($p = 2$)}} & {\bf{30.85}} & 6.439 & 4.088 & \underline{0.8118} & \underline{30.11} & 6.426 & 4.379 & \underline{0.7876} & \underline{29.58} & 6.448 & 4.661 & \underline{0.7693} \\ \hline
	\end{tabular}}
\end{center}
\end{table*}

\begin{table*}[tp]
\begin{center}
	\caption{The quality measures  of the results on HS pansharpening experiments with $r = 8$ and $16$ (boldface: the highest performance, underline: the second performance).}
	\label{tab:HSPan_816}
	\begin{tabular}{|c|c||c|c|c|c||c|c|c|c|} \hline
	\multirow{2}{*}{image} & \multirow{2}{*}{method} & \multicolumn{4}{c||}{$r = 8 ,~\sigma_{\v} = 0.1,~\sigma_{\g} = 0.02$} & \multicolumn{4}{c|}{$r = 16,~\sigma_{\v} = 0.1,~\sigma_{\g} = 0.02$} \\ \cline{3-10}
	& & PSNR[dB] & SAM & ERGAS & $Q2^n$ & PSNR[dB] & SAM & ERGAS & $Q2^n$ \\ \hline
	 \multirow{10}{*}{Reno} & GSA~\cite{GSA} & 28.72 & 8.009 & 3.065 & 0.6098 & 25.08 & \underline{8.664} & 2.140 & 0.2874 \\  
	 & MTF-GLP~\cite{MTF_GLP} & 26.09 & 8.009 & 3.655 & 0.3104 & 21.76 & \underline{8.664} & 3.037 & 0.2813 \\ 
	 & GFPCA~\cite{GFPCA} & 23.65 & 7.628 & 4.581 & 0.2943 & 20.96 & 9.795 & 3.091 & 0.2813\\ 
	 & CNMF~\cite{CNMF} & 26.84 & 10.28 & 3.516 & 0.6586 & 24.68 & 12.61 & 2.146 & 0.6791 \\ 
	 & HySure~\cite{HySure} & {\bf{30.51}} & \underline{6.849} & {\bf{2.374}} & 0.7645 & {\bf{29.45}} & {\bf{8.026}} & {\bf{1.398}} & 0.7006 \\ 
	 & Lanaras's~\cite{Lanaras_fusion} & 29.54 & 8.439 & 2.630 & 0.8066 & 28.24 & 10.55 & \underline{1.534} & 0.7603 \\ 
	 & LTMR~\cite{LTMR} & 29.38 & 8.635 & 2.710 & 0.7807 & 28.28 & 12.51 & 1.943 & 0.5142 \\ 
	 & CNN-Fus~\cite{CNN_Fus} & \underline{30.31} & {\bf{6.702}} & {\underline{2.386}} & {\bf{0.8740}} & \underline{28.56} & 9.748 & 1.544 & {\bf{0.8244}} \\ 
	 & {\bf{proposed ($p = 1$)}} & 29.47 & 8.908 & 2.958 & 0.8190 & 28.20 & 10.68 & 1.728 & \underline{0.7841} \\ 
	 & {\bf{proposed ($p = 2$)}} & 29.59 & 8.865 & 2.904 & \underline{0.8209} & 28.32 & 10.23 & 1.662 & 0.7826 \\  \hline
	 \multirow{10}{*}{Pavia U} & GSA~\cite{GSA} & 22.82 & 14.72 & 5.262 & 0.3687 & 21.43 & 14.86 & 3.568 & 0.1771 \\  
	 & MTF-GLP~\cite{MTF_GLP} & 22.02 & 14.72 & 6.236 & 0.1964 & 20.82 & 14.86 & 3.868 & 0.1719 \\ 
	 & GFPCA~\cite{GFPCA} & 22.67 & 10.69 & 5.483 & 0.1719 & 21.23 & 13.58 & 3.309 & 0.1719 \\ 
	 & CNMF~\cite{CNMF} & 23.08 & 14.99 & 4.580 & 0.1577 & 22.90 & 12.91 & 2.426 & 0.1826 \\ 
	 & HySure~\cite{HySure} & 26.09 & 10.40 & 3.107 & 0.3739 & 24.44 & 11.99 & 1.911 & 0.3525 \\ 
	 & Lanaras's~\cite{Lanaras_fusion} & 25.50 & 10.81 & 3.576 & 0.4066 & 24.55 & 12.00 & 1.891 & 0.4203 \\ 
	 & LTMR~\cite{LTMR} & 26.06 & 10.19 & 3.395 & 0.4125 & 24.93 & 13.68 & 1.893 & 0.3256 \\ 
	 & CNN-Fus~\cite{CNN_Fus} & 26.00 & 9.111 & 3.386 & 0.8120 & 24.93 & 11.27 & 1.791 & {\bf{0.7920}} \\ 
	 & {\bf{proposed ($p = 1$)}} & \underline{26.65} & \underline{9.336} & \underline{2.965} & \underline{0.8232} & \underline{24.98} & \underline{11.80} & \underline{1.780} & 0.7747 \\
	 & {\bf{proposed ($p = 2$)}} & {\bf{26.68}} & {\bf{9.333}} & {\bf{2.950}} & {\bf{0.8240}} & {\bf{25.05}} & {\bf{11.43}} & {\bf{1.724}} & \underline{0.7759} \\  \hline
	\end{tabular}
\end{center}
\end{table*}

\begin{figure*}[t]
\begin{center}
  
  \begin{minipage}[t]{0.128\hsize} 
\includegraphics[width=\hsize]{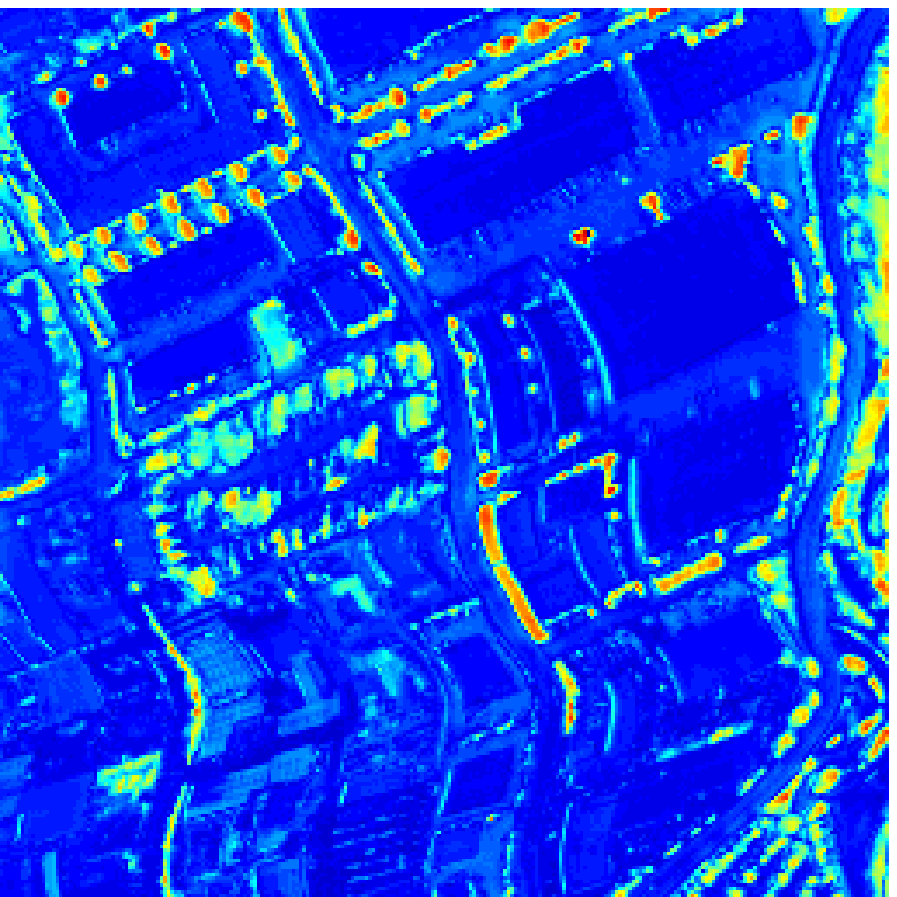}
\end{minipage}
\begin{minipage}[t]{0.128\hsize}
\includegraphics[width=\hsize]{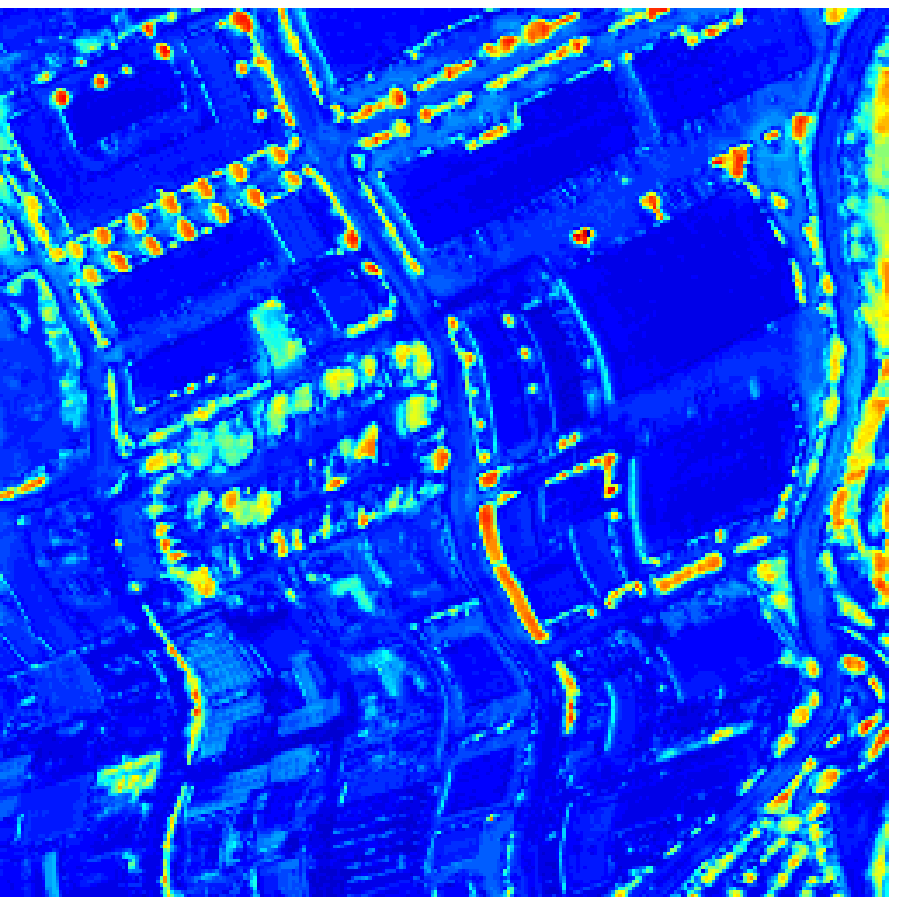}
\end{minipage}
\begin{minipage}[t]{0.128\hsize}
\includegraphics[width=\hsize]{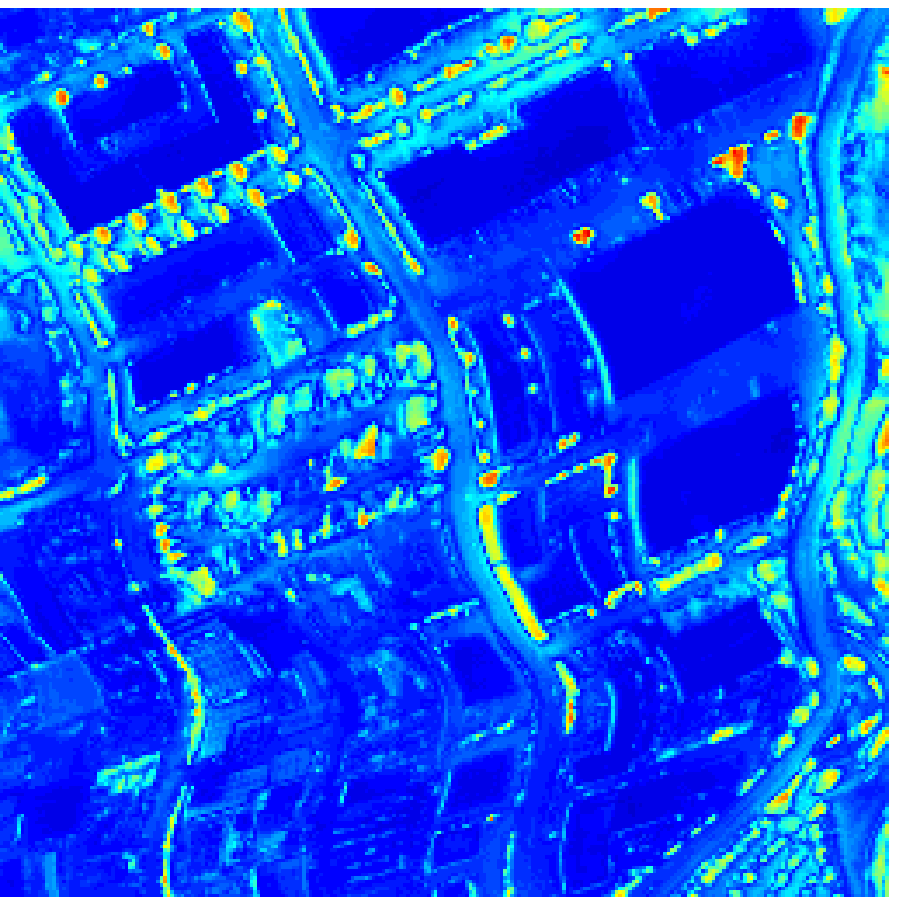}
\end{minipage}
  \begin{minipage}[t]{0.128\hsize}
\includegraphics[width=\hsize]{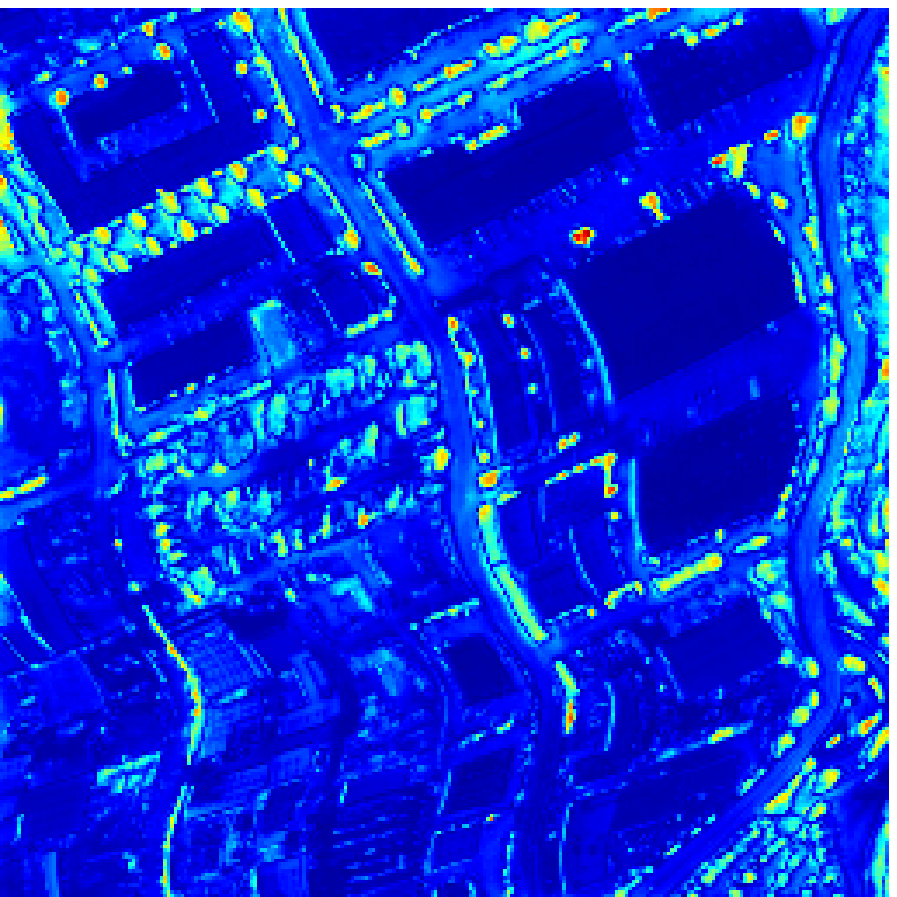}
\end{minipage}
 \begin{minipage}[t]{0.128\hsize}
\includegraphics[width=\hsize]{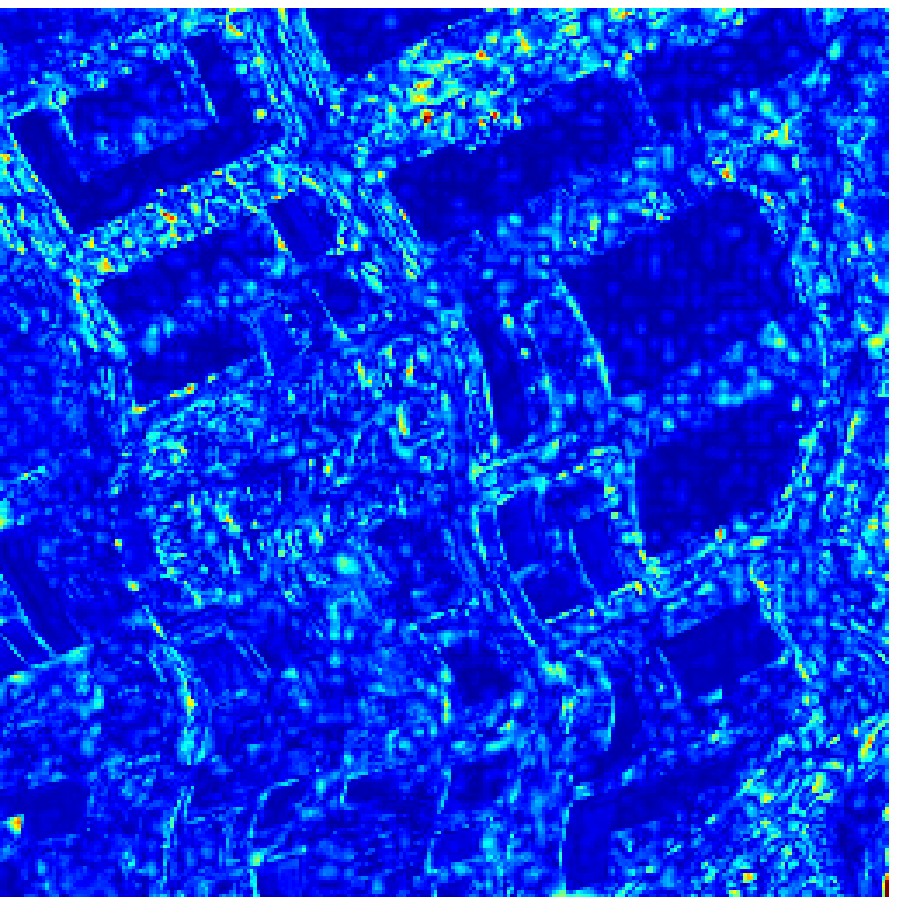}
\end{minipage}
  \begin{minipage}[t]{0.128\hsize}
\includegraphics[width=\hsize]{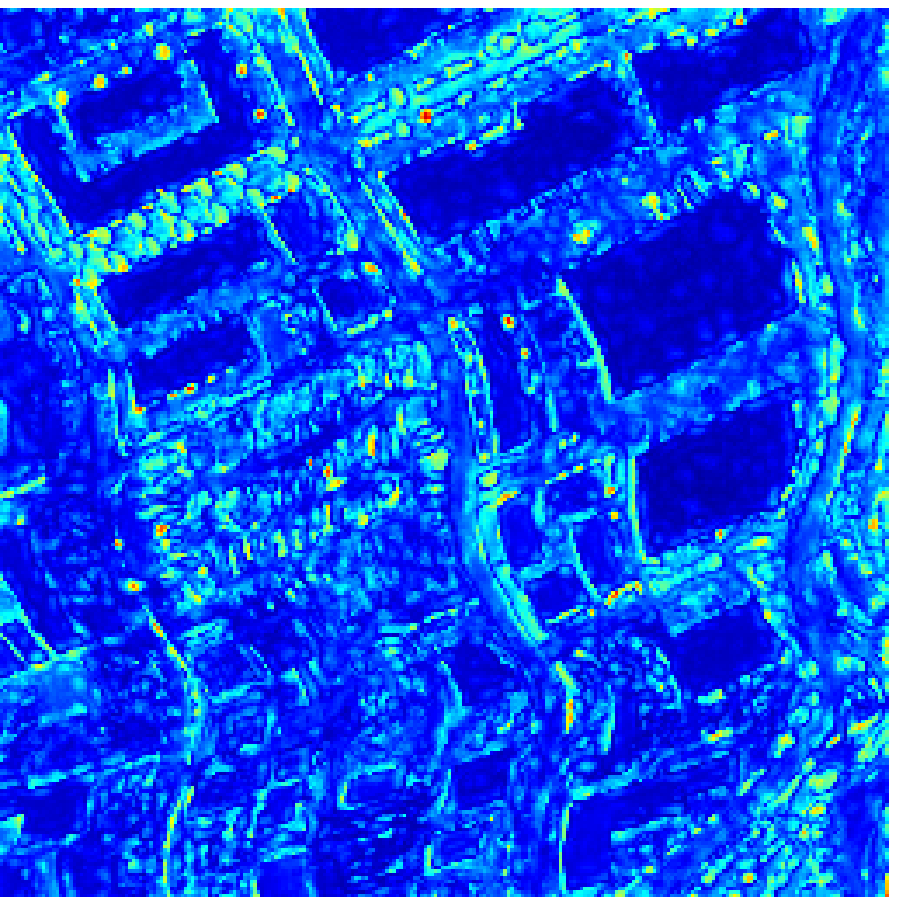}
\end{minipage}
  \begin{minipage}[t]{0.128\hsize}
\includegraphics[width=\hsize]{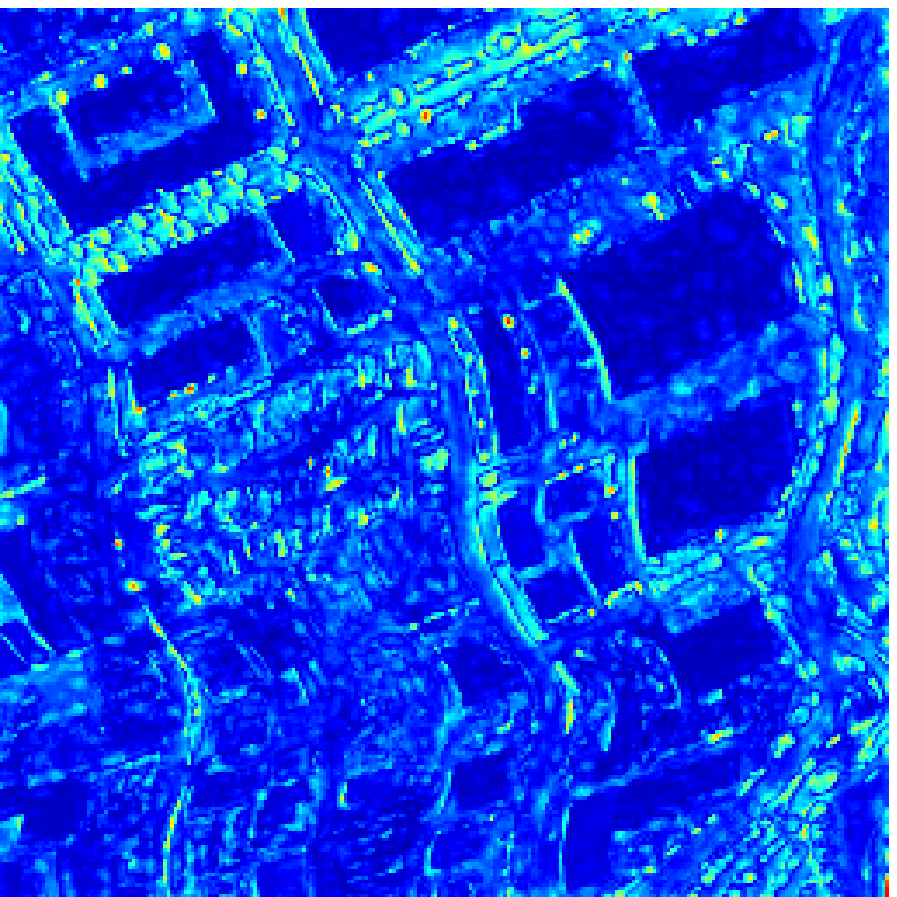}
\end{minipage}
\begin{minipage}[t]{0.06\hsize} 
\includegraphics[width=0.65\hsize]{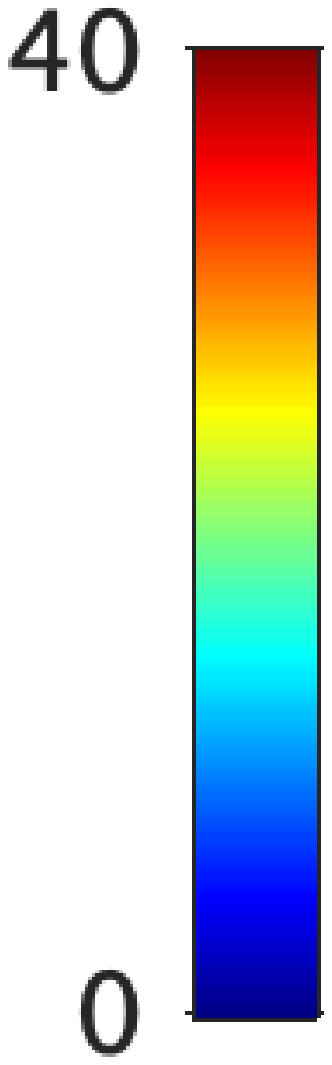}
\end{minipage}

  \begin{minipage}[t]{0.128\hsize}
\centerline{\footnotesize{ GSA }}
\end{minipage}
\begin{minipage}[t]{0.128\hsize}
\centerline{\footnotesize{ MTF-GLP }}
\end{minipage}
  \begin{minipage}[t]{0.128\hsize}
\centerline{\footnotesize{ GFPCA }}
\end{minipage}
  \begin{minipage}[t]{0.128\hsize}
\centerline{\footnotesize{ HySure }}
\end{minipage}
\begin{minipage}[t]{0.128\hsize}
\centerline{\footnotesize{ CNN-Fus }}
\end{minipage}
  \begin{minipage}[t]{0.128\hsize}
\centerline{\footnotesize{ proposed ($p = 1$) }}
\end{minipage}
  \begin{minipage}[t]{0.128\hsize}
\centerline{\footnotesize{ proposed ($p = 2$) }}
\end{minipage}
\begin{minipage}[t]{0.06\hsize}
\centerline{\footnotesize{  }}
\end{minipage}


   \caption{SAM map on HS pansharpening experiments 
   (Reno, $r = 4$, and $\sigma_\g = 0.02$).}
 \label{fig:SAMmap_HSpan}
\end{center}
\end{figure*}


\begin{figure*}[t]
    \centering
    \begin{minipage}[t]{0.32\hsize}
        \includegraphics[width=\hsize]{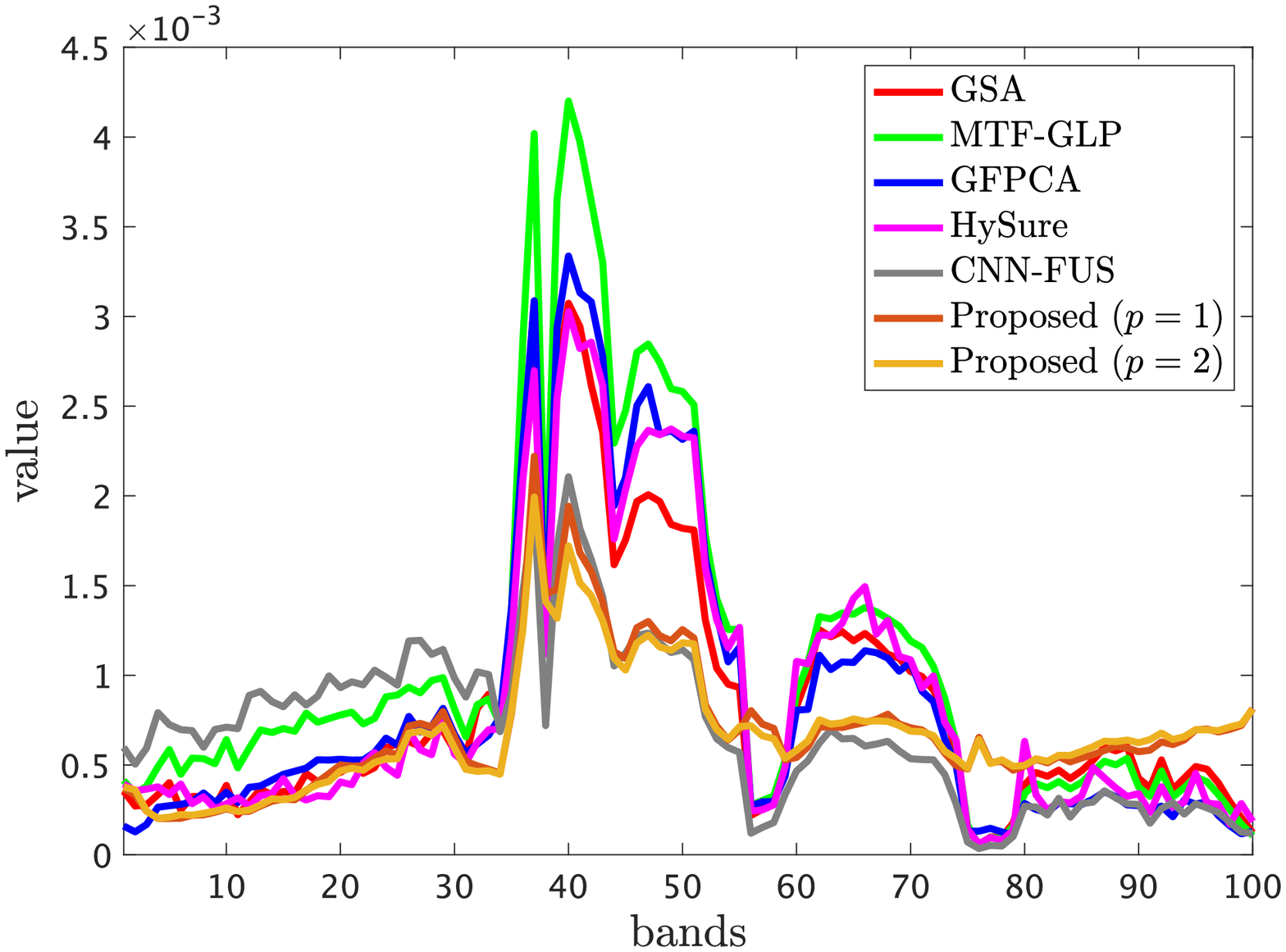}
    \end{minipage}
    \begin{minipage}[t]{0.32\hsize}
        \includegraphics[width=\hsize]{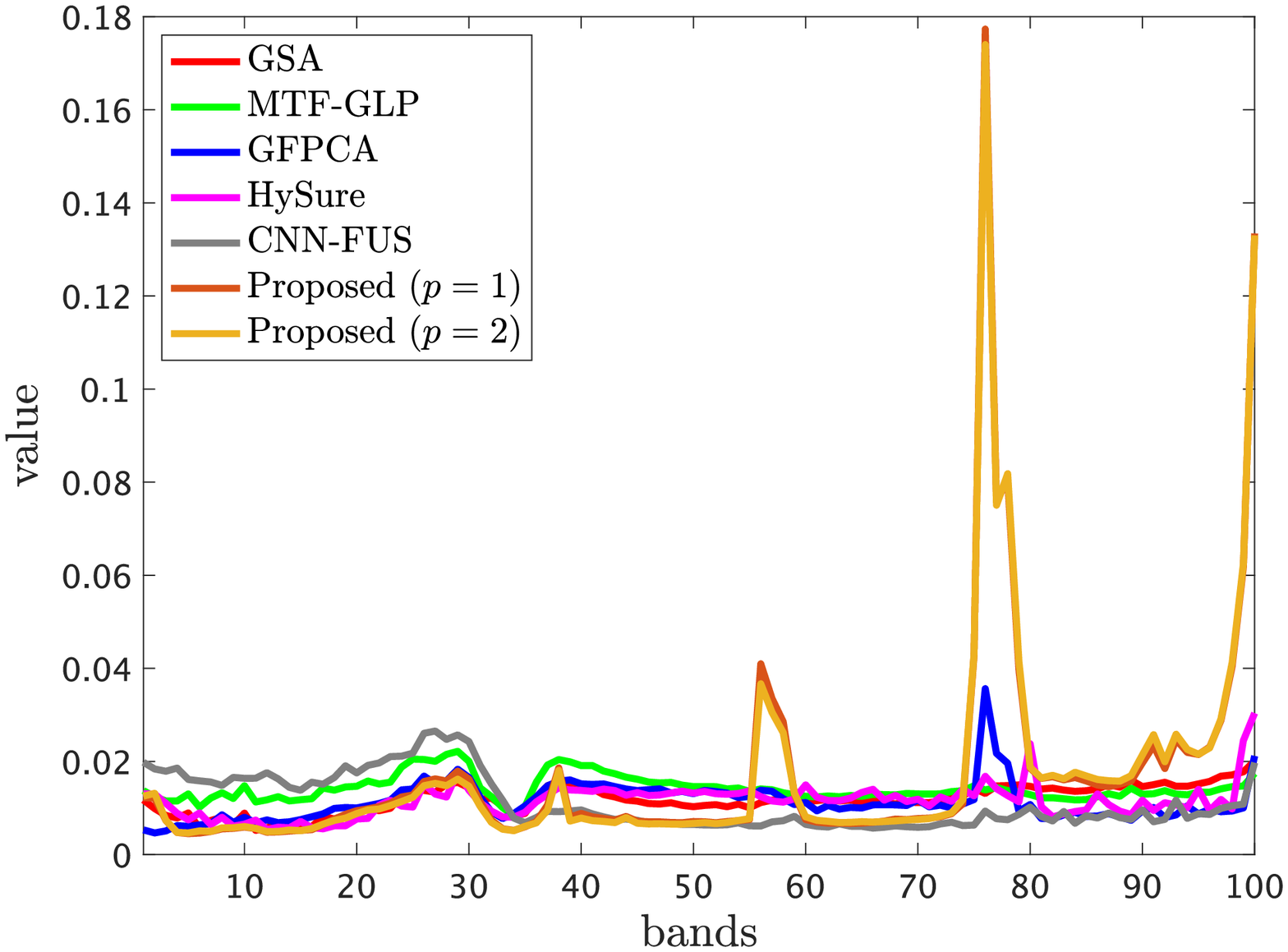}
    \end{minipage}
    \begin{minipage}[t]{0.32\hsize}
        \includegraphics[width=\hsize]{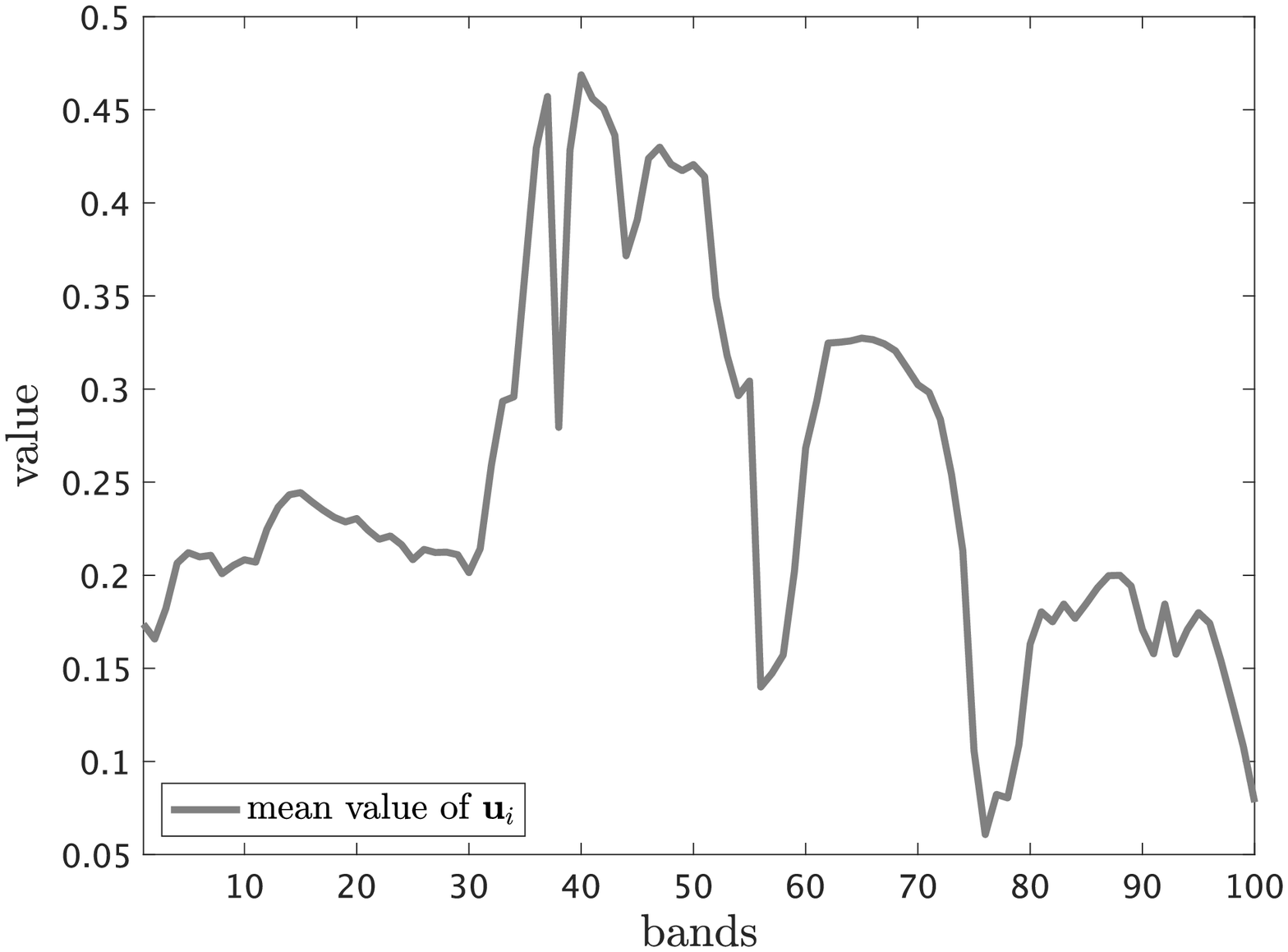}
    \end{minipage}
    
    \begin{minipage}[t]{0.32\hsize}
        \centerline{\footnotesize{MSE}}
    \end{minipage}
    \begin{minipage}[t]{0.32\hsize}
        \centerline{\footnotesize{band-wise normalized MSE}}
    \end{minipage}
    \begin{minipage}[t]{0.32\hsize}
        \centerline{\footnotesize{mean luminance}}
    \end{minipage}
    \caption{MSE and band-wise normalized MSE of the HS pansharpening results by GSA, MTF-GLP, GFPCA, HySure, CNN-Fus, and proposed ($p = 1$ and $2$) and mean luminance of the test HS image (Salinas, $r = 4$, and $\sigma_{\g} = 0.04$). }
    \label{fig:bandwiseMSE_pan}
\end{figure*}

\begin{figure*}[t]
\begin{center}
  \begin{minipage}[t]{0.135\hsize}
\includegraphics[width=\hsize]{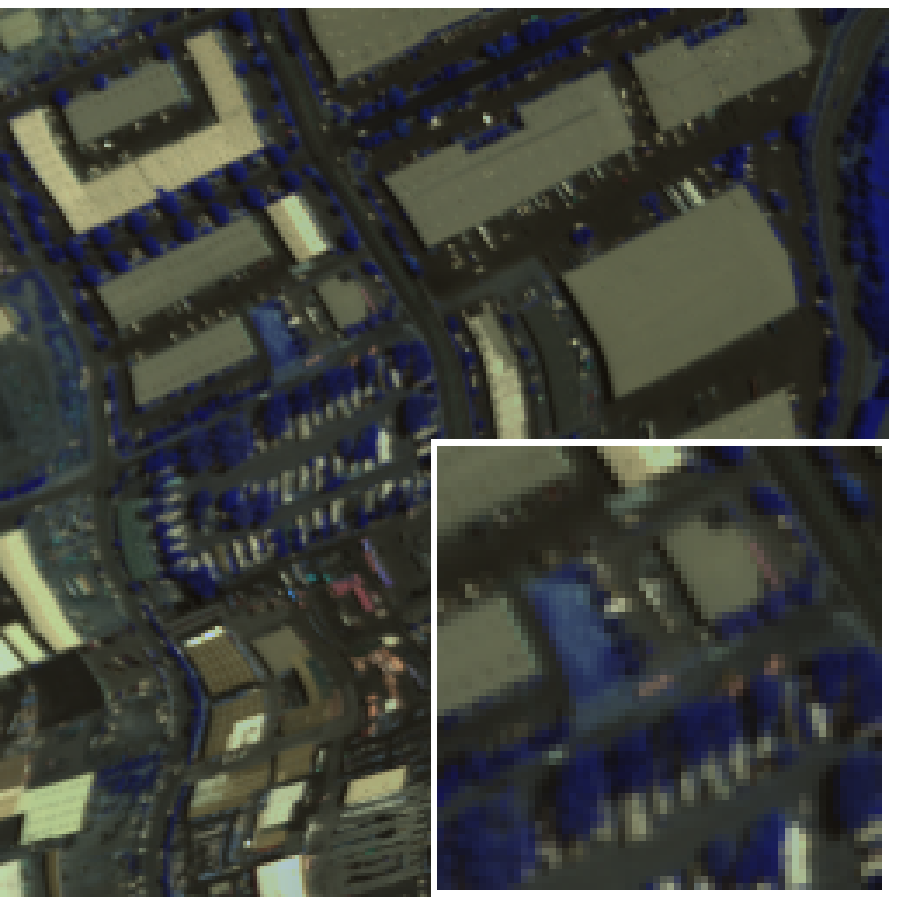}
\end{minipage}
  \begin{minipage}[t]{0.135\hsize}
\includegraphics[width=\hsize]{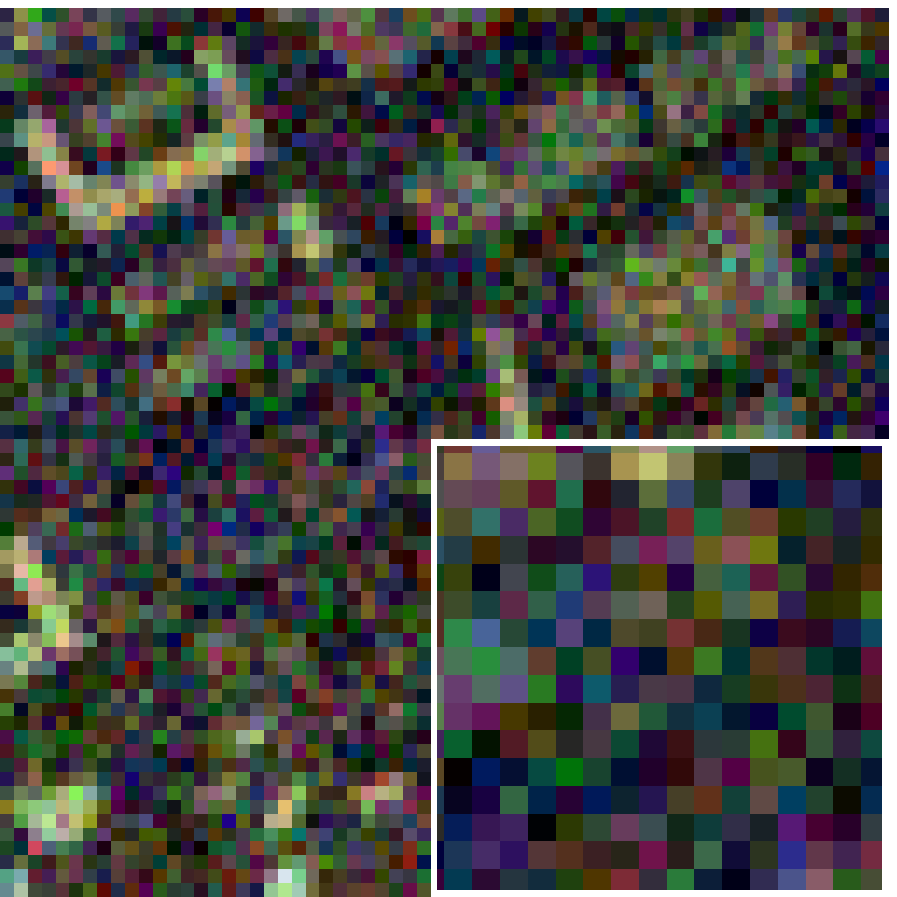}
\end{minipage}
\begin{minipage}[t]{0.135\hsize}
\includegraphics[width=\hsize]{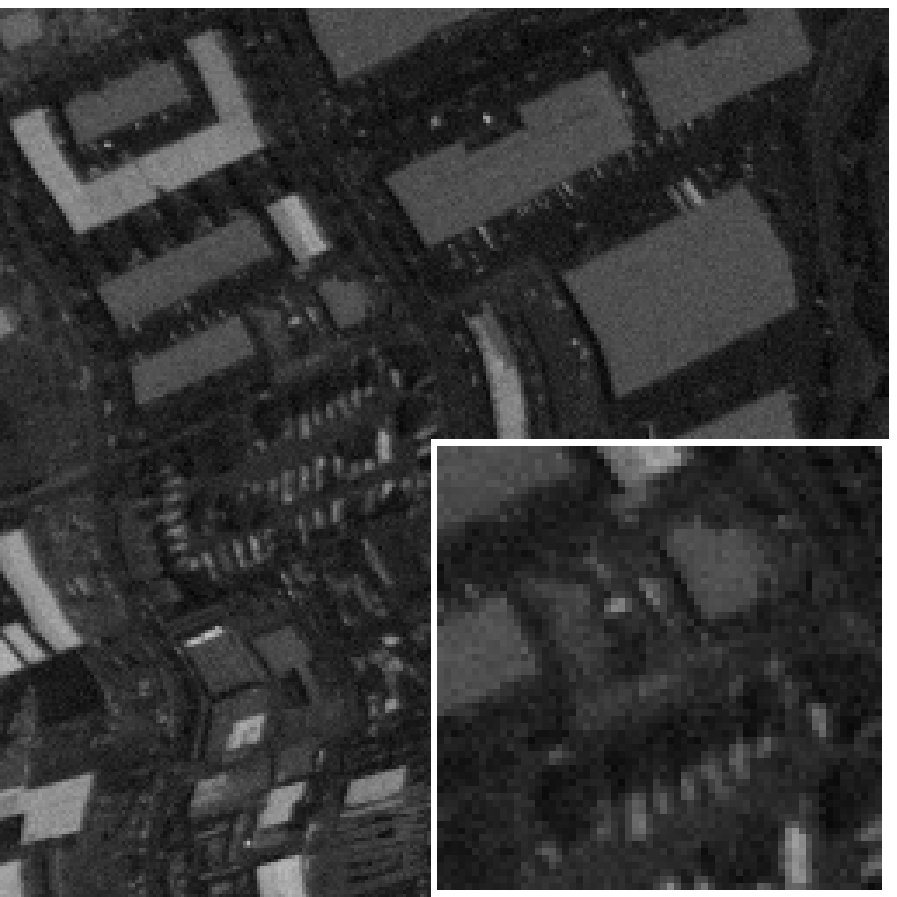}
\end{minipage}
  \begin{minipage}[t]{0.135\hsize}
\includegraphics[width=\hsize]{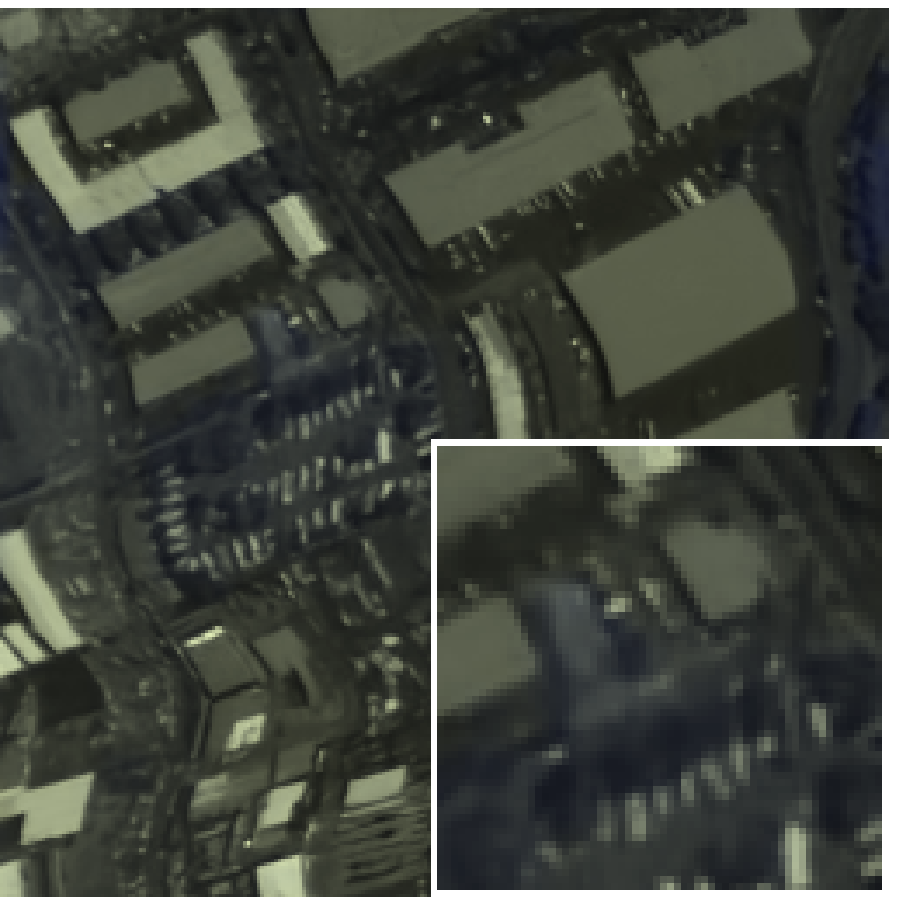}
\end{minipage}
  \begin{minipage}[t]{0.135\hsize}
\includegraphics[width=\hsize]{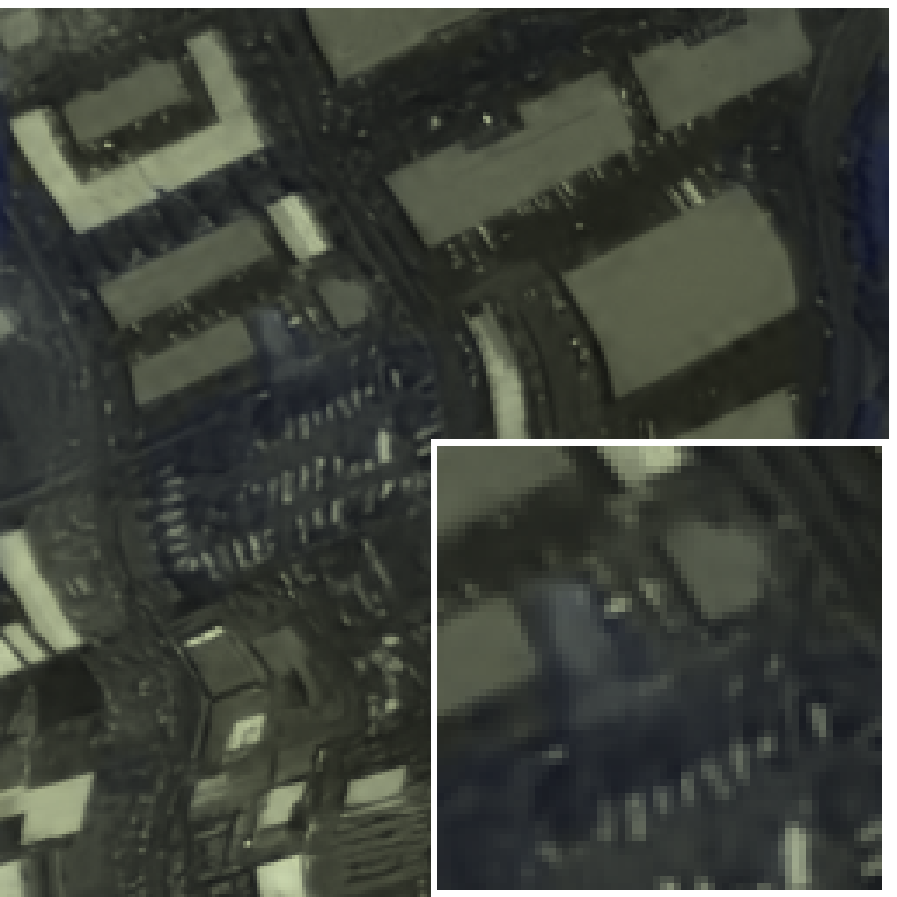}
\end{minipage}
  \begin{minipage}[t]{0.135\hsize}
\includegraphics[width=\hsize]{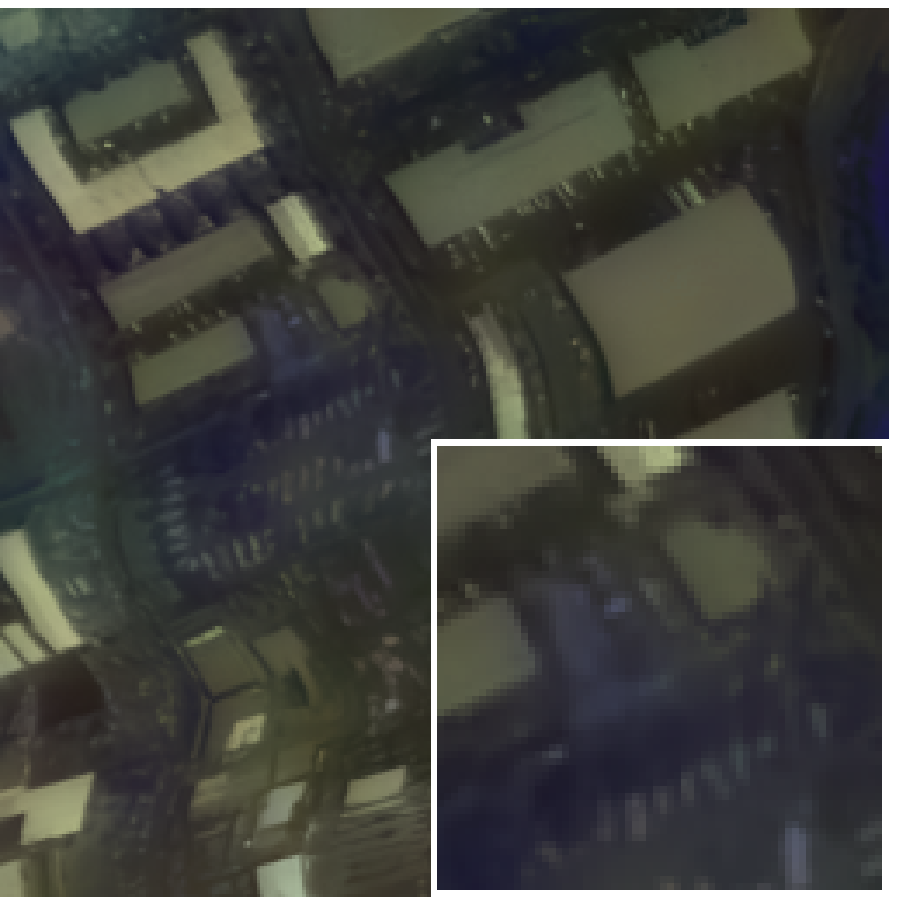}
\end{minipage}
\begin{minipage}[t]{0.135\hsize}
\includegraphics[width=\hsize]{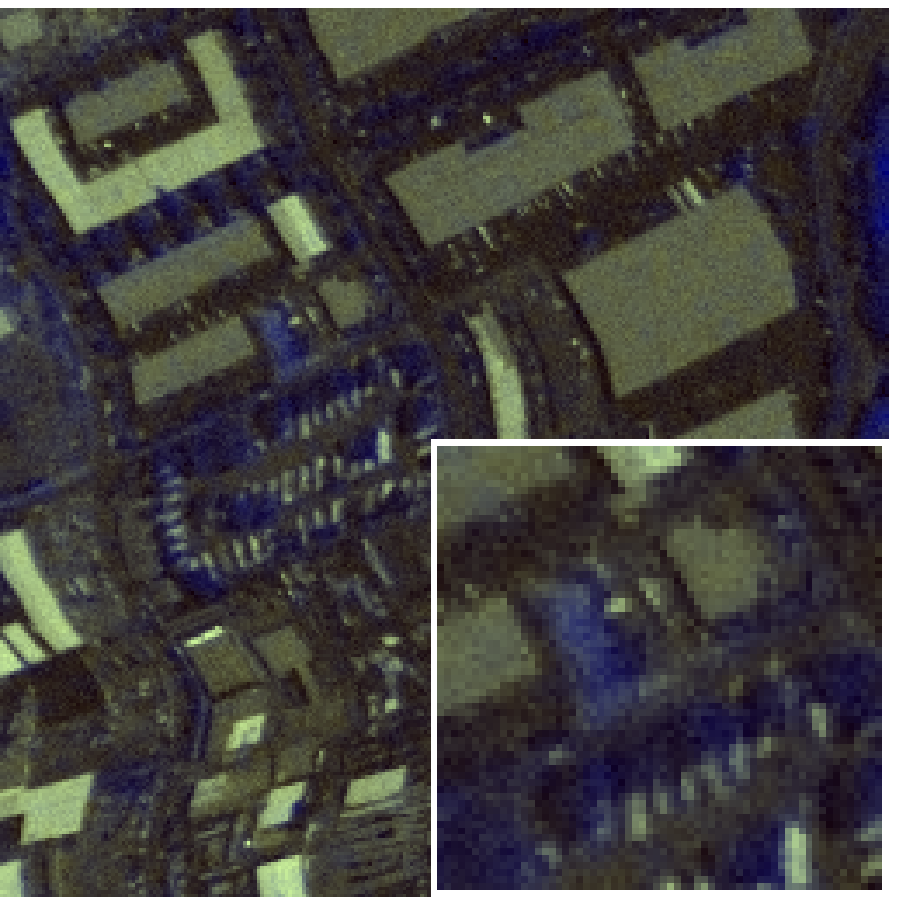}
\end{minipage}

  \begin{minipage}[t]{0.135\hsize}
\centerline{\footnotesize{ ground-truth }}
\end{minipage}
  \begin{minipage}[t]{0.135\hsize}
\centerline{\footnotesize{ observed HS image }}
\end{minipage}
\begin{minipage}[t]{0.135\hsize}
\centerline{\footnotesize{ observed PAN image }}
\end{minipage}
  \begin{minipage}[t]{0.135\hsize}
\centerline{\footnotesize{ GSA }}
\end{minipage}
  \begin{minipage}[t]{0.135\hsize}
\centerline{\footnotesize{ MTF-GLP }}
\end{minipage}
  \begin{minipage}[t]{0.135\hsize}
\centerline{\footnotesize{ GFPCA }}
\end{minipage}
\begin{minipage}[t]{0.135\hsize}
\centerline{\footnotesize{ CNMF }}
\end{minipage}

\vspace{4pt}
  \begin{minipage}[t]{0.135\hsize}
\includegraphics[width=\hsize]{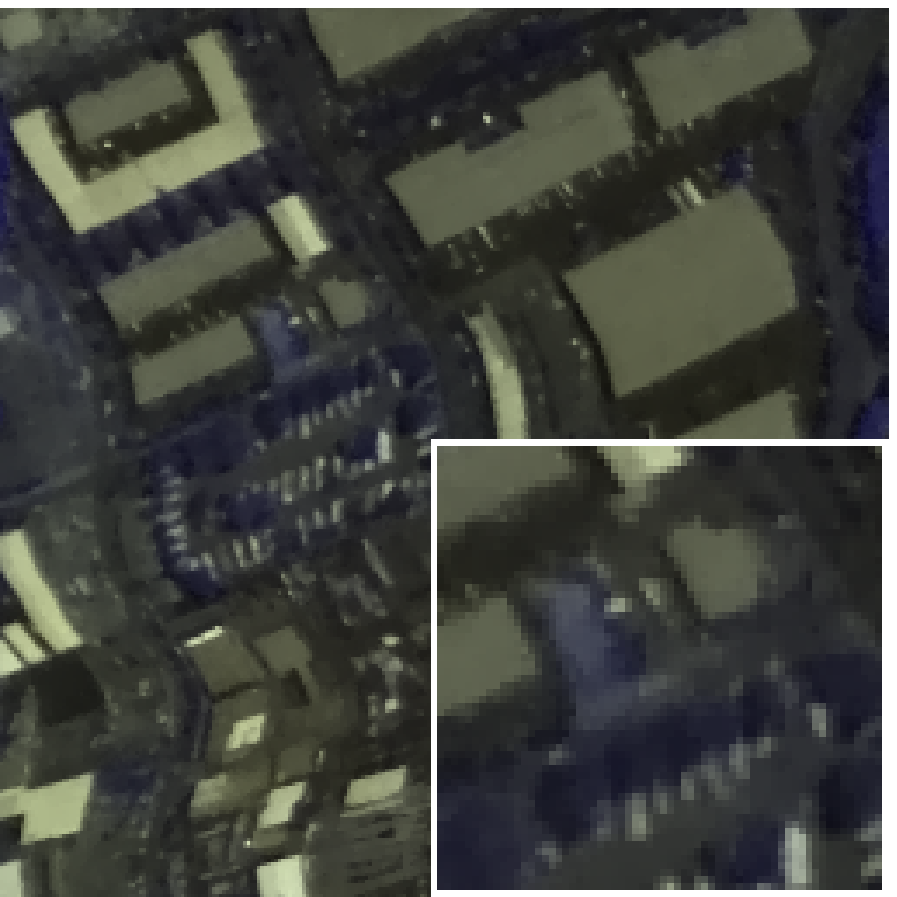}
\end{minipage}
  \begin{minipage}[t]{0.135\hsize}
\includegraphics[width=\hsize]{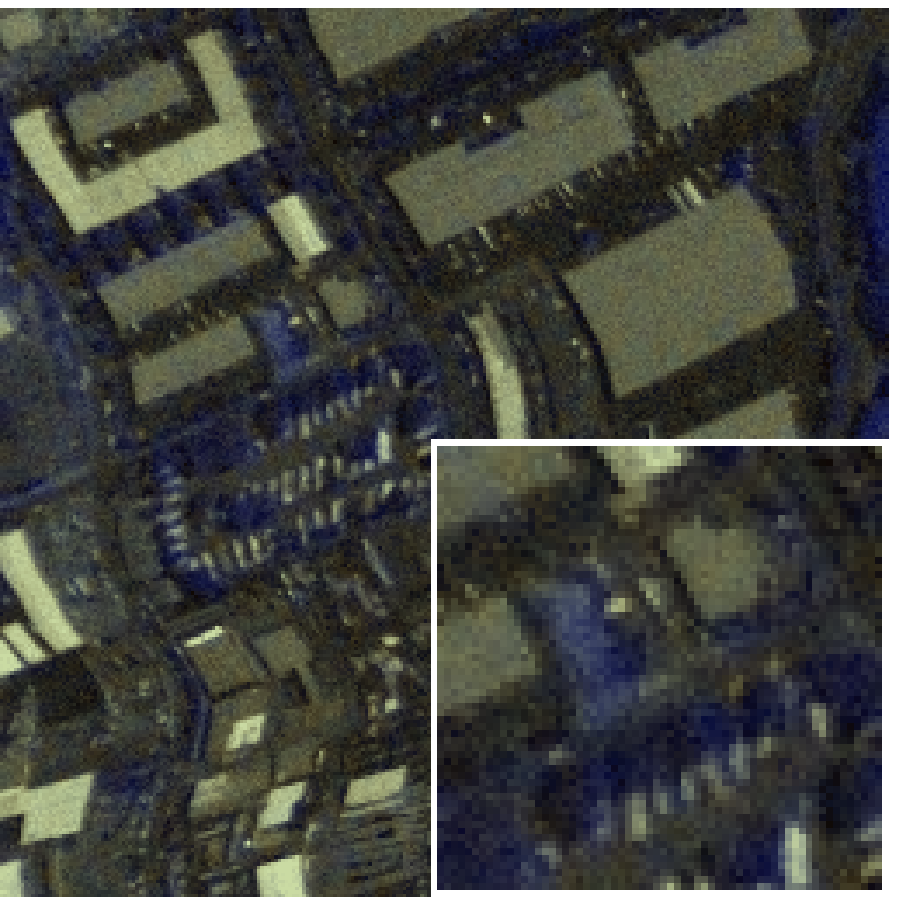}
\end{minipage}
  \begin{minipage}[t]{0.135\hsize}
\includegraphics[width=\hsize]{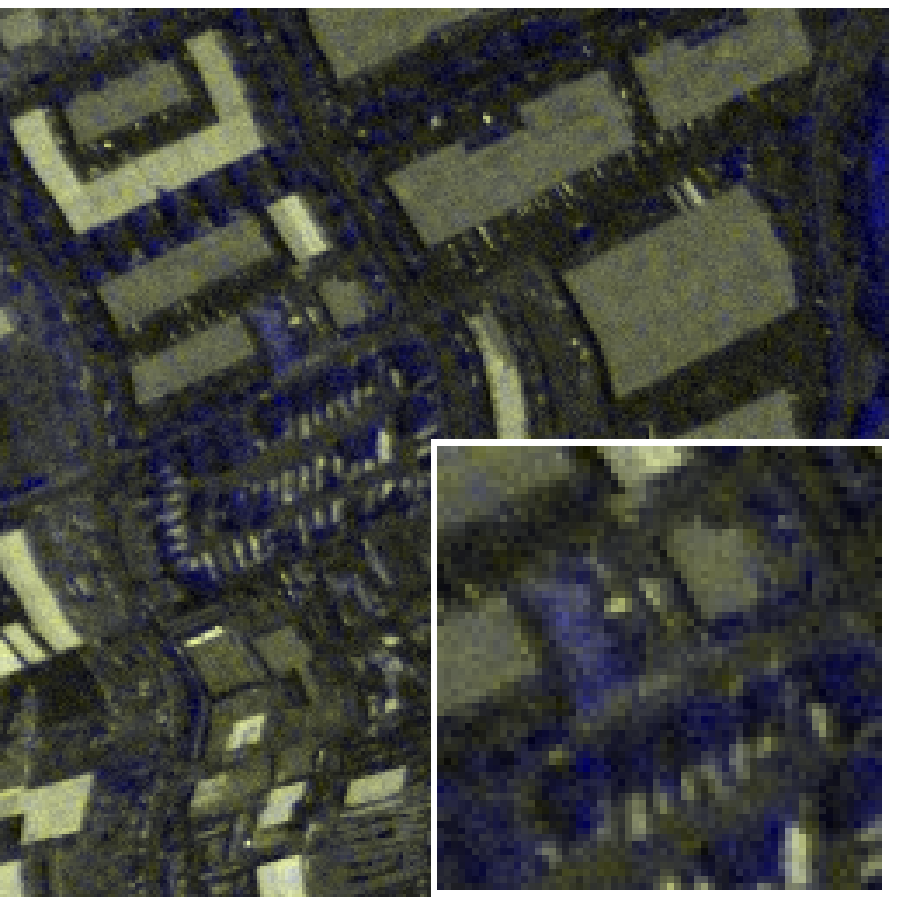}
\end{minipage}
\begin{minipage}[t]{0.135\hsize}
\includegraphics[width=\hsize]{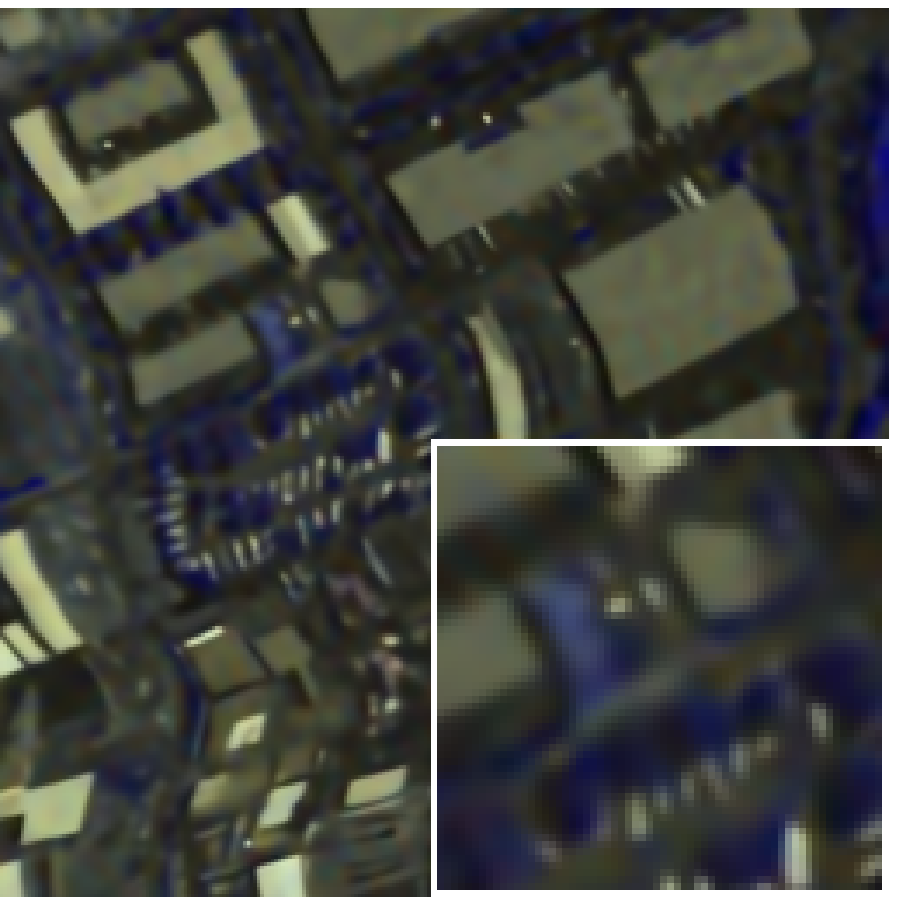}
\end{minipage}
  \begin{minipage}[t]{0.135\hsize}
\includegraphics[width=\hsize]{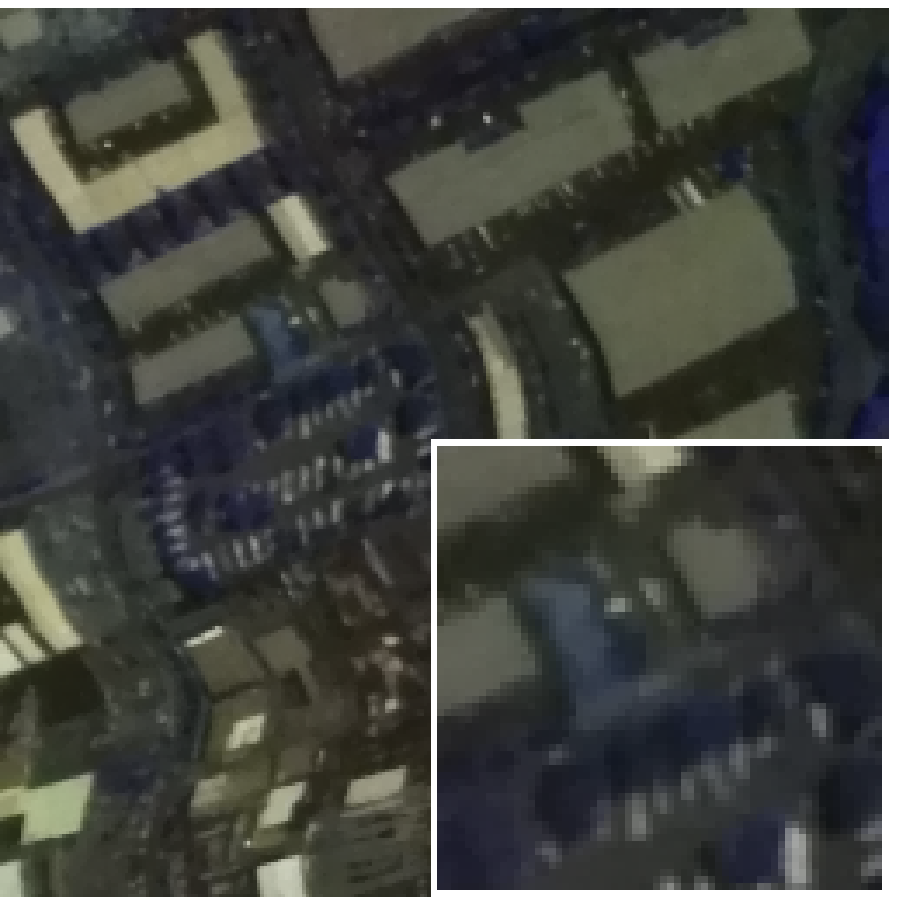}
\end{minipage}
  \begin{minipage}[t]{0.135\hsize}
\includegraphics[width=\hsize]{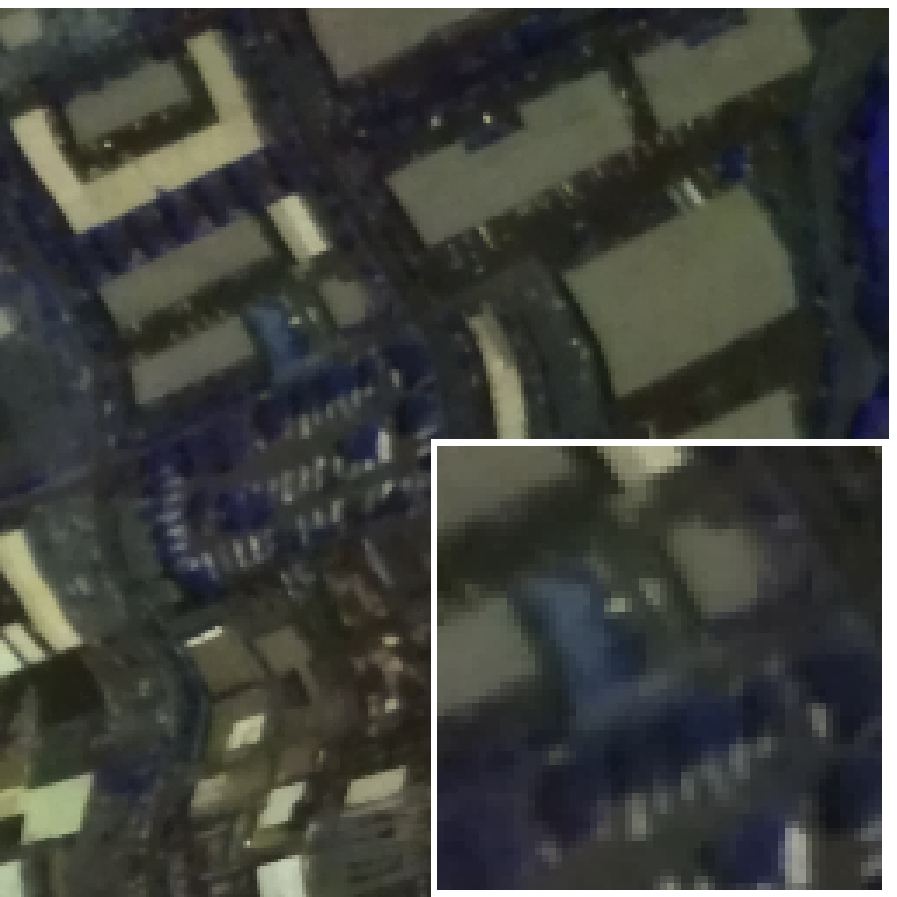}
\end{minipage}
\begin{minipage}[t]{0.135\hsize}
\centerline{\footnotesize{  }}
\end{minipage}

  \begin{minipage}[t]{0.135\hsize}
\centerline{\footnotesize{ HySure }}
\end{minipage}
  \begin{minipage}[t]{0.135\hsize}
\centerline{\footnotesize{ Lanaras's }}
\end{minipage}
  \begin{minipage}[t]{0.135\hsize}
\centerline{\footnotesize{ LTMR }}
\end{minipage}
\begin{minipage}[t]{0.135\hsize}
\centerline{\footnotesize{ CNN-Fus }}
\end{minipage}
  \begin{minipage}[t]{0.135\hsize}
\centerline{\footnotesize{ \bf{proposed ($p = 1$)} }}
\end{minipage}
  \begin{minipage}[t]{0.135\hsize}
\centerline{\footnotesize{ \bf{proposed ($p = 2$)} }}
\end{minipage}
\begin{minipage}[t]{0.135\hsize}
\centerline{\footnotesize{  }}
\end{minipage}

\vspace{4pt}
  \begin{minipage}[t]{0.135\hsize}
\includegraphics[width=\hsize]{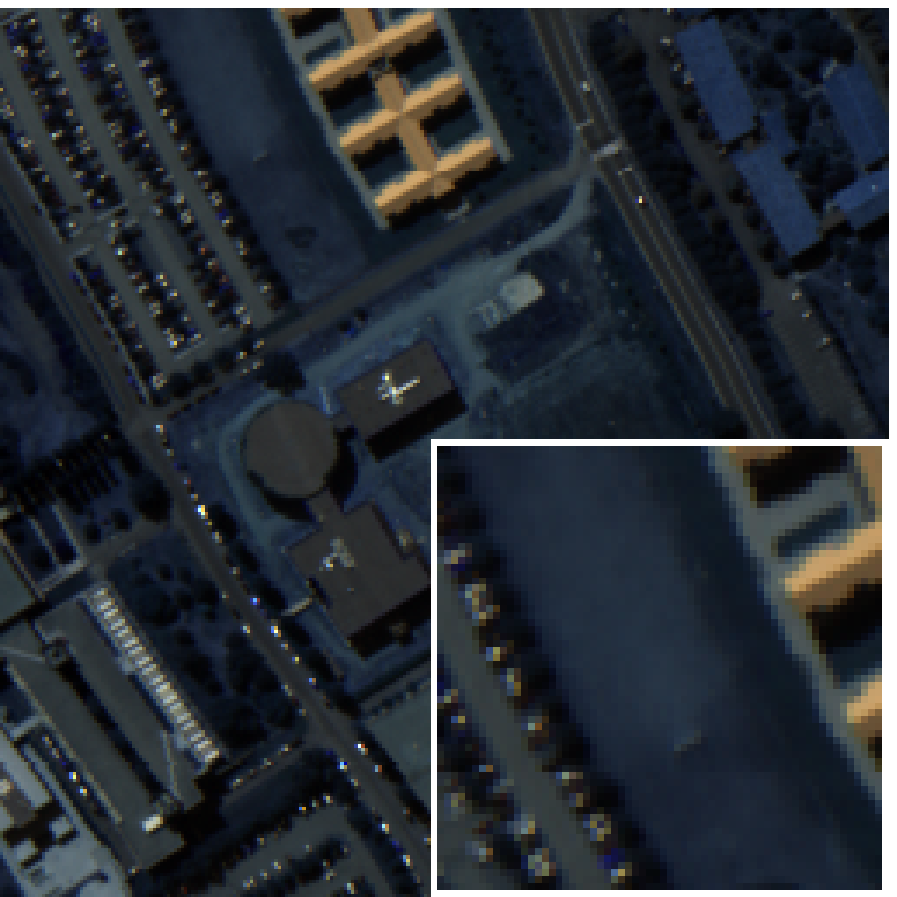}
\end{minipage}
  \begin{minipage}[t]{0.135\hsize}
\includegraphics[width=\hsize]{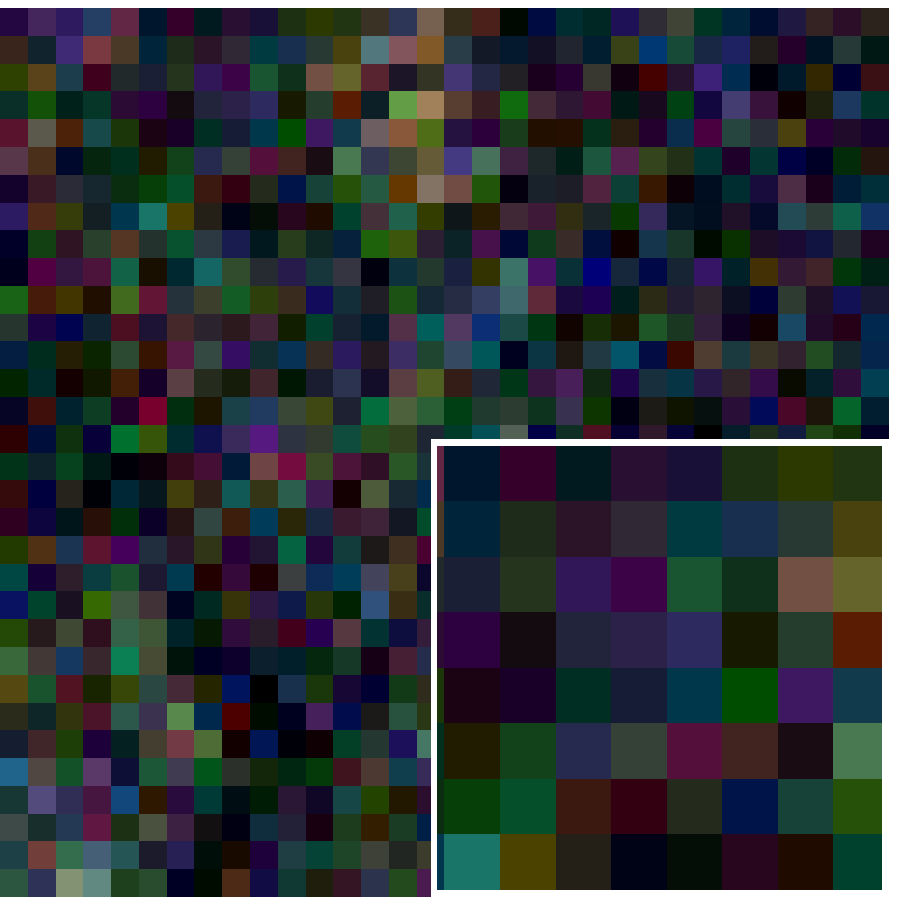}
\end{minipage}
\begin{minipage}[t]{0.135\hsize}
\includegraphics[width=\hsize]{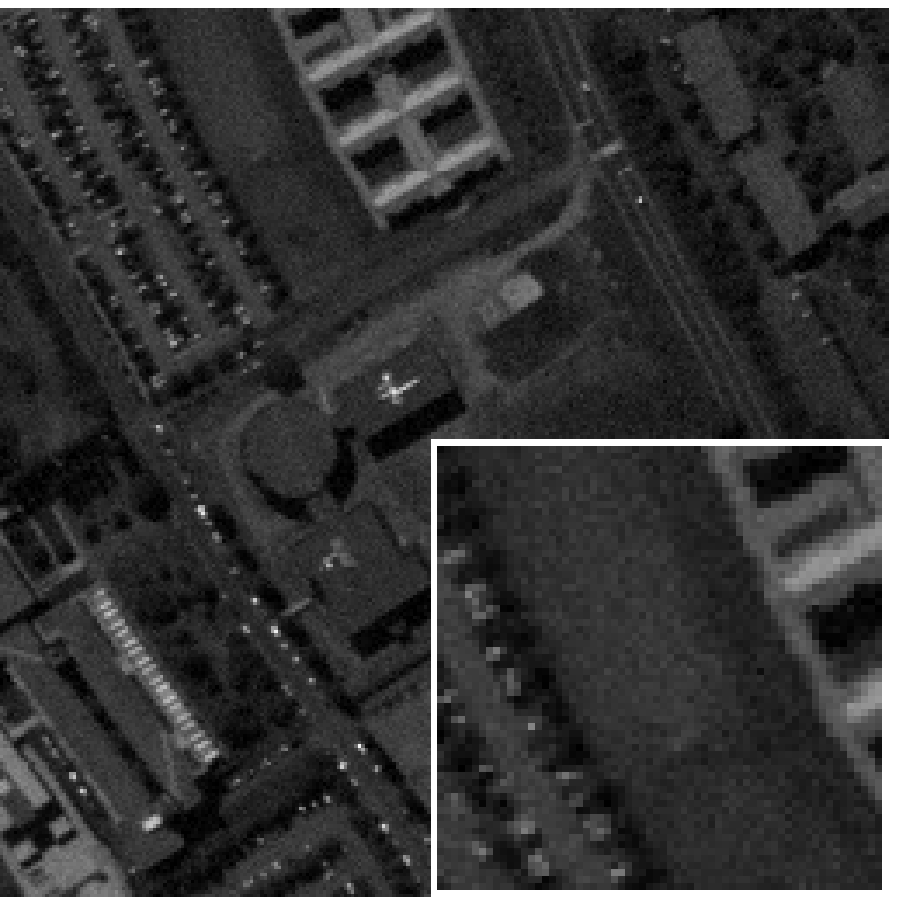}
\end{minipage}
  \begin{minipage}[t]{0.135\hsize}
\includegraphics[width=\hsize]{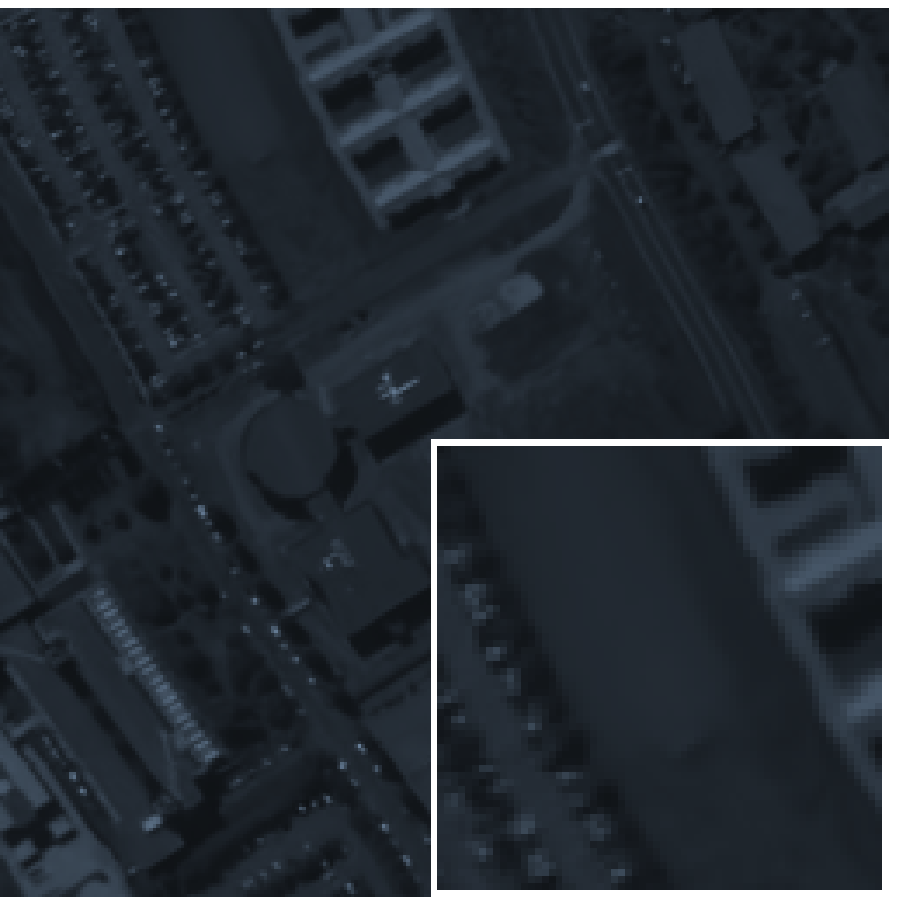}
\end{minipage}
  \begin{minipage}[t]{0.135\hsize}
\includegraphics[width=\hsize]{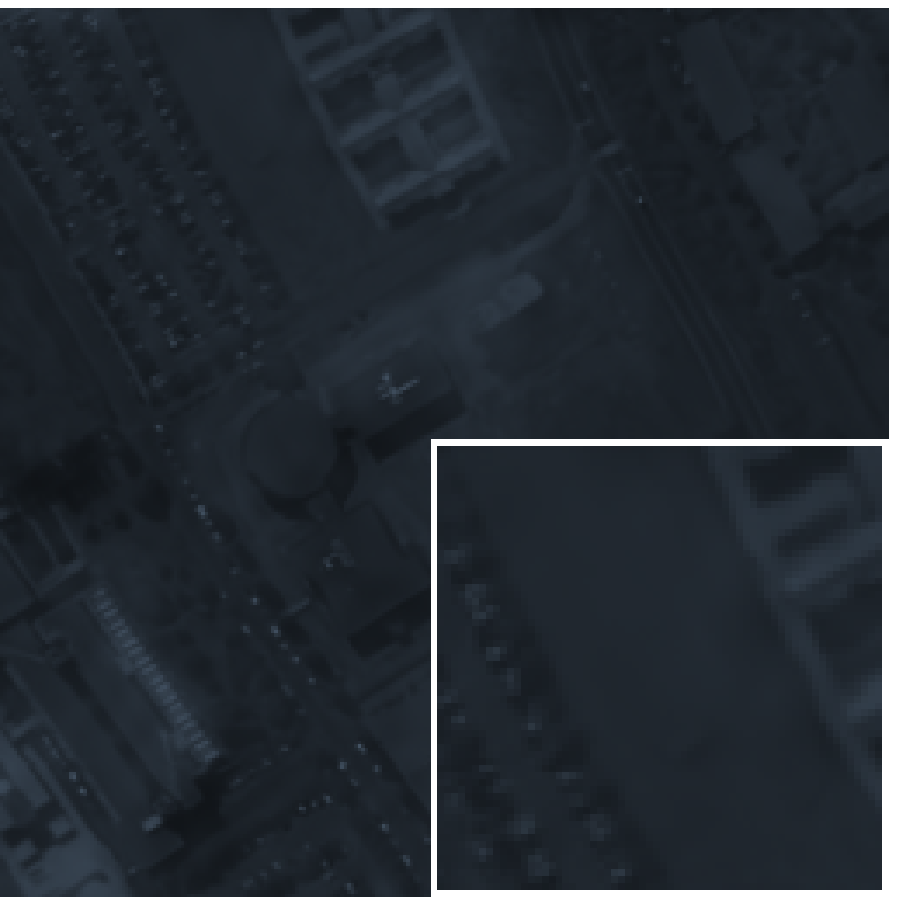}
\end{minipage}
  \begin{minipage}[t]{0.135\hsize}
\includegraphics[width=\hsize]{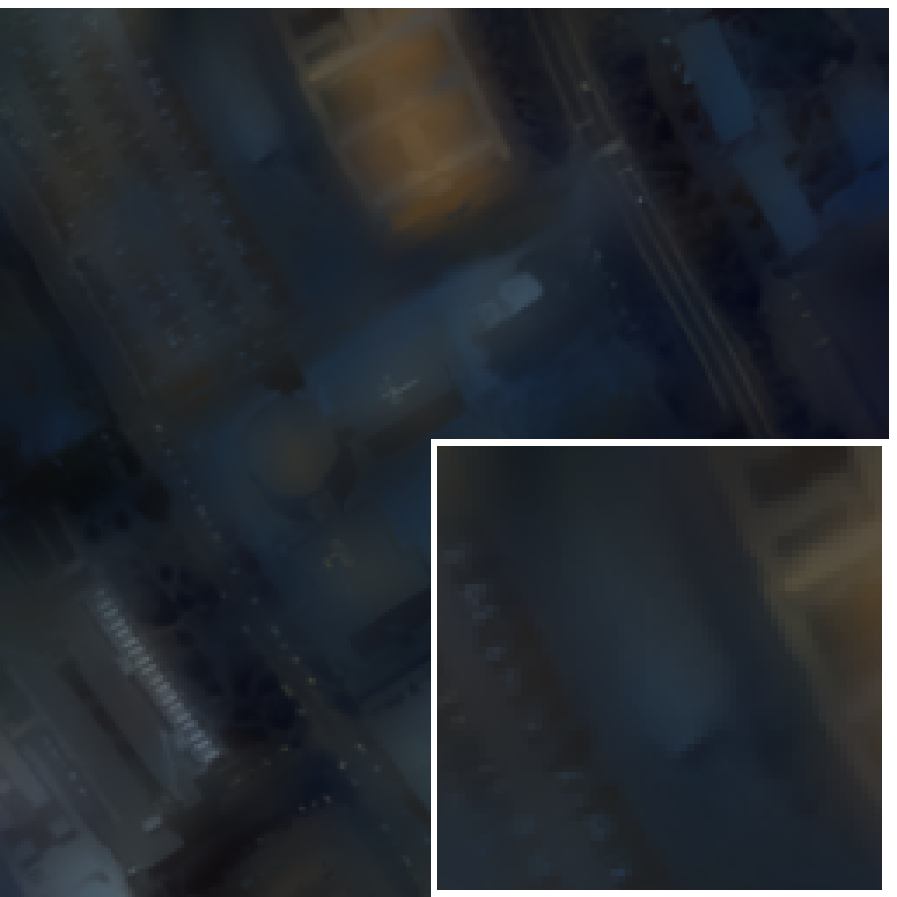}
\end{minipage}
\begin{minipage}[t]{0.135\hsize}
\includegraphics[width=\hsize]{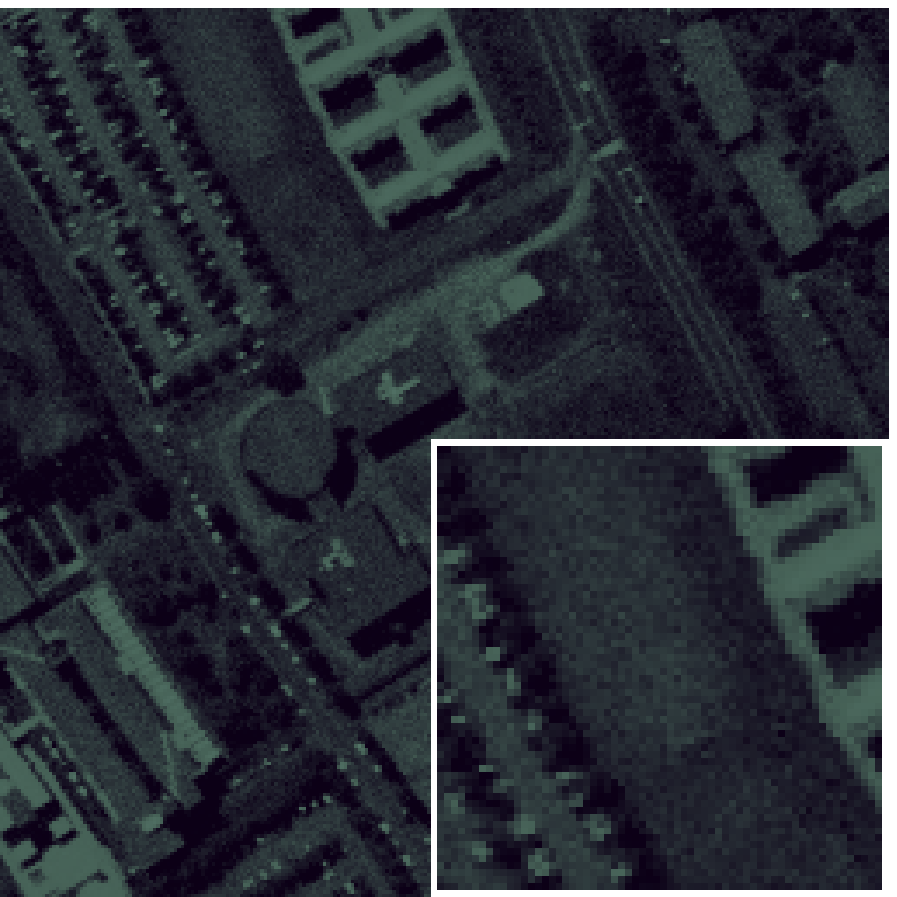}
\end{minipage}

  \begin{minipage}[t]{0.135\hsize}
\centerline{\footnotesize{ ground-truth }}
\end{minipage}
  \begin{minipage}[t]{0.135\hsize}
\centerline{\footnotesize{ observed HS image }}
\end{minipage}
\begin{minipage}[t]{0.135\hsize}
\centerline{\footnotesize{ observed PAN image }}
\end{minipage}
  \begin{minipage}[t]{0.135\hsize}
\centerline{\footnotesize{ GSA }}
\end{minipage}
  \begin{minipage}[t]{0.135\hsize}
\centerline{\footnotesize{ MTF-GLP }}
\end{minipage}
  \begin{minipage}[t]{0.135\hsize}
\centerline{\footnotesize{ GFPCA }}
\end{minipage}
  \begin{minipage}[t]{0.135\hsize}
\centerline{\footnotesize{ CNMF }}
\end{minipage}

\vspace{4pt}
  \begin{minipage}[t]{0.135\hsize}
\includegraphics[width=\hsize]{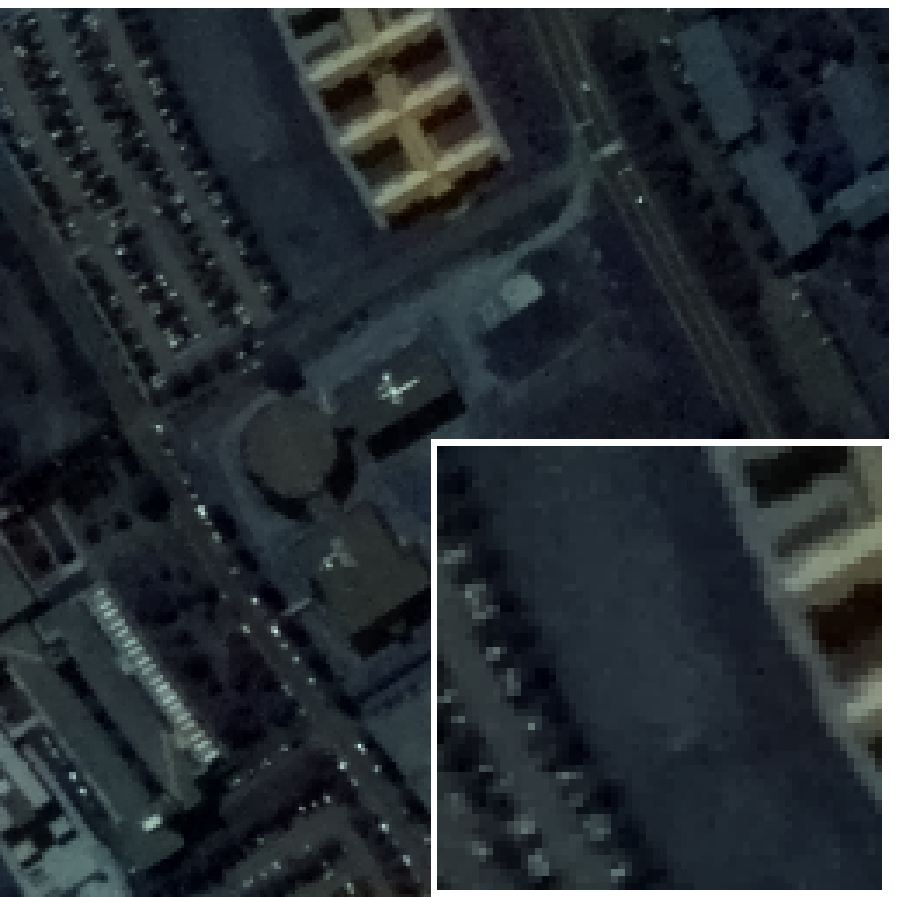}
\end{minipage}
  \begin{minipage}[t]{0.135\hsize}
\includegraphics[width=\hsize]{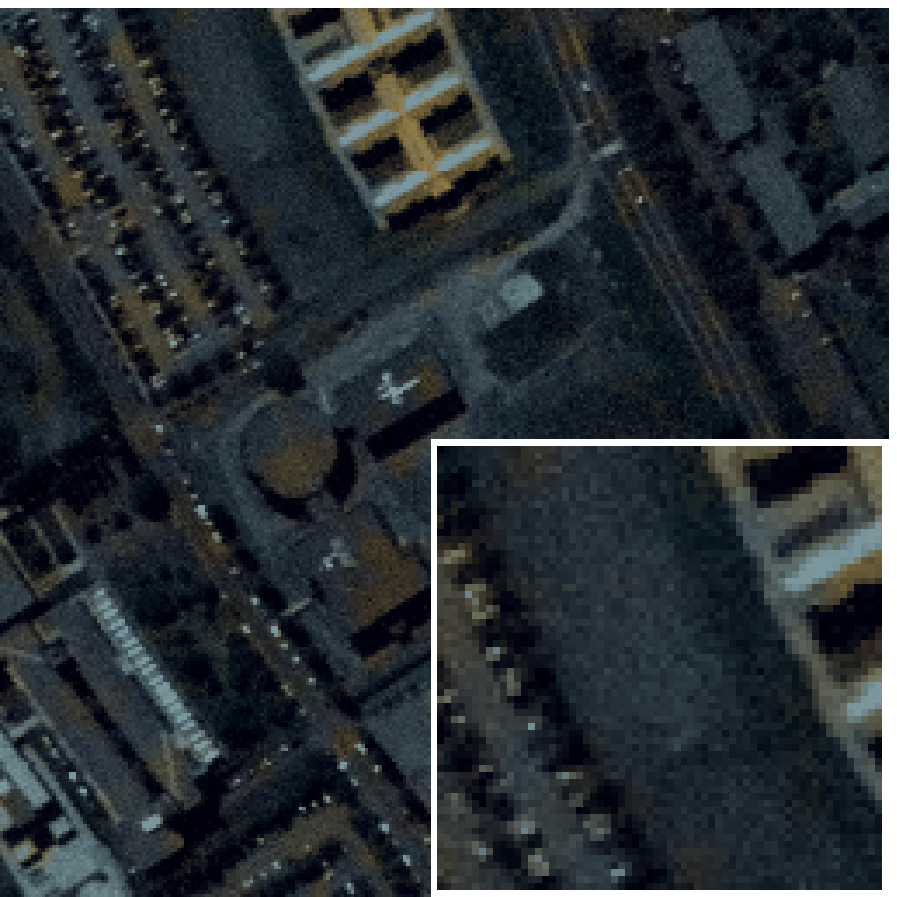}
\end{minipage}
  \begin{minipage}[t]{0.135\hsize}
\includegraphics[width=\hsize]{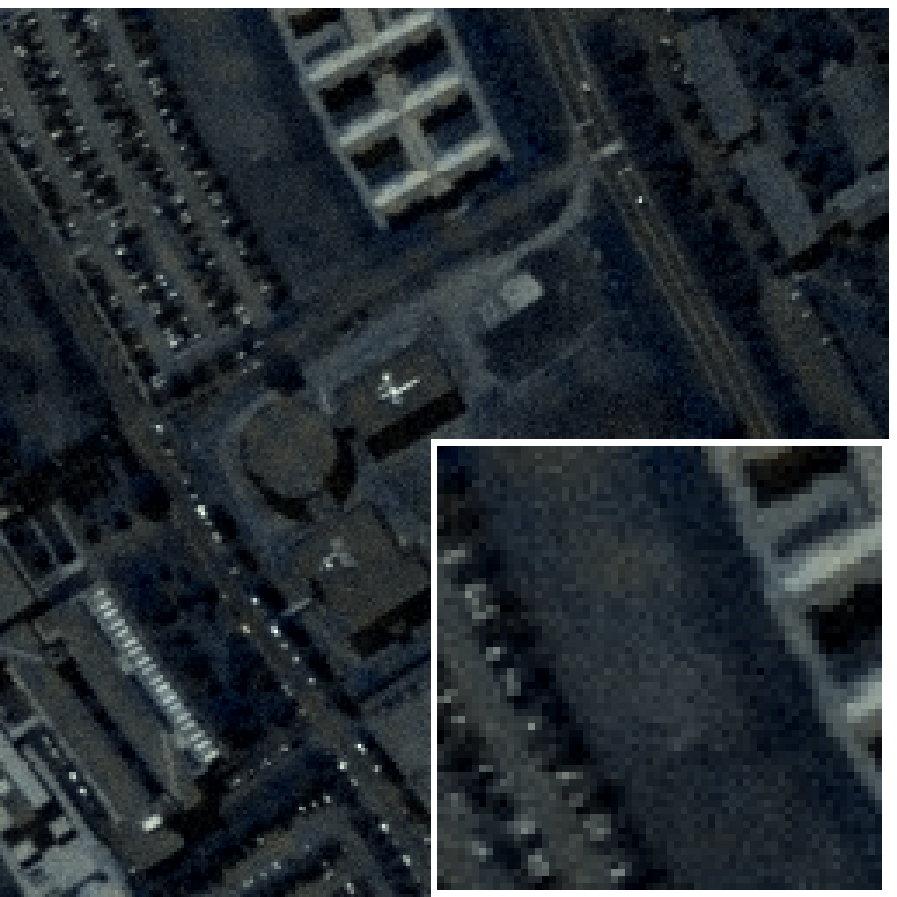}
\end{minipage}
\begin{minipage}[t]{0.135\hsize}
\includegraphics[width=\hsize]{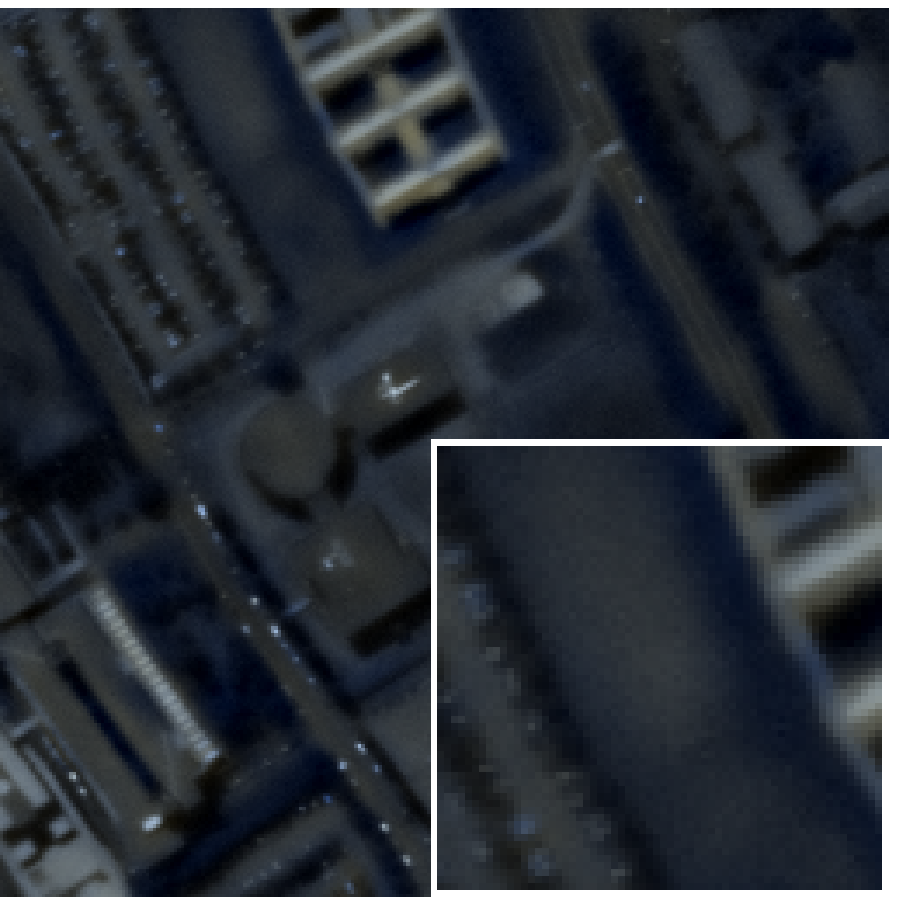}
\end{minipage}
  \begin{minipage}[t]{0.135\hsize}
\includegraphics[width=\hsize]{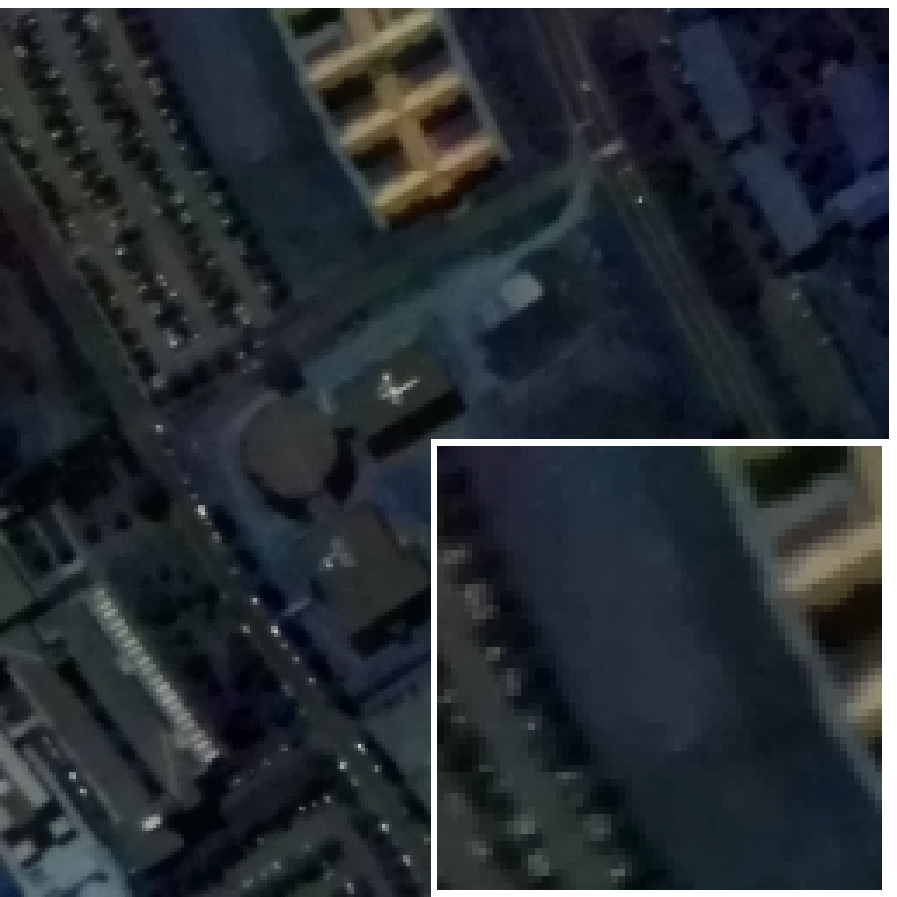}
\end{minipage}
  \begin{minipage}[t]{0.135\hsize}
\includegraphics[width=\hsize]{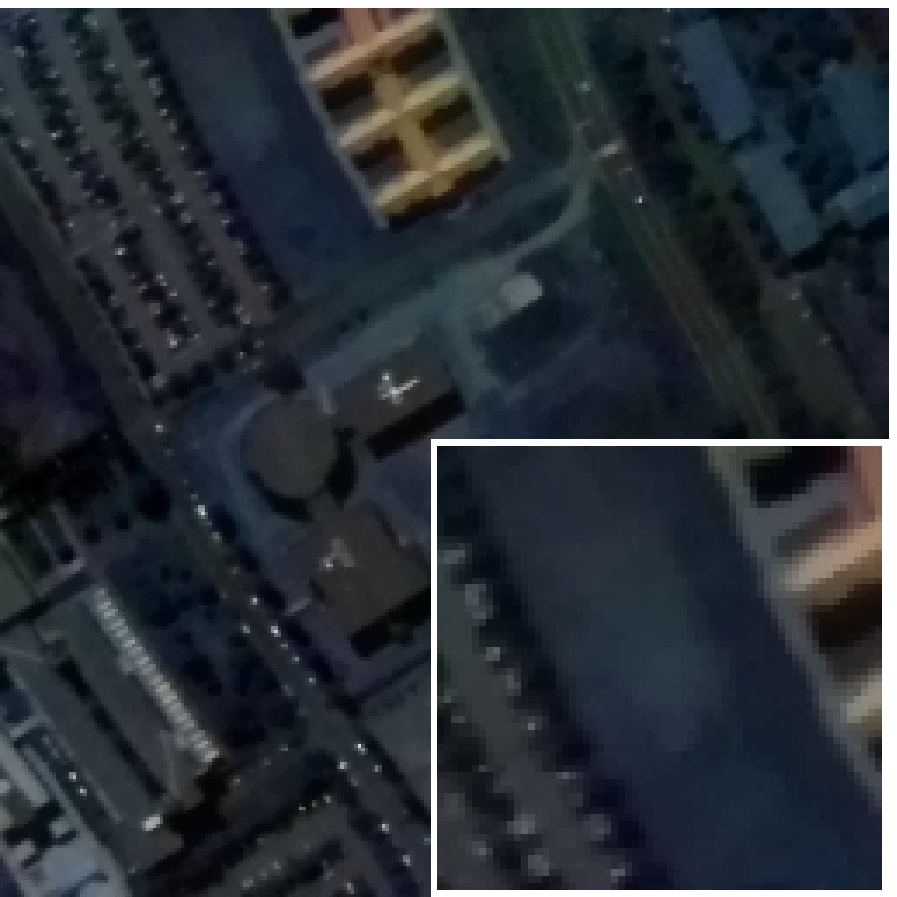}
\end{minipage}
\begin{minipage}[t]{0.135\hsize}
\centerline{\footnotesize{  }}
\end{minipage}

  \begin{minipage}[t]{0.135\hsize}
\centerline{\footnotesize{ HySure }}
\end{minipage}
  \begin{minipage}[t]{0.135\hsize}
\centerline{\footnotesize{ Lanaras's }}
\end{minipage}
  \begin{minipage}[t]{0.135\hsize}
\centerline{\footnotesize{ LTMR }}
\end{minipage}
 \begin{minipage}[t]{0.135\hsize}
\centerline{\footnotesize{ CNN-Fus }}
\end{minipage}
  \begin{minipage}[t]{0.135\hsize}
\centerline{\footnotesize{ \bf{proposed ($p = 1$)} }}
\end{minipage}
  \begin{minipage}[t]{0.135\hsize}
\centerline{\footnotesize{ \bf{proposed ($p = 2$)} }}
\end{minipage}
 \begin{minipage}[t]{0.135\hsize}
\centerline{\footnotesize{  }}
\end{minipage}

   \caption{Results on HS pansharpening experiments (top: Reno, $r = 4$, and $\sigma_\g = 0.02$, bottom: PaviaU, $r = 8$, and $\sigma_\g = 0.02$).}
 \label{fig:img_HSpan}
\end{center}
\end{figure*}

\begin{figure*}[t]
\begin{center}
	\begin{minipage}[t]{0.24\hsize}
	\includegraphics[width=1.0\hsize]{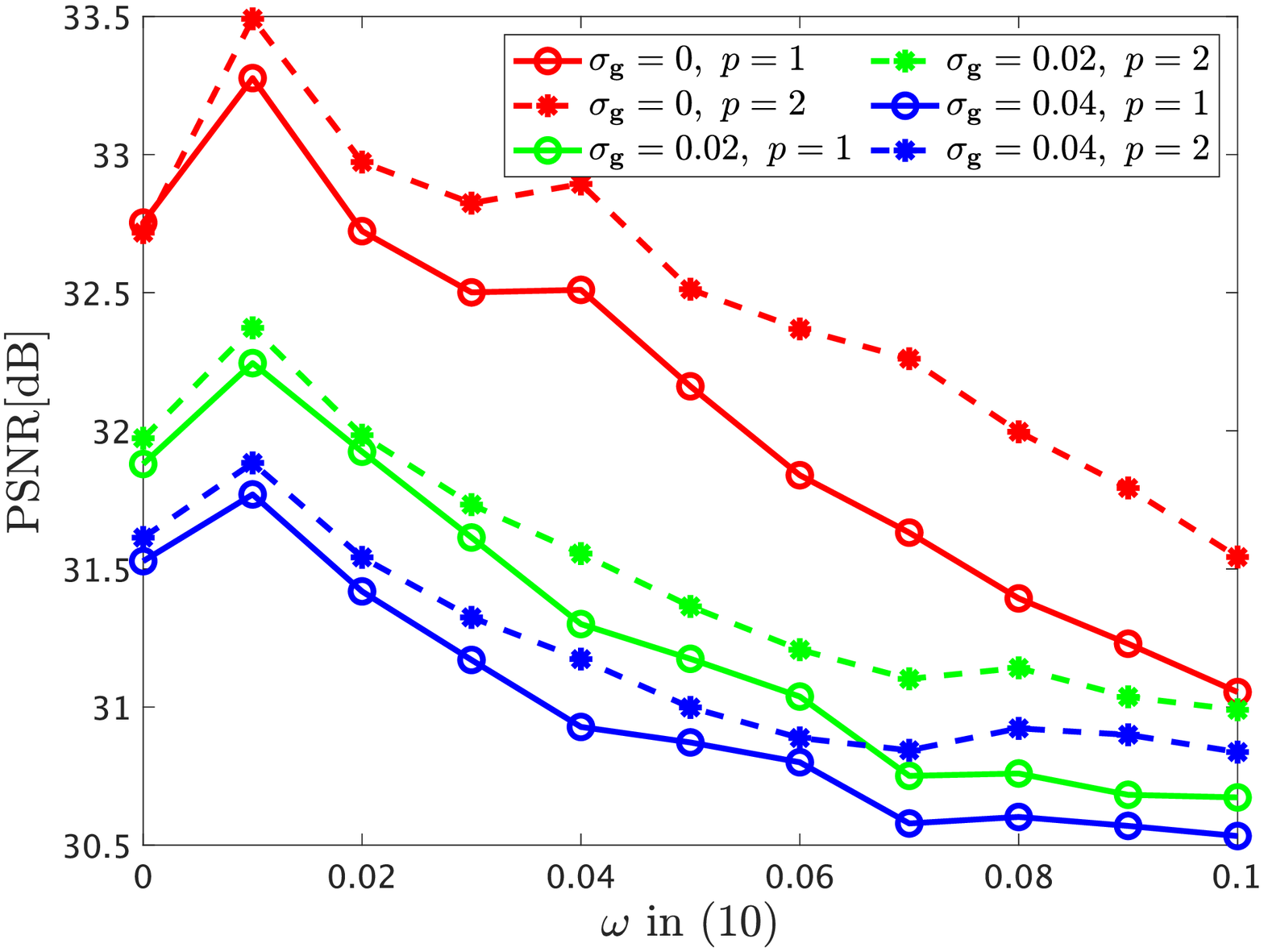}
	\end{minipage}
	\begin{minipage}[t]{0.24\hsize}
	\includegraphics[width=1.0\hsize]{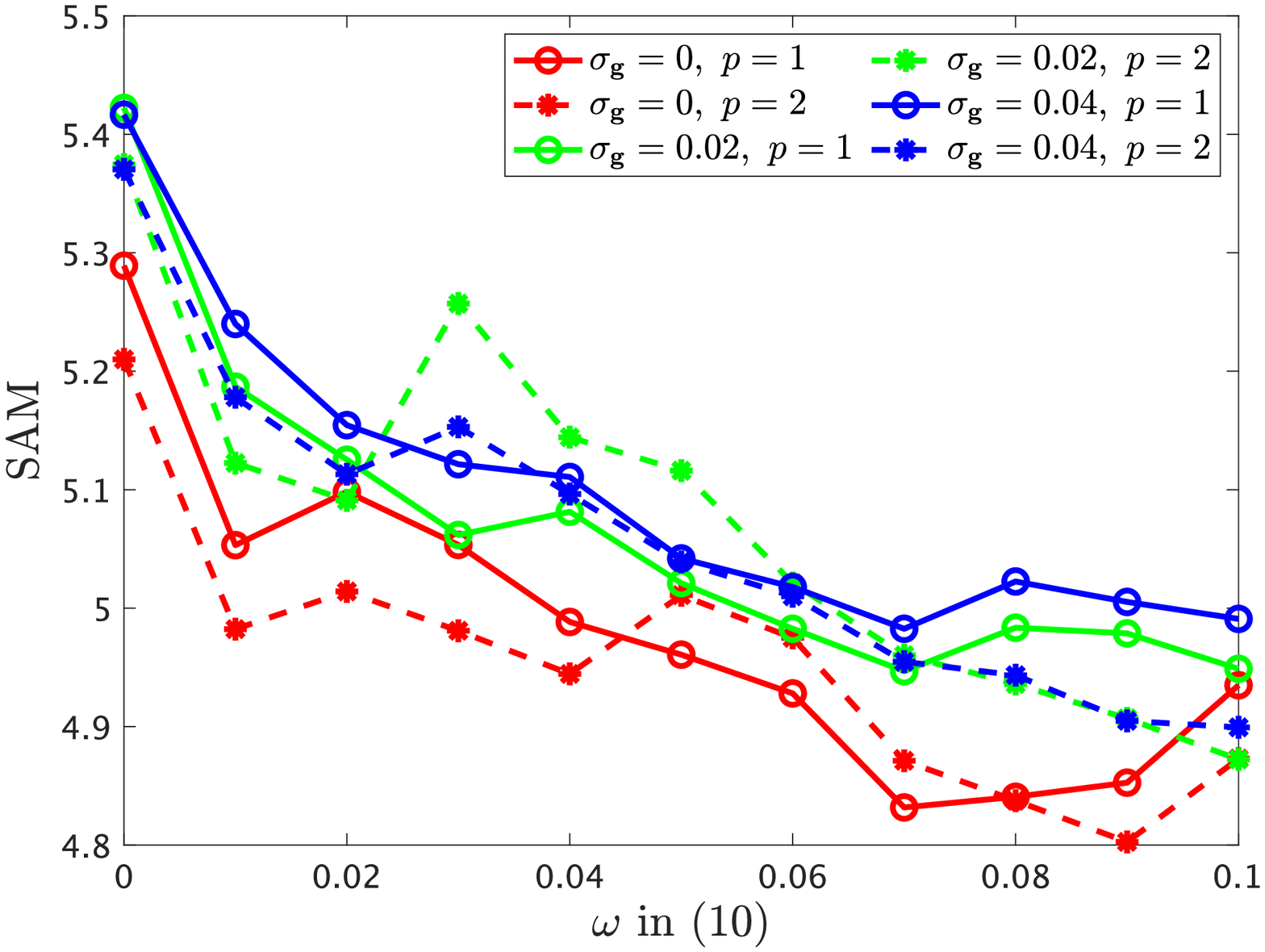}
	\end{minipage}
	\begin{minipage}[t]{0.24\hsize}
	\includegraphics[width=1.0\hsize]{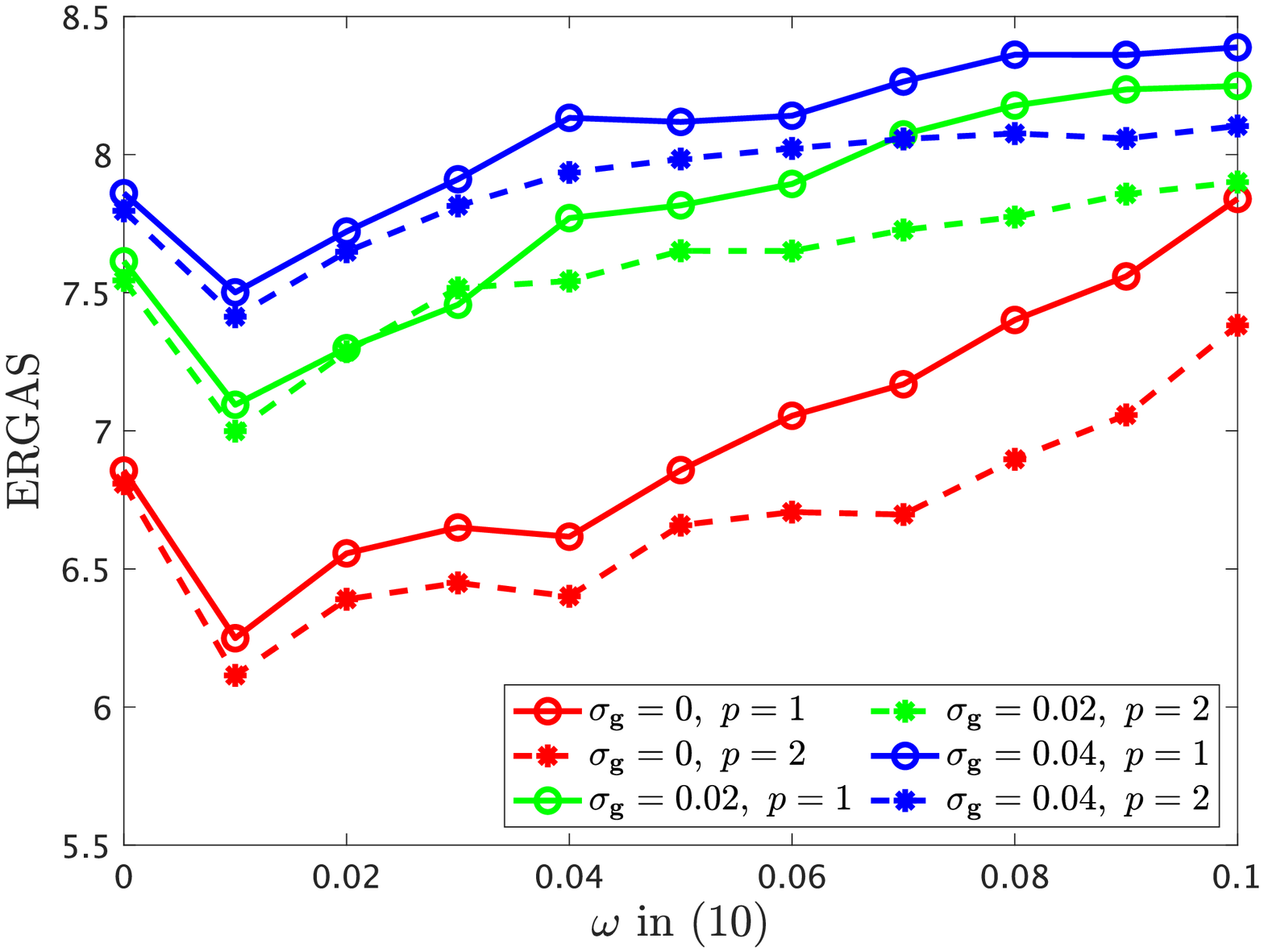}
	\end{minipage}
	\begin{minipage}[t]{0.24\hsize}
	\includegraphics[width=1.0\hsize]{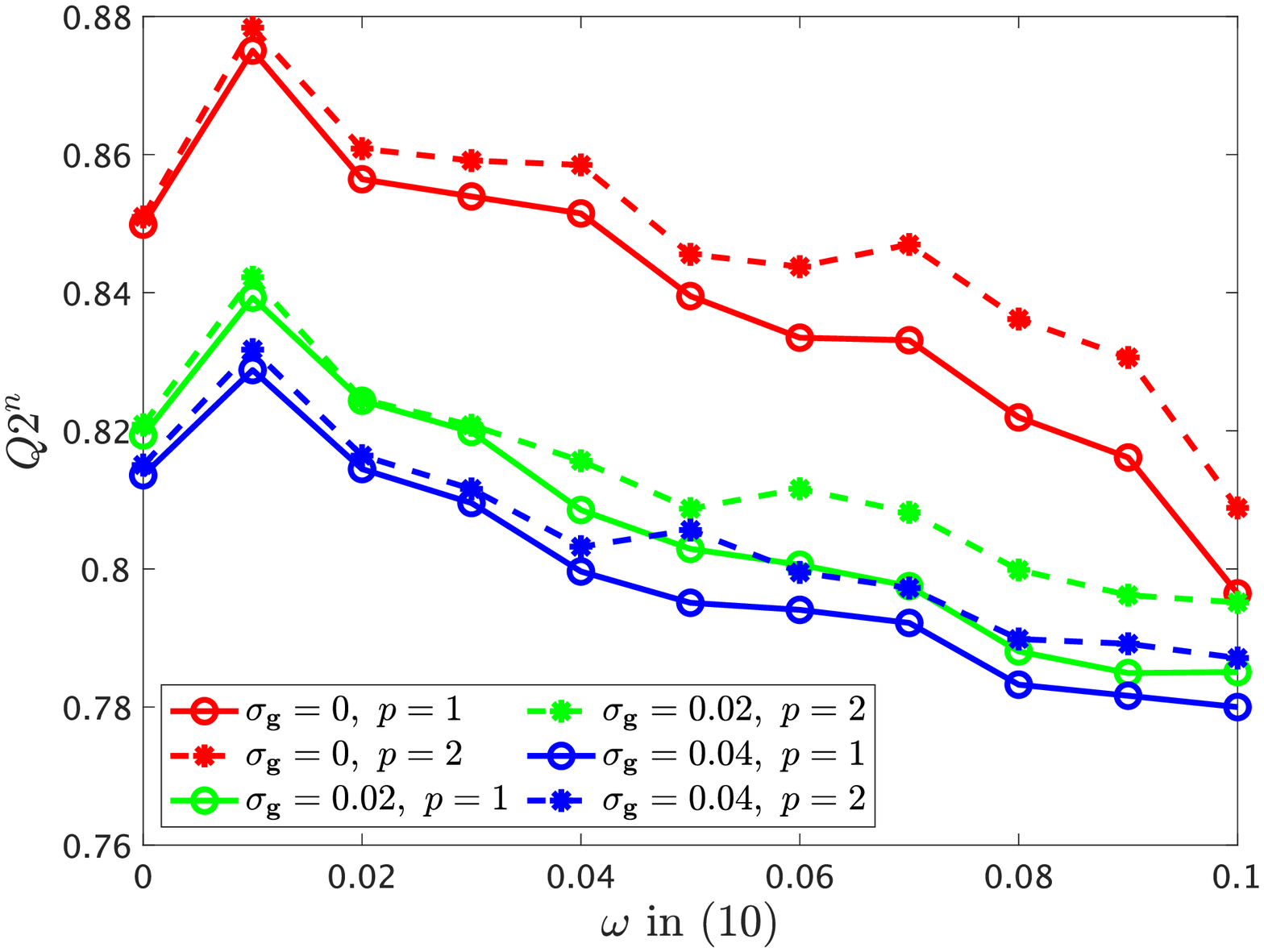}
	\end{minipage}
	
	
	\begin{minipage}[t]{0.24\hsize}
	\includegraphics[width=1.0\hsize]{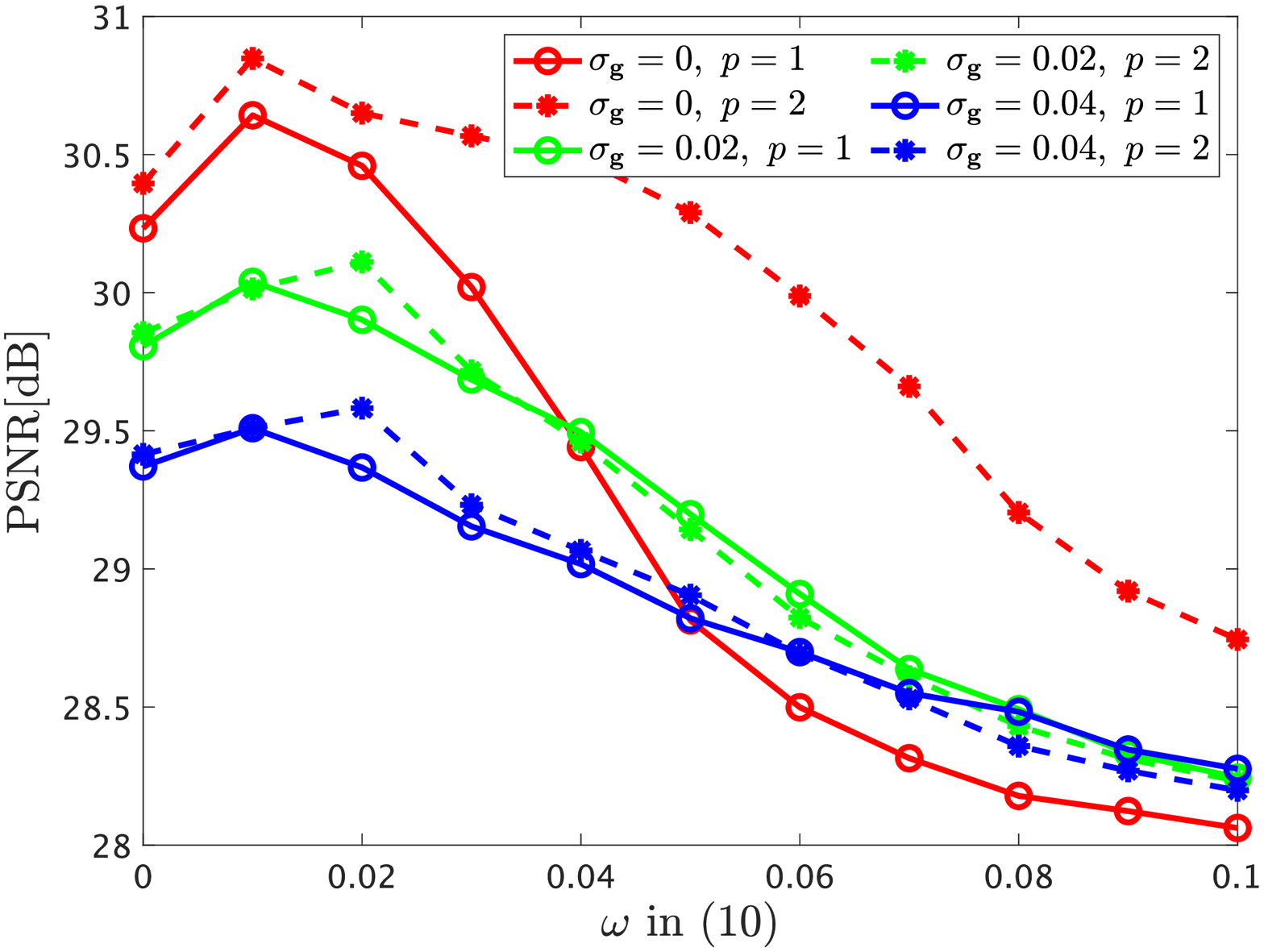}
	\end{minipage}
	\begin{minipage}[t]{0.24\hsize}
	\includegraphics[width=1.0\hsize]{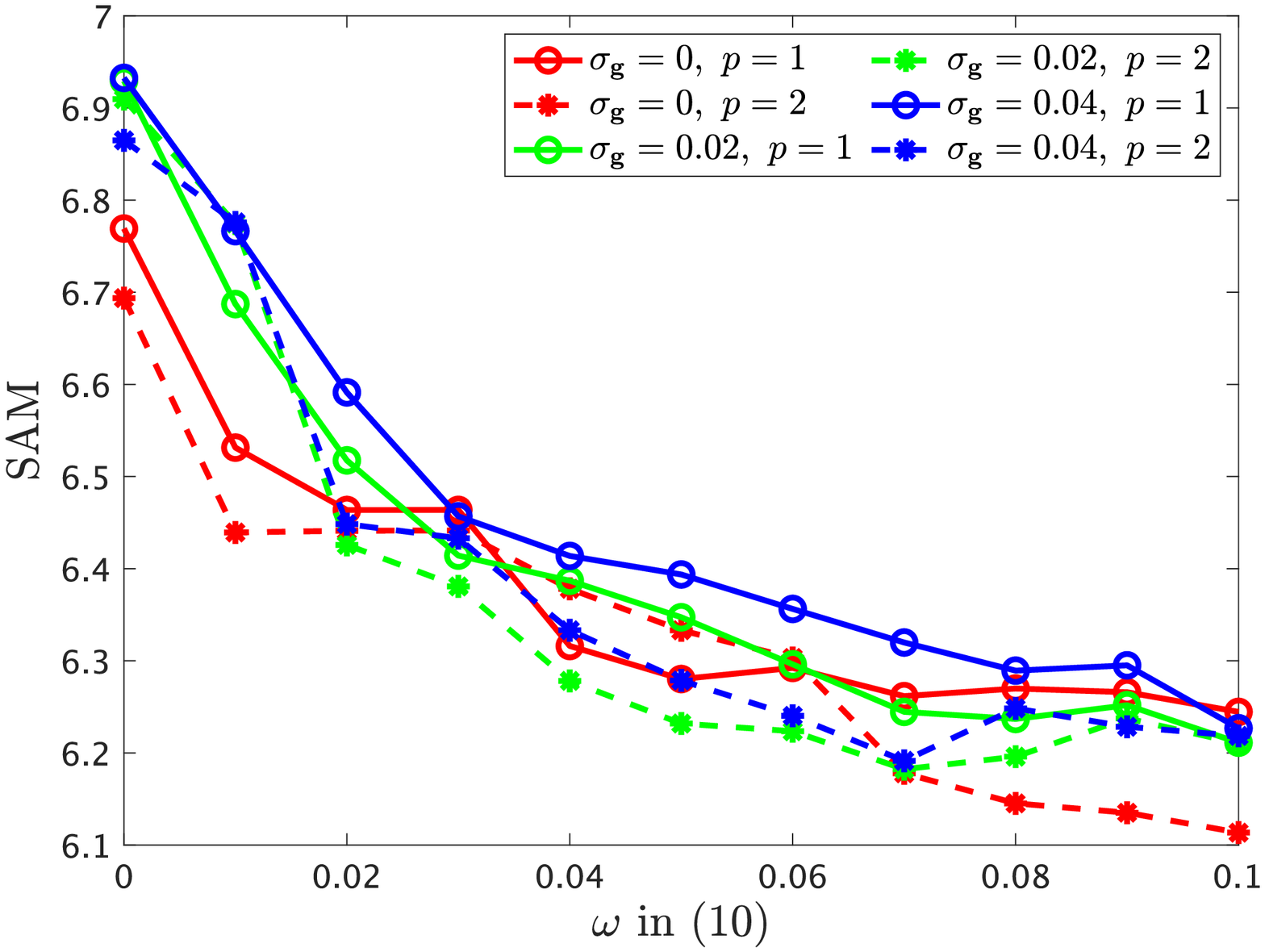}
	\end{minipage}
	\begin{minipage}[t]{0.24\hsize}
	\includegraphics[width=1.0\hsize]{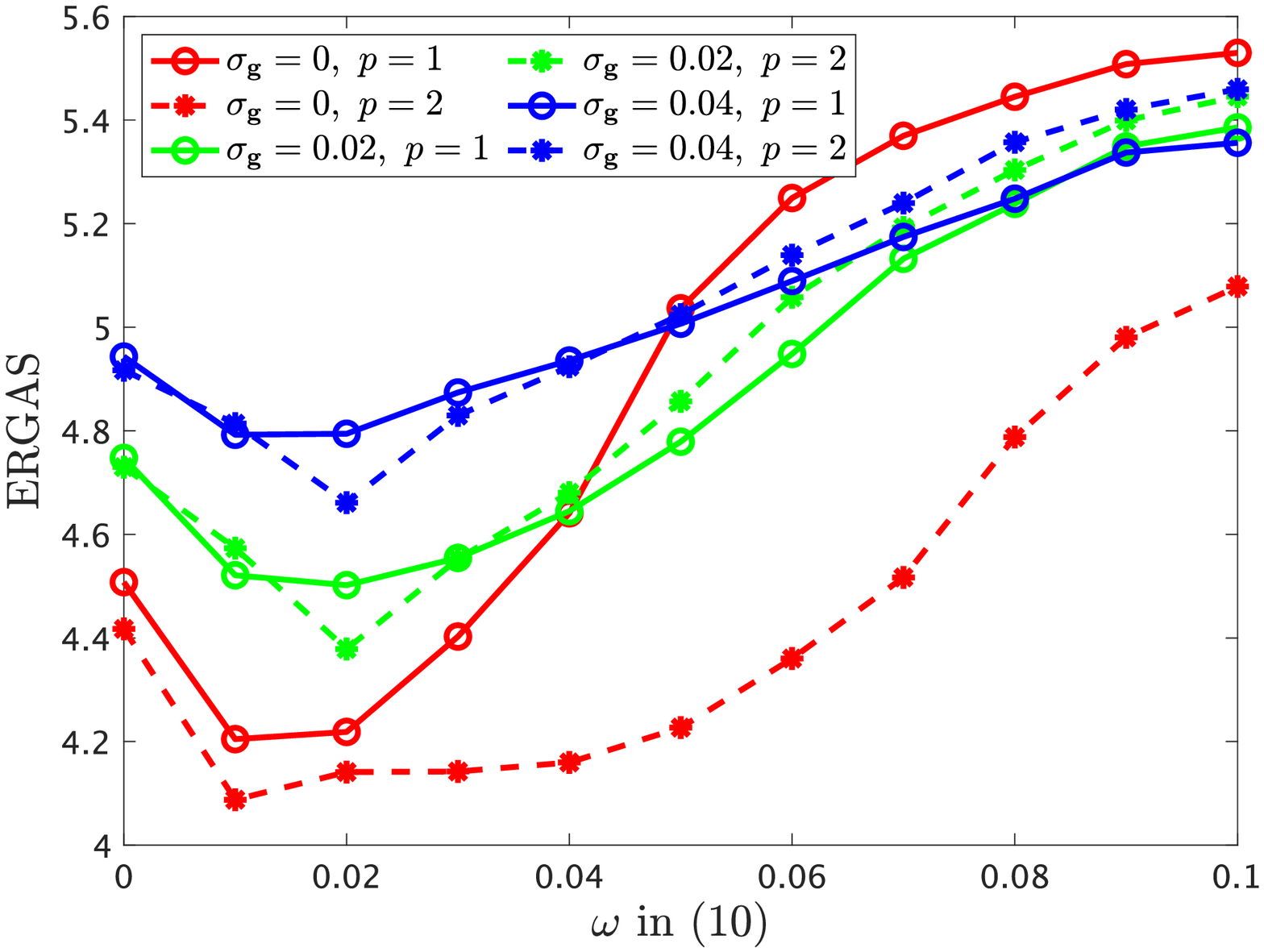}
	\end{minipage}
	\begin{minipage}[t]{0.24\hsize}
	\includegraphics[width=1.0\hsize]{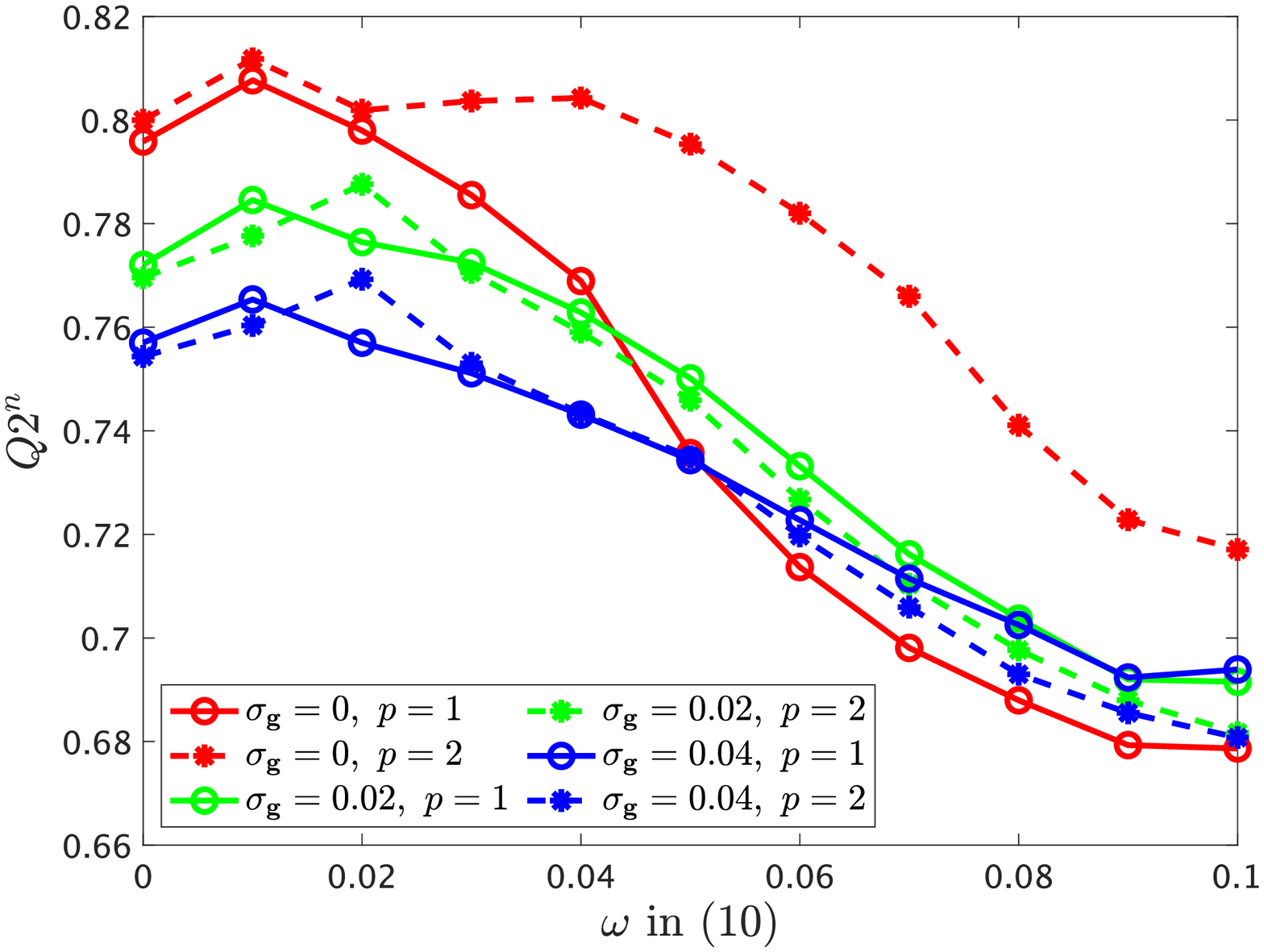}
	\end{minipage}
	
	
	\vspace{-3pt}
	\begin{minipage}[t]{0.24\hsize}
	\centerline{\footnotesize{PSNR[dB]}}
	\end{minipage}
	\begin{minipage}[t]{0.24\hsize}
	\centerline{\footnotesize{SAM}}
	\end{minipage}
	\begin{minipage}[t]{0.24\hsize}
	\centerline{\footnotesize{ERGAS}}
	\end{minipage}
	\begin{minipage}[t]{0.24\hsize}
	\centerline{\footnotesize{$Q2^n$}}
	\end{minipage}
	\caption{Four quality measures versus $\omega$ in \eqref{def:HSSTV} on HS pansharpening (top: $r = 2$, bottom: $r = 4$).}
	\label{omega_graph_HSpan}
\end{center}
\end{figure*}

\begin{figure*}[t]
\begin{center}
	\begin{minipage}[t]{0.24\hsize}
	\includegraphics[width=1.0\hsize]{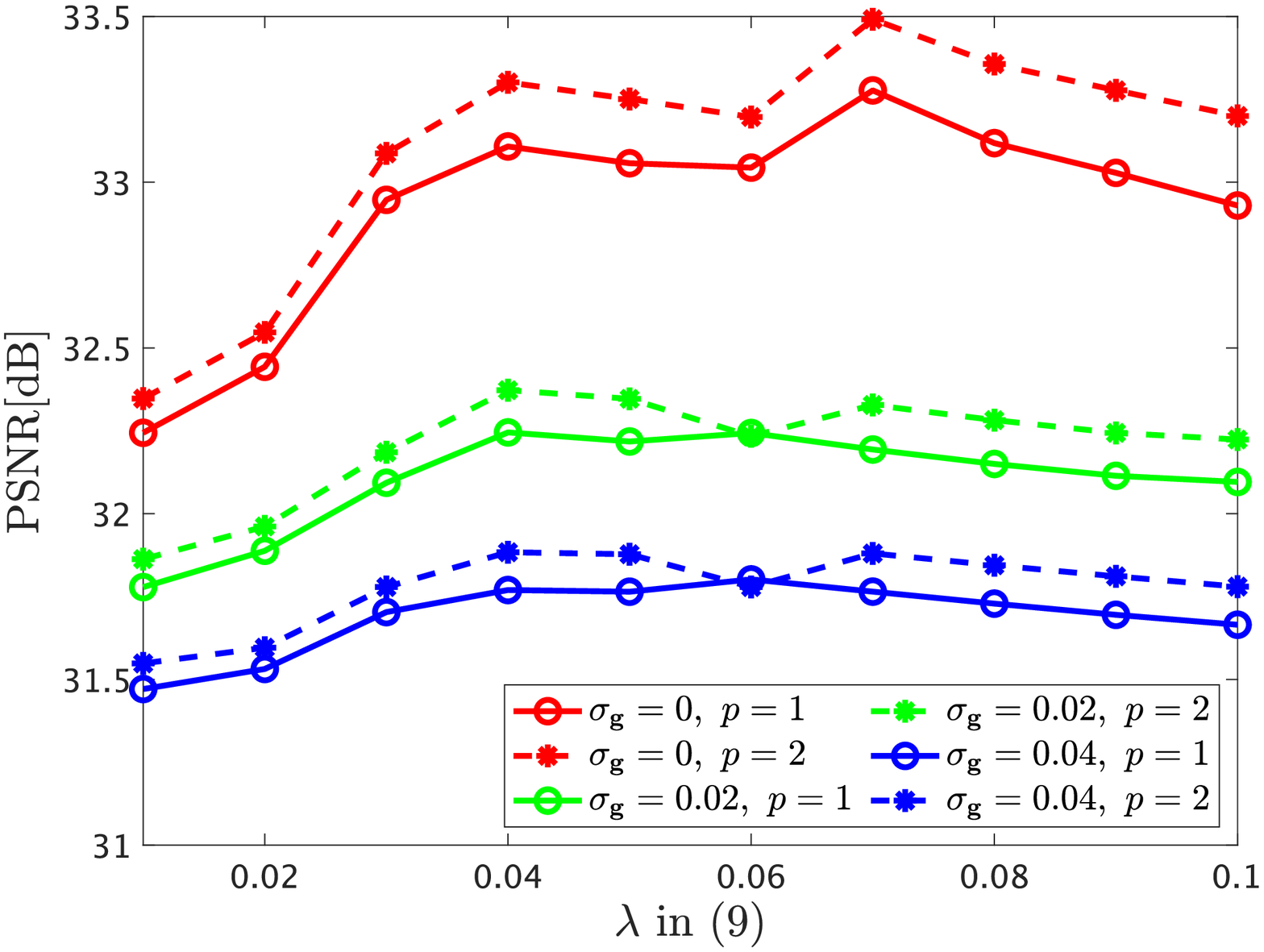}
	\end{minipage}
	\begin{minipage}[t]{0.24\hsize}
	\includegraphics[width=1.0\hsize]{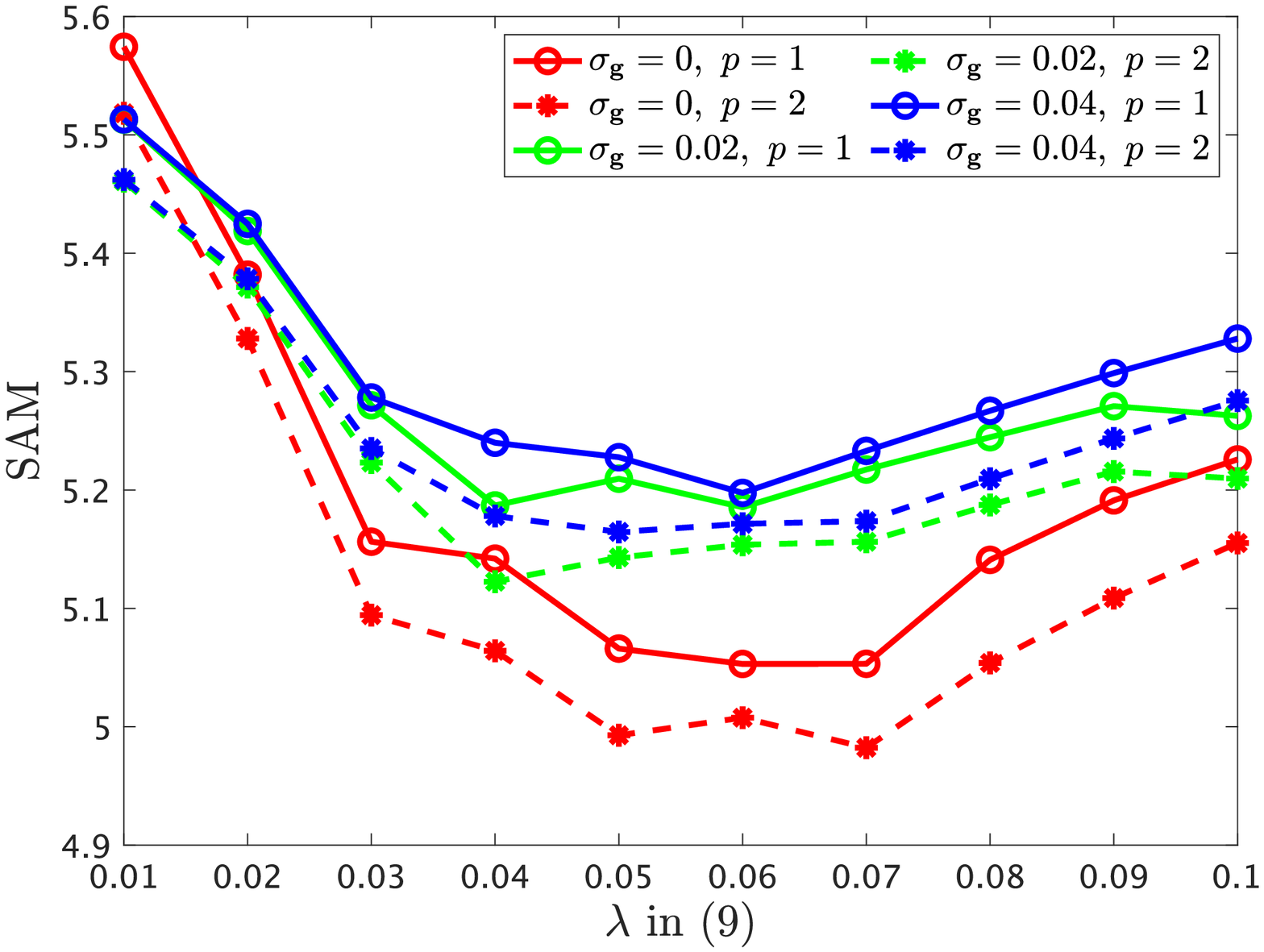}
	\end{minipage}
	\begin{minipage}[t]{0.24\hsize}
	\includegraphics[width=1.0\hsize]{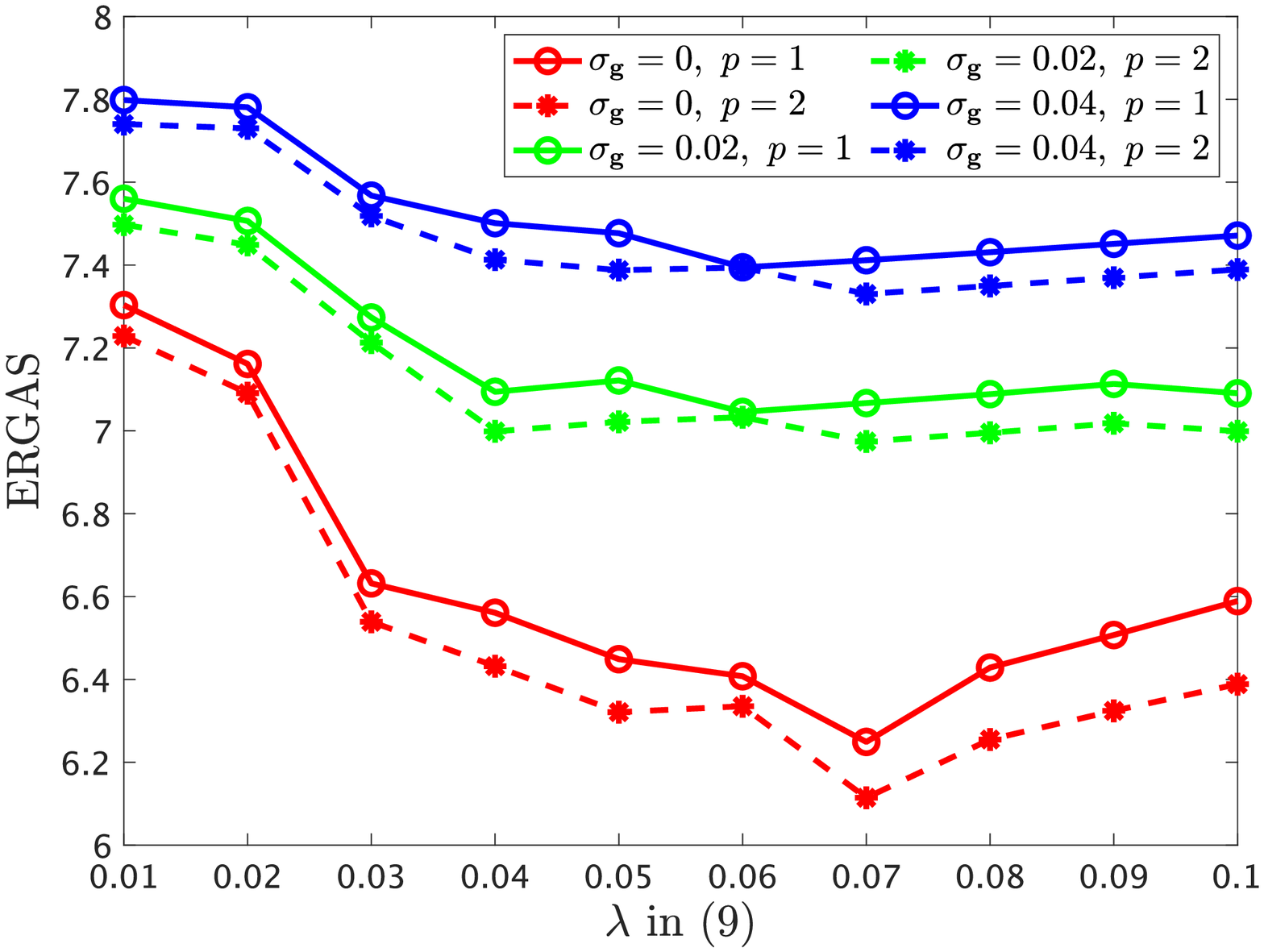}
	\end{minipage}
	\begin{minipage}[t]{0.24\hsize}
	\includegraphics[width=1.0\hsize]{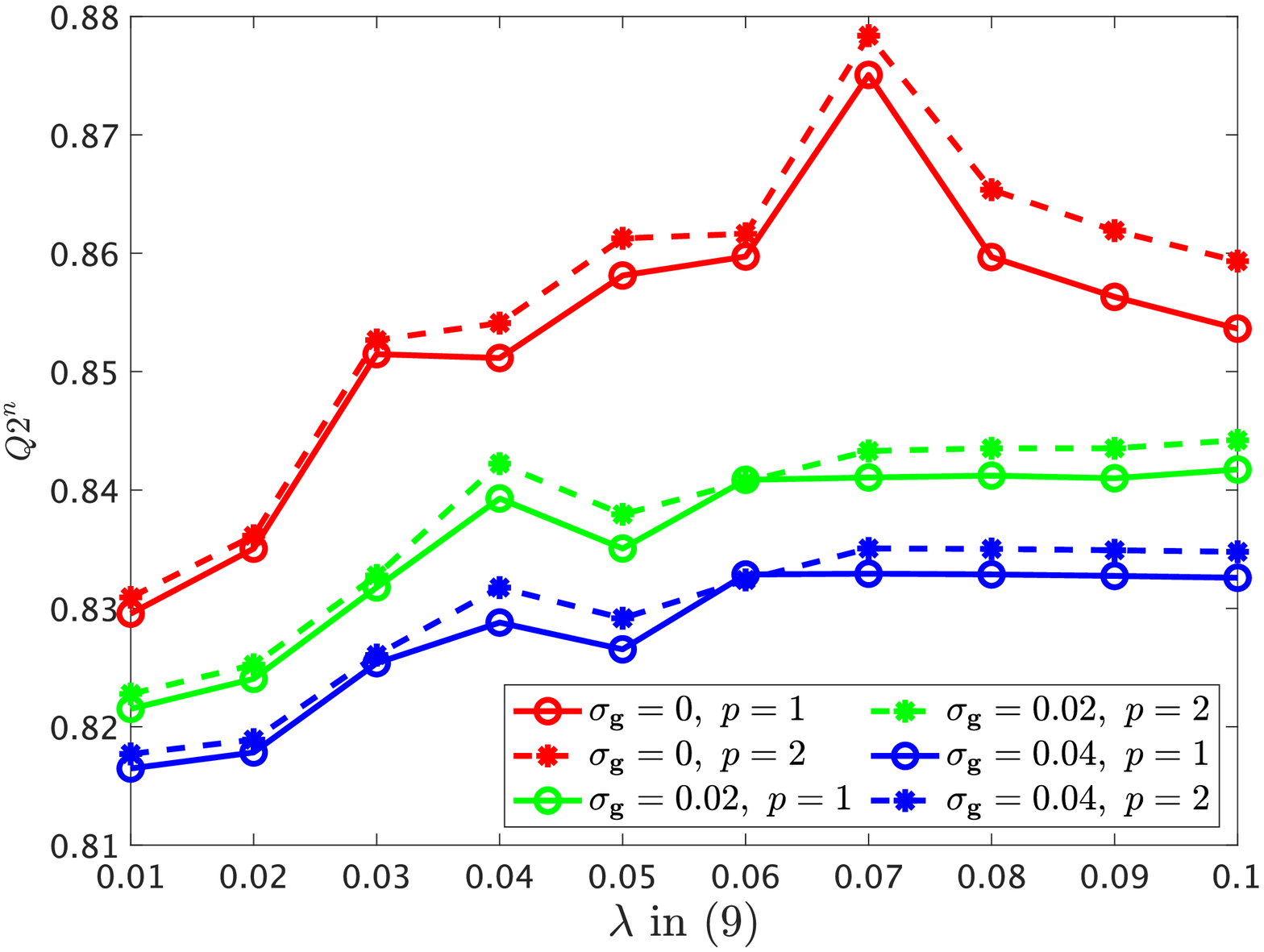}
	\end{minipage}
	
	
	\begin{minipage}[t]{0.24\hsize}
	\includegraphics[width=1.0\hsize]{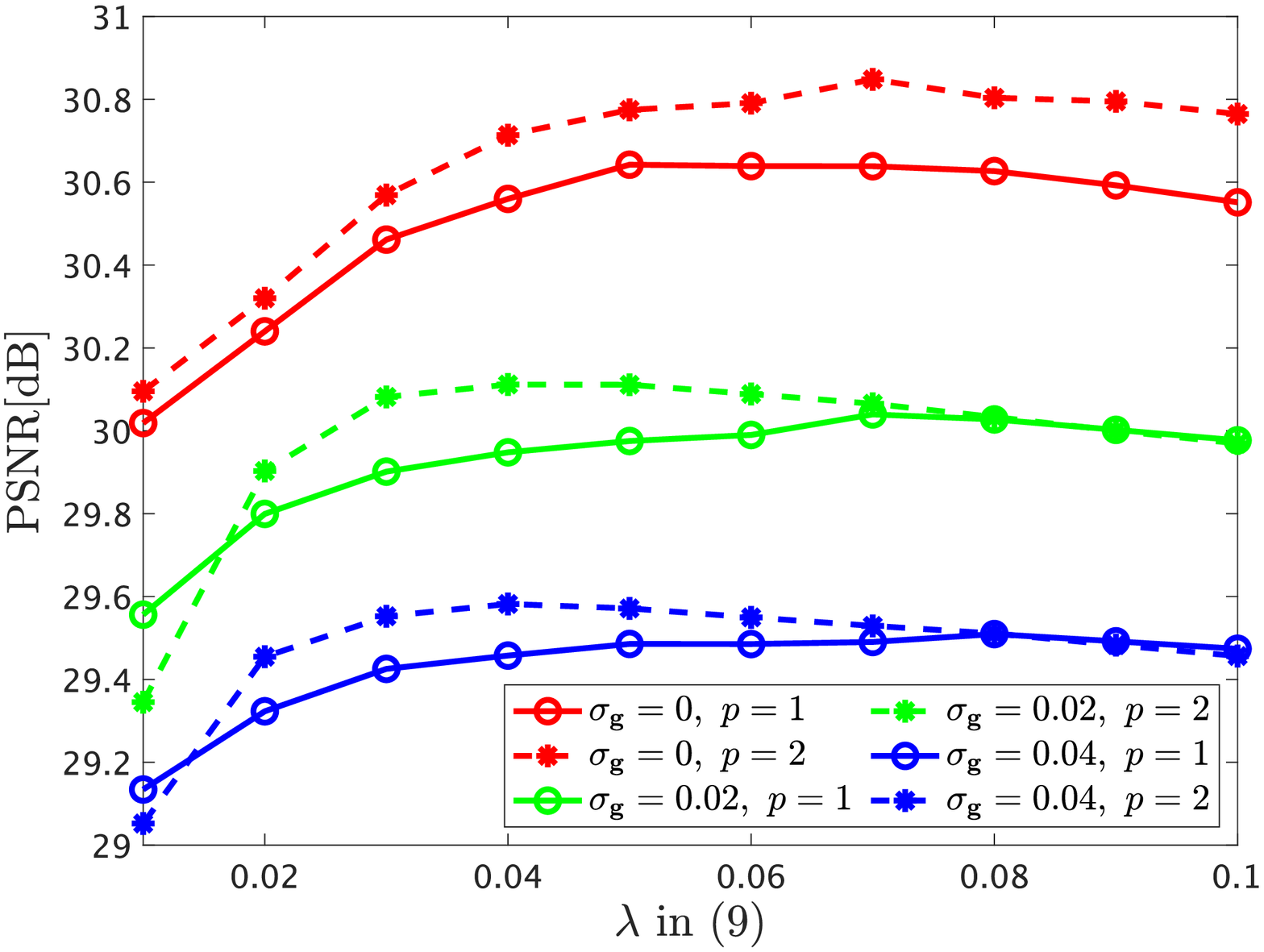}
	\end{minipage}
	\begin{minipage}[t]{0.24\hsize}
	\includegraphics[width=1.0\hsize]{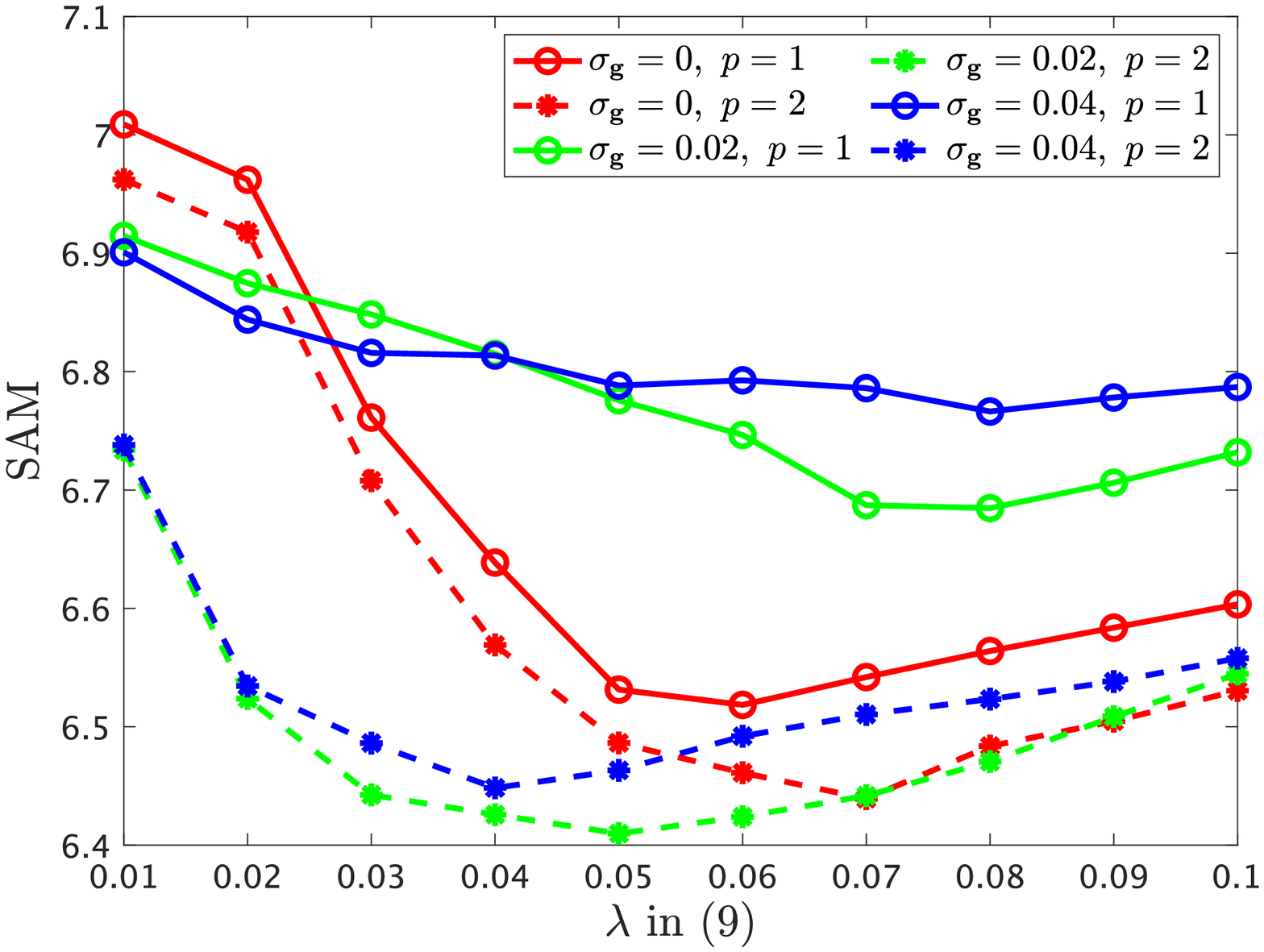}
	\end{minipage}
	\begin{minipage}[t]{0.24\hsize}
	\includegraphics[width=1.0\hsize]{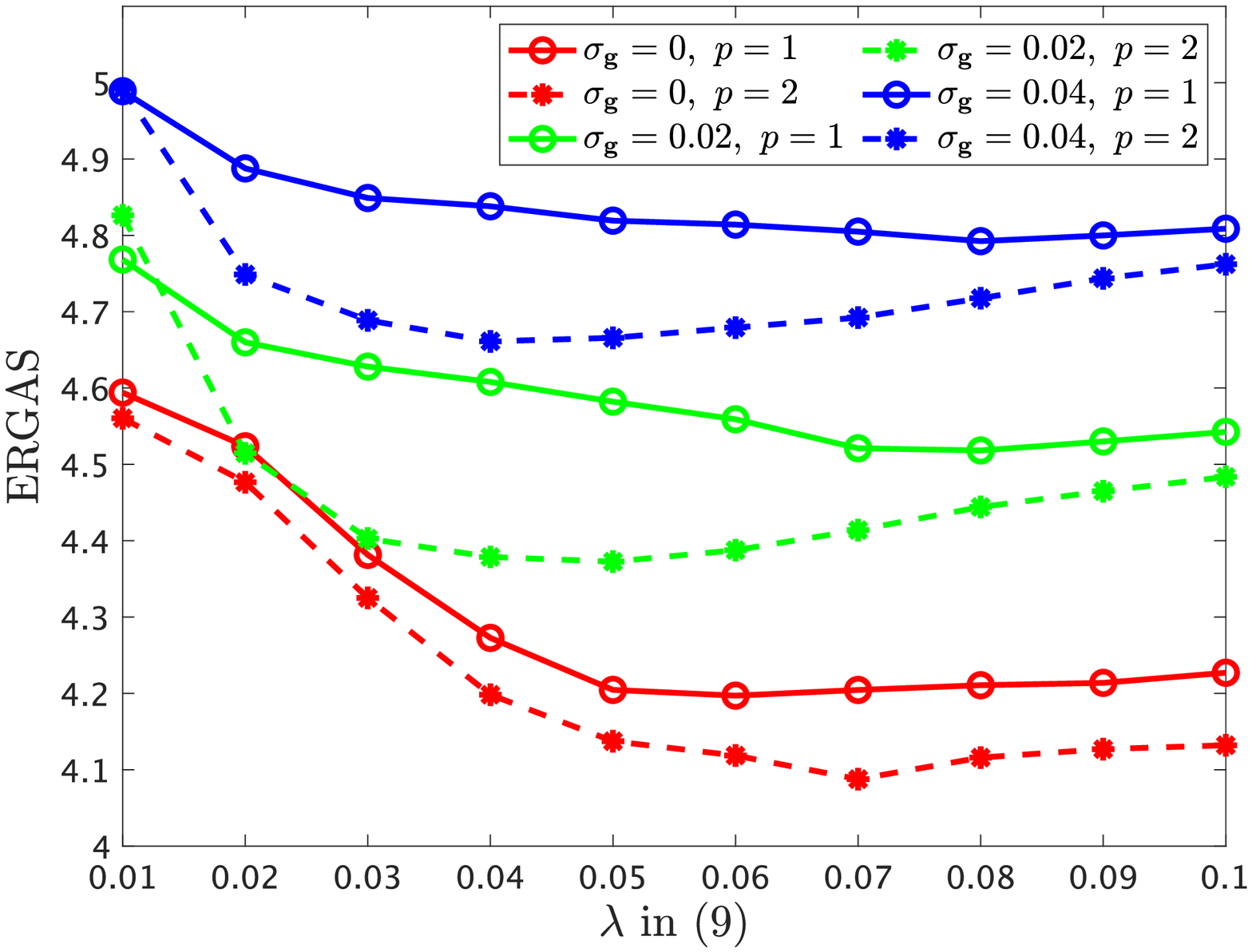}
	\end{minipage}
	\begin{minipage}[t]{0.24\hsize}
	\includegraphics[width=1.0\hsize]{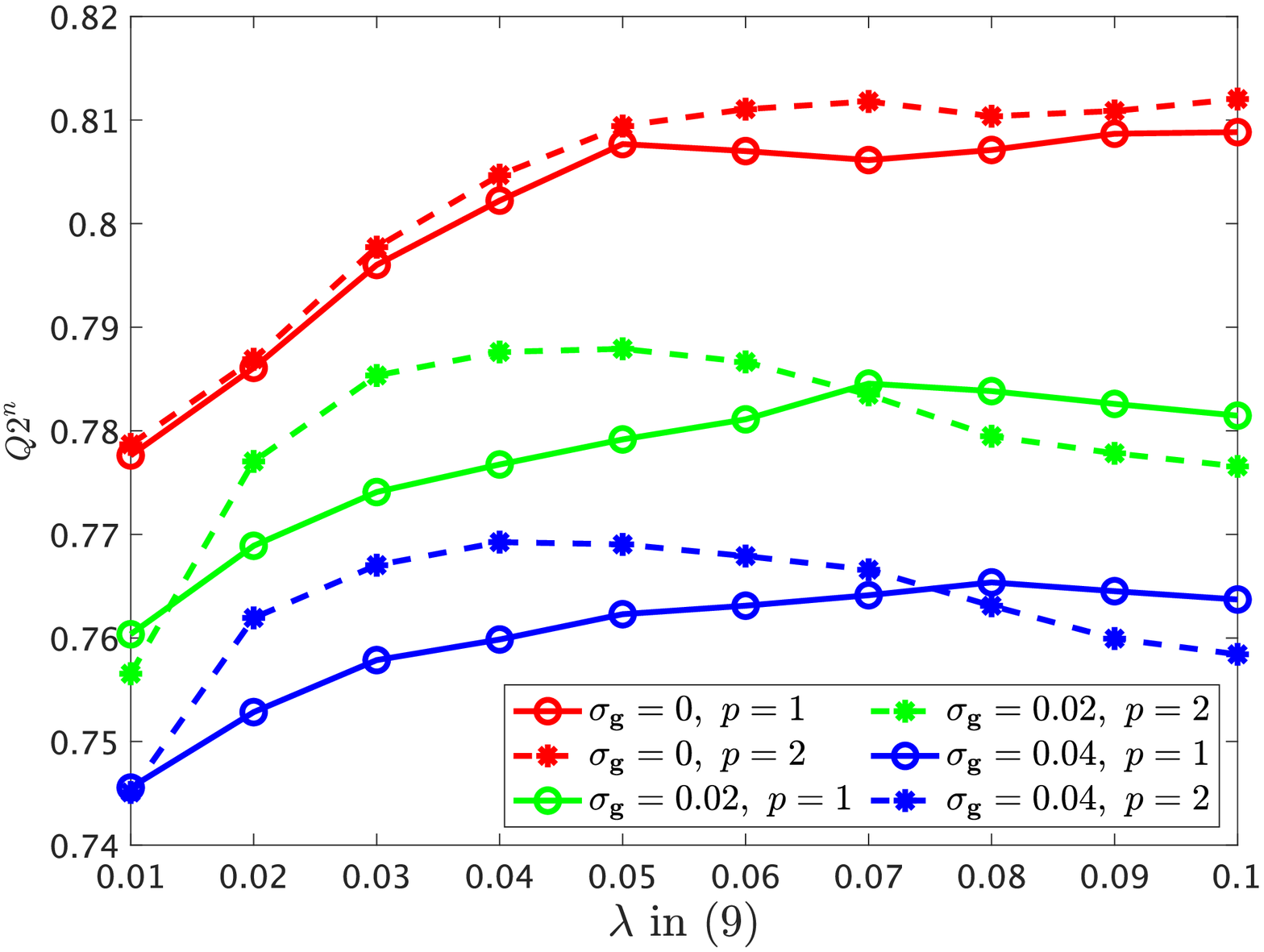}
	\end{minipage}
	
	
	\vspace{-3pt}
	\begin{minipage}[t]{0.24\hsize}
	\centerline{\footnotesize{PSNR[dB]}}
	\end{minipage}
	\begin{minipage}[t]{0.24\hsize}
	\centerline{\footnotesize{SAM}}
	\end{minipage}
	\begin{minipage}[t]{0.24\hsize}
	\centerline{\footnotesize{ERGAS}}
	\end{minipage}
	\begin{minipage}[t]{0.24\hsize}
	\centerline{\footnotesize{$Q2^n$}}
	\end{minipage}
	\caption{Four quality measures versus $\lambda$ in \eqref{prob:HRHSestimation} on HS pansharpening (top: $r = 2$, bottom: $r = 4$).}
	\label{lambda_graph_HSpan}
\end{center}
\end{figure*}

\begin{figure}[t]
\begin{center}
	\begin{minipage}[t]{0.48\hsize}
	\includegraphics[width=1.0\hsize]{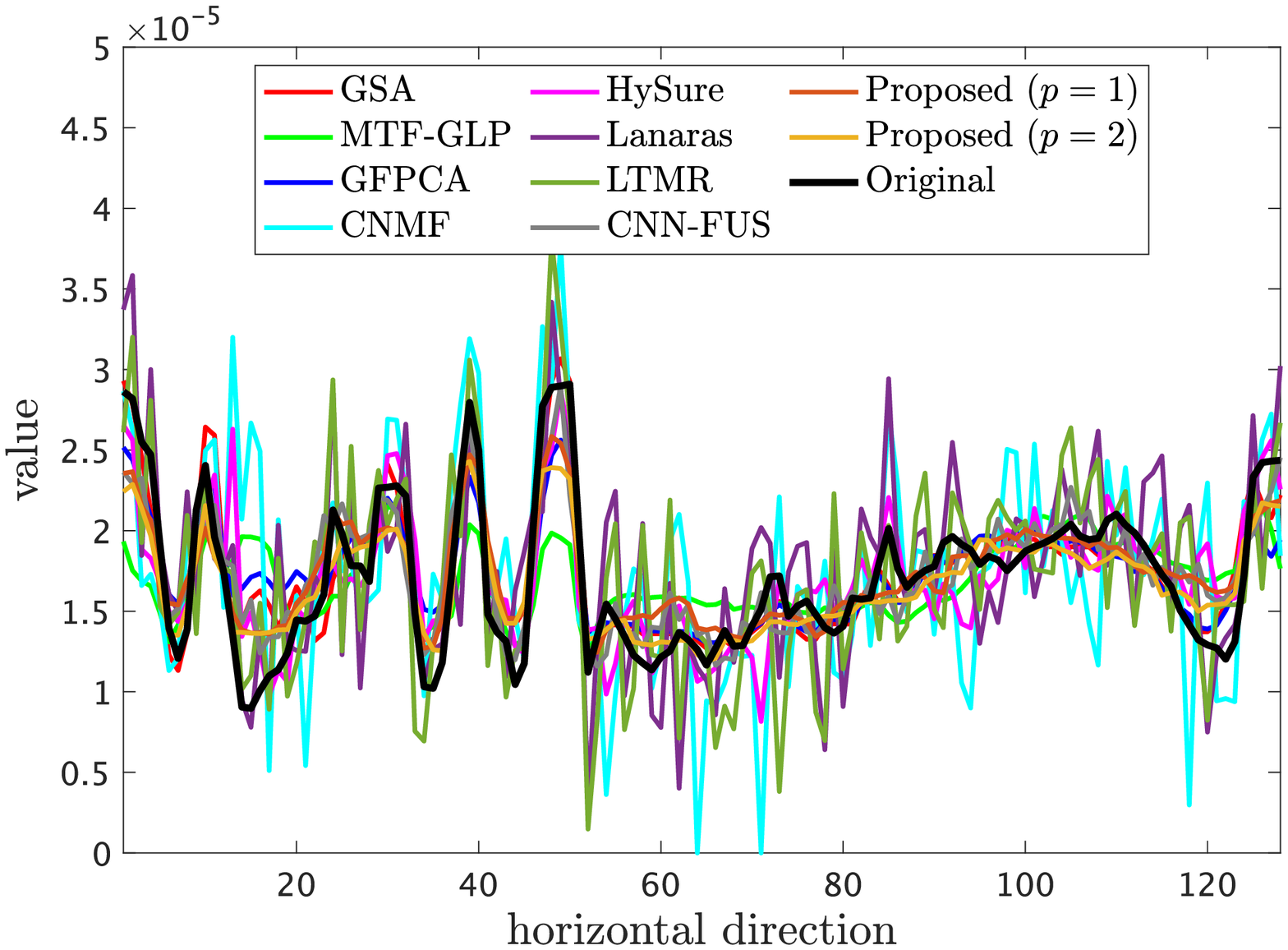}
	\end{minipage}
	\begin{minipage}[t]{0.48\hsize}
	\includegraphics[width=1.0\hsize]{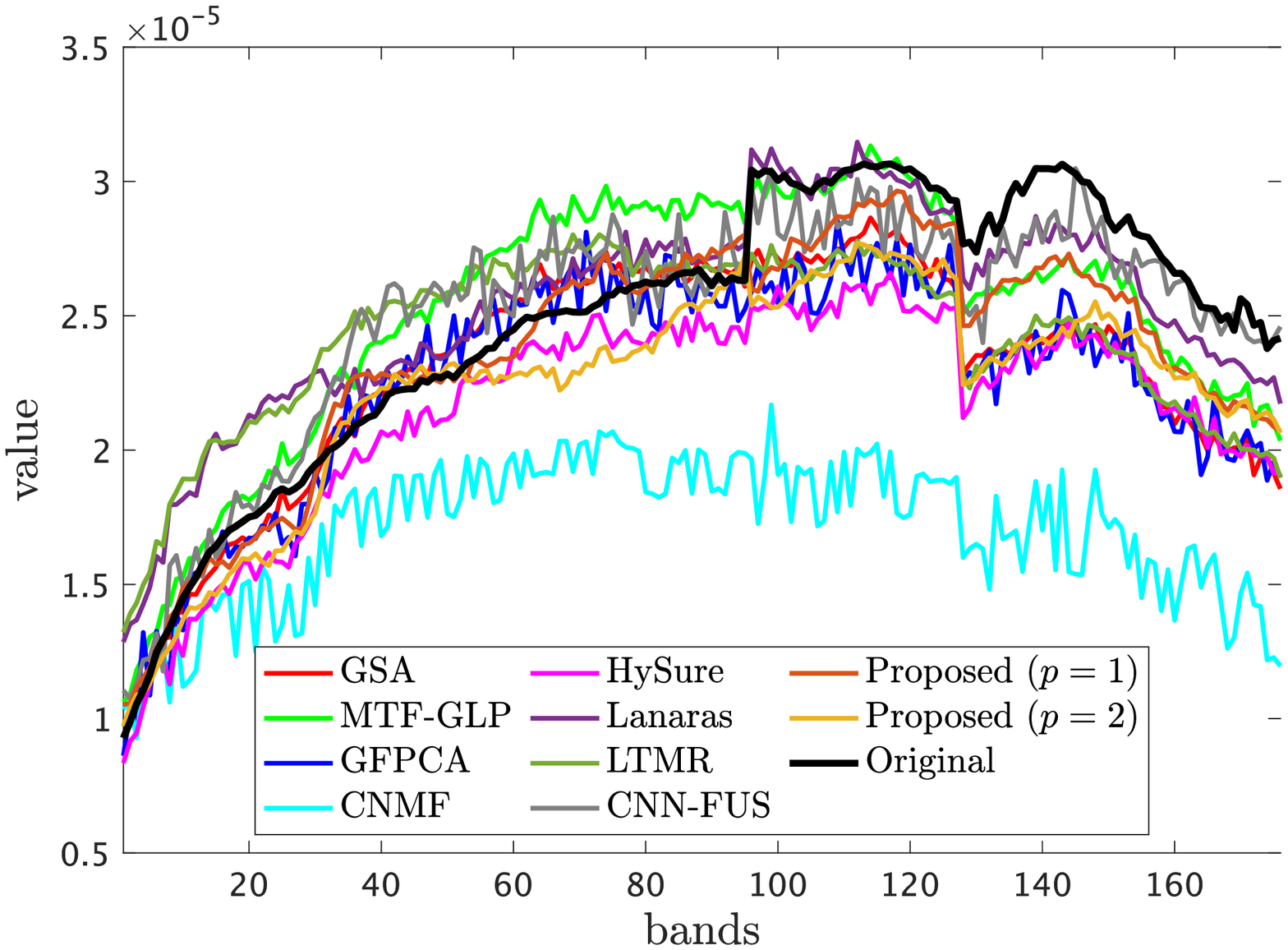}
	\end{minipage}

	\caption{ Spatial (left) and spectral (right) response of the results on HS pansharpening experiments (Moffett field, $r = 2$, and $\sigma_{\g} = 0.04$).}
	\label{graph:response_HSpan}
	\vspace{-5pt}
\end{center}
\end{figure}

\begin{table}[t]
\begin{center}
    \caption{ Computational time (s) of all processes and endmember/abundance estimation with the average times of each iteration and iteration number (PaviaU, $r = 2$, and $\sigma_{\g} = 0.02$).}
	\label{tab:comp_time}
	\scalebox{0.9}{
    \begin{tabular}{|c||c|c|c|c|} \hline
	\multirow{2}{*}{method} & computational & iteration & endmember & abundance \\ 
	& time (s) & numbers & \multicolumn{2}{c|}{(average time of each iteration)} \\ \hline
	GSA & 0.3483 & - & - & - \\ \hline
	MTF-GLP & 0.3273  & - & - & - \\ \hline 
	GFPCA & 0.4234 & - & - & - \\ \hline
	CNMF & 23.09 & 2 & \multicolumn{2}{c|}{0.5375 (0.2688)} \\ \hline
	HySure & 26.50 & 383 & 0.4431 & 25.82 (0.06742) \\ \hline
	Lanaras's & 19.02 & 219 & \multicolumn{2}{c|}{18.84 (0.08604)} \\ \hline
	LTMR & 7.342 & 11 & 0.02488 & 6.601 (0.6001) \\ \hline
	CNN-Fus & 12.13 & 25 & 0.1611 & 11.65 (0.4659) \\ \hline 
	\begin{tabular}{c}{\bf{proposed}}\\{\bf{($p = 1$)}}\end{tabular} & 94.58 & 1702 & \multicolumn{2}{c|}{94.54 (0.05555)} \\ \hline
	\begin{tabular}{c}{\bf{proposed}}\\{\bf{($p = 2$)}}\end{tabular} & 98.38 & 1697 & \multicolumn{2}{c|}{98.37 (0.05796)} \\ \hline
	\end{tabular}}
\end{center}
\end{table}

We demonstrate the advantages of our proposed method on HS pansharpening and HS and MS image fusion experiments.
In the experiments, we artificially generated an LR-HS image $\v$ and a noisy guide image $\g$ from a test HS image based on \eqref{model_v} and \eqref{model_g}, and then we estimated an HR-HS image from them.
Here, the down-sampling rate of $\S$ was set as $r = 2$, $4$, $8$, and $16$, and $B$ was a blur matrix with the size of the kernel $2r + 1 \times 2r + 1$.
For test HS images, we took four HS images from SpecTIR~\cite{SpecTIR}, GIC~\cite{GIC}, and Moffett field dataset and normalized all test HS images.
Fig.~\ref{fig:groundtruth} is the test images with their size, where they were depicted as RGB images with (R, G, B) = (16th, 32nd, 64th) bands.

We adopted several existing methods as compared methods: GSA~\cite{GSA}, MTF-GLP~\cite{MTF_GLP}, GFPCA~\cite{GFPCA}, CNMF~\cite{CNMF}, HySure~\cite{HySure}, Lanaras's~\cite{Lanaras_fusion}, LTMR~\cite{LTMR}, and CNN-Fus~\cite{CNN_Fus}.
GSA, MTF-GLP, and GFPCA are component analysis-based methods, CNMF, HySure, Lanaras's, and LTMR are optimization-based methods, and CNN-Fus is a deep learning-based method.
We used the MATLAB code presented by the authors of \cite{Pansharpening1, Pansharpening2, Lanaras_fusion, LTMR, CNN_Fus}.
GSA and MTF-GLP do not consider noisy observations, so we estimated an HR-HS image after pre-denoising.
Since GFPCA assumes a noisy LR-HS image and a noiseless guide image, we denoised an observed guide image, and then we estimate an HR-HS image.
We adopted FGSLR~\cite{FGSLR} and FFDNet~\cite{FFDNet} as pre-denoising methods for an LR-HS image and a guide image, respectively.
HySure, Lanaras's, LTMR, and CNN-Fus require the hyperparameter settings with each situation, so we experimentally set the value that achieves the highest performance in several settings.
In addition, CNN-Fus requires setting the number of iterations $T$, so we decide it in the same way as the original paper.
Tab.~\ref{parameter_settings} shows the suitable parameters for each situation.

In the proposed method, $\varepsilon$ and $\eta$ in Prob.~\eqref{prob:HRHSestimation} were set as $\varepsilon = \|\S\B\bar{\u} - \v\|$ and $\varepsilon = \|\bR\bar{\u} - \g\|$.
The hyperparameters $\omega$ in \eqref{def:HSSTV}, $\lambda$, and $\rho$ in \eqref{prob:HRHSestimation} were set like Tab.~\ref{parameter_settings}, where the values achieve the highest performance in $\omega \in [0, 0.1]$, $\lambda \in [0.01, 0.1]$, and $\rho \in [0.5, 1.5]$.

To quantitatively evaluate the estimation accuracy, we used four quality measures: the peak signal-to-noise ratio (PSNR) [dB] index, spectral angle mapper (SAM)~\cite{SAM}, erreur relative globale adimensionnelle de synth\`{e}se (ERGAS)~\cite{ERGAS}, and $Q2^n$~\cite{q2n}.
PSNR is defined by $10\log_{10}(NB/\|\u-\bar\u\|^2)$.
It evaluates the pixel-wise similarity between $\u$ and $\bar\u$, and the larger the PSNR value is, the more similar the two HS images are.
SAM measures the angle between spectral vectors of an estimated HS image and a ground truth, and it is defined by 
$\cos^{-1}(\u_{\mathrm{spe},j}\cdot \bar\u_{\mathrm{spe},j}/\|\u_{\mathrm{spe},j}\|\|\bar\u_{\mathrm{spe},j}\|)$, 
where $\u_{\mathrm{spe},j}$ and $\bar\u_{\mathrm{spe},j}$ are the $j$th spectral vectors in an estimated HS image and a ground-truth ($j = 1, 2, \ldots, N$).
For the definition, SAM is not affected by gain, so it can evaluate inherent spectral similarity.
If SAM is equal to 0, the two spectral vectors are the same, and the small value means that they are similar.
ERGAS is a global statistical measure for evaluating fuse data with the best value at $0$ and is defined by
\begin{equation}\label{def_ERGAS}
    \mathrm{ERGAS}(\u,\bar\u) = \frac{100}{r} \sqrt{\frac{1}{B}\sum_{i = 1}^B\frac{\|\u_{\mathrm{spa},i}-\bar\u_{\mathrm{spa},i}\|^2}{\left(\frac{1}{N}\mathbf{1}^{\top}\bar\u_{\mathrm{spa},i}\right)^2}},
\end{equation}
where $\u_{\mathrm{spa},i} \in \R^{N}$ and $\bar\u_{\mathrm{spa},i} \in \R^{N}$ ($i = 1, \ldots, B$) are the images of $i$th band in $\u$ and $\bar\u$, respectively, and $\mathbf{1} \in \R^{N}$ is all-one vector. 
ERGAS mainly calculates the band-wise normalized root-mean-square error (RMSE) and considers the difficulty in the fusion problem by multiplying the ground sampling distance ratio between LR-HS and guide images.
$Q2^n$ is a generalization of the universal image quality index~\cite{UIQI} and evaluates three factors: correlation coefficient, the change of the mean bias, and the change in contrast.
The value is from 0 to 1, and the two images are similar when it is close to 1.

We set the max iteration number and the stopping criterion of the primal-dual splitting method to $10000$ and $\|\u^{(n)} - \u^{(n+1)}\|/\|\u^{(n+1)}\| < 1.0 \times 10^{-4}$, respectively.

\begin{table*}[tp]
\begin{center}
	\caption{The quality measures  of the results on HS and MS image fusion experiments with $r = 2$ (boldface: the highest performance, underline: the second performance).}
	\label{tab:HSMSfusion_2}
	\scalebox{0.88}{
	\begin{tabular}{|c|c||c|c|c|c||c|c|c|c||c|c|c|c|} \hline
	\multirow{2}{*}{image} & \multirow{2}{*}{method} & \multicolumn{4}{c||}{$\sigma_{\g} = 0$} & \multicolumn{4}{c||}{$\sigma_{\g} = 0.05$} & \multicolumn{4}{c|}{$\sigma_{\g} = 0.1$} \\ \cline{3-14}
	& & PSNR[dB] & SAM & ERGAS & $Q2^n$ & PSNR[dB] & SAM & ERGAS & $Q2^n$ & PSNR[dB] & SAM & ERGAS & $Q2^n$ \\ \hline
	 \multirow{10}{*}{Reno} & GSA~\cite{GSA} & \underline{36.36} & \underline{3.916} & \underline{4.834} & 0.8617 & {\bf{32.73}} & \underline{5.356} & {\bf{6.815}} & 0.7589 & 30.36 & \underline{5.624} & {\bf{8.783}} & 0.6988 \\ 
	 & MTF-GLP~\cite{MTF_GLP} & 31.50 & 5.148 & 7.842 & 0.6598 & 30.59 & {\bf{5.341}} & 8.594 & 0.6410 & 29.84 & {\bf{5.471}} & 9.285 & 0.6254\\ 
	 & GFPCA~\cite{GFPCA} & 31.52 & 6.754 & 8.256 & 0.6296 & 29.87 & 7.592 & 10.02 & 0.6367 & 29.17 & 7.787 & 10.59 & 0.6259 \\ 
	 & CNMF~\cite{CNMF} & 29.55 & 6.561 & 10.78 & 0.8180 & 26.48 & 10.56 & 13.96 & 0.5350 & 22.79 & 14.73 & 20.33 & 0.3785 \\ 
	 & HySure~\cite{HySure} & 33.23 & 6.460 & 7.028 & 0.8111 & 31.44 & 6.833 & 8.388 & 0.7212 & 29.89 & 6.682 & 9.471 & 0.6736 \\ 
	 & Lanaras's~\cite{Lanaras_fusion} & {\bf{40.27}} & {\bf{2.758}} & {\bf{3.660}} & 0.8918 & 28.90 & 9.141 & 11.45 & 0.7656 & 25.13 & 13.10 & 16.86 & 0.5615 \\
	 & LTMR~\cite{LTMR} & 29.72 & 8.175 & 10.55 & 0.7311 & 26.45 & 12.25 & 15.51 & 0.6364 & 22.31 & 19.50 & 24.97 & 0.4269 \\
	 & CNN-Fus~\cite{CNN_Fus} & 32.17 & 5.335 & 7.285 & 0.9335 & 30.60 & 7.069 & 8.790 & 0.8936 & 30.45 & 7.160 & 8.911 & 0.8908 \\
	 & {\bf{proposed ($p = 1$)}} & 35.07 & 5.411 & 7.464 & \underline{0.9416} & 31.79 & 6.510 & 8.281 & \underline{0.9175} & \underline{30.86} & 6.703 & 8.921 & \underline{0.9015} \\ 
	 & {\bf{proposed ($p = 2$)}} & 35.58 & 5.246 & 6.923 & {\bf{0.9486}} & \underline{31.80} & 6.498 & \underline{8.258} & {\bf{0.9177}} & {\bf{30.88}} & 6.688 & \underline{8.898} & {\bf{0.9016}} \\ \hline
	 \multirow{10}{*}{Pavia U} & GSA~\cite{GSA} & 29.41 & 6.783 & 10.10 & 0.5244 & 28.17 & 7.697 & 11.62 & 0.4662 & 26.96 & 7.529 & 14.29 & 0.3340 \\ 
	 & MTF-GLP~\cite{MTF_GLP} & 28.47 & 6.854 & 11.40 & 0.3998 & 27.85 & 7.154 & 12.17 & 0.3825 & 27.20 & 7.309 & 13.09 & 0.3385 \\ 
	 & GFPCA~\cite{GFPCA} & 27.87 & 7.046 & 10.41 & 0.3491 & 26.68 & 7.661 & 11.65 & 0.3245 & 26.21 & 7.791 & 12.74 & 0.3146 \\ 
	 & CNMF~\cite{CNMF} & 29.86 & 7.453 & 8.254 & 0.5758 & 23.78 & 14.03 & 15.11 & 0.3015 & 20.20 & 19.38 & 22.67 & 0.2054 \\ 
	 & HySure~\cite{HySure} & 29.67 & 8.135 & 8.951 & 0.5084 & 28.48 & 8.341 & 10.24 & 0.4637 & 27.15 & 8.075 & 11.90 & 0.3743\\ 
	 & Lanaras's~\cite{Lanaras_fusion} & {\bf{36.80}} & {\bf{3.108}} & {\bf{3.817}} & 0.7879 & 25.17 & 11.18 & 14.02 & 0.3933 & 21.68 & 15.48 & 21.26 & 0.3815 \\
	 & LTMR~\cite{LTMR} & 27.05 & 9.803 & 12.60 & 0.4664 & 23.85 & 14.34 & 16.72 & 0.3374 & 19.77 & 21.95 & 25.24 & 0.2183 \\
	 & CNN-Fus~\cite{CNN_Fus} & 29.62 & 6.205 & 9.788 & 0.8910 & 28.56 & 8.009 & 11.35 & 0.8429 & 28.41 & 8.178 & 11.65 & 0.8370 \\
	 & {\bf{proposed ($p = 1$)}} & \underline{32.87} & 5.346 & 5.674 & {\bf{0.9618}} & \underline{29.82} & \underline{6.531} & \underline{9.117} & \underline{0.8921} & \underline{28.94} & \underline{6.784} & \underline{10.30} & \underline{0.8680}\\ 
	 & {\bf{proposed ($p = 2$)}} & 32.83 & \underline{5.338} & \underline{5.615} & \underline{0.9612} & {\bf{29.90}} & {\bf{6.495}} & {\bf{9.045}} & {\bf{0.8932}} & {\bf{29.02}} & {\bf{6.726}} & {\bf{10.23}} & {\bf{0.8695}} \\ \hline
	 \multirow{10}{*}{Salinas} & GSA~\cite{GSA} & 34.28 & 3.536 & 4.367 & 0.6624 & \underline{32.76} & \underline{3.834} & \underline{4.922} & 0.5139 & 31.89 & 4.003 & 5.337 & 0.4738\\ 
	 & MTF-GLP~\cite{MTF_GLP} & 32.85 & 3.953 & 4.816 & 0.5503 & 32.09 & 4.067 & 5.142 & 0.4845 & 31.83 & 4.086 & \underline{5.273} & 0.4736 \\ 
	 & GFPCA~\cite{GFPCA} & 33.36 & 3.593 & 4.689 & 0.4931 & 31.99 & 3.877 & 5.194 & 0.4508 & 31.64 & \underline{3.944} & 5.461 & 0.4697 \\ 
	 & CNMF~\cite{CNMF} & 31.09 & 4.495 & 7.122 & 0.5530 & 25.61 & 7.329 & 10.37 & 0.2292 & 21.09 & 11.65 & 15.91 & 0.1438 \\ 
	 & HySure~\cite{HySure} & 34.00 & 3.107 & 4.317 & 0.5185 & {\bf{32.92}} & {\bf{3.101}} & {\bf{4.611}} & 0.4461 & 31.88 & {\bf{3.514}} & {\bf{5.174}} & 0.4464\\ 
	 & Lanaras's~\cite{Lanaras_fusion} & {\bf{37.41}} & {\bf{1.783}} & {\bf{3.056}} & {\bf{0.8328}} & 27.17 & 6.108 & 7.674 & 0.2533 & 22.16 & 9.991 & 13.15 & 0.1562 \\
	 & LTMR~\cite{LTMR} & 28.72 & 5.467 & 6.913 & 0.2872 & 25.49 & 7.497 & 9.539 & 0.2241 & 21.33 & 11.54 & 14.81 & 0.1545 \\
	 & CNN-Fus~\cite{CNN_Fus} & \underline{34.52} & \underline{2.819} & \underline{3.832} & \underline{0.7452} & 31.34 & 4.376 & 5.613 & \underline{0.5945} & 30.92 & 4.574 & 5.922 & {\bf{0.5807}} \\
	 & {\bf{proposed ($p = 1$)}} & 34.41 & 3.343 & 4.280 & 0.7045 & 32.54 & 3.945 & 5.506 & 0.5941 & \underline{32.05} & 4.103 & 5.916 & 0.5741 \\ 
	 & {\bf{proposed ($p = 2$)}} & 34.48 & 3.301 & 4.231 & 0.7055 & 32.63 & 3.905 & 5.456 & {\bf{0.5946}} & {\bf{32.16}} & 4.049 & 5.847 & \underline{0.5758} \\ \hline
	 \multirow{10}{*}{\begin{tabular}{c} Moffett\\field \end{tabular}} & GSA~\cite{GSA} & 25.41 & 10.13 & 14.78 & 0.4853 & 24.93 & 9.953 & 15.50 & 0.5260 & 24.21 & 9.935 & 16.65 & 0.5745 \\ 
	 & MTF-GLP~\cite{MTF_GLP} & 27.45 & 9.935 & 12.37 & 0.5923 & 26.94 & 9.935 & 12.96 & 0.6538 & 26.58 & 9.935 & 13.44 & 0.5893\\ 
	 & GFPCA~\cite{GFPCA} & \underline{29.55} & 7.242 & \underline{8.688} & 0.5867 & 27.58 & 8.033 & 10.82 & 0.5775 & 26.86 & 8.341 & 12.01 & 0.5850 \\ 
	 & CNMF~\cite{CNMF} & 26.43 & 8.596 & 12.26 & 0.2952 & 22.09 & 14.44 & 19.50 & 0.1469 & 19.52 & 17.66 & 25.77 & 0.1071 \\ 
	 & HySure~\cite{HySure} & 28.72 & 7.717 & 9.632 & 0.5449 & 28.00 & 7.617 & 10.41 & 0.4277 & 27.53 & 7.940 & 11.08 & 0.5902 \\ 
	 & Lanaras's~\cite{Lanaras_fusion} & 28.92 & \underline{6.494} & 9.352 & 0.5303 & 23.74 & 10.54 & 16.46 & 0.5264 & 21.65 & 11.90 & 21.38 & 0.3866 \\
	 & LTMR~\cite{LTMR} & 27.56 & 10.17 & 11.95 & 0.2349 & 23.63 & 15.89 & 18.26 & 0.2069 & 18.88 & 25.09 & 31.15 & 0.1185 \\
	 & CNN-Fus~\cite{CNN_Fus} & {\bf{30.71}} & {\bf{6.005}} & {\bf{7.931}} & {\bf{0.8902}} & {\bf{29.70}} & 7.073 & 9.049 & \underline{0.8475} & {\bf{29.66}} & 7.301 & 9.252 & 0.8427 \\
	 & {\bf{proposed ($p = 1$)}} & 29.35 & 7.347 & 8.896 & \underline{0.8679} & \underline{29.55} & \underline{6.802} & \underline{8.974} & 0.8474 & 29.62 & \underline{6.795} & \underline{9.009} & \underline{0.8473} \\ 
	 & {\bf{proposed ($p = 2$)}} & 29.22 & 7.351 & 9.007 & 0.8635 & \underline{29.55} & {\bf{6.794}} & {\bf{8.967}} & {\bf{0.8476}} & \underline{29.63} & {\bf{6.779}} & {\bf{8.996}} & {\bf{0.8476}} \\ \hline
	 \multirow{10}{*}{Average} & GSA~\cite{GSA} & 31.36 & 6.091 & 8.520 & 0.6335 & 29.65 & 6.710 & 9.716 & 0.5662 & 28.35 & 6.773 & 11.27 & 0.5203 \\ 
	 & MTF-GLP~\cite{MTF_GLP} & 30.07 & 6.473 & 9.107 & 0.5506 & 29.37 & 6.624 & 9.717 & 0.5405 & 28.86 & 6.700 & 10.27 & 0.5067 \\
	 & GFPCA~\cite{GFPCA} & 30.57 & 6.159 & 8.011 & 0.5146 & 29.03 & 6.791 & 9.421 & 0.4974 & 28.47 & 6.966 & 10.20 & 0.4988 \\
	 & CNMF~\cite{CNMF} & 29.23 & 6.776 & 9.605 & 0.5605 & 24.49 & 11.59 & 14.74 & 0.3032 & 20.90 & 15.86 & 21.17 & 0.2087 \\
	 & HySure~\cite{HySure} & 31.40 & 6.355 & 7.482 & 0.5957 & 30.21 & 6.473 & 8.412 & 0.5147 & 29.11 & 6.553 & 9.406 & 0.5211 \\
	 & Lanaras's~\cite{Lanaras_fusion} & {\bf{35.85}} & {\bf{3.536}} & {\bf{4.971}} & 0.7607 & 26.24 & 9.241 & 12.40 & 0.4846 & 22.66 & 12.62 & 18.16 & 0.3715 \\
	 & LTMR~\cite{LTMR} & 28.26 & 8.404 & 10.50 & 0.4299 & 24.85 & 12.49 & 15.01 & 0.3512 & 20.57 & 19.52 & 24.04 & 0.2296 \\
	 & CNN-Fus~\cite{CNN_Fus} & 31.75 & \underline{5.091} & 7.209 & 0.8650 & 30.05 & 6.632 & 8.701 & 0.7946 & 29.86 & 6.803 & 8.934 & 0.7878 \\
	 & {\bf{proposed ($p = 1$)}} & 32.93 & 5.362 & 6.578 & \underline{0.8690} & \underline{30.92} & \underline{5.947} & \underline{7.970} & \underline{0.8128} & \underline{30.37} & \underline{6.097} & \underline{8.537} & \underline{0.7977} \\
	 & {\bf{proposed ($p = 2$)}} & \underline{33.03} & 5.309 & \underline{6.444} & {\bf{0.8697}} & {\bf{30.97}} & {\bf{5.923}} & {\bf{7.932}} & {\bf{0.8133}} & {\bf{30.42}} & {\bf{6.061}} & {\bf{8.493}} & {\bf{0.7986}} \\ \hline
	\end{tabular}}
\end{center}
\end{table*}
\begin{table*}[tp]
\begin{center}
	\caption{The quality measures  of the results on HS and MS image fusion experiments with $r = 4$ (boldface: the highest performance, underline: the second performance).}
	\label{tab:HSMSfusion_4}
	\scalebox{0.88}{
	\begin{tabular}{|c|c||c|c|c|c||c|c|c|c||c|c|c|c|} \hline
	\multirow{2}{*}{image} & \multirow{2}{*}{method} & \multicolumn{4}{c||}{$\sigma_{\g} = 0$} & \multicolumn{4}{c||}{$\sigma_{\g} = 0.05$} & \multicolumn{4}{c|}{$\sigma_{\g} = 0.1$} \\ \cline{3-14}
	& & PSNR[dB] & SAM & ERGAS & $Q2^n$ & PSNR[dB] & SAM & ERGAS & $Q2^n$ & PSNR[dB] & SAM & ERGAS & $Q2^n$ \\ \hline
	 \multirow{10}{*}{Reno} & GSA~\cite{GSA} & 28.52 & 7.943 & 6.391 & 0.6115 & 27.92 & 7.943 & 6.651 & 0.5843 & 27.30 & 7.943 & 6.928 & 0.5576 \\ 
	 & MTF-GLP~\cite{MTF_GLP} & 26.36 & 7.943 & 7.235 & 0.4737 & 26.06 & 7.943 & 7.438 & 0.4648 & 25.68 & 7.943 & 7.716 & 0.4224 \\ 
	 & GFPCA~\cite{GFPCA} & 27.22 & 8.218 & 6.353 & 0.4586 & 26.89 & 8.525 & 6.682 & 0.4450 & 26.69 & 8.649 & 6.784 & 0.4541 \\ 
	 & CNMF~\cite{CNMF} & 30.20 & 6.082 & 5.038 & 0.7864 & 25.74 & 10.85 & 7.714 & 0.5042 & 22.33 & 14.69 & 10.85 & 0.3726 \\ 
	 & HySure~\cite{HySure} & 31.29 & 8.316 & 4.574 & 0.7605 & 30.02 & {\bf{7.754}} & {\bf{5.082}} & 0.7796 & 28.87 & {\bf{7.765}} & {\bf{5.461}} & 0.7289 \\ 
	 & Lanaras's~\cite{Lanaras_fusion} & {\bf{38.78}} & {\bf{3.375}} & {\bf{2.102}} & 0.8914 & 28.57 & 9.684 & 5.964 & 0.7899 & 25.88 & 9.439 & 7.415 & 0.6001 \\
	 & LTMR~\cite{LTMR} & 32.85 & 6.410 & 3.888 & 0.8440 & 26.79 & 12.04 & 7.679 & 0.6558 & 21.78 & 20.76 & 13.51 & 0.3781 \\
	 & CNN-Fus~\cite{CNN_Fus} & \underline{33.31} & \underline{5.834} & \underline{3.528} & {\bf{0.9377}} & 28.33 & 7.912 & 5.699 & 0.8026 & 28.35 & \underline{7.865} & 5.663 & 0.8084 \\
	 & {\bf{proposed ($p = 1$)}} & 33.17 & 6.866 & 4.588 & 0.9217 & {\bf{30.37}} & \underline{7.847} & \underline{5.097} & {\bf{0.8895}} & {\bf{28.99}} & 8.301 & \underline{5.549} & {\bf{0.8468}} \\ 
	 & {\bf{proposed ($p = 2$)}} & 33.22 & 6.967 & 4.512 & \underline{0.9220} & \underline{30.34} & 7.923 & {\bf{5.082}} & \underline{0.8884} & \underline{28.98} & 8.329 & 5.552 & \underline{0.8465} \\ \hline
	 \multirow{10}{*}{Pavia U} & GSA~\cite{GSA} & 26.95 & 9.777 & 6.743 & 0.3997 & 26.14 & 10.29 & 7.360 & 0.3588 & 25.14 & 10.78 & 8.252 & 0.3203 \\ 
	 & MTF-GLP~\cite{MTF_GLP} & 24.56 & 10.54 & 8.988 & 0.2310 & 24.25 & 10.74 & 9.287 & 0.2369 & 23.86 & 10.89 & 9.688 & 0.1986 \\ 
	 & GFPCA~\cite{GFPCA} & 25.06 & 8.769 & 7.950 & 0.2190 & 24.72 & 9.123 & 8.000 & 0.2323 & 24.53 & 9.348 & 8.31 & 0.2224 \\ 
	 & CNMF~\cite{CNMF} & \underline{30.63} & \underline{4.884} & 3.697 & 0.6466 & 22.99 & 15.13 & 8.304 & 0.2664 & 19.50 & 20.19 & 12.24 & 0.1834 \\ 
	 & HySure~\cite{HySure} & 27.84 & 10.66 & 5.556 & 0.4548 & 26.08 & 10.15 & 6.231 & 0.4224 & 25.12 & 10.49 & 7.024 & 0.3631 \\ 
	 & Lanaras's~\cite{Lanaras_fusion} & {\bf{35.78}} & {\bf{4.058}} & {\bf{2.192}} & 0.7412 & 24.98 & 11.18 & 7.058 & 0.3106 & 22.61 & 12.64 & 10.04 & 0.3716 \\
	 & LTMR~\cite{LTMR} & 28.71 & 8.279 & 5.089 & 0.5697 & 24.01 & 14.19 & 7.915 & 0.3416 & 19.14 & 23.38 & 13.23 & 0.2031 \\
	 & CNN-Fus~\cite{CNN_Fus} & 29.07 & 7.564 & 5.046 & 0.9063 & 26.09 & 8.358 & 7.080 & 0.7269 & 26.08 & {\bf{8.411}} & 7.115 & 0.7247 \\
	 & {\bf{proposed ($p = 1$)}} & 30.48 & 7.025 & 3.621 & {\bf{0.9418}} & \underline{28.08} & {\bf{8.155}} & {\bf{5.333}} & \underline{0.8472} & \underline{26.91} & \underline{8.563} & \underline{6.355} & \underline{0.7861} \\ 
	 & {\bf{proposed ($p = 2$)}} & 30.58 & 7.059 & \underline{3.607} & \underline{0.9416} & {\bf{28.12}} & \underline{8.182} & \underline{5.337} & {\bf{0.8477}} & {\bf{26.93}} & 8.578 & {\bf{6.346}} & {\bf{0.7866}} \\ \hline
	 \multirow{10}{*}{Salinas} & GSA~\cite{GSA} & 32.14 & 4.427 & 2.672 & 0.5744 & 31.17 & 4.670 & 2.907 & 0.4998 & 30.55 & 4.812 & 3.092 & 0.4653  \\ 
	 & MTF-GLP~\cite{MTF_GLP} & 29.88 & 5.189 & 3.286 & 0.4559 & 29.39 & 5.312 & 3.439 & 0.4588 & 29.15 & 5.353 & 3.521 & 0.4416 \\ 
	 & GFPCA~\cite{GFPCA} & 30.46 & 4.723 & 3.228 & 0.4277 & 29.91 & 4.98 & 3.409 & 0.3749 & 29.61 & 5.140 & 3.470 & 0.4225 \\ 
	 & CNMF~\cite{CNMF} & 28.26 & 9.156 & 4.928 & 0.5030 & 25.49 & 7.565 & 5.274 & 0.2329 & 21.01 & 11.55 & 8.086 & 0.1431 \\ 
	 & HySure~\cite{HySure} & \underline{32.48} & \underline{4.022} & 2.791 & 0.5151 & {\bf{31.15}} & {\bf{4.232}} & \underline{2.981} & 0.4790 & 30.25 & {\bf{4.355}} & \underline{3.121} & 0.3705 \\ 
	 & Lanaras's~\cite{Lanaras_fusion} & {\bf{36.28}} & {\bf{2.481}} & {\bf{1.834}} & {\bf{0.7838}} & 27.00 & 6.318 & 3.932 & 0.2500 & 19.55 & 10.68 & 8.927 & 0.1633 \\
	 & LTMR~\cite{LTMR} & 31.54 & 4.431 & 2.652 & 0.4253 & 25.93 & 7.223 & 4.493 & 0.2312 & 20.93 & 11.94 & 7.692 & 0.1496 \\
	 & CNN-Fus~\cite{CNN_Fus} & 32.26 & 4.171 & \underline{2.507} & {\underline{0.7253}} & 30.55 & \underline{4.483} & {\bf{2.975}} & \underline{0.5628} & {\bf{30.55}} & \underline{4.433} & {\bf{2.992}} & {\bf{0.5676}} \\
	 & {\bf{proposed ($p = 1$)}} & 32.10 & 4.426 & 2.797 & 0.6531 & \underline{31.04} & 4.722 & 3.155 & {\bf{0.5653}} & 30.38 & 5.083 & 3.477 & \underline{0.5204} \\ 
	 & {\bf{proposed ($p = 2$)}} & 31.95 & 4.571 & 2.905 & 0.6249 & 30.98 & 4.807 & 3.219 & 0.5530 & \underline{30.43} & 5.071 & 3.469 & 0.5189 \\ \hline
	 \multirow{10}{*}{Moffett field} & GSA~\cite{GSA} & 25.47 & 10.22 & 7.575 & 0.4821 & 25.01 & 10.15 & 7.930 & 0.5244 & 24.36 & 10.15 & 8.454 & 0.5758 \\ 
	 & MTF-GLP~\cite{MTF_GLP} & 25.89 & 10.15 & 7.401 & 0.6449 & 25.43 & 10.15 & 7.738 & 0.6419 & 25.01 & 10.15 & 8.084 & 0.6415 \\ 
	 & GFPCA~\cite{GFPCA} & 26.21 & \underline{9.417} & 6.701 & 0.5408 & 25.62 & 9.703 & 7.08 & 0.5507 & 25.32 & 9.934 & 7.302 & 0.4947 \\ 
	 & CNMF~\cite{CNMF} & 24.96 & 9.901 & 7.045 & 0.2128 & 22.20 & 13.74 & 9.509 & 0.1380 & 19.84 & 16.71 & 12.21 & 0.1031 \\ 
	 & HySure~\cite{HySure} & 25.48 & 11.17 & 6.935 & 0.3952 & 25.51 & 9.393 & 6.711 & 0.4584 & 25.38 & 9.334 & 6.764 & 0.4575 \\ 
	 & Lanaras's~\cite{Lanaras_fusion} & \underline{27.13} & {\bf{8.303}} & {\bf{5.710}} & 0.5863 & 23.84 & 10.58 & 8.297 & 0.2993 & 22.46 & 10.53 & 9.948 & 0.3884 \\
	 & LTMR~\cite{LTMR} & 26.55 & 10.86 & 6.567 & 0.2155 & 22.33 & 17.57 & 10.41 & 0.1762 & 17.37 & 28.52 & 18.66 & 0.08507 \\
	 & CNN-Fus~\cite{CNN_Fus} & {\bf{27.55}} & 9.766 & \underline{5.870} & {\bf{0.8172}} & \underline{26.39} & {\bf{8.673}} & 6.411 & 0.6784 & 26.32 & {\bf{8.766}} & 6.488 & 0.6784 \\
	 & {\bf{proposed ($p = 1$)}} & 26.49 & 9.577 & 6.147 & 0.7704 & \underline{26.39} & 9.026 & \underline{6.352} & \underline{0.6972} & {\bf{26.54}} & \underline{8.846} & {\bf{6.343}} & {\bf{0.6908}} \\ 
	 & {\bf{proposed ($p = 2$)}} & 26.67 & 9.521 & 6.033 & \underline{0.7774} & {\bf{26.49}} & \underline{8.999} & {\bf{6.293}} & {\bf{0.7045}} & \underline{26.53} & 8.863 & \underline{6.349} & \underline{0.6905}\\ \hline
	 \multirow{10}{*}{Average} & GSA~\cite{GSA} & 28.27 & 8.092 & 5.846 & 0.5169 & 27.56 & 8.264 & 6.212 & 0.4918 & 26.84 & 8.422 & 6.682 & 0.4798 \\ 
	 & MTF-GLP~\cite{MTF_GLP} & 26.68 & 8.456 & 6.728 & 0.4514 & 26.28 & 8.537 & 6.976 & 0.4506 & 25.93 & 8.585 & 7.252 & 0.4260 \\
	 & GFPCA~\cite{GFPCA} & 27.24 & 7.782 & 6.058 & 0.4115 & 26.79 & 8.083 & 6.293 & 0.4007 & 26.54 & 8.268 & 6.467 & 0.3984 \\
	 & CNMF~\cite{CNMF} & 28.51 & 7.506 & 5.177 & 0.5372 & 23.77 & 12.27 & 7.955 & 0.2765 & 20.62 & 16.01 & 10.88 & 0.1978 \\
	 & HySure~\cite{HySure} & 29.27 & 8.542 & 4.964 & 0.5314 & 28.19 & 7.882 & 5.251 & 0.5349 & 27.41 & 7.986 & 5.593 & 0.4800 \\
	 & Lanaras's~\cite{Lanaras_fusion} & {\bf{34.49}} & {\bf{4.554}} & {\bf{2.960}} & 0.7507 & 26.10 & 9.441 & 6.313 & 0.4124 & 22.62 & 10.82 & 9.084 & 0.3808 \\
	 & LTMR~\cite{LTMR} & 29.92 & 7.495 & 4.549 & 0.5136 & 24.77 & 12.76 & 7.625 & 0.3512 & 19.80 & 21.15 & 13.27 & 0.2040 \\
	 & CNN-Fus~\cite{CNN_Fus} & 30.55 & \underline{6.834} & \underline{4.238} & {\bf{0.8466}} & 27.84 & {\bf{7.357}} & 5.541 & 0.6927 & 27.83 & {\bf{7.369}} & 5.565 & 0.6948 \\
	 & {\bf{proposed ($p = 1$)}} & 30.56 & 6.974 & 4.288 & \underline{0.8217} & \underline{28.97} & \underline{7.437} & \underline{4.984} & {\bf{0.7498}} & \underline{28.20} & \underline{7.698} & \underline{5.431} & {\bf{0.7110}} \\
	 & {\bf{proposed ($p = 2$)}} & \underline{30.61} & 7.030 & 4.264 & 0.8165 & {\bf{28.98}} & 7.478 & {\bf{4.983}} & \underline{0.7484} & {\bf{28.22}} & 7.710 & {\bf{5.429}} & \underline{0.7106} \\ \hline
	\end{tabular}}
\end{center}
\end{table*}

\begin{table*}[tp]
\begin{center}
	\caption{The quality measures of the results on HS and MS image fusion experiments with $r = 8$ and $16$ (boldface: the highest performance, underline: the second performance).}
	\label{tab:HSMSfusion_816}
	\begin{tabular}{|c|c||c|c|c|c||c|c|c|c|} \hline
	\multirow{2}{*}{image} & \multirow{2}{*}{method} & \multicolumn{4}{c||}{$r = 8 ,~\sigma_{\v} = 0.1,~\sigma_{\g} = 0.05$} & \multicolumn{4}{c|}{$r = 16,\sigma_{\v} = 0.1,~\sigma_{\g} = 0.05$} \\ \cline{3-10}
	& & PSNR[dB] & SAM & ERGAS & $Q2^n$ & PSNR[dB] & SAM & ERGAS & $Q2^n$ \\ \hline
	 \multirow{10}{*}{Reno} & GSA~\cite{GSA} & 27.87 & 8.009 & 3.340 & 0.5810 & 24.66 & \underline{8.664} & 2.257 & 0.2813 \\  
	 & MTF-GLP~\cite{MTF_GLP} & 24.62 & 8.009 & 4.273 & 0.3432 & 21.86 & \underline{8.664} & 3.007 & 0.2813 \\ 
	 & GFPCA~\cite{GFPCA} & 23.55 & 7.604 & 4.633 & 0.2939 & 20.97 & 9.960 & 3.105 & 0.2813 \\ 
	 & CNMF~\cite{CNMF} & 27.15 & 10.79 & 3.633 & 0.5824 & 26.87 & 10.44 & 1.943 & 0.7390 \\ 
	 & HySure~\cite{HySure} & 29.27 & \underline{7.387} & 2.717 & 0.7587 & 27.55 & 9.887 & 1.727 & 0.7487 \\ 
	 & Lanaras's~\cite{Lanaras_fusion} & 28.49 & 9.876 & 3.009 & 0.7818 & 27.68 & 11.61 & 1.624 & 0.6918 \\ 
	 & LTMR~\cite{LTMR} & 27.29 & 11.44 & 3.655 & 0.6514 & 26.39 & 13.18 & 2.103 & 0.5485 \\ 
	 & CNN-Fus~\cite{CNN_Fus} & {\bf{32.69}} & {\bf{6.053}} & {\bf{1.837}} & {\bf{0.9388}} & {\bf{31.27}} & {\bf{7.872}} & {\bf{1.113}} & {\bf{0.9253}} \\ 
	 & {\bf{proposed ($p = 1$)}} & \underline{30.22} & 7.603 & \underline{2.646} & \underline{0.8901} & 28.69 & 9.495 & 1.739 & 0.8649 \\ 
	 & {\bf{proposed ($p = 2$)}} & \underline{30.22} & 7.666 & 2.648 & 0.8896 & \underline{29.00} & 9.345 & \underline{1.605} & \underline{0.8729} \\  \hline 
	 \multirow{10}{*}{Pavia U} & GSA~\cite{GSA} & 23.26 & 14.72 & 5.832 & 0.2902 & 21.13 & 14.86 & 3.641 & 0.1744 \\  
	 & MTF-GLP~\cite{MTF_GLP} & 22.17 & 14.72 & 6.175 & 0.2213 & 20.88 & 14.86 & 3.860 & 0.1719  \\ 
	 & GFPCA~\cite{GFPCA} & 22.69 & 10.30 & 5.469 & 0.1720 & 21.23 & 13.65 & 3.286 & 0.1719 \\ 
	 & CNMF~\cite{CNMF} & 23.39 & 15.44 & 4.897 & 0.3153 & 22.61 & 16.35 & 2.339 & 0.2646 \\ 
	 & HySure~\cite{HySure} & 26.74 & 9.496 & 3.077 & 0.3972 & 25.10 & 12.27 & 1.770 & 0.3025 \\ 
	 & Lanaras's~\cite{Lanaras_fusion} & 24.92 & 11.59 & 3.562 & 0.3015 & 24.75 & 12.46 & 1.859 & 0.2948 \\ 
	 & LTMR~\cite{LTMR} & 24.38 & 13.60 & 3.764 & 0.3416 & 24.06 & 14.40 & 1.966 & 0.3055 \\ 
	 & CNN-Fus~\cite{CNN_Fus} & {\bf{30.04}} & {\bf{7.519}} & {\bf{2.255}} & {\bf{0.8970}} & {\bf{28.64}} & {\bf{9.285}} & {\bf{1.342}} & {\bf{0.8871}} \\ 
	 & {\bf{proposed ($p = 1$)}} & 27.55 & \underline{8.412} & 2.763 & 0.8342 & \underline{26.46} & \underline{10.01} & \underline{1.568} & \underline{0.8111} \\ 
	 & {\bf{proposed ($p = 2$)}} & \underline{27.58} & 8.428 & \underline{2.754} & \underline{0.8348} & 26.25 & 10.27 & 1.619 & 0.7991 \\  \hline
	\end{tabular}
\end{center}
\end{table*}

\begin{figure*}[t]
\begin{center}
  \begin{minipage}[t]{0.128\hsize}
\includegraphics[width=\hsize]{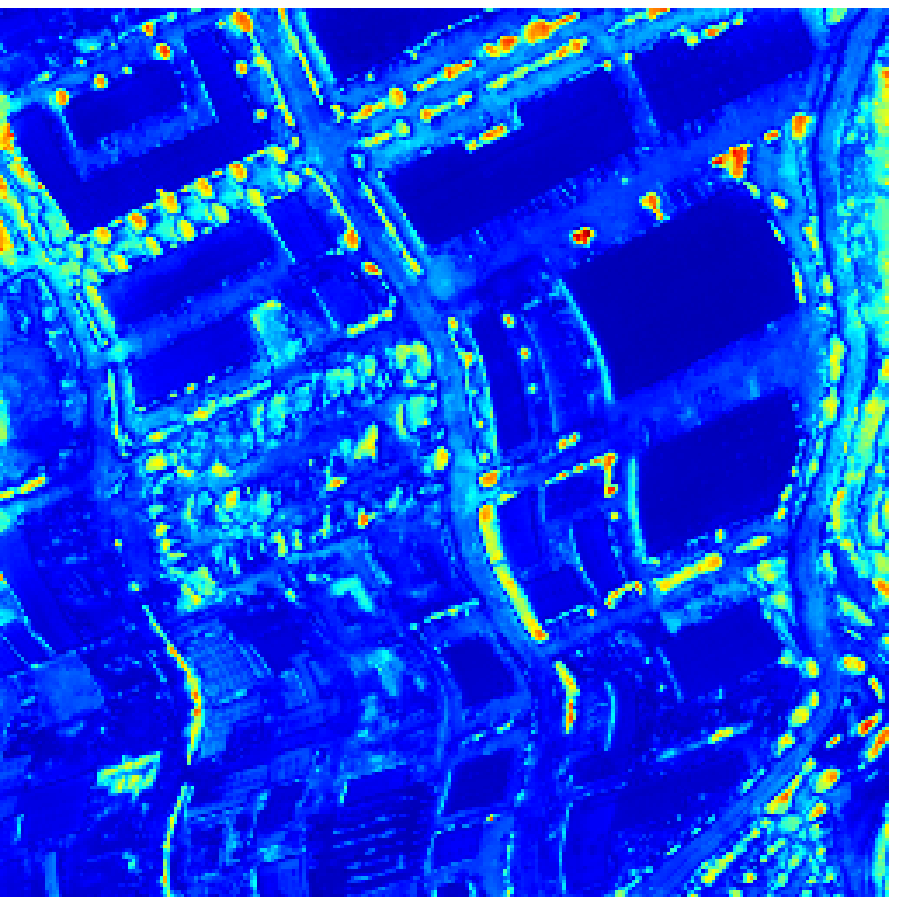}
\end{minipage}
\begin{minipage}[t]{0.128\hsize}
\includegraphics[width=\hsize]{./img/fusion_SAMmap_HySure.eps}
\end{minipage}
\begin{minipage}[t]{0.128\hsize}
\includegraphics[width=\hsize]{./img/fusion_SAMmap_HySure.eps}
\end{minipage}
  \begin{minipage}[t]{0.128\hsize}
\includegraphics[width=\hsize]{./img/fusion_SAMmap_HySure.eps}
\end{minipage}
\begin{minipage}[t]{0.128\hsize}
\includegraphics[width=\hsize]{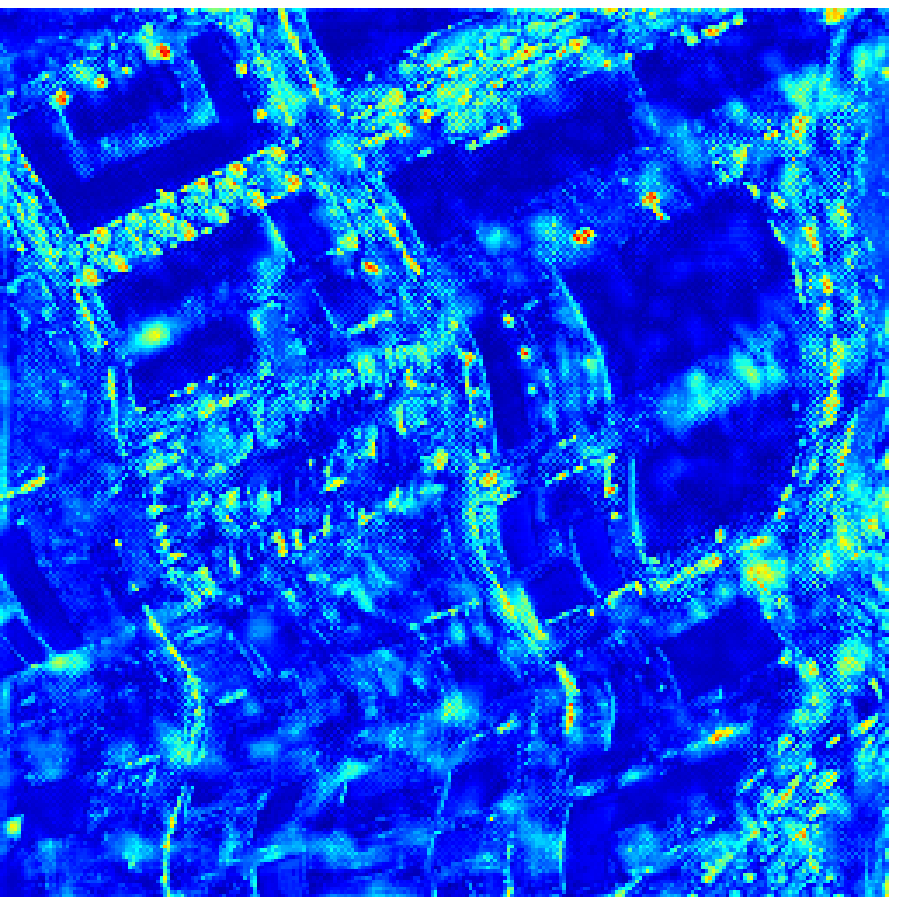}
\end{minipage}
  \begin{minipage}[t]{0.128\hsize}
\includegraphics[width=\hsize]{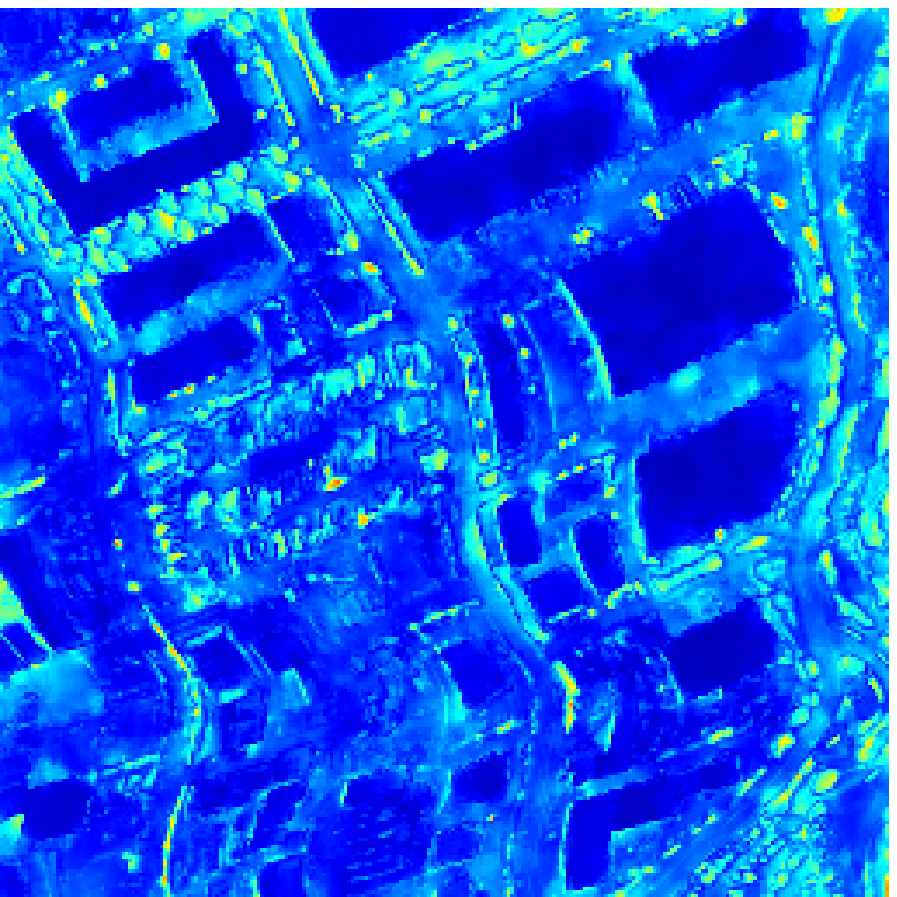}
\end{minipage}
  \begin{minipage}[t]{0.128\hsize}
\includegraphics[width=\hsize]{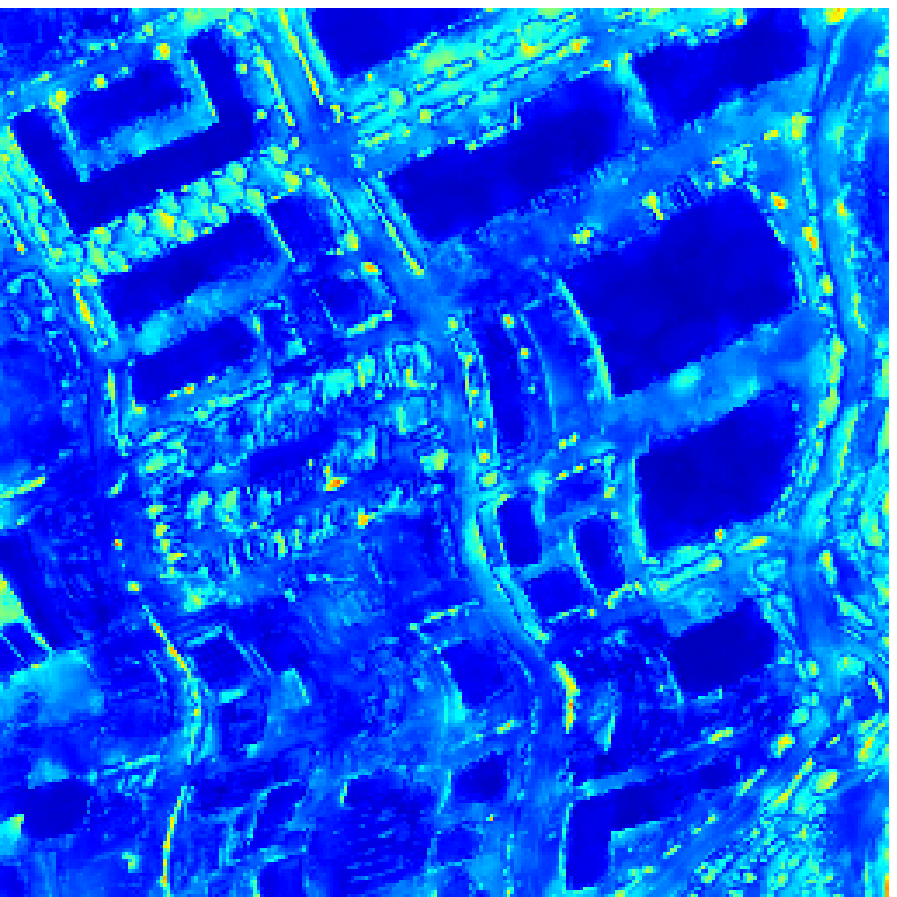}
\end{minipage}
\begin{minipage}[t]{0.06\hsize}
\includegraphics[width=0.65\hsize]{./img/colorbar_jet.eps}
\end{minipage}

  \begin{minipage}[t]{0.128\hsize}
\centerline{\footnotesize{ GSA }}
\end{minipage}
\begin{minipage}[t]{0.128\hsize}
\centerline{\footnotesize{ MTF-GLP }}
\end{minipage}
\begin{minipage}[t]{0.128\hsize}
\centerline{\footnotesize{ GFPCA }}
\end{minipage}
  \begin{minipage}[t]{0.128\hsize}
\centerline{\footnotesize{ HySure }}
\end{minipage}
\begin{minipage}[t]{0.128\hsize}
\centerline{\footnotesize{ CNN-Fus }}
\end{minipage}
  \begin{minipage}[t]{0.128\hsize}
\centerline{\footnotesize{ proposed ($p = 1$) }}
\end{minipage}
  \begin{minipage}[t]{0.128\hsize}
\centerline{\footnotesize{ proposed ($p = 2$) }}
\end{minipage}
\begin{minipage}[t]{0.06\hsize}
\centerline{\footnotesize{  }}
\end{minipage}

   \caption{SAM map of the results by some existing methods and proposed ($p = 1$ and $2$) on HS and MS image fusion experiments 
   (Reno, $r = 4$, and $\sigma_\g = 0.1$).}
 \label{fig:SAM_MSfusion}
\end{center}
\end{figure*}


\begin{figure*}[t]
    \centering
    \begin{minipage}[t]{0.32\hsize}
        \includegraphics[width=\hsize]{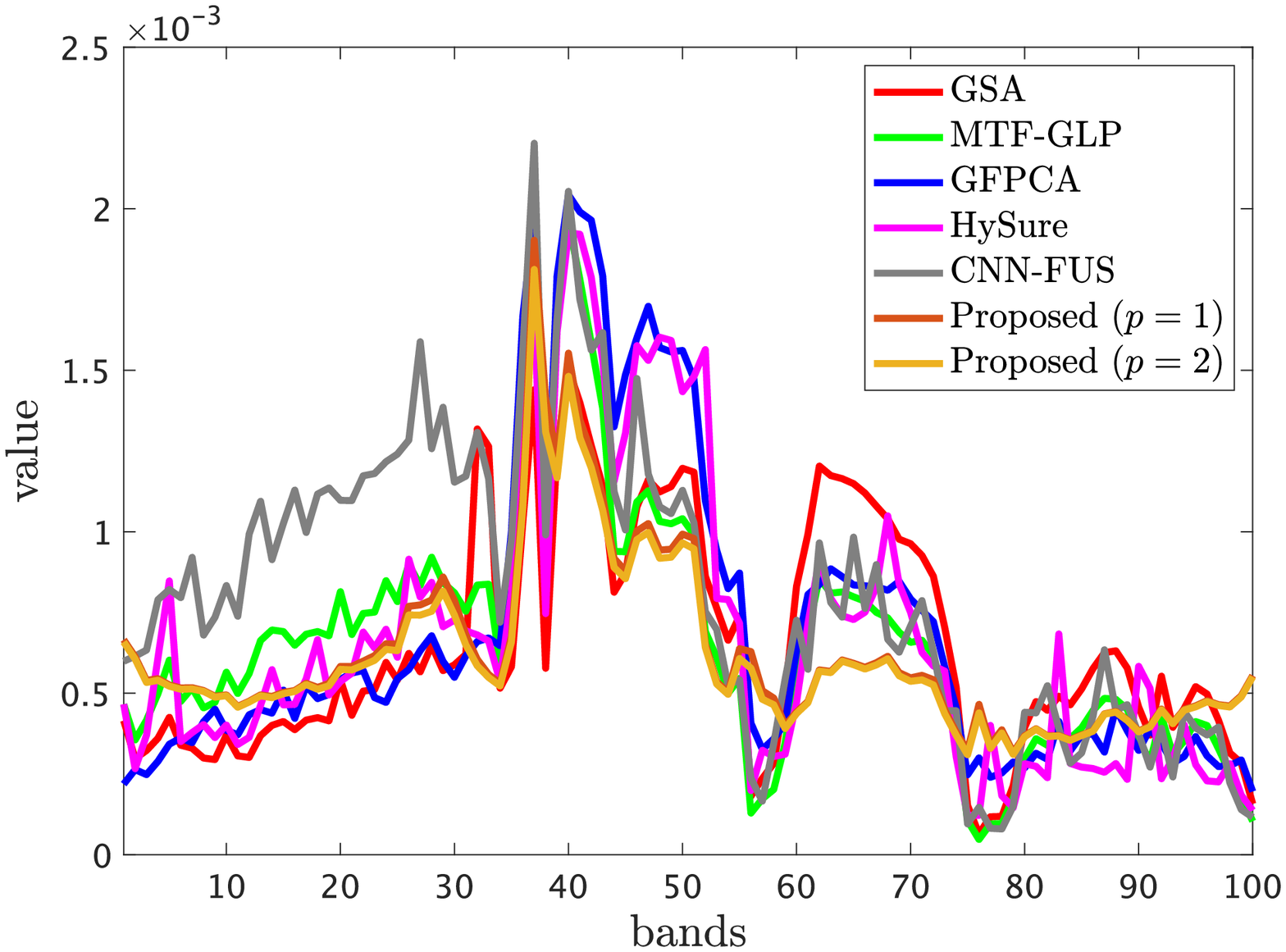}
    \end{minipage}
    \begin{minipage}[t]{0.32\hsize}
        \includegraphics[width=\hsize]{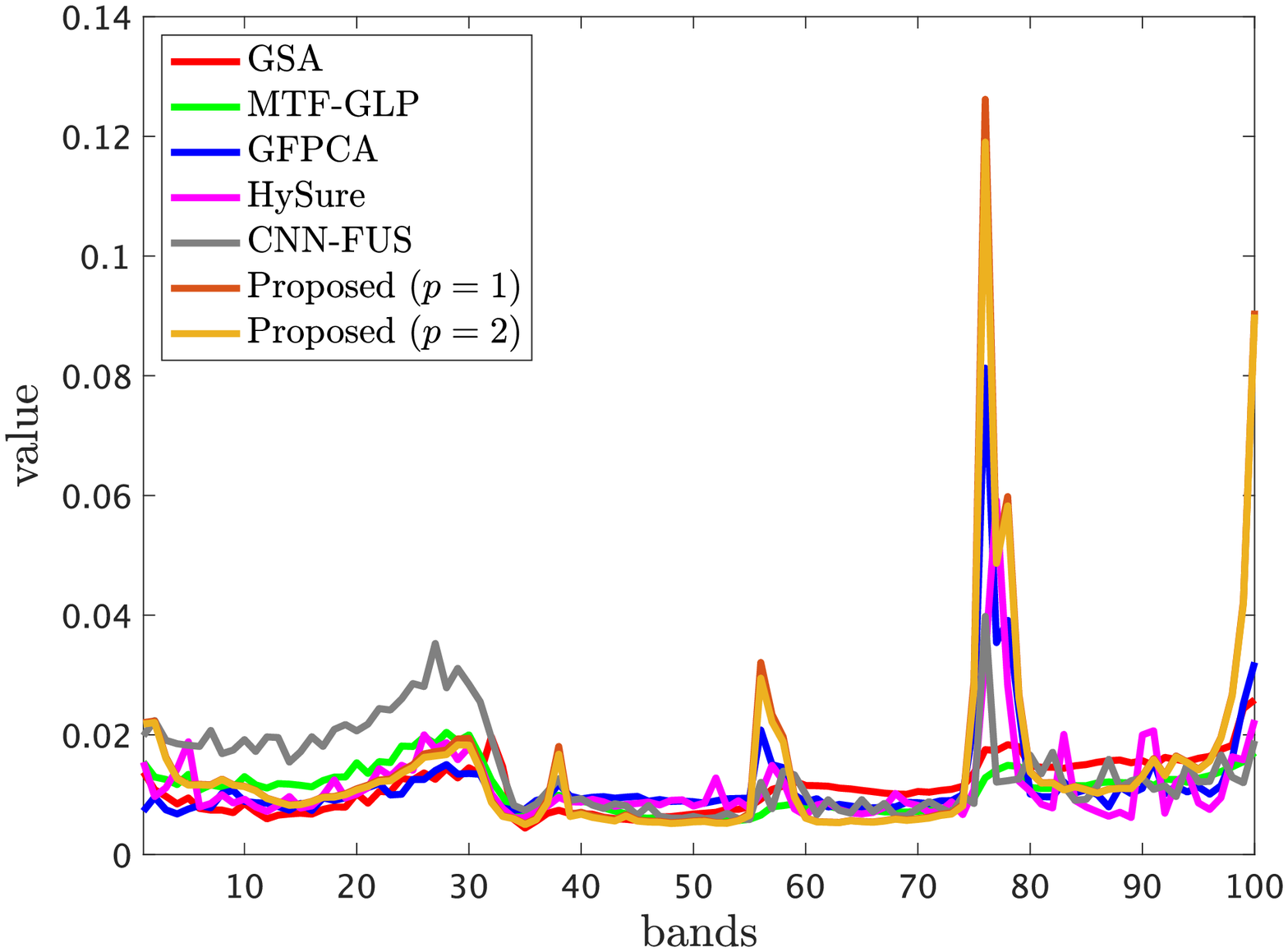}
    \end{minipage}
    \begin{minipage}[t]{0.32\hsize}
        \includegraphics[width=\hsize]{./img/mean_value.eps}
    \end{minipage}
    
    \begin{minipage}[t]{0.32\hsize}
        \centerline{\footnotesize{MSE}}
    \end{minipage}
    \begin{minipage}[t]{0.32\hsize}
        \centerline{\footnotesize{band-wise normalized MSE}}
    \end{minipage}
    \begin{minipage}[t]{0.32\hsize}
        \centerline{\footnotesize{mean luminance}}
    \end{minipage}
    \caption{MSE and band-wise normalized MSE of the HS and MS fusion results by some existing methods and proposed ($p = 1$ and $2$) and mean luminance of the test HS image (Salinas, $r = 2$, and $\sigma_{\g} = 0.1$). }
    \label{fig:bandwiseMSE_MSfusion}
\end{figure*}

\begin{figure*}[t]
\begin{center}
  \begin{minipage}[t]{0.135\hsize}
\includegraphics[width=\hsize]{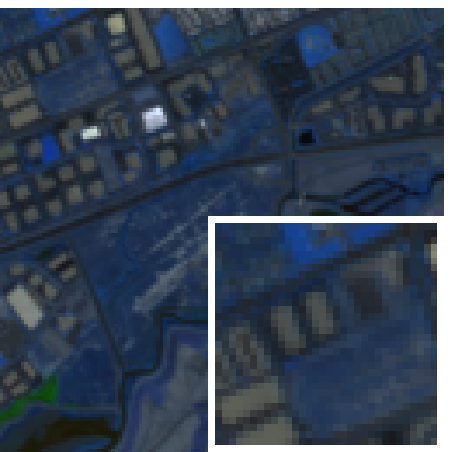}
\end{minipage}
  \begin{minipage}[t]{0.135\hsize}
\includegraphics[width=\hsize]{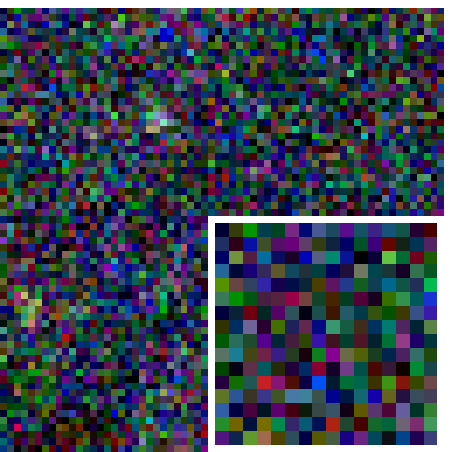}
\end{minipage}
  \begin{minipage}[t]{0.135\hsize}
\includegraphics[width=\hsize]{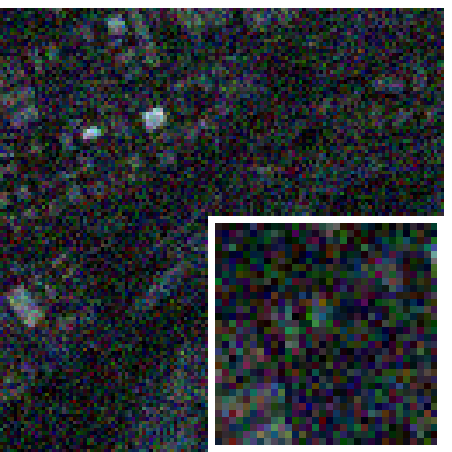}
\end{minipage}
  \begin{minipage}[t]{0.135\hsize}
\includegraphics[width=\hsize]{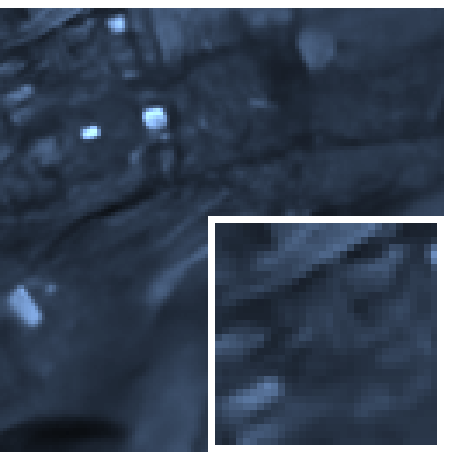}
\end{minipage}
  \begin{minipage}[t]{0.135\hsize}
\includegraphics[width=\hsize]{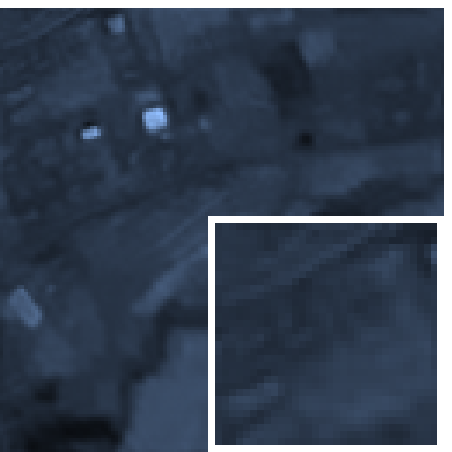}
\end{minipage}
  \begin{minipage}[t]{0.135\hsize}
\includegraphics[width=\hsize]{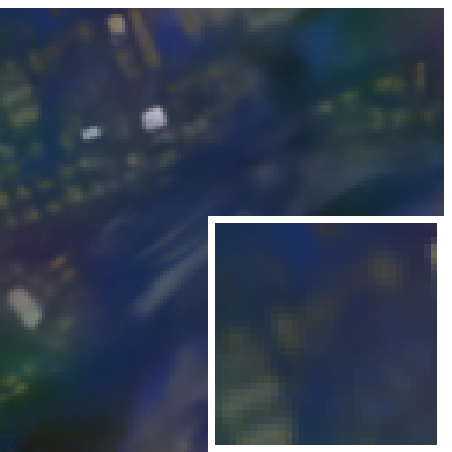}
\end{minipage}
  \begin{minipage}[t]{0.135\hsize}
\includegraphics[width=\hsize]{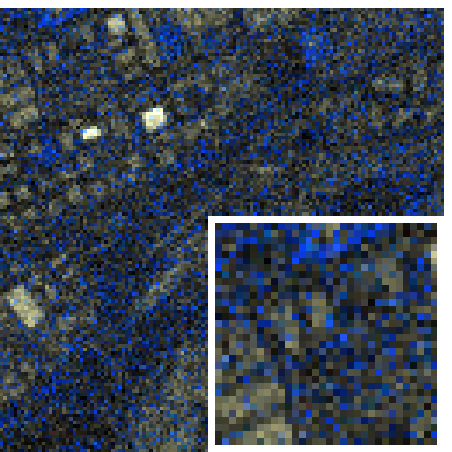}
\end{minipage}

  \begin{minipage}[t]{0.135\hsize}
\centerline{\footnotesize{ ground-truth }}
\end{minipage}
  \begin{minipage}[t]{0.135\hsize}
\centerline{\footnotesize{ observed HS image }}
\end{minipage}
\begin{minipage}[t]{0.135\hsize}
\centerline{\footnotesize{ observed MS image }}
\end{minipage}
  \begin{minipage}[t]{0.135\hsize}
\centerline{\footnotesize{ GSA }}
\end{minipage}
  \begin{minipage}[t]{0.135\hsize}
\centerline{\footnotesize{ MTF-GLP }}
\end{minipage}
  \begin{minipage}[t]{0.135\hsize}
\centerline{\footnotesize{ GFPCA }}
\end{minipage}
\begin{minipage}[t]{0.135\hsize}
\centerline{\footnotesize{ CNMF }}
\end{minipage}

\vspace{4pt}
  \begin{minipage}[t]{0.135\hsize}
\includegraphics[width=\hsize]{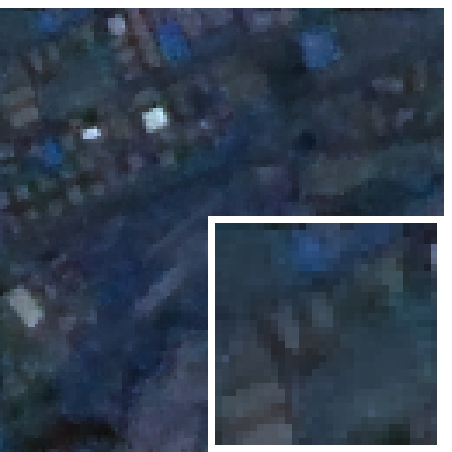}
\end{minipage}
  \begin{minipage}[t]{0.135\hsize}
\includegraphics[width=\hsize]{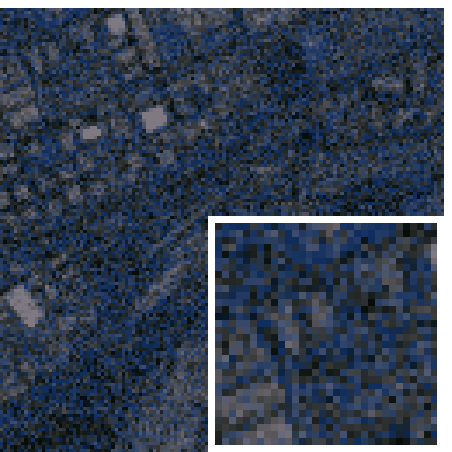}
\end{minipage}
  \begin{minipage}[t]{0.135\hsize}
\includegraphics[width=\hsize]{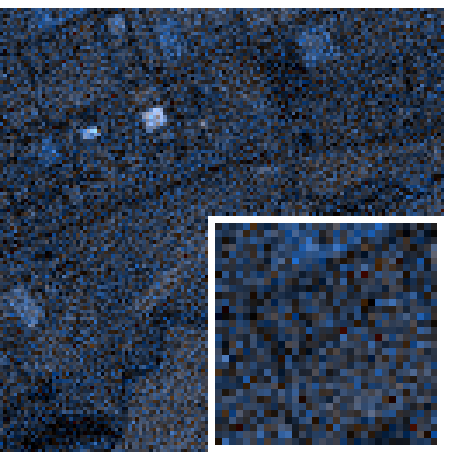}
\end{minipage}
  \begin{minipage}[t]{0.135\hsize}
\includegraphics[width=\hsize]{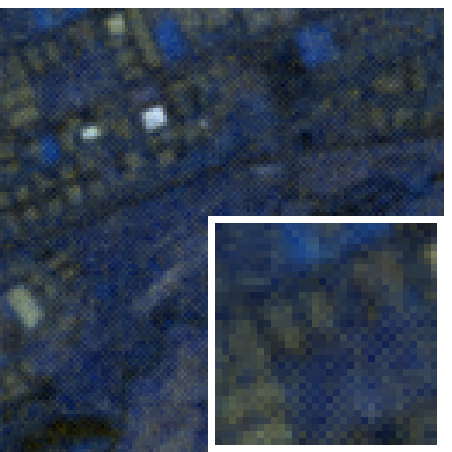}
\end{minipage}
  \begin{minipage}[t]{0.135\hsize}
\includegraphics[width=\hsize]{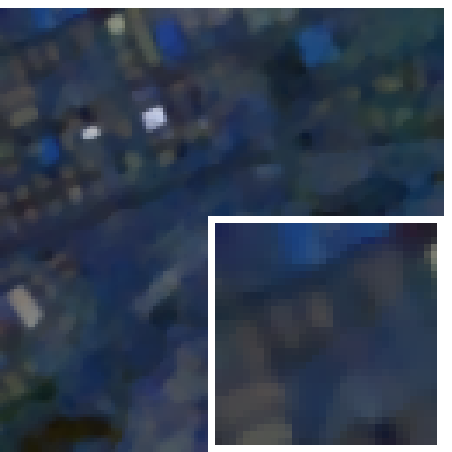}
\end{minipage}
\begin{minipage}[t]{0.135\hsize}
\includegraphics[width=\hsize]{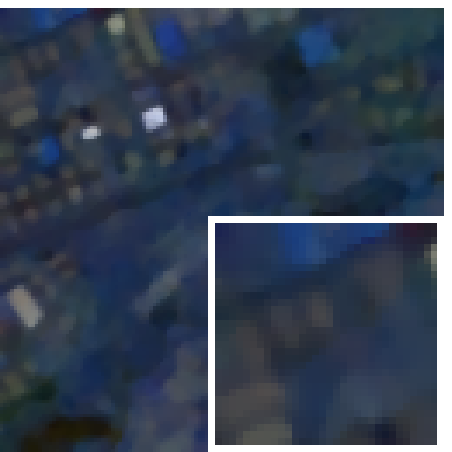}
\end{minipage}
\begin{minipage}[t]{0.135\hsize}
\centerline{\footnotesize{  }}
\end{minipage}

\begin{minipage}[t]{0.135\hsize}
\centerline{\footnotesize{ HySure }}
\end{minipage}
  \begin{minipage}[t]{0.135\hsize}
\centerline{\footnotesize{ Lanaras's }}
\end{minipage}
  \begin{minipage}[t]{0.135\hsize}
\centerline{\footnotesize{ LTMR }}
\end{minipage}
\begin{minipage}[t]{0.135\hsize}
\centerline{\footnotesize{ CNN-Fus }}
\end{minipage}
  \begin{minipage}[t]{0.135\hsize}
\centerline{\footnotesize{ \bf{proposed ($p = 1$)} }}
\end{minipage}
  \begin{minipage}[t]{0.135\hsize}
\centerline{\footnotesize{ \bf{proposed ($p = 2$)} }}
\end{minipage}
\begin{minipage}[t]{0.135\hsize}
\centerline{\footnotesize{  }}
\end{minipage}

\vspace{4pt}
  \begin{minipage}[t]{0.135\hsize}
\includegraphics[width=\hsize]{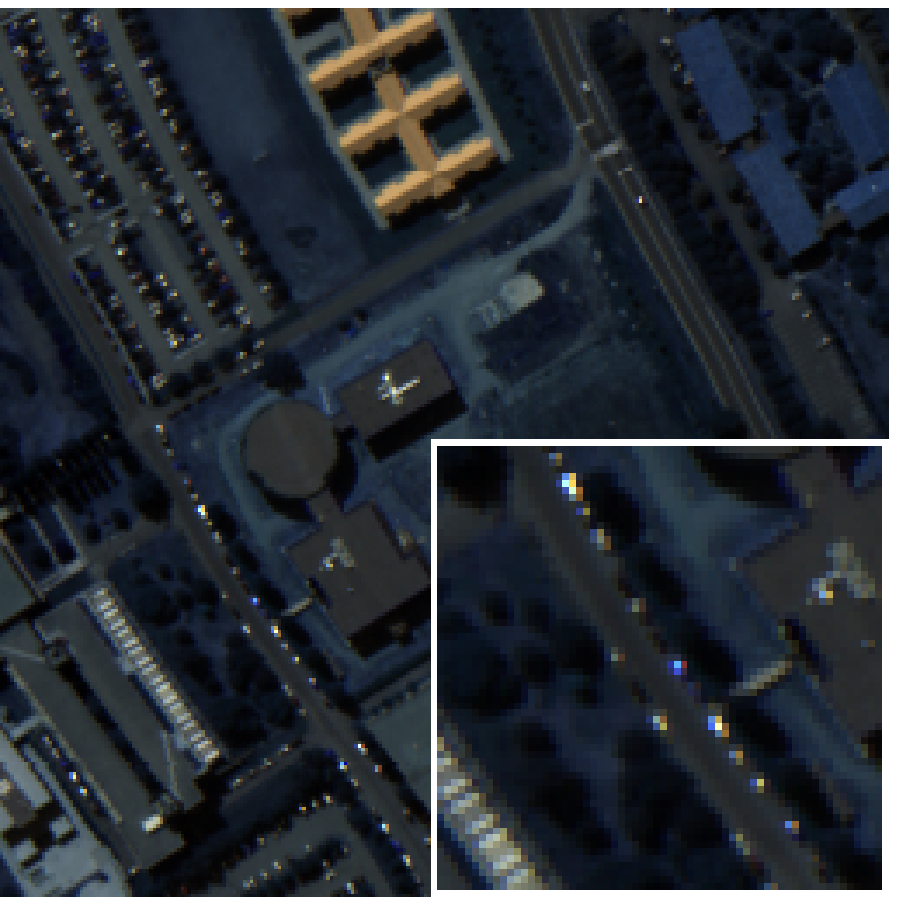}
\end{minipage}
  \begin{minipage}[t]{0.135\hsize}
\includegraphics[width=\hsize]{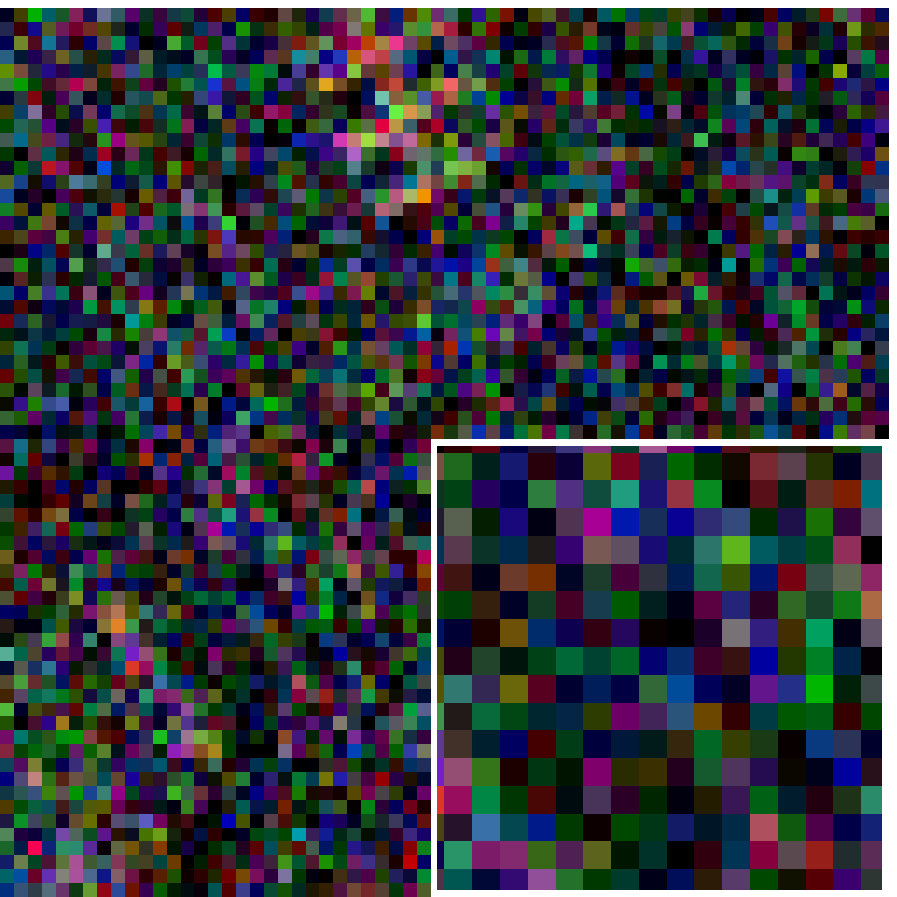}
\end{minipage}
  \begin{minipage}[t]{0.135\hsize}
\includegraphics[width=\hsize]{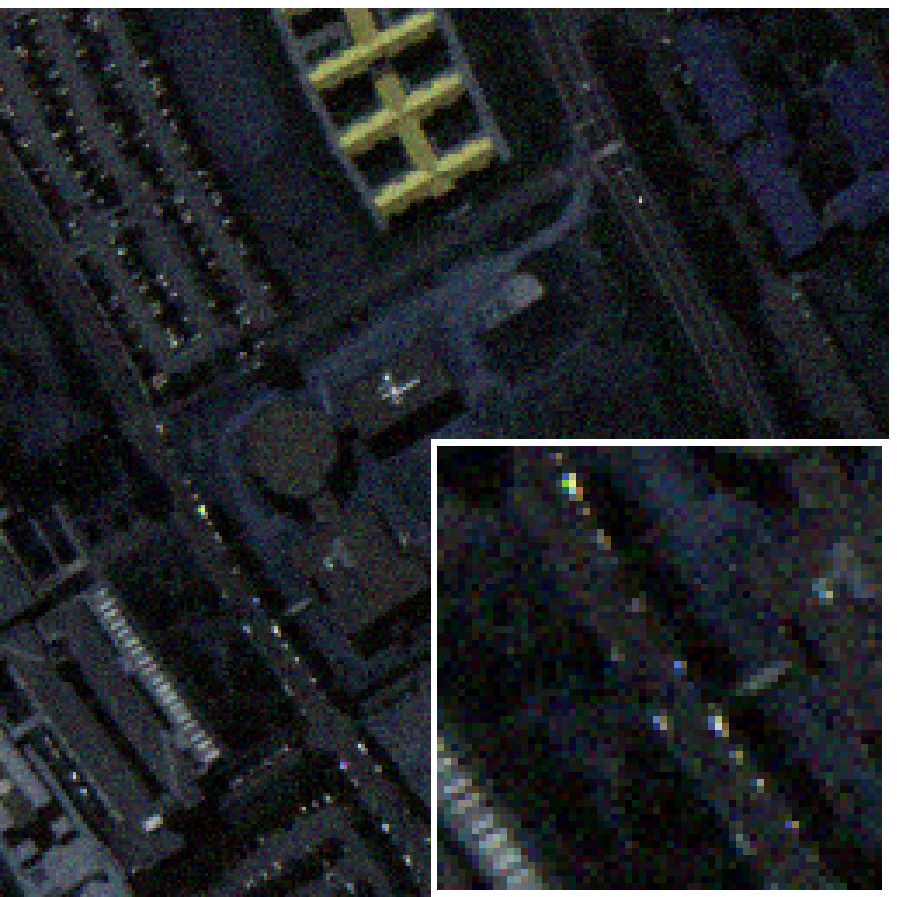}
\end{minipage}
  \begin{minipage}[t]{0.135\hsize}
\includegraphics[width=\hsize]{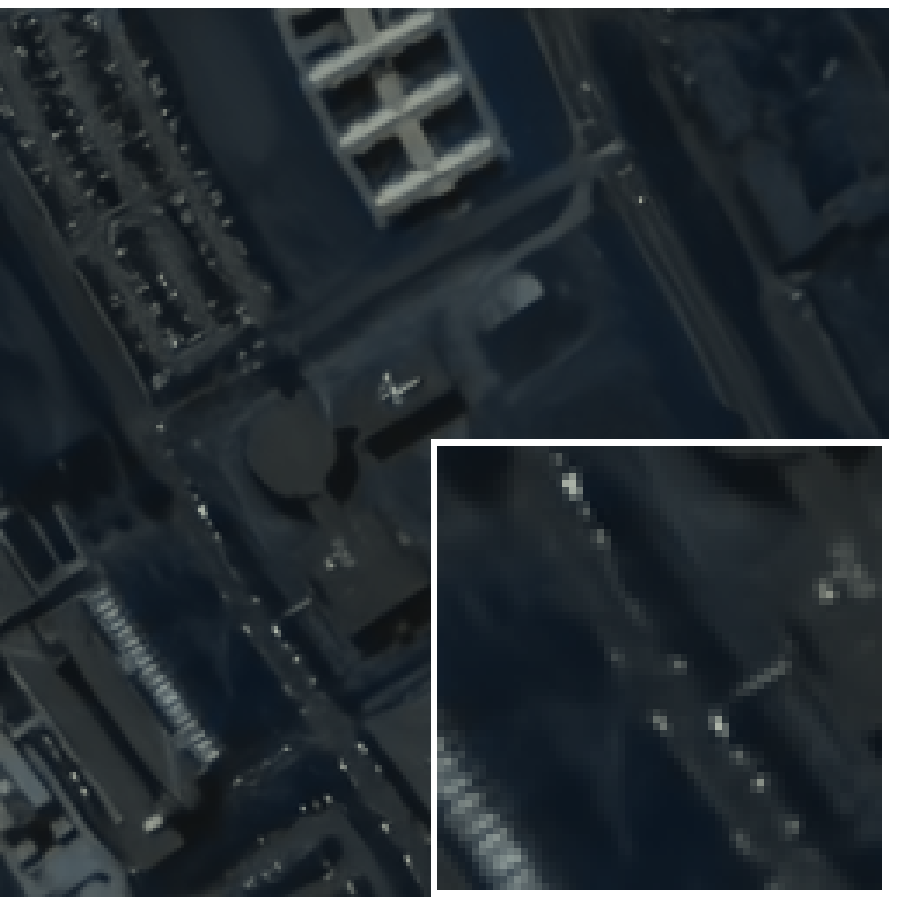}
\end{minipage}
  \begin{minipage}[t]{0.135\hsize}
\includegraphics[width=\hsize]{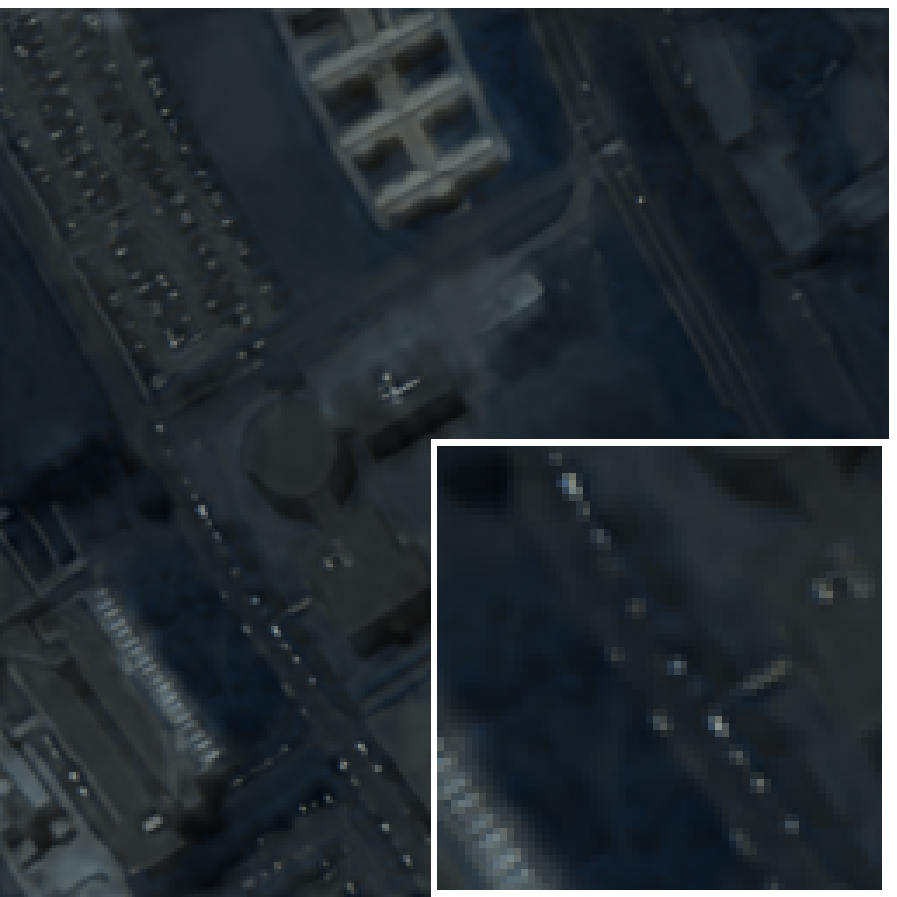}
\end{minipage}
  \begin{minipage}[t]{0.135\hsize}
\includegraphics[width=\hsize]{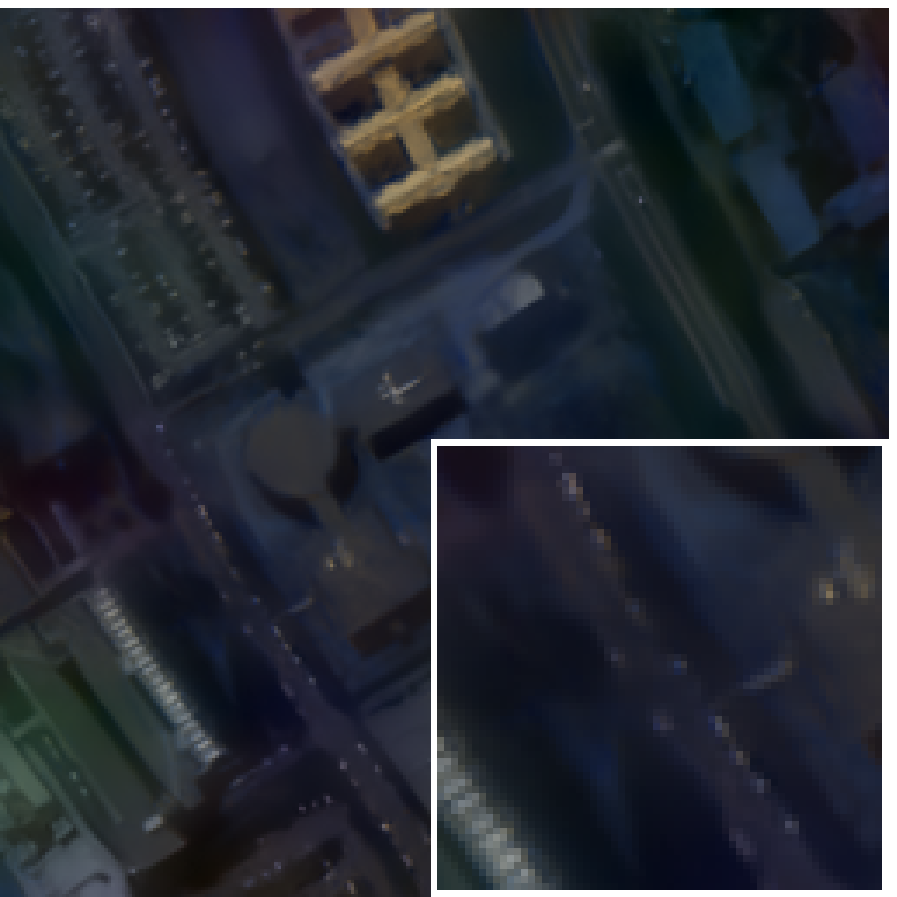}
\end{minipage}
\begin{minipage}[t]{0.135\hsize}
\includegraphics[width=\hsize]{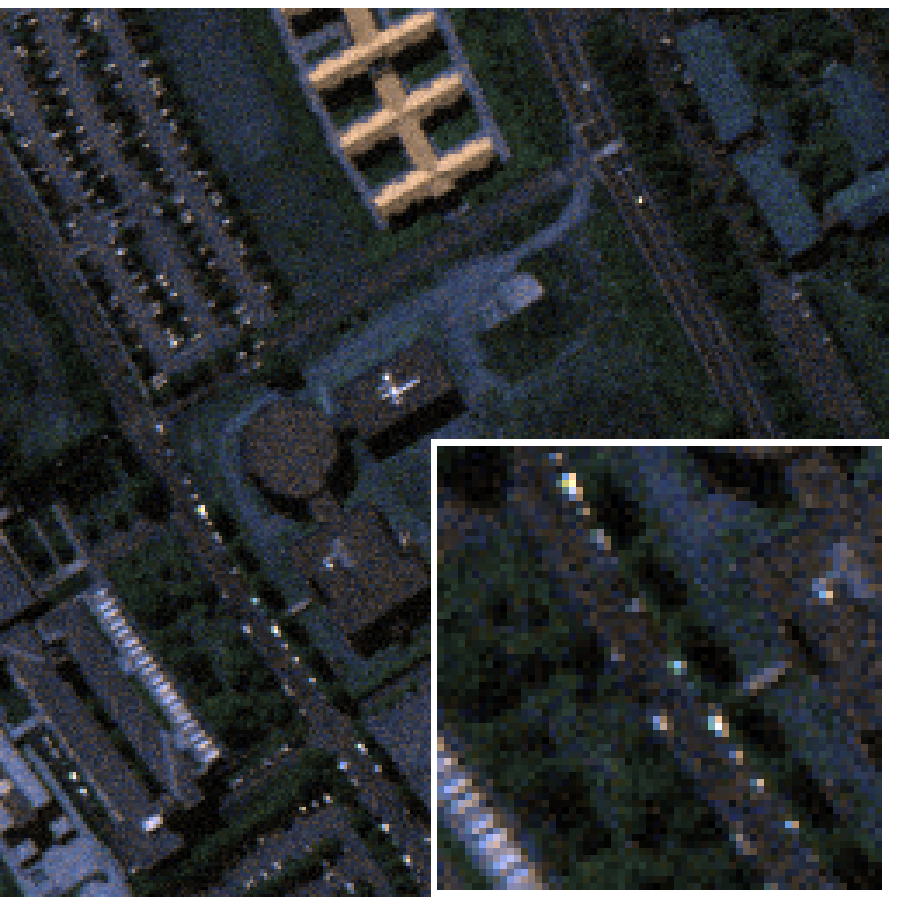}
\end{minipage}

  \begin{minipage}[t]{0.135\hsize}
\centerline{\footnotesize{ ground-truth }}
\end{minipage}
  \begin{minipage}[t]{0.135\hsize}
\centerline{\footnotesize{ observed HS image }}
\end{minipage}
\begin{minipage}[t]{0.135\hsize}
\centerline{\footnotesize{ observed MS image }}
\end{minipage}
  \begin{minipage}[t]{0.135\hsize}
\centerline{\footnotesize{ GSA }}
\end{minipage}
  \begin{minipage}[t]{0.135\hsize}
\centerline{\footnotesize{ MTF-GLP }}
\end{minipage}
  \begin{minipage}[t]{0.135\hsize}
\centerline{\footnotesize{ GFPCA }}
\end{minipage}
\begin{minipage}[t]{0.135\hsize}
\centerline{\footnotesize{ CNMF }}
\end{minipage}

\vspace{4pt}
  \begin{minipage}[t]{0.135\hsize}
\includegraphics[width=\hsize]{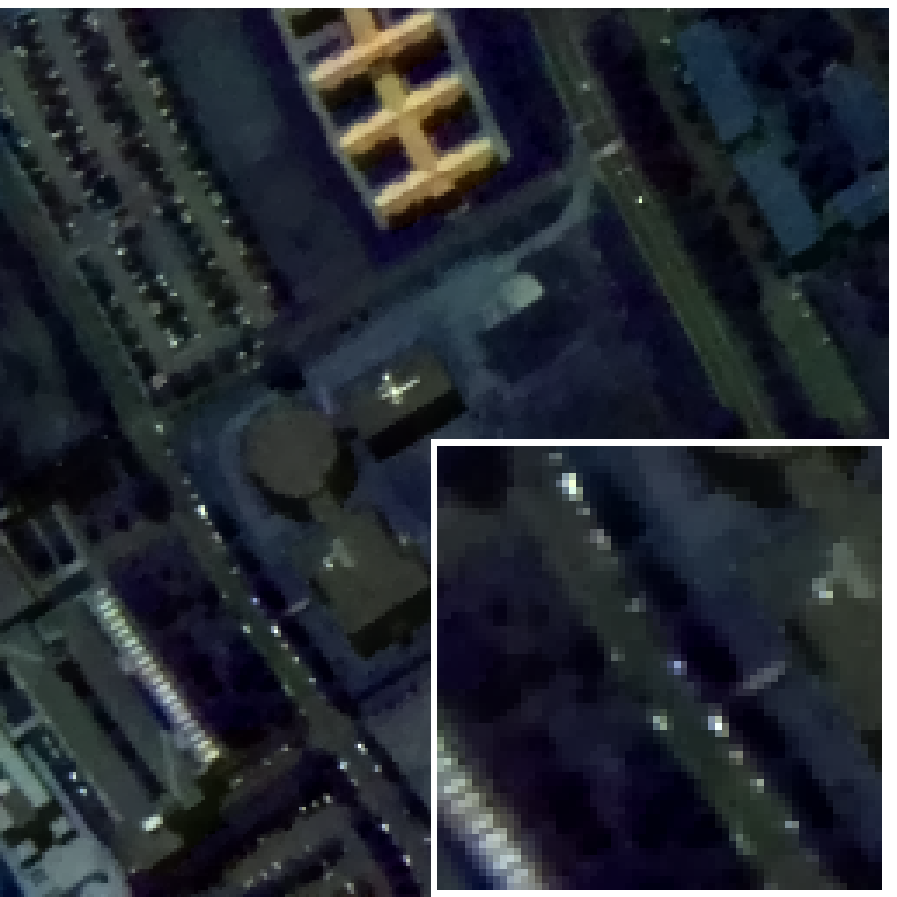}
\end{minipage}
  \begin{minipage}[t]{0.135\hsize}
\includegraphics[width=\hsize]{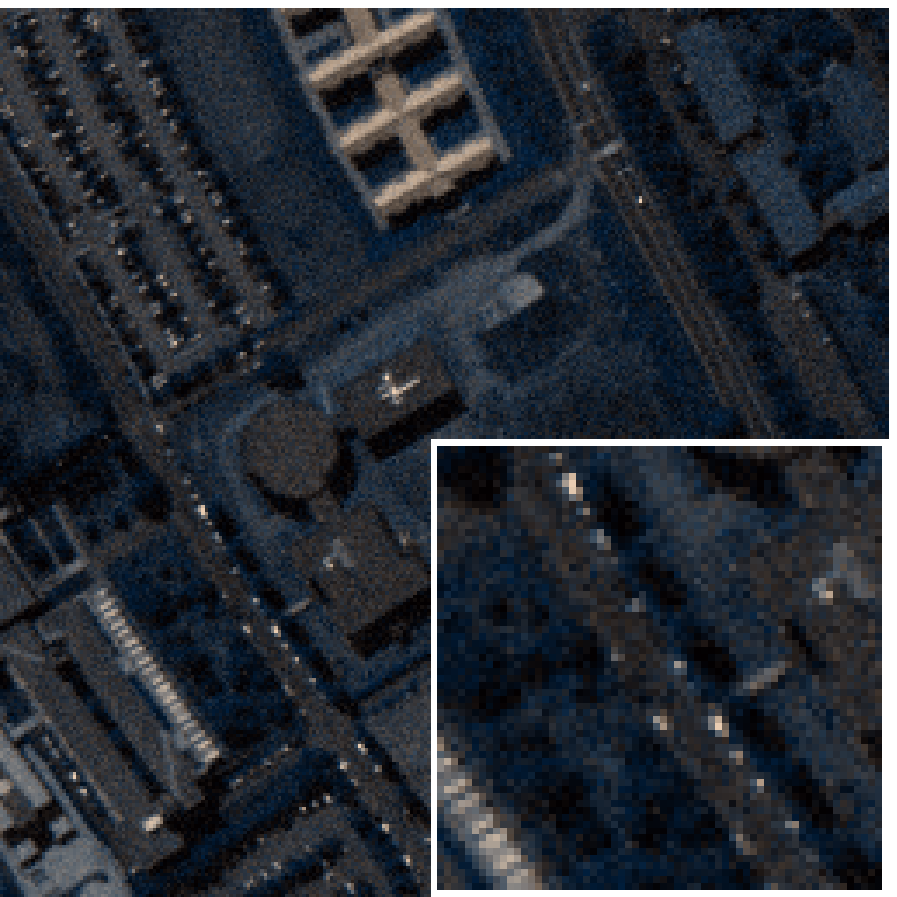}
\end{minipage}
  \begin{minipage}[t]{0.135\hsize}
\includegraphics[width=\hsize]{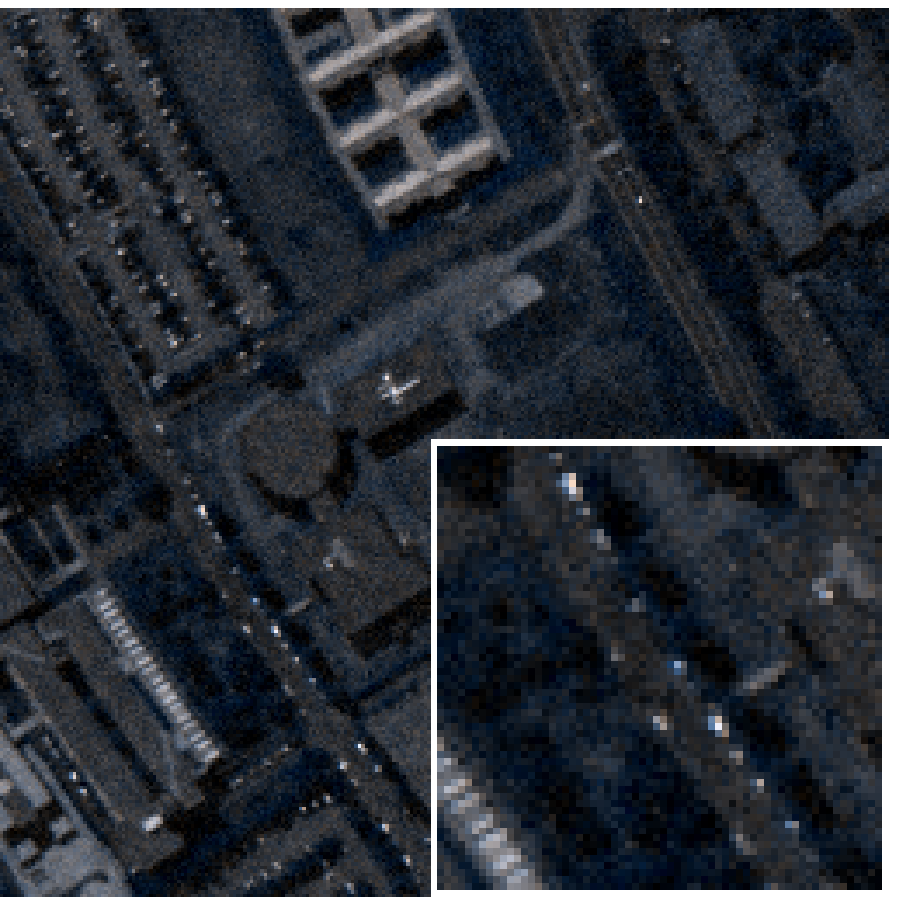}
\end{minipage}
\begin{minipage}[t]{0.135\hsize}
\includegraphics[width=\hsize]{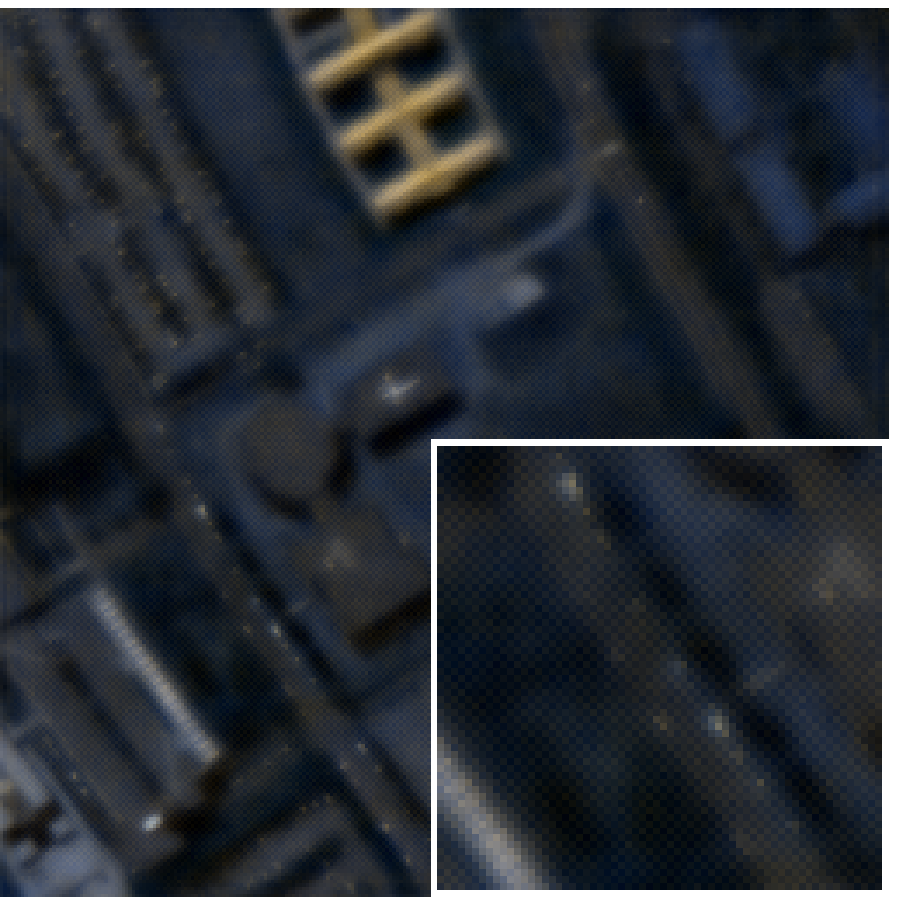}
\end{minipage}
  \begin{minipage}[t]{0.135\hsize}
\includegraphics[width=\hsize]{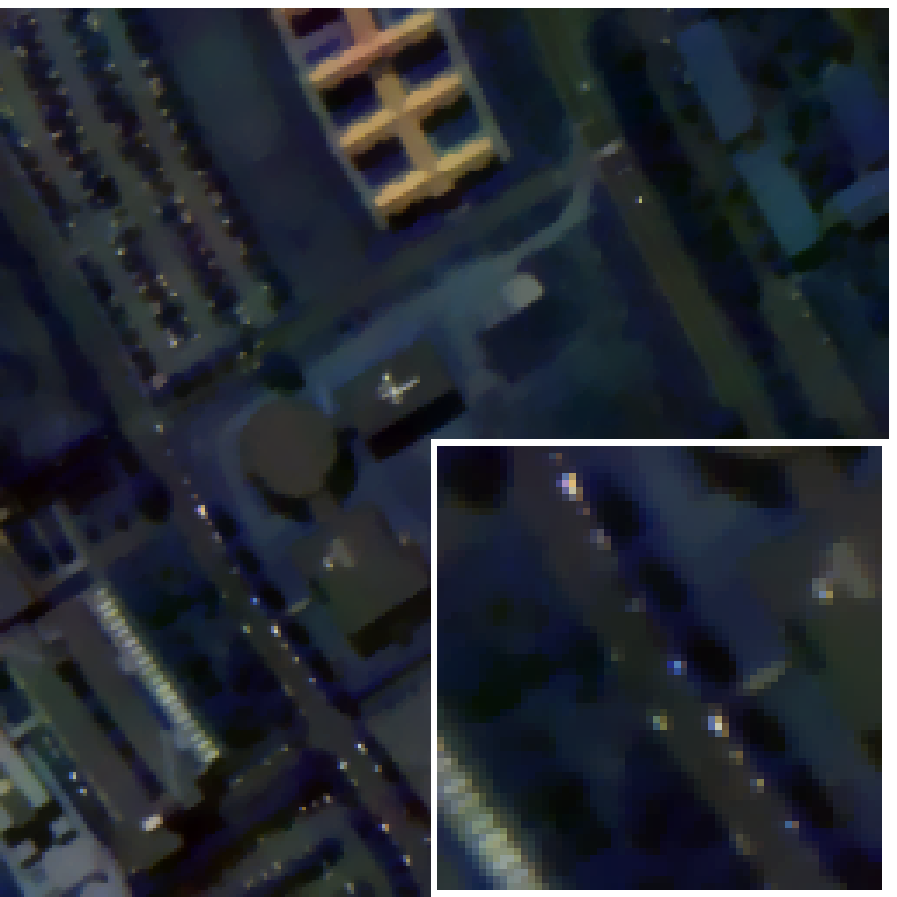}
\end{minipage}
\begin{minipage}[t]{0.135\hsize}
\includegraphics[width=\hsize]{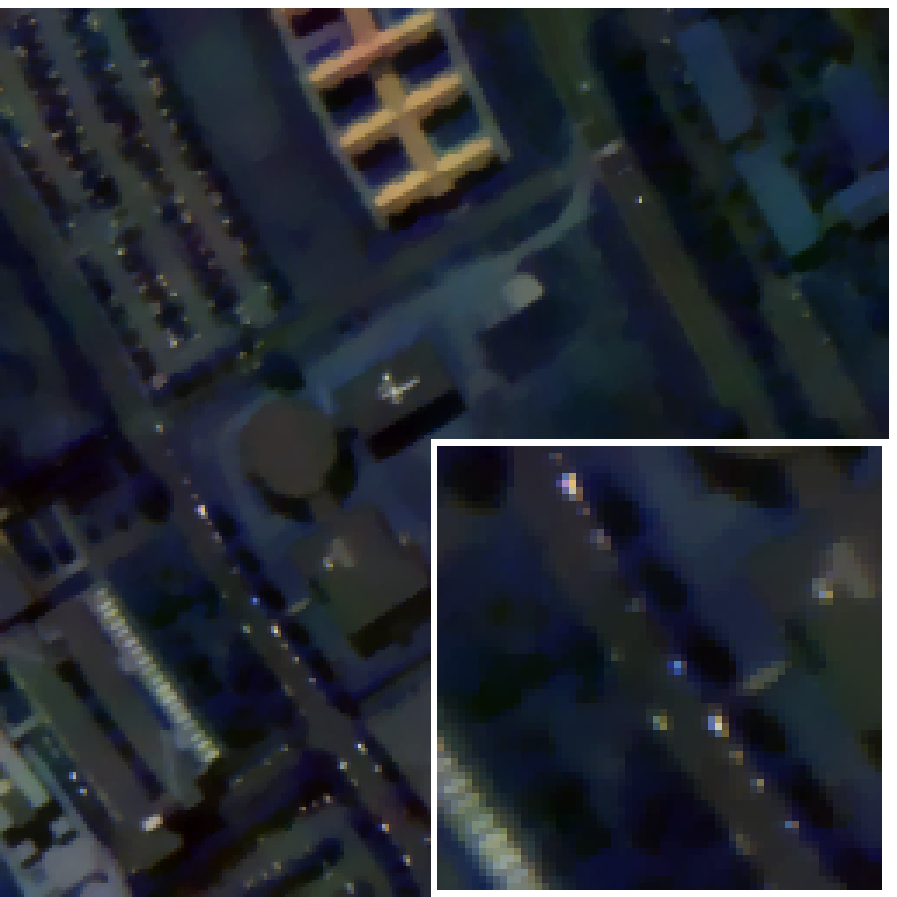}
\end{minipage}
  \begin{minipage}[t]{0.135\hsize}
\centerline{\footnotesize{  }}
\end{minipage}

  \begin{minipage}[t]{0.135\hsize}
\centerline{\footnotesize{ HySure }}
\end{minipage}
  \begin{minipage}[t]{0.135\hsize}
\centerline{\footnotesize{ Lanaras's }}
\end{minipage}
  \begin{minipage}[t]{0.135\hsize}
\centerline{\footnotesize{ LTMR }}
\end{minipage}
  \begin{minipage}[t]{0.135\hsize}
\centerline{\footnotesize{ CNN-Fus }}
\end{minipage}
  \begin{minipage}[t]{0.135\hsize}
\centerline{\footnotesize{ \bf{proposed ($p = 1$)} }}
\end{minipage}
  \begin{minipage}[t]{0.135\hsize}
\centerline{\footnotesize{ \bf{proposed ($p = 2$)} }}
\end{minipage}
  \begin{minipage}[t]{0.135\hsize}
\centerline{\footnotesize{  }}
\end{minipage}

   \caption{Results on HS and MS image fusion experiments (top: Reno, $r = 2$, and $\sigma_{\g} = 0.1$, bottom: Pavia U, $r = 4$, and $\sigma_{\g} = 0.05$).}
 \label{fig:img_MSfusion}
\end{center}
\end{figure*}

\begin{figure*}[t]
\begin{center}
	
	\begin{minipage}[t]{0.24\hsize}
	\includegraphics[width=1.0\hsize]{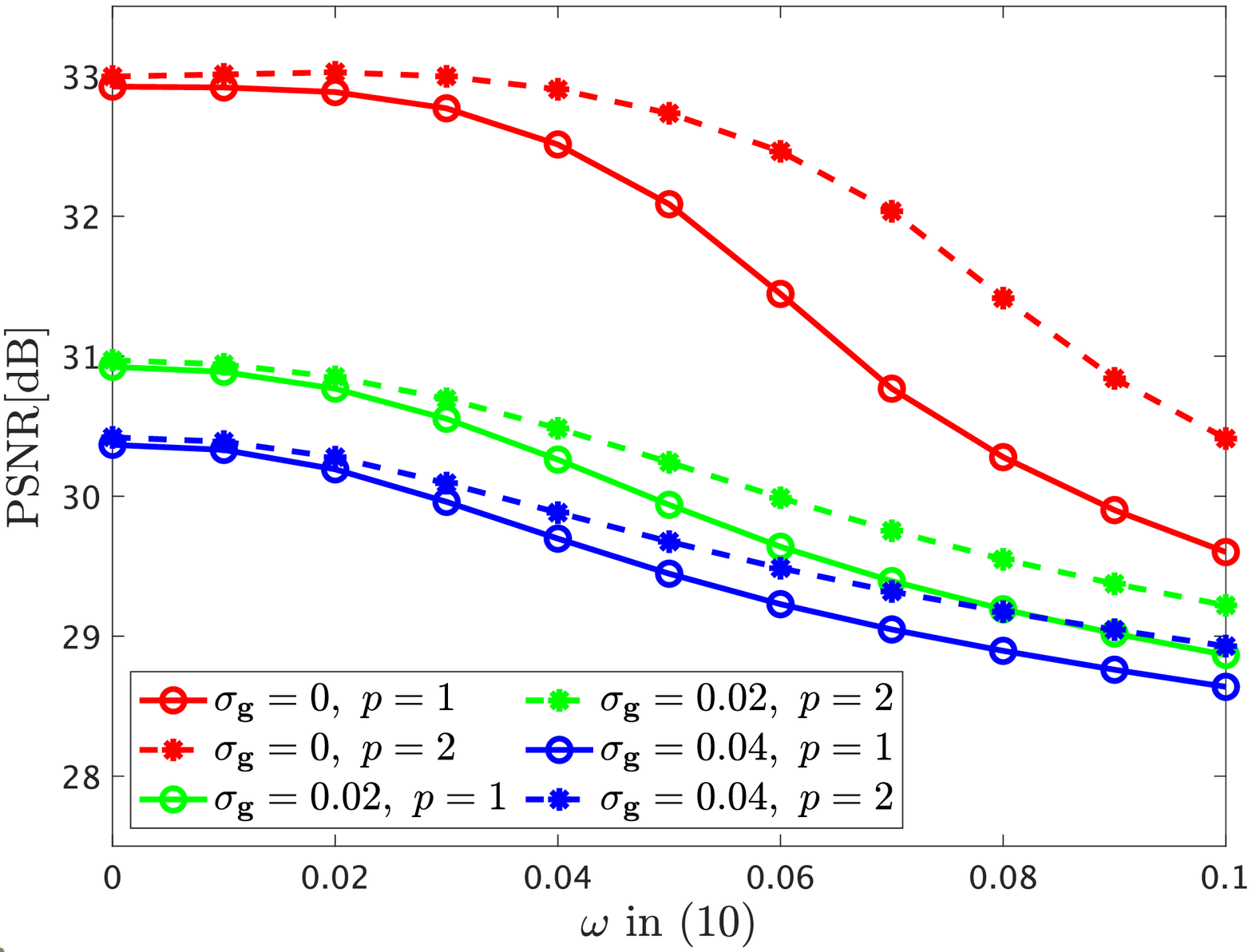}
	\end{minipage}
	\begin{minipage}[t]{0.24\hsize}
	\includegraphics[width=1.0\hsize]{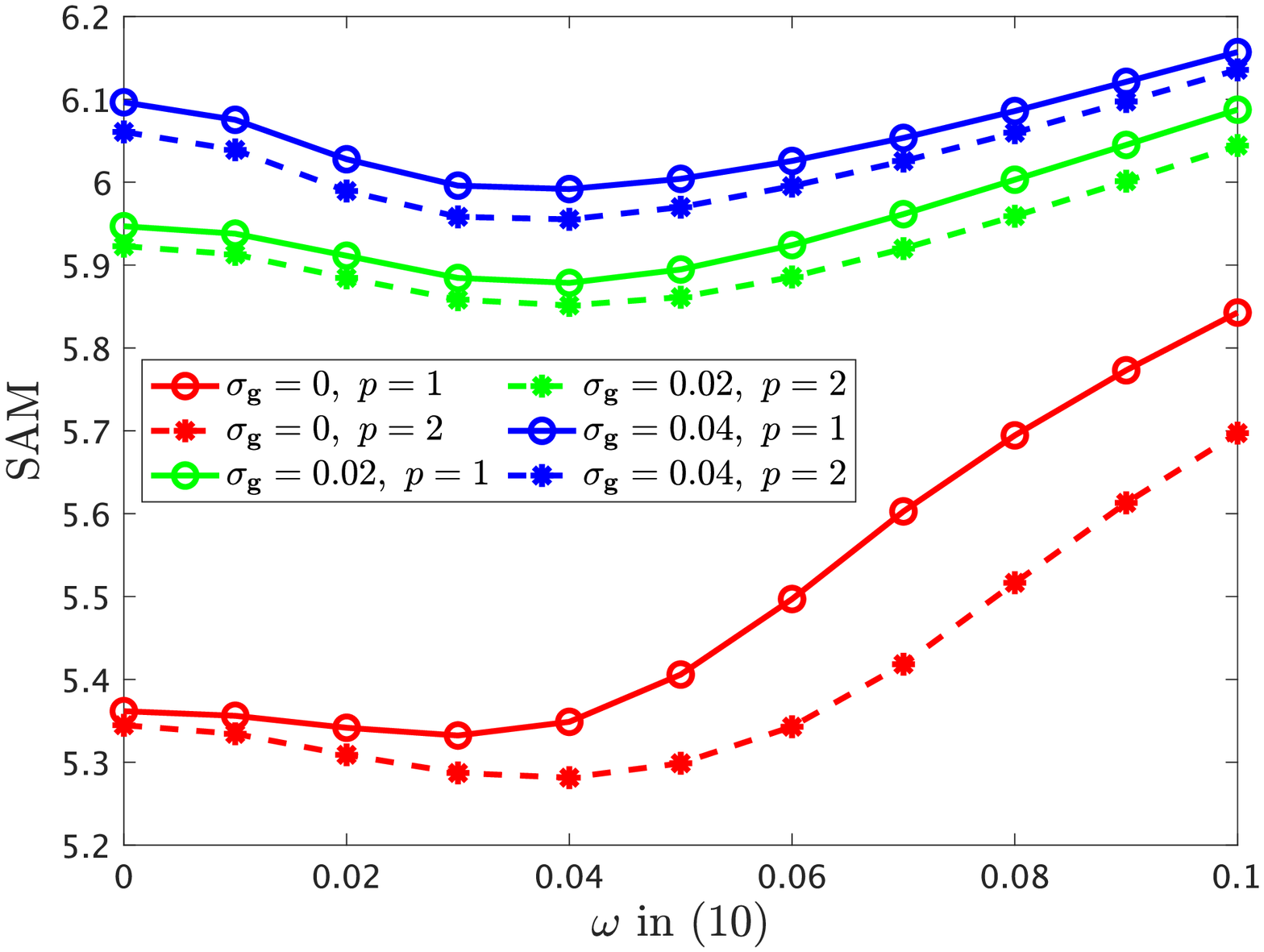}
	\end{minipage}
	\begin{minipage}[t]{0.24\hsize}
	\includegraphics[width=1.0\hsize]{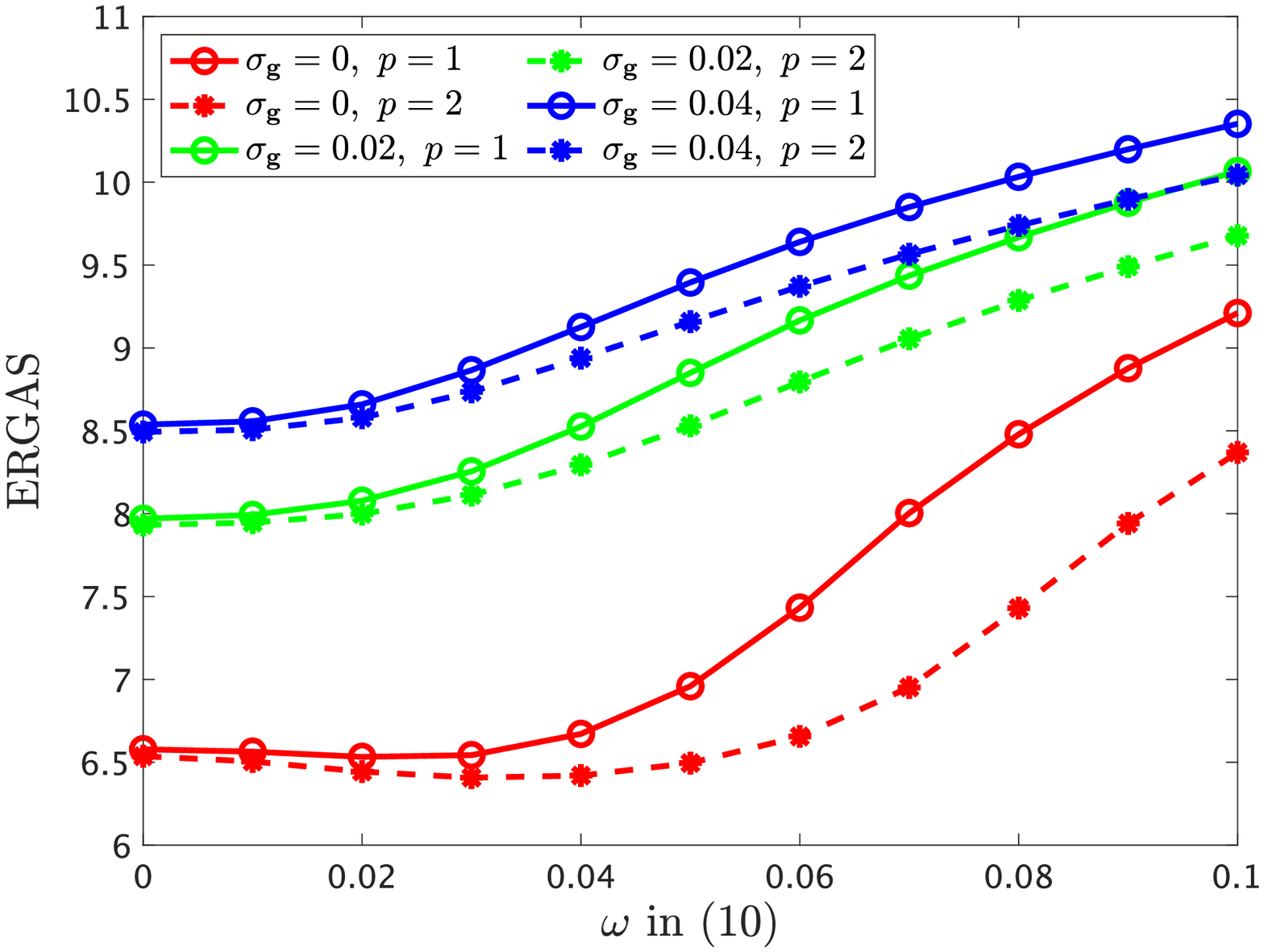}
	\end{minipage}
	\begin{minipage}[t]{0.24\hsize}
	\includegraphics[width=1.0\hsize]{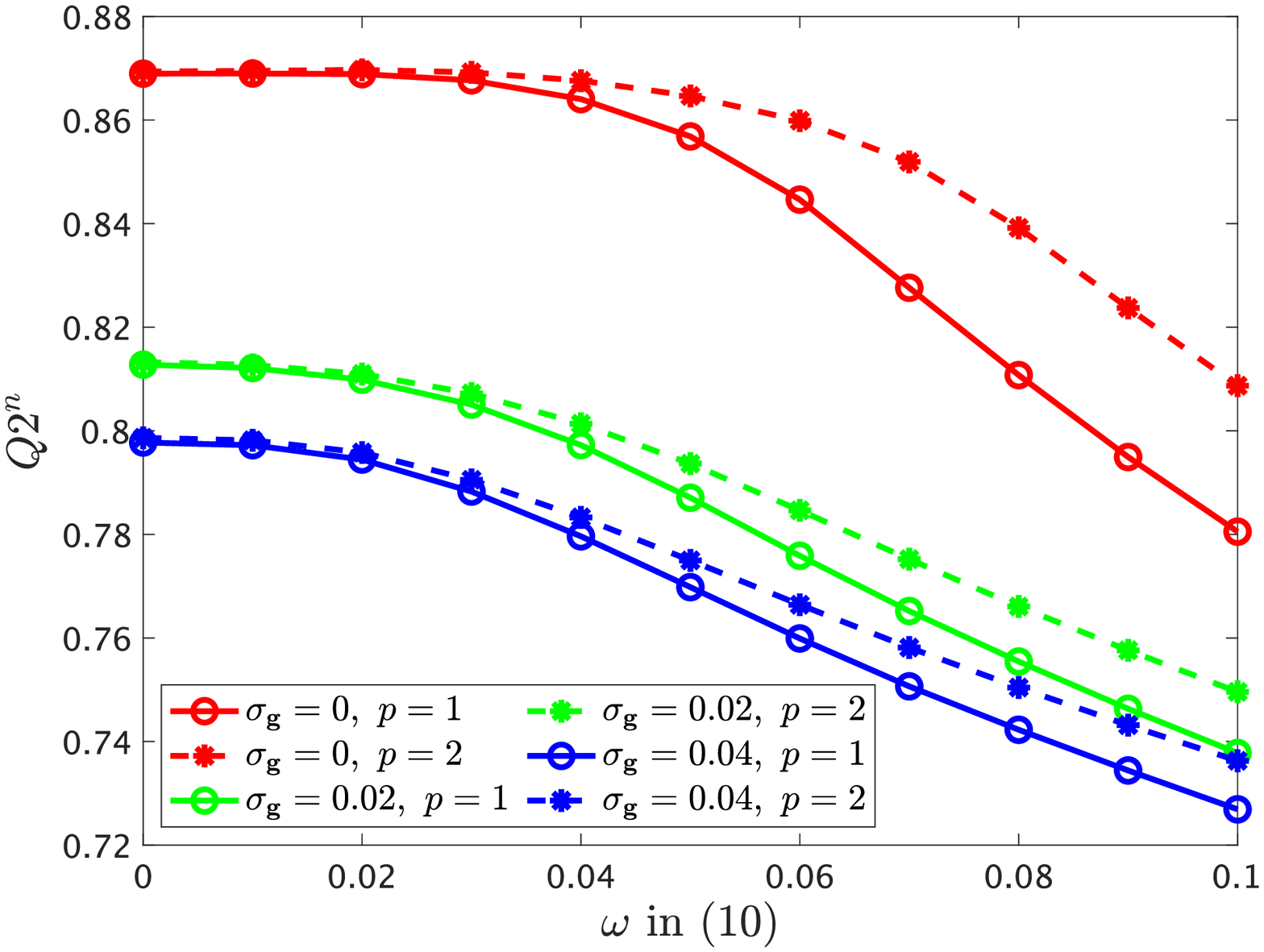}
	\end{minipage}
	
	
	\begin{minipage}[t]{0.24\hsize}
	\includegraphics[width=1.0\hsize]{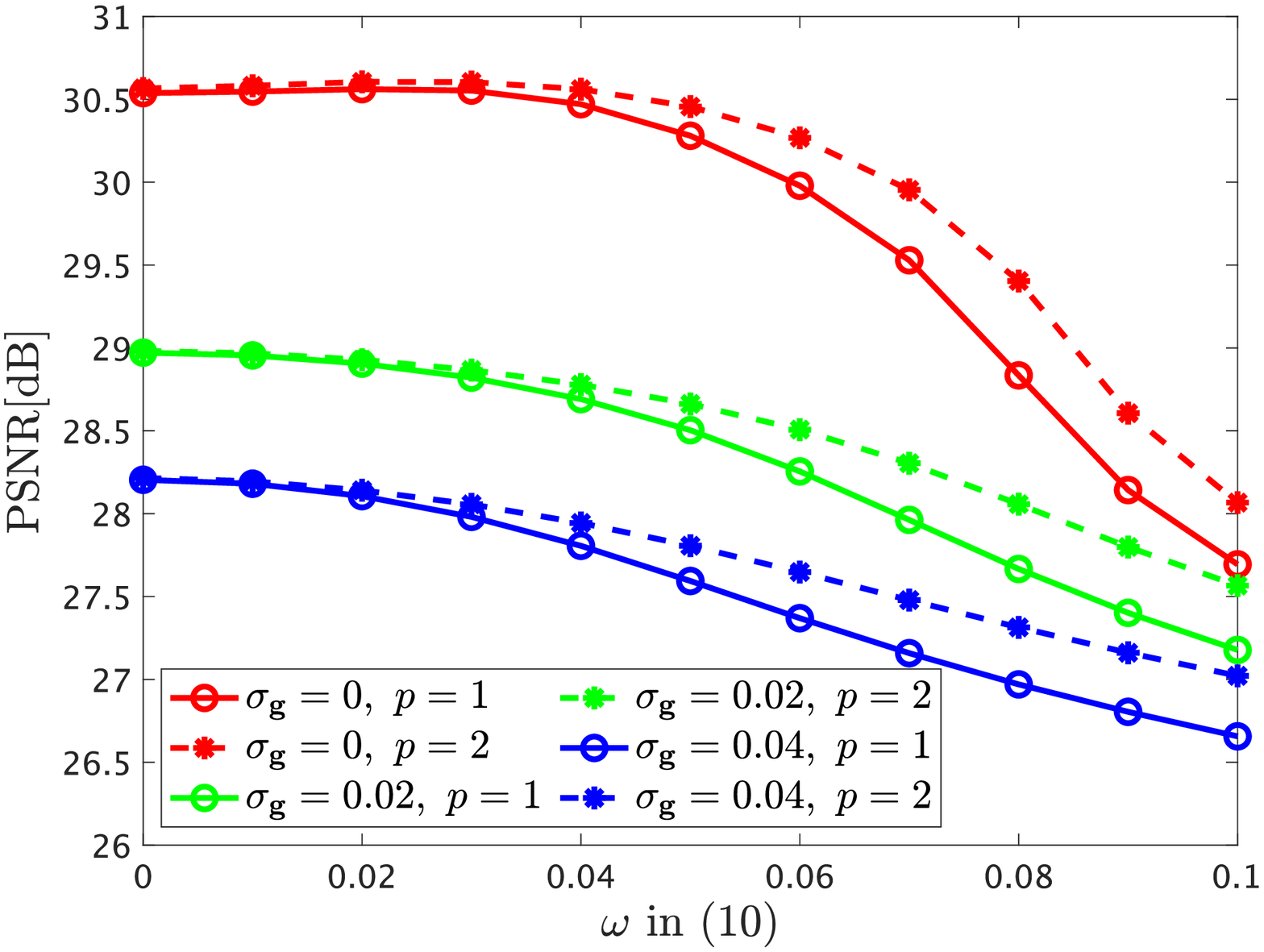}
	\end{minipage}
	\begin{minipage}[t]{0.24\hsize}
	\includegraphics[width=1.0\hsize]{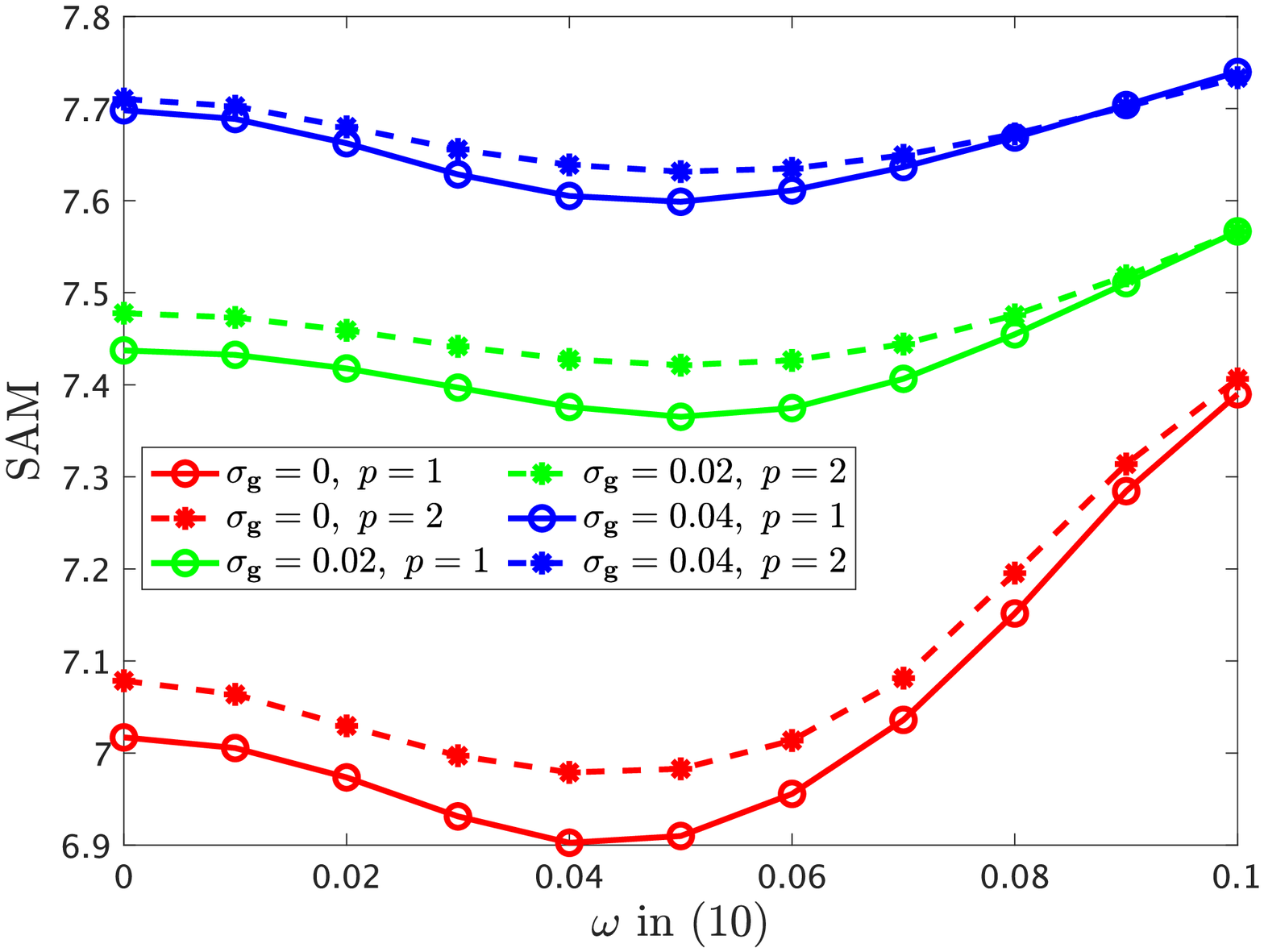}
	\end{minipage}
	\begin{minipage}[t]{0.24\hsize}
	\includegraphics[width=1.0\hsize]{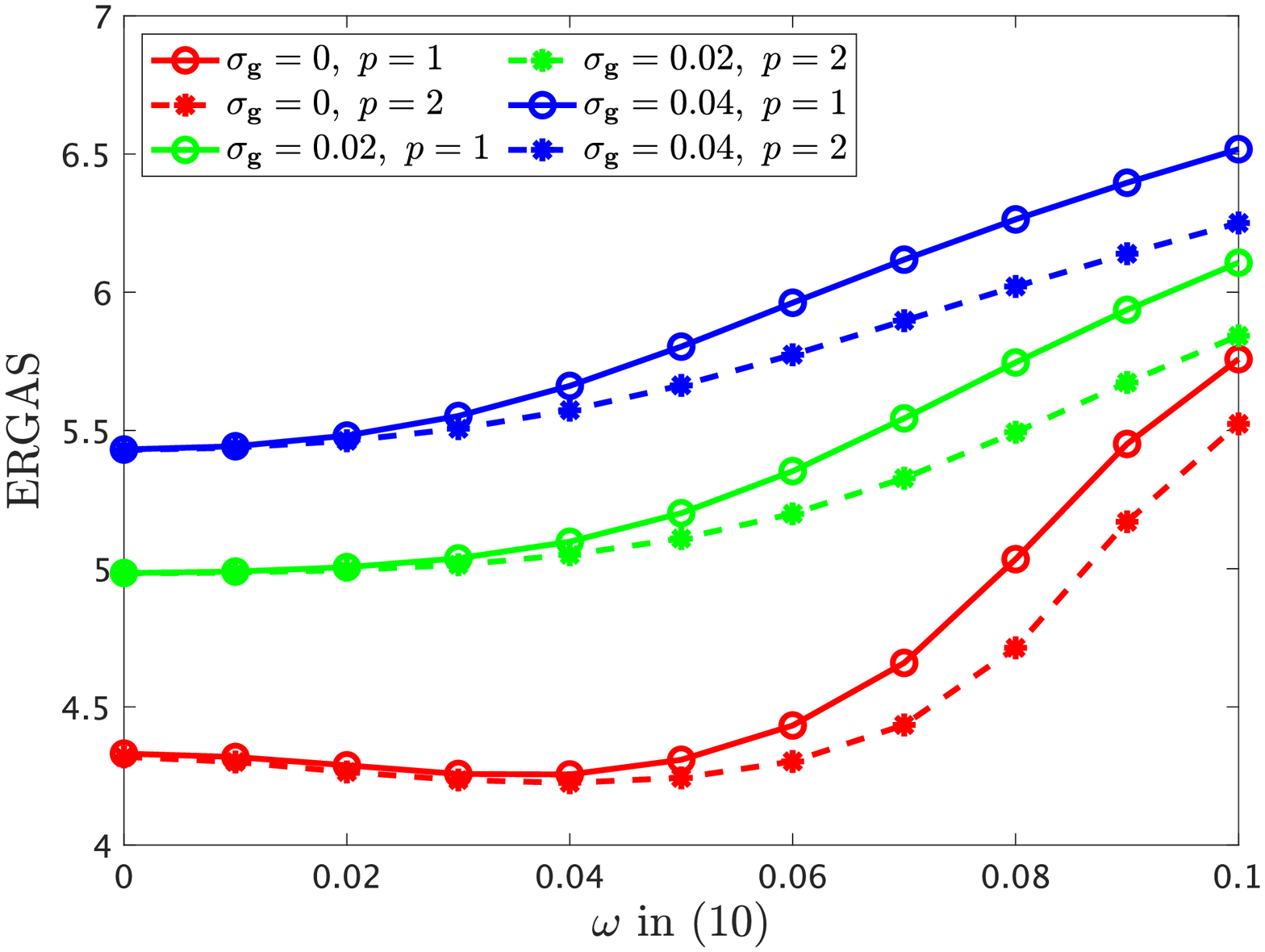}
	\end{minipage}
	\begin{minipage}[t]{0.24\hsize}
	\includegraphics[width=1.0\hsize]{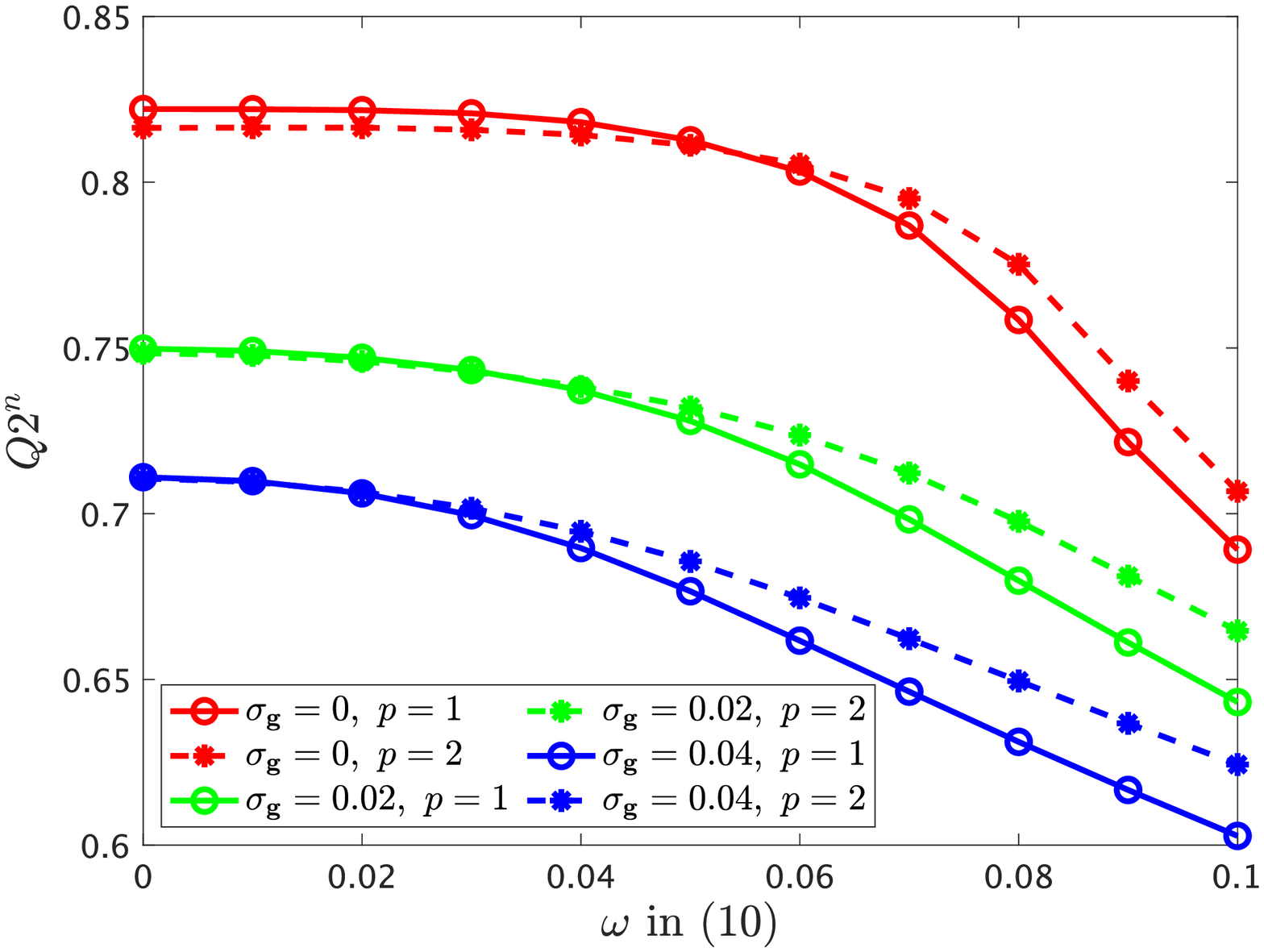}
	\end{minipage}
	
	\vspace{-3pt}
	\begin{minipage}[t]{0.24\hsize}
	\centerline{\footnotesize{PSNR[dB]}}
	\end{minipage}
	\begin{minipage}[t]{0.24\hsize}
	\centerline{\footnotesize{SAM}}
	\end{minipage}
	\begin{minipage}[t]{0.24\hsize}
	\centerline{\footnotesize{ERGAS}}
	\end{minipage}
	\begin{minipage}[t]{0.24\hsize}
	\centerline{\footnotesize{$Q2^n$}}
	\end{minipage}
	\caption{Four quality measures versus $\omega$ in \eqref{def:HSSTV} on HS and MS image fusion (top: $r = 2$, bottom: $r = 4$).}
	\label{omega_graph_MSfusion}
\end{center}
\end{figure*}

\begin{figure*}[t]
\begin{center}
	\begin{minipage}[t]{0.24\hsize}
	\includegraphics[width=1.0\hsize]{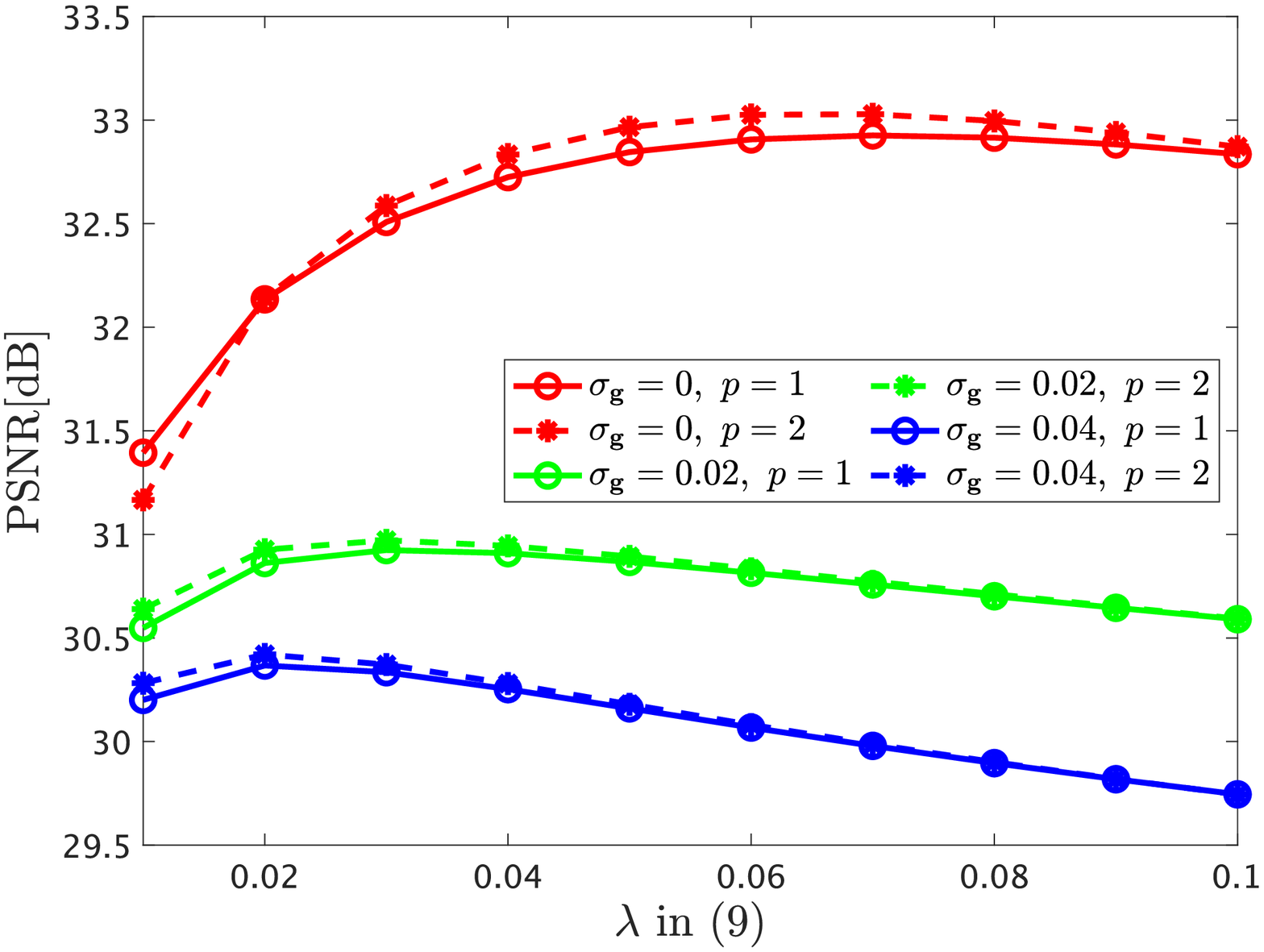}
	\end{minipage}
	\begin{minipage}[t]{0.24\hsize}
	\includegraphics[width=1.0\hsize]{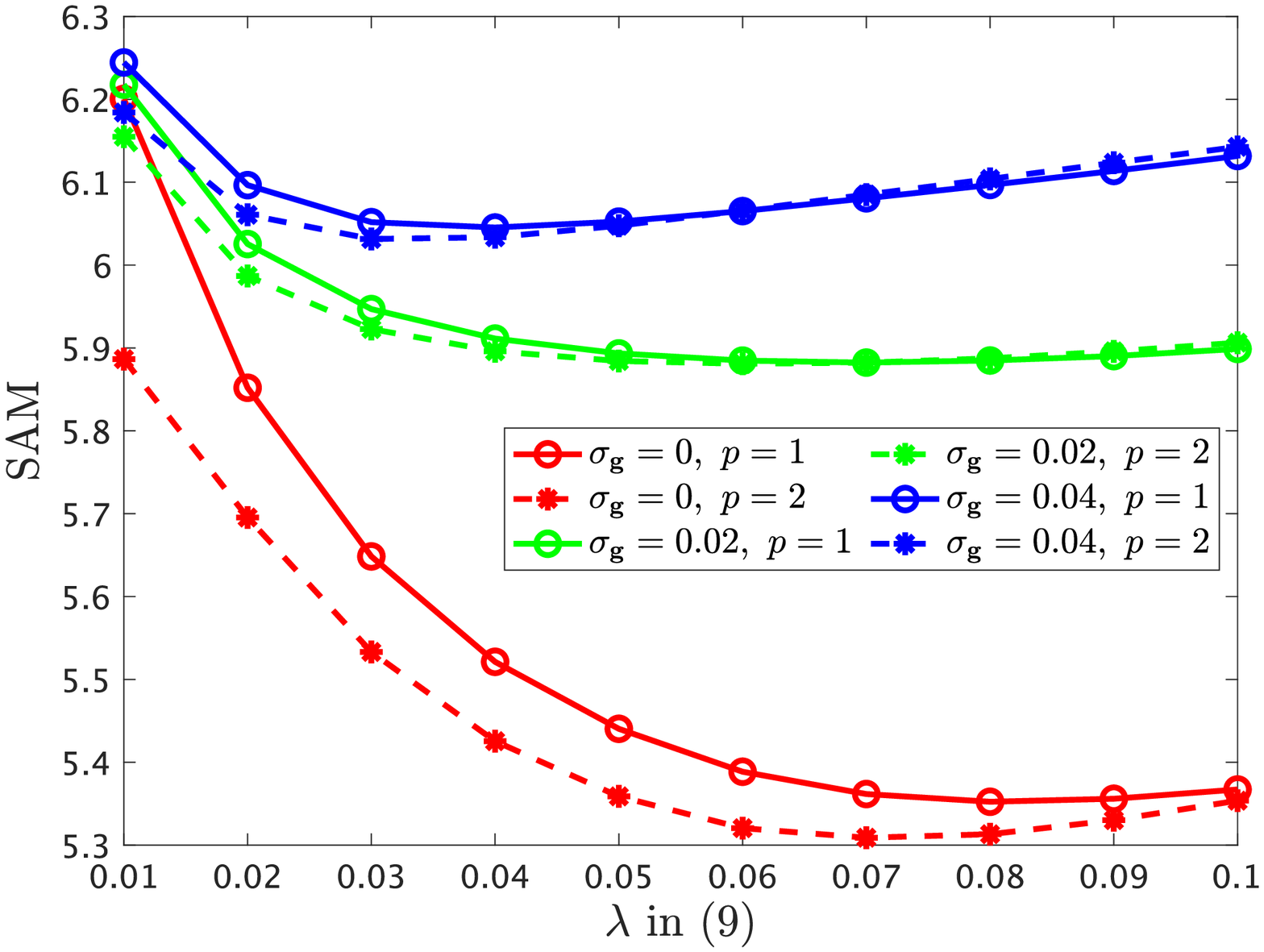}
	\end{minipage}
	\begin{minipage}[t]{0.24\hsize}
	\includegraphics[width=1.0\hsize]{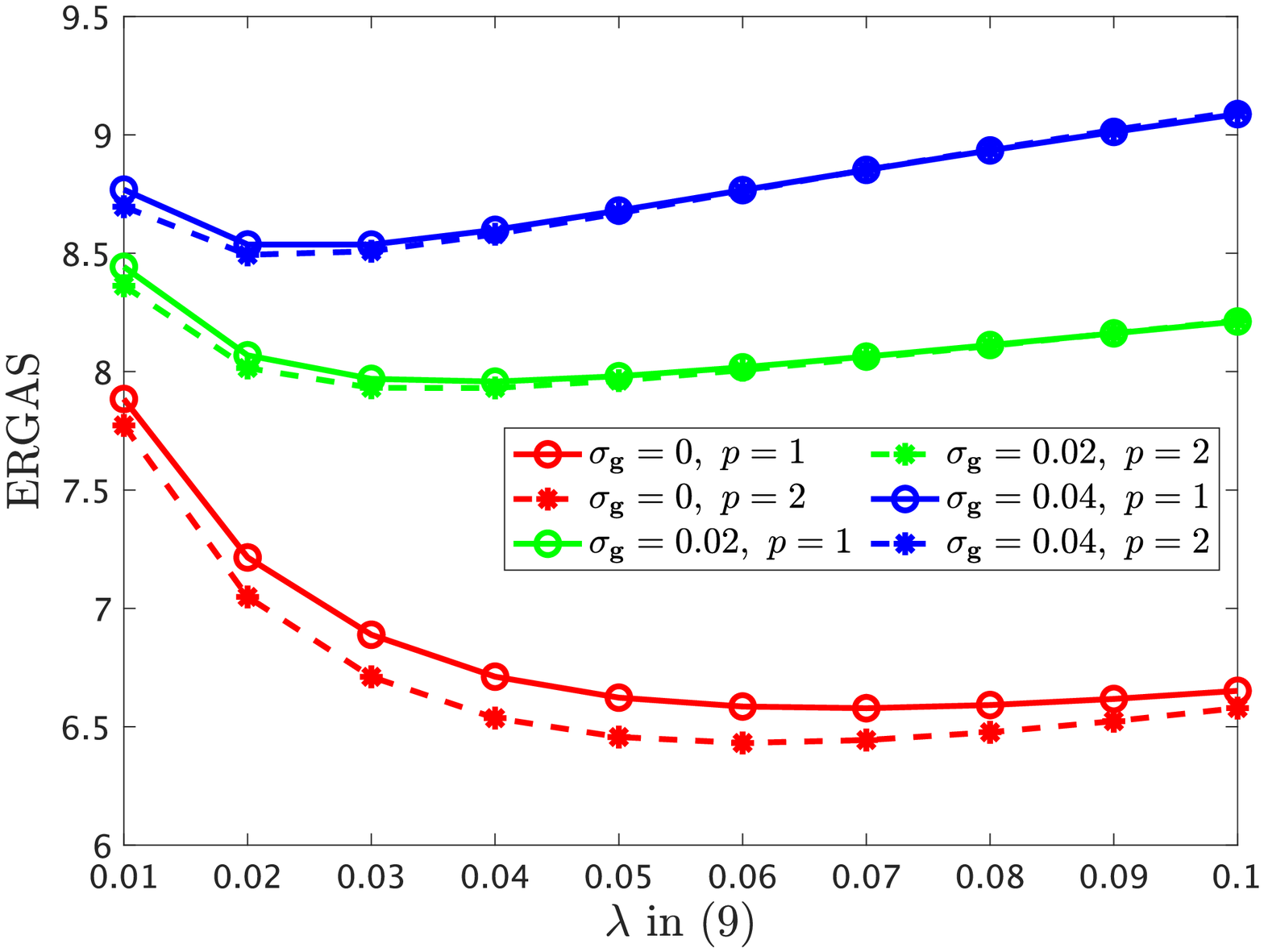}
	\end{minipage}
	\begin{minipage}[t]{0.24\hsize}
	\includegraphics[width=1.0\hsize]{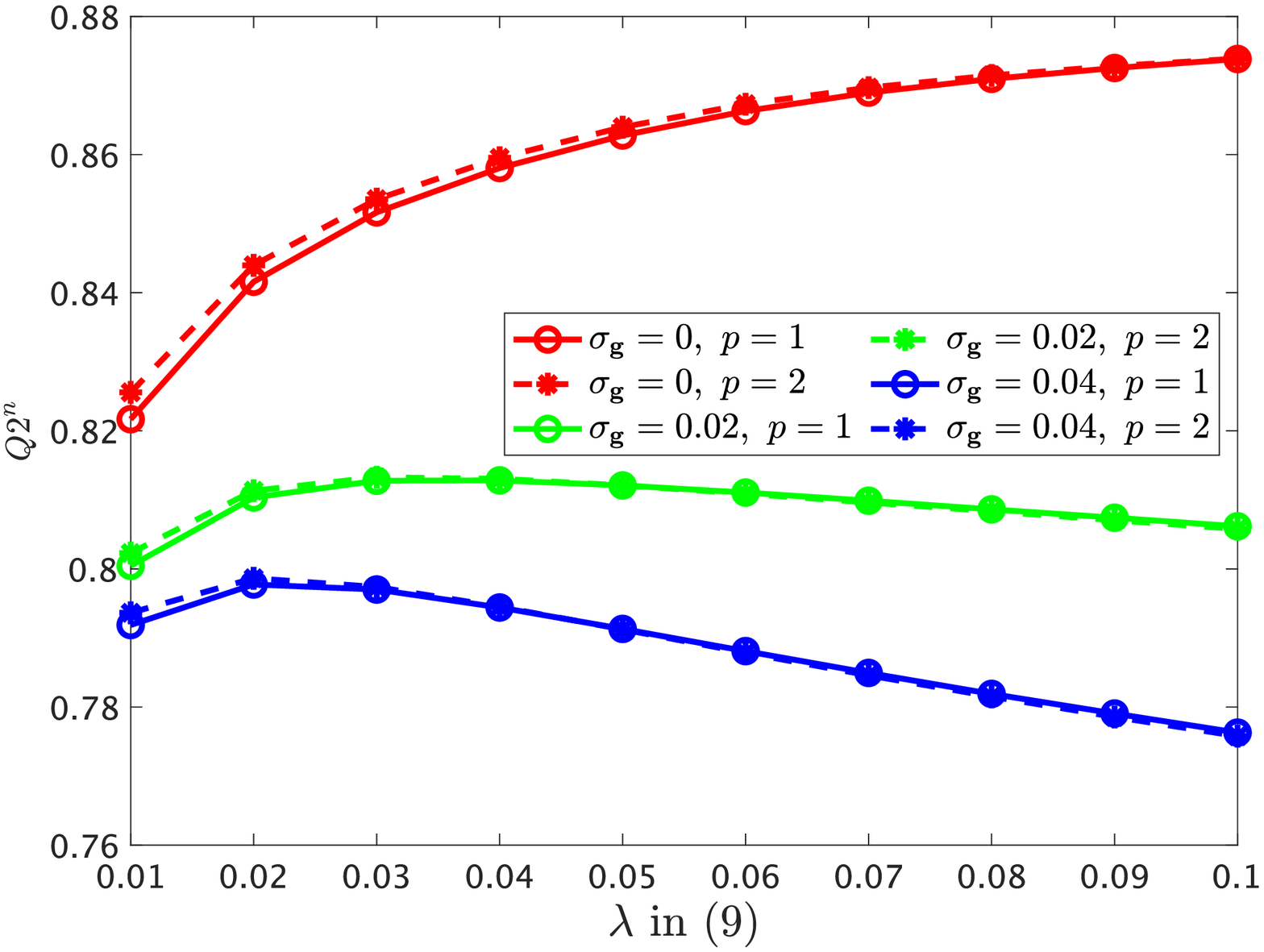}
	\end{minipage}
	
	
	
	\begin{minipage}[t]{0.24\hsize}
	\includegraphics[width=1.0\hsize]{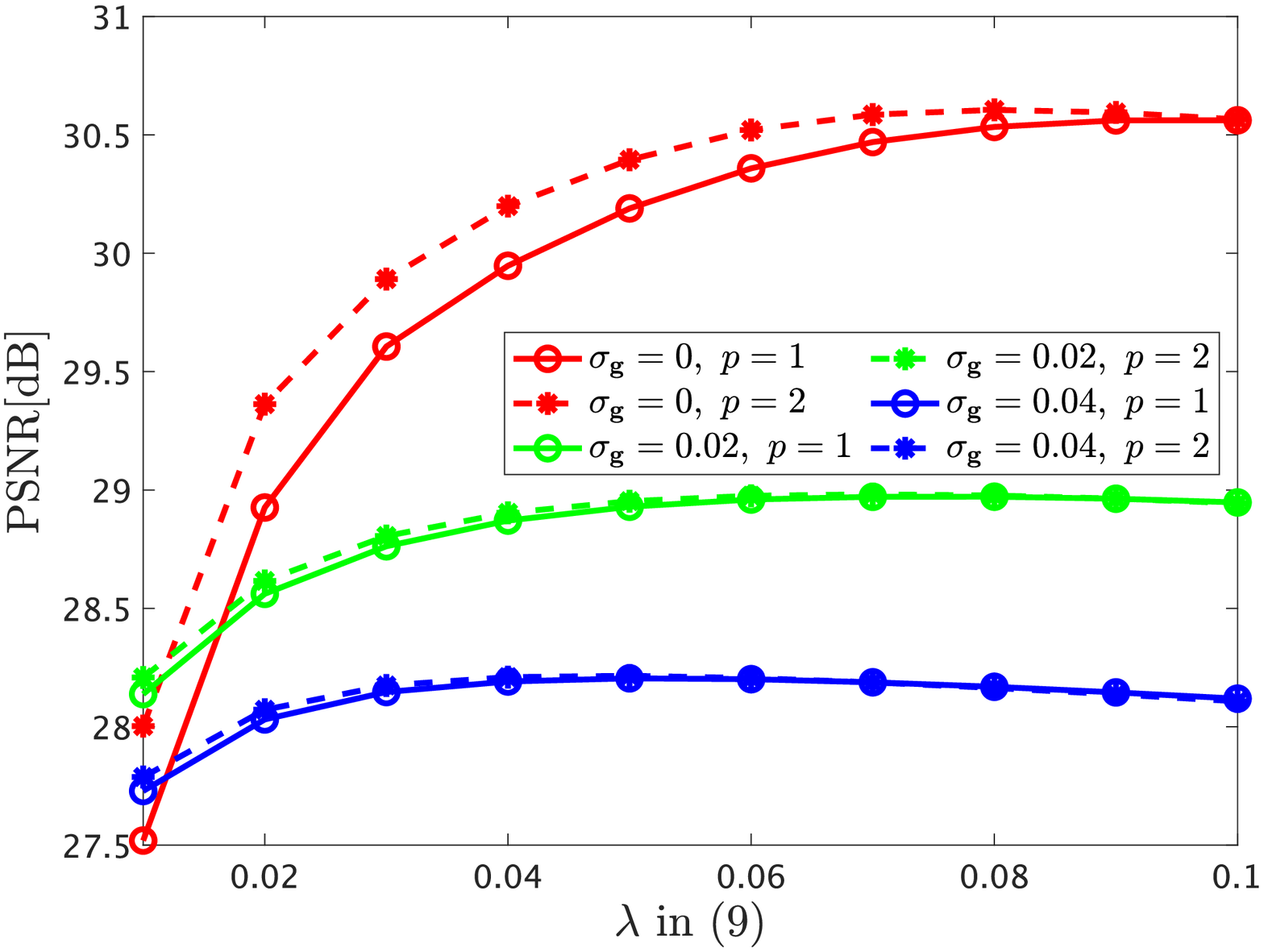}
	\end{minipage}
	\begin{minipage}[t]{0.24\hsize}
	\includegraphics[width=1.0\hsize]{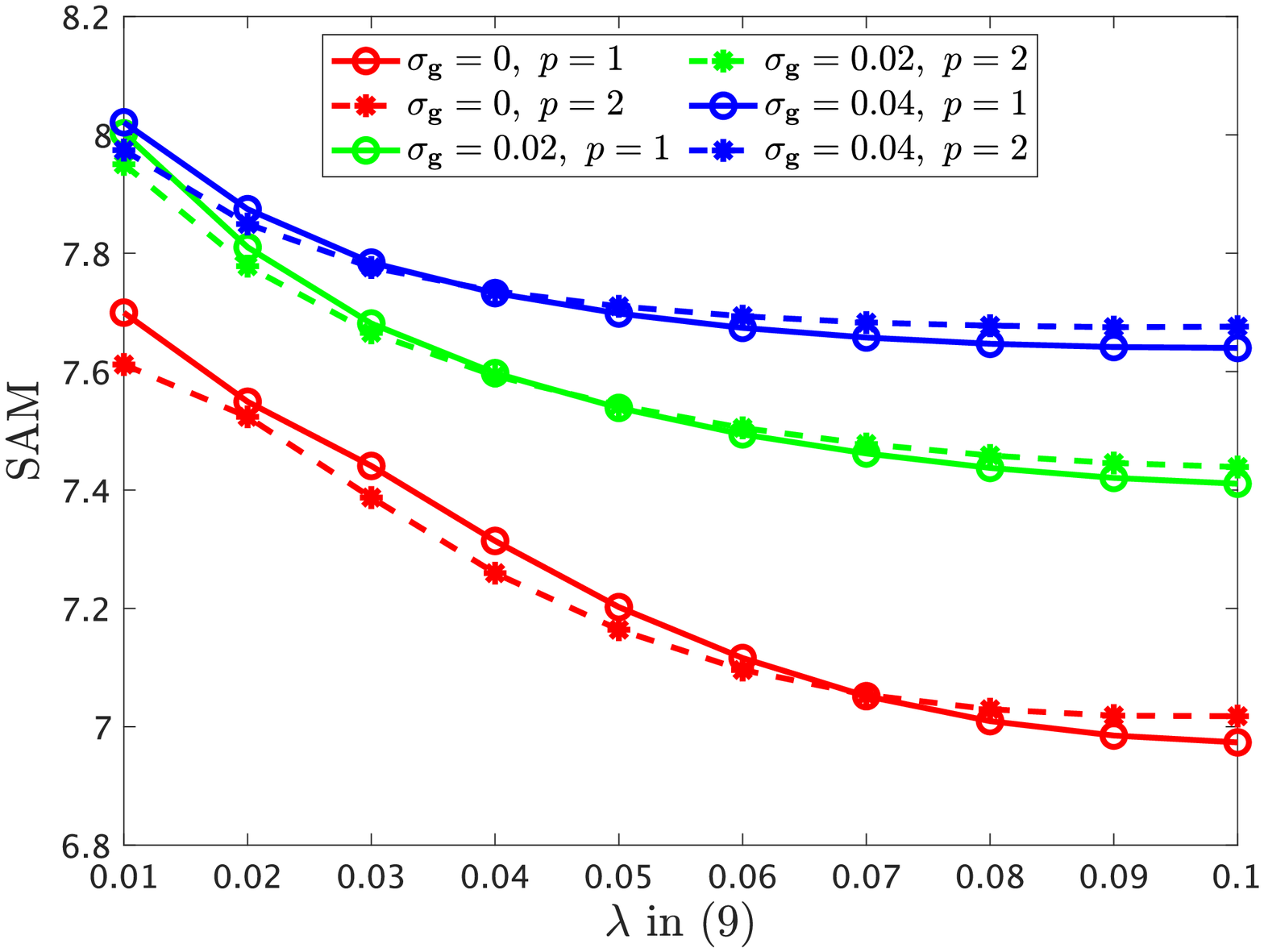}
	\end{minipage}
	\begin{minipage}[t]{0.24\hsize}
	\includegraphics[width=1.0\hsize]{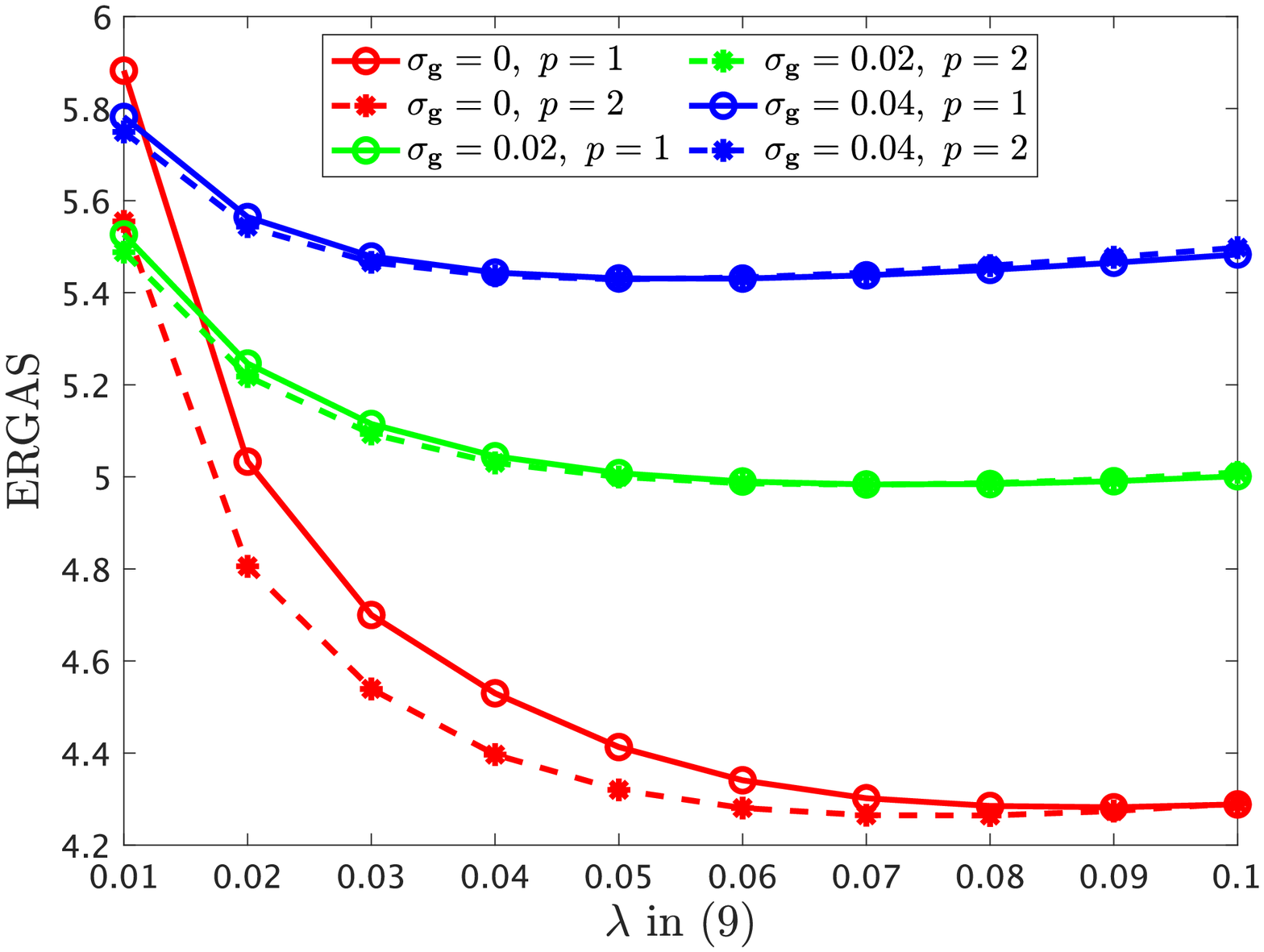}
	\end{minipage}
	\begin{minipage}[t]{0.24\hsize}
	\includegraphics[width=1.0\hsize]{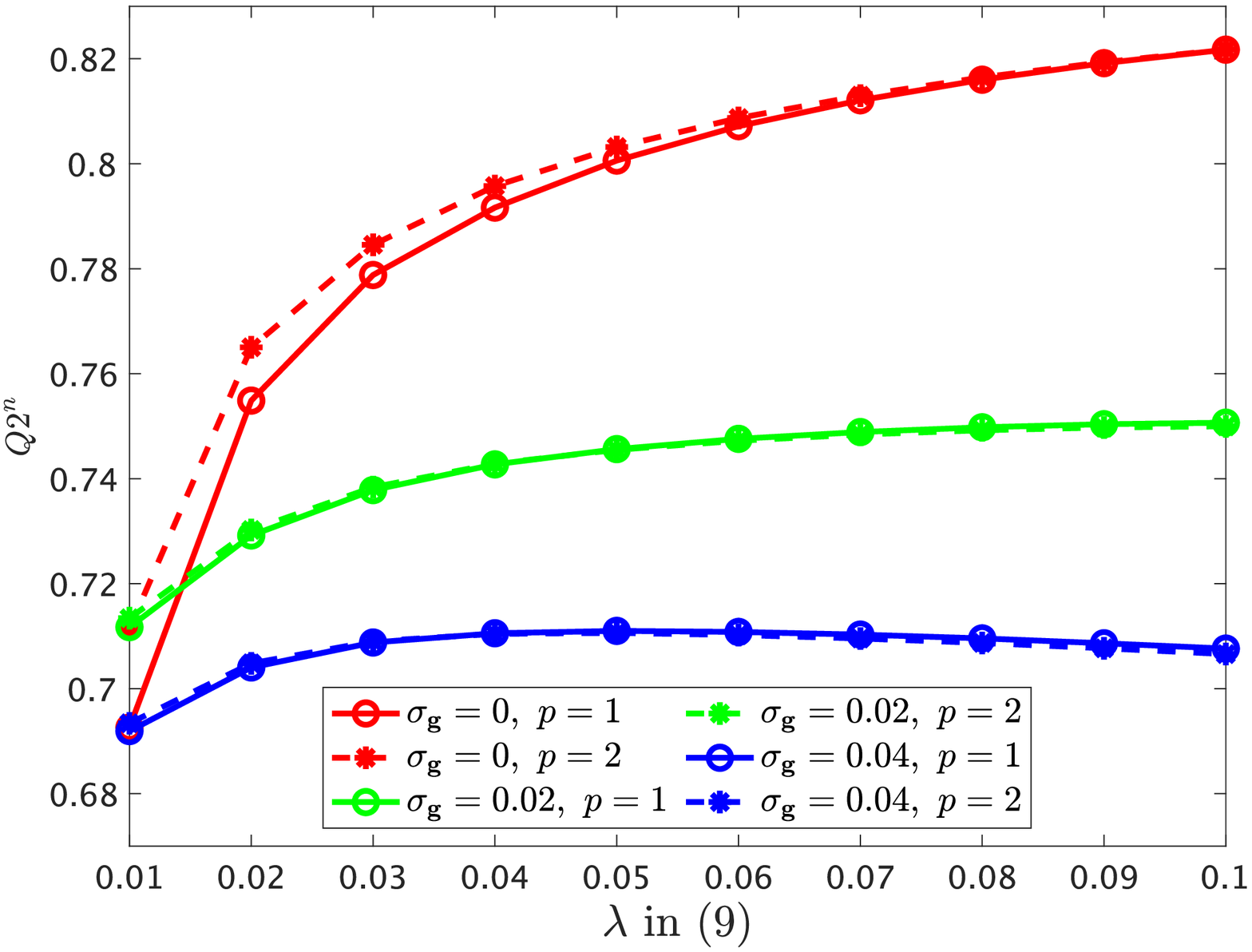}
	\end{minipage}
	
	\vspace{-3pt}
	\begin{minipage}[t]{0.24\hsize}
	\centerline{\footnotesize{PSNR[dB]}}
	\end{minipage}
	\begin{minipage}[t]{0.24\hsize}
	\centerline{\footnotesize{SAM}}
	\end{minipage}
	\begin{minipage}[t]{0.24\hsize}
	\centerline{\footnotesize{ERGAS}}
	\end{minipage}
	\begin{minipage}[t]{0.24\hsize}
	\centerline{\footnotesize{$Q2^n$}}
	\end{minipage}
	\caption{Four quality measures versus $\lambda$ in \eqref{prob:HRHSestimation} on HS and MS image fusion (top: $r = 2$, bottom: $r = 4$).}
	\label{lambda_graph_MSfusion}
\end{center}
\end{figure*}

\begin{figure}[t]
\begin{center}
	\begin{minipage}[t]{0.48\hsize}
	\includegraphics[width=1.0\hsize]{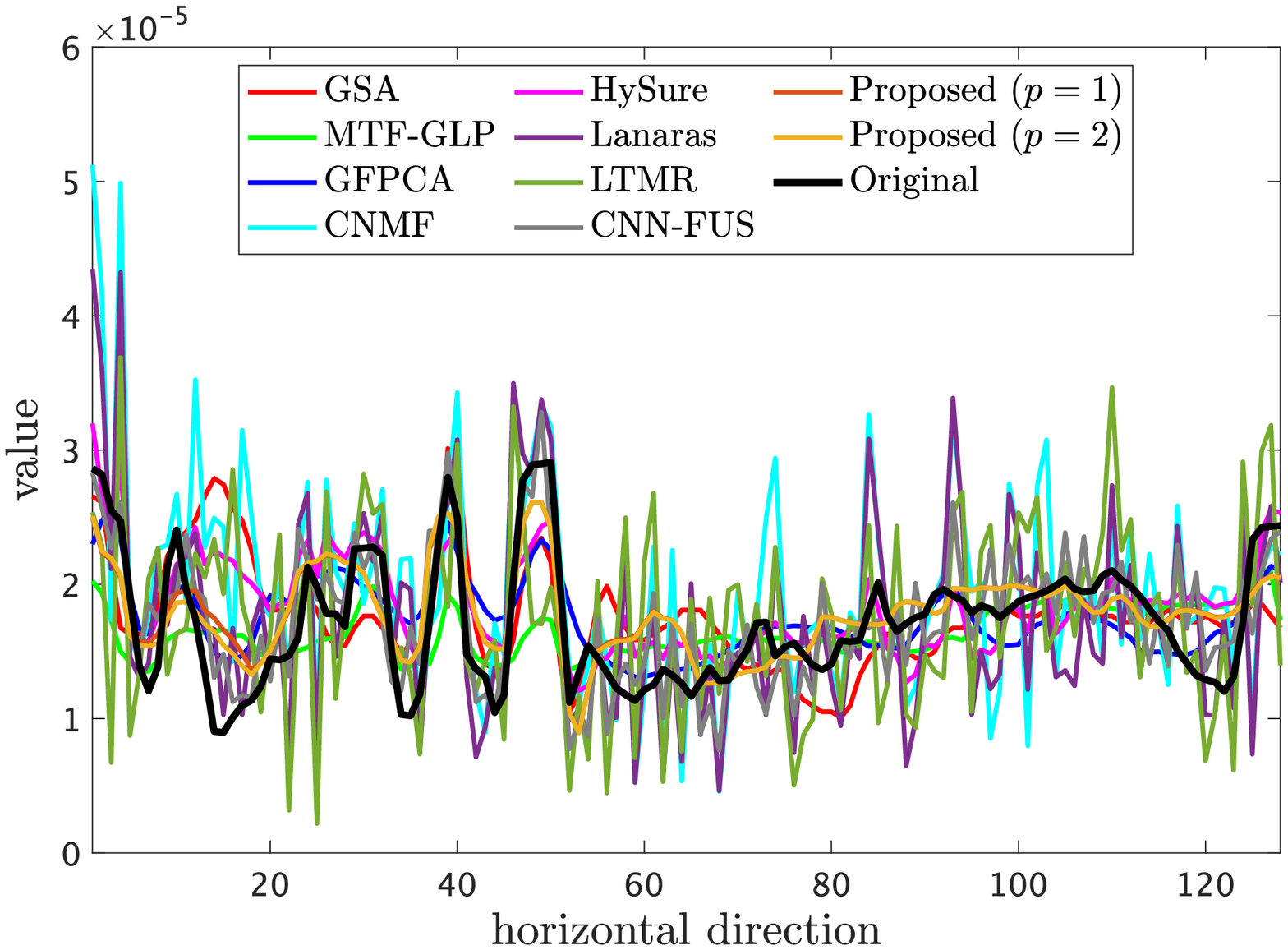}
	\end{minipage}
	\begin{minipage}[t]{0.48\hsize}
	\includegraphics[width=1.0\hsize]{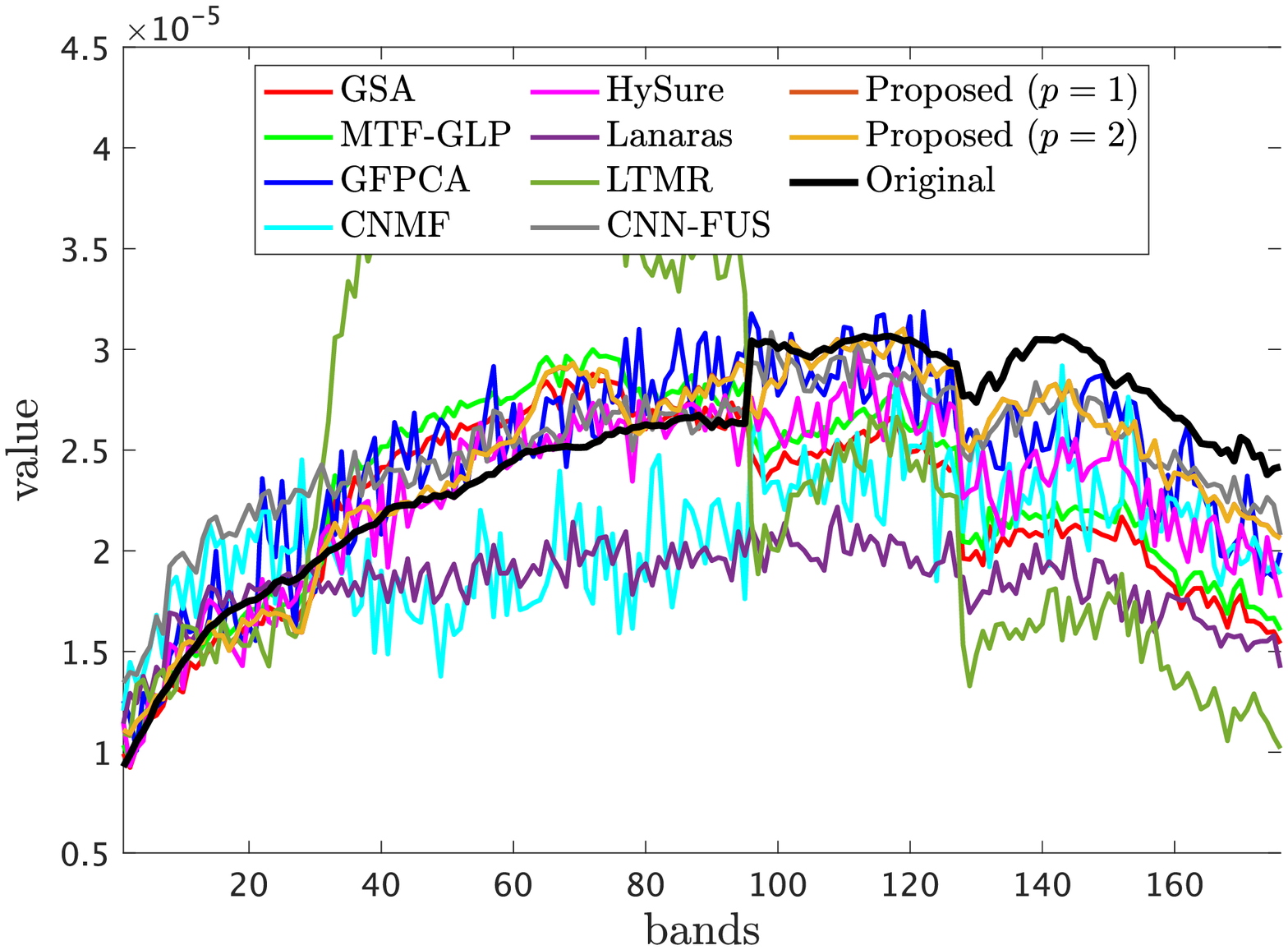}
	\end{minipage}

	\caption{Spatial (left) and spectral (right) response of the results on HS and MS image fusion experiments (Moffett field, $r = 2$, and $\sigma_{\g} = 0.1$).}
	\label{graph:response_MSfusion}
\end{center}
\end{figure}

\subsection{HS Pansharpening}\label{subsec:ex_pan}

We conducted HS pansharpening experiments, where the guide image $\g \in \R^{N}$ has only single-channel information, i.e., a grayscale image.
In the experiments, the spectral response operator $\bR \in \R^{N \times NB}$ in \eqref{model_g} averages information of the visible-light band, and we set the standard deviations of $\n_{\v}$ in \eqref{model_v} and $\n_{\g}$ in \eqref{model_g} as $\sigma_{\v} = 0.1$ and $\sigma_{\g} = 0$, $0.02$ and $0.04$ in $r = 2$ and $4$ cases, respectively.
For $r = 8$ and $16$, $\sigma_{\v}$ and $\sigma_{\g}$ are set as $0.1$ and $0.02$, and we use \textit{Reno} and \textit{Pavia U} as ground-truth.
To satisfy the convergence condition of the primal-dual splitting method, the stepsizes were set to $\gamma_1 = 0.005$ and $\gamma_2 = 0.1818$.

Tab.~\ref{tab:HSPan_2} and \ref{tab:HSPan_4} are the results of the quality measures for $r = 2$ and $r = 4$, respectively.
Here, the tables show the average SAM values, and boldface and underline are the best and the second performances, respectively.
In many $r = 2$ cases, GSA and MTF-GLP achieve high performance.
The input images are pre-denoised by state-of-the-art denoising methods, so they can effectively extract spatial and spectral characteristics from the images.
However, since for the \textit{Moffett field}, GSA is significantly low performance, one can see that the performance is affected by scene objects.
Moreover, from Tab.~\ref{tab:HSPan_2} and \ref{tab:HSPan_4}, the performance of GSA, MTF-GLP, and GFPCA is more sensitive to down-sampling ratio $r$ than other methods.
Existing optimization-based methods can estimate a high-performance HR-HS image, but the performance significantly decreases with increasing the noise intensity of a guide image, even though the methods consider noisy observations.
This would be because they do not denoise a guide image and do not sufficiently use spatial information in the image. 
The tables show that CNN-Fus achieves the highest-estimation performance of all methods in some cases and quality measures, so one can see that CNN-Fus is an effective estimation method.
However, CNN-Fus requires setting the iteration number because it cannot converge and the performance is affected by the iteration number such as Fig~\ref{fig:conv_CNN_Fus}.
The suitable iteration number is different by noise intensity and down-sampling ratio like Tab.~\ref{parameter_settings}, so the setting is a very difficult task when HR-HS image estimation from real observations.
One can see that our proposed methods achieve the highest and/or second estimation performance in some cases and comparable performance in other situations.
Specifically, for \textit{Pavia U} and \textit{Moffet field} that have more spatial details, our proposed methods can estimate more desirable HS images than compared methods, but for \textit{Reno} and \textit{Salinas}, SAM and ERGAS show that some existing methods achieve great performance.

Tab.~\ref{tab:HSPan_816} shows the results in strong down-sampling situations, i.e., $r = 8$ and $16$.
From the table, one can see that some methods and our proposed methods can effectively estimate an HR-HS image.
Like the $r = 2$ and $4$ cases, our methods can estimate a more desirable HR-HS image for \textit{Pavia U}, but quality measures especially SAM and ERGAS show that some existing methods achieve more high performance than our methods for \textit{Reno}.

To verify SAM performance in detail, we plot SAM maps of the results by GSA, MTF-GLP, GFPCA, HySure, CNN-Fus, and proposed (Reno, $r=4$, and $\sigma_{\g} = 0.02$) in Fig.~\ref{fig:SAMmap_HSpan} that achieve high SAM performance.
The SAM maps show that GSA, MTF-GLP, GFPCA, and HySure can estimate desirable spectral information in smooth areas, but the performance in high-variation areas is low.
The methods represent an HR-HS image by several bases (endmember) less than the number of the bands and evaluate the smoothness of the coefficient (abundance) in the basis space.
Therefore, they can promote low rankness of HS images but would lose spectral information that cannot be represented by the basis.
On the other hand, in Fig.~\ref{fig:SAMmap_HSpan}, one can see that SAM maps of the results by CNN-Fus and our methods have more spectral error than GSA, MTF-GLP, GFPCA, and HySure in smooth areas, but the methods suppress the outstanding error in high-variation areas.
One of the applications of HS images is the detection of extraordinary objects, so preserving information on high-variation areas is important.
Based on the above discussion, one can see that CNN-Fus and our methods can remain desirable information.

Both ERGAS and PSNR are based on MSE, but in Tab.~\ref{tab:HSPan_2} and \ref{tab:HSPan_4}, the two measures show different evaluations for some situations.
Specifically, in Salinas, $r = 4$, and $\sigma_{\g} = 0.04$, one can see that the proposed method ($p = 2$) achieves the highest PSNR for all methods but is inferior to GSA, GFPCA, HySure, and CNN-Fus for ERGAS and almost the same to MTF-GLP.
To verify the reason, we confirmed MSE, band-wise normalized MSE, and the band-wise mean luminance in the above situation.
This is because PSNR averages MSE by all pixels, and in contrast, ERGAS normalizes MSE by the mean of the spatial vector $\bar\u_{\mathrm{spa},i}$ at band-wise.
Fig.~\ref{fig:bandwiseMSE_pan} shows the MSE, band-wise normalized MSE of the results by GSA, MTF-GLP, GFPCA, HySure, CNN-Fus, and proposed methods, and the band-wise mean luminance of Salinas.
Here, band-wise normalized MSE means $\mathrm{\|\u_{\mathrm{spa},i}-\bar\u_{\mathrm{spa},i}\|^2/(\mathrm{1}^{\top}\bar\u_{\mathrm{spa},i}/N)^2}$ included in \eqref{def_ERGAS}. 
For MSE, one can see that our proposed method almost improves the estimation performance from the 1st to the 75th bands, but over the 76th bands, the proposed method is inferior to other methods in many bands.
In the bands of the higher MSE by proposed than others, the mean luminance is relatively low, so one can see that the proposed method is a little weak in the estimation of the low luminance bands.
The graph of band-wise normalized MSE shows that the value by the proposed method is significantly high in the low luminance bands.
Especially, from 75th to 80 bands, the mean luminance is less than 0.1, and the band-wise normalized MSE is very high.
From the results, ERGAS regards MSE in low luminance bands as more important, and our proposed method is weak in the estimation of the bands including low spatial information.
However, one can see that our method achieves high performance in most bands.

We show some results in Fig.~\ref{fig:img_HSpan}, which depict the resulting HS images as RGB images with $R = 16$th, $G = 32$nd, and $B = 64$th bands.
One can see that (i) GFPCA produces spatial artifacts, (ii) CNMF, Lanaras's, and LTMR remain noise included in the observed PAN images, (iii) GSA, MTF-GLP, CNMF, HySure, and LTMR cannot fully estimate spectral information, (iv) CNN-Fus produces oversmoothing, and (v) our proposed methods can effectively estimate HR-HS images.
From the enlarged area of the top of Fig.~\ref{fig:img_HSpan} and the top center area in the bottom, the results of our proposed methods have more similar colors to ground-truth than that by HySure (especially blue and yellow architectures), so our methods can preserve spectral information of small areas.

Fig~\ref{omega_graph_HSpan} and \ref{lambda_graph_HSpan} show $\omega$ and $\lambda$ versus four quality measures, respectively, where the graphs plot the average.
In Fig.~\ref{omega_graph_HSpan}, one can see that the suitable $\omega$ is rarely affected by $r$ and $\sigma_{\g}$.
In PSNR, ERGAS, and $Q2^n$, it is good that $\omega$ is set as less than 0.03, and For SAM, $\omega \in [0.07 0.1]$ is suitable.
The two results are contrastive, so $\omega$ requires to be adjusted by what information one requires.
In Fig.~\ref{lambda_graph_HSpan}, one can see that the suitable $\lambda$ changes between SAM and the other quality measures.
Since the hyperparameter $\lambda$ is the weight of the edge-similarity evaluation, the results are appropriate.

We plot the spatial and spectral response of the results in a Moffett field, $r = 2$, and $\sigma_{\g} = 0.04$ case in Fig.~\ref{graph:response_HSpan}.
Here, we picked up the spatial vectors at the 43rd row and the 30th band of the results and the spectral vectors at the 43rd row and the 107th column of them.
The graph of the spatial response shows that many methods produce artifacts, and the result by MTF-GLP has oversmoothing. 
Especially, the result by CNMF has some less than 0 value that cannot be actually observed value.
GSA, GFPCA, and HySure suppress spatial artifacts compared with other existing methods, but the value of some pixels is significantly different, e.g., 13th, 53rd, and 120th elements. 
On the other hand, CNN-Fus and our proposed method can estimate spectral response without artifacts and oversmoothing compared with other methods.
From the spectral response graph, one can see that CNMF suppresses intrinsic spectral response, and the results by GFPCA, CNMF, and CNN-FUS have spectral artifacts.
In addition, one can also see that the values estimated by GSA, GFPCA, HySure, LTMR, and the proposed ($p = 2$) method are far apart with the large of the band number, and they are not good at estimating information of large wavelength.
MTF-GLP, LTMR, and Lanaras's 
produce the increase of spectral response that does not exist in the original in small bands.
In contrast, the proposed ($p = 1$) method can estimate a more similar spectral response to the original one than the other methods.
Based on the results of the spatial and spectral response, one can see that our method ($p = 1$) achieves effective spatial and spectral estimation. 

We measured computational time, where we use Pavia U as a test HS image and set $r = 2$ and $\sigma_{\g} = 0.02$.
We conducted the experiments with MATLAB 2022a and a computer equipped with 64-GB random access memory and 12th Gen Intel(R) Core(TM) i9-12900K CPU with 3.2 GHz or NVIDIA GeForce RTX 3090.
In Tab.~\ref{tab:comp_time}, we align the computational time of all processing.
In terms of the optimization-based methods and CNN-Fus, the table also shows the iteration number and endmember/abundance estimation time with the average time of each iteration.
In this table, our methods spend much time compared with existing methods, but the iteration number is changed by image, noise intensity, degradation operator, and so on.
From the time of each iteration, one can see that our methods achieve less computational time than existing optimization-based methods and CNN-Fus.
Moreover, one can see that GSA, MTF-GLP, and GFPCA achieve less computational time than other methods, but they require a pre-denoising step, so one needs to consider the time of the step.
In the pre-denoising process, FFDNet and FGSLR spend 0.1210s 
and 246.4s for denoising a guide image and an LR-HS image, respectively.

\subsection{HS and MS Image Fusion}\label{subsec:ex_HSMSFusion}
To verify the performance of our proposed method for HS and MS image fusion, we did an experiment that estimates an HR-HS image from a pair of LR-HS and MS images.
Here, we assumed a 4-bands MS image, and $\bR \in \R^{4N \times NB}$ in \eqref{model_g} was the spectral response of the IKONOS satellite.
In addition, we set the standard deviations of $\n_{\v}$ in \eqref{model_v} and $\n_{\g}$ in \eqref{model_g} as $\sigma_{\v} = 0.2$ and $\sigma_{\g} = 0$, $0.05$ and $0.1$, respectively, for $r = 2$ and $4$ cases.
In the experiments on large downsampling ratio $r = 8$ and $16$, we set $\sigma_{\v} = 0.1$ and $\sigma_{\g} = 0.05$ and use only Reno and Pavia U because of the spatial size of the test image. 
We set the stepsizes $\gamma_1$ and $\gamma_2$ to $0.01$ and $0.5$, respectively, as satisfied with the convergence condition.

In Tab.~\ref{tab:HSMSfusion_2} and \ref{tab:HSMSfusion_4}, we show four quality measures of the results on $r = 2$ and $r = 4$, respectively.
When the noiseless guide image ($\sigma_{\g} = 0$), Lanaras's achieves the highest performance of all methods.
However, the performance of Lanaras's significantly decrease as the noise intensity is large.
Even though the method considers noisy observations, one can see that the result is sensitive to noise in a guide image.
In some cases, the results of CNN-Fus are better than the other methods.
As mentioned in Sec.~\ref{subsec:ex_pan}, CNN-Fus has a serious problem, so in a real situation, the achievement of the performance is very difficult.
GSA, MTF-GLP, and HySure outperform our methods in some cases, but the performance is affected by image.
From the average of the results, one can see that our methods achieve the highest performance and the stable estimation of all methods from a noisy guide image.

Tab.~\ref{tab:HSMSfusion_816} shows the quality measures of the results, and one can see that CNN-Fus achieves the best performance.
However, from the discussion in the above experiments, the setting of the iteration number is very difficult but essential.
Therefore, in fact, it is difficult that CNN-Fus realizes the performance.
For Reno, SAM of GSA, MTF-GLP, and HySure is superior to that of our methods about 0.3 ~ 0.7, but our methods greatly outperform them about from 1 to 5 in Pavia U cases.
In PSNR, ERGAS, and $Q2^n$, the table shows that our methods achieve the second performance, i.e., one can see that our methods can estimate a desirable HR-HS image in the stable estimation.

From the results of quality measures, our methods can estimate a higher-performance HR-HS image on average, but SAM shows that some methods achieve the higher of almost the same performance as our methods for Reno and Salinas.
To deeply analyze the reason, we depict the SAM map of the results by GSA, MTF-GLP, GFPCA, HySure, CNN-Fus, and the proposed methods (Salinas, $r = 4$, and $\sigma_{\g} = 0.05$) in Fig.~\ref{fig:SAM_MSfusion}.
One can see that GSA, MTF-GLP, GFPCA, and HySure suppress the error of spectral angle in the smooth area, but in the high-variation area, e.g., circle objects in the upper of the result, they produce a large gap of spectral angle.
CNN-Fus achieves reduced spectral angle gaps compared with them in the small area, but still remains several gaps (red area in the SAM map).
In contrast, the SAM map of our methods rarely has red areas, so our proposed method can estimate spectral information with a smaller gap of spectral angles than the other methods in the high-variation area like the results on HS pansharpening.

Fig.~\ref{fig:bandwiseMSE_MSfusion} plots MSE, band-wise normalized MSE, and the band-wise mean luminance in Salinas, $r = 2$, and $\sigma_{\g} = 0.05$ cases to verify ERGAS of the results by GSA, MTF-GLP, GFPCA, HySure, CNN-Fus, and the proposed method in more detail.
The graph of MSE shows that our proposed method almost improves the estimation performance from existing methods, especially the MSEs from 60th to 70th bands are significantly reduced.
However, in over 90th bands, the proposed method is inferior to existing methods in many bands.
From the band-wise mean luminance graph, since these bands have low luminance, one can see that the proposed method is weak in the estimation of the low luminance bands like the HS pansharpening experiments.
For this reason, the band-wise normalized MSEs of some bands are significantly high, so the ERGAS of the result by the proposed method is larger than the results by the existing methods.
The graph of the band-wise normalized MSE shows that the proposed method overcomes the existing methods in the many bands and has a high estimation ability.

We show two results in Fig.~\ref{fig:img_MSfusion}, where they are depicted by RGB images with $R = 16$th $G = 32$nd $B = 64$th bands.
The upper two rows are the results in Moffett fileld, $r = 2$, and $\sigma_{\g} = 0.1$ case, and the lower two are that in Pavia U, $r = 4$, and $\sigma_{\g} = 0.05$ case.
One can see that GSA, MTF-GLP, GFPCA, CNMF, Lanaras's, and LTMR produce spatial and/or spectral artifacts, and the results by CNN-Fus are spatial oversmoothing or artifacts.
One can also see that HySure can estimate spatial information, but the results by HySure have a few different colors, e.g. square architecture of Moffett field and orange architecture in the top center of Pavia U.
Therefore, one can see that HySure produces a little distortion in the spectral direction.
In contrast, our methods can estimate an HR-HS image without spatial and spectral artifacts compared with existing methods.

In Fig.~\ref{omega_graph_MSfusion} and \ref{lambda_graph_MSfusion}, we plot $\omega$ and $\lambda$ versus four quality measures, where the top and bottom are the results in $r = 2$ and $4$ cases, respectively.
Fig.~\ref{omega_graph_MSfusion} shows that the suitable $\omega$ rarely changes by $r$ and $\sigma_{\g}$ in most cases.
In PSNR, ERGAS, and $Q2^n$, one can see that it is good that $\omega$ is set as less than 0.03.
On the other hand, the graph shows that in SAM, the suitable $\omega$ value is from 0.03 to 0.05 different from the above quality measures.
Fig.~\ref{lambda_graph_MSfusion} shows that the suitable $\lambda$ is almost the same in different $\sigma_{\g}$ and $r$ cases, and $\lambda \in [0.07,~0.1]$ is good in most cases.

In Fig.~\ref{graph:response_MSfusion}, we plot the spatial and spectral response of the results in a Moffett field, $r = 2$, and $\sigma_{\g} = 0.1$ case in like manner to HS pansharpening experiments.
From the spatial response graph, one can see that other than GFPCA, HySure, and the proposed methods produce spatial artifacts.
GFPCA, HySure, and our proposed method can reduce spatial artifacts and estimate some rapid spatial change.
Especially, around the 40th elements and from the 45th to the 50th elements, the proposed methods can estimate a similar response to the original one than other methods.
The spectral response graph shows that GFPCA, CNMF, HySure, Lanaras's, and LTMR produce huge spectral artifacts. 
Moreover, one can see that the values estimated by GSA and MTF-GLP are far apart from the original after the 95th band.
CNN-Fus can estimate similar spectral vectors to the original, but from the 10th to 40th bands, the spectral value significantly increases from the original.
In contrast, our proposed methods can estimate more similar spectral responses to the original without spectral artifacts than other methods.
From the above results, one can see that our method can estimate desirable spatial and spectral information.

\section{Conclusion} \label{sec:C}
We have proposed a new HR-HS estimation method based on convex optimization robust to noise in observations.
Our method simultaneously estimates an HR-HS image and a noiseless guide image to suppress the effect of noise.
The method can utilize the spatial detailed information in a guide image because it evaluates the edge-similarity between an HR-HS image and an estimated guide image. 
Moreover, it utilizes HSSTV to evaluate spatial and spectral piecewise smoothness of an HS image, which is TV-based regularization and achieves high performance for HS image denoising and compressed sensing reconstruction. 
Thanks to the design, the proposed method achieves robust HR-HS image estimation.
To efficiently solve the problem, we adopt the primal-dual splitting method.
In the experiments, we verify the performance of our method and illustrate the advantages over several existing methods.


%



\section*{Acknowledgment}
This work was supported in part by JST PRESTO under Grant JPMJPR21C4, and in part by JSPS KAKENHI under Grant 22H03610, 22H00512, 21K21312, 20H02145, and 18H05413.

\ifCLASSOPTIONcaptionsoff
  \newpage
\fi


\bibliographystyle{IEEEtran}
\bibliography{TGRS_pansharpening}

%

\begin{IEEEbiography}{Saori Takeyama}
Saori Takeyama (S'17-M'21) received a B.E. degree in Computer Science in 2016 and M.E. and Ph.D. degrees in Information and Communications Engineering in 2018 and 2021 from the Tokyo Institute of Technology, respectively.

From April 2018 to March 2021, she was a Research Fellow (DC1) of the Japan Society for the Promotion of Science (JSPS). She is currently an Assistant Professor in the Department of Information and Communications Engineering, School of Engineering, Tokyo Institute of Technology. Her research interests include signal processing, hyperspectral imaging and processing, and mathematical optimization.

Dr. Takeyama received the Young Researchers’ Award from the IEICE in 2019, the Student Conference Paper Award from the IEEE SPS Japan Chapter in 2020, and the Telecomsystem Technology Student Award from the Telecommunications Advancement Foundation in 2021.
\end{IEEEbiography}

\begin{IEEEbiography}{Shunsuke Ono}
Shunsuke Ono (S'11-M'15) received a B.E. degree in Computer Science in 2010 and M.E. and Ph.D. degrees in Communications and Computer Engineering in 2012 and 2014 from the Tokyo Institute of Technology, respectively.

From April 2012 to September 2014, he was a Research Fellow (DC1) of the Japan Society for the Promotion of Science (JSPS). He is currently an Associate Professor in the Department of Computer Science, School of Computing, Tokyo Institute of Technology. From October 2016 to March 2020 and from October 2021 to present, he was/is a Researcher of Precursory Research for Embryonic Science and Technology (PRESTO), Japan Science and Technology Corporation (JST), Tokyo, Japan. His research interests include signal processing, image analysis, remote sensing, mathematical optimization, and data science.

Dr. Ono received the Young Researchers’ Award and the Excellent Paper Award from the IEICE in 2013 and 2014, respectively, the Outstanding Student Journal Paper Award and the Young Author Best Paper Award from the IEEE SPS Japan Chapter in 2014 and 2020, respectively, the Funai Research Award from the Funai Foundation in 2017, the Ando Incentive Prize from the Foundation of Ando Laboratory in 2021, and the Young Scientists’ Award from MEXT in 2022. He has been an Associate Editor of IEEE TRANSACTIONS ON SIGNAL AND INFORMATION PROCESSING OVER NETWORKS since 2019.
\end{IEEEbiography}





\end{document}